\documentclass[twoside,12pt]{article}
\usepackage{epsfig,color}
\usepackage{bm}
\usepackage{amsmath}
\usepackage{url}

\def\Journal#1#2#3#4{{#1} {#2} (#4) #3 }

\def\NPA{{\em Nucl. Phys.} A}

\def\NPB{{\em Nucl. Phys.} B}

\def\PLB{{\em Phys. Lett.} B}

\def\PRL{\em Phys. Rev. Lett.}

\def\PREP{\em Phys. Rep.}

\def\PRD{{\em Phys. Rev.} D}
\def\PRC{{\em Phys. Rev.} C}

\def\ZPC{{\em Z. Phys.} C}

\newcommand{\be}{\begin{equation}}
\newcommand{\ee}{\end{equation}}
\newcommand{\bea}{\begin{eqnarray}}
\newcommand{\eea}{\end{eqnarray}}

\newcommand{\lsim}{\mbox{\raisebox{-0.6ex}{$\stackrel{<}{\sim}$}}\:}

\newcommand{\Blue}[1]{\textcolor[named]{Blue}{#1}}

\newcommand{\commentold}[1]{}

\begin{document}

\title{ \vspace{1cm} Integrated Dynamical Approach to\\
Relativistic Heavy Ion Collisions}
\author{T.\ Hirano,$^{1,2}$ P.\ Huovinen,$^{3,4}$ K.\ Murase,$^{2,1}$ 
Y.\ Nara,$^{5}$\\
\\
$^1$Department of Physics, Sophia University, Tokyo 102-8554, Japan\\
$^2$Department of Physics, The University of Tokyo, Tokyo 113-0033, Japan\\
$^3$Frankfurt Institute for Advanced Studies, 60438 Frankfurt am Main, 
Germany\\
$^4$Institut f\"ur Theoretische Physik, Johann Wolfgang Goethe-Universit\"at,\\
60438 Frankfurt am Main, Germany \\
$^5$Akita International University, Yuwa, Akita-city 010-1292, Japan}
\maketitle
\begin{abstract}
We review integrated dynamical approaches
to describe heavy ion
reaction as a whole at ultrarelativistic energies.  
Since final observables result from all the history of the reaction,
it is important to describe all the stages of the reaction to obtain the
properties of the quark gluon plasma from experimental data.
As an example of these approaches, we develop an integrated dynamical
model, which is composed of a fully (3+1) dimensional ideal
hydrodynamic  model with the
state-of-the-art equation of state  based on lattice
QCD, and subsequent hadronic cascade in the late stage.
Initial conditions are obtained employing Monte Carlo versions of the
Kharzeev-Levin-Nardi model (MC-KLN) or the Glauber
model (MC-Glauber).
Using this integrated model, we first simulate relativistic heavy ion
collisions at the RHIC and LHC energies starting from conventional
smooth initial conditions.  We next utilise each Monte Carlo samples
of initial conditions on an event-by-event basis and perform
event-by-event dynamical simulations to accumulate a large number of
minimum bias events.  A special attention is paid to performing the
flow analysis as in experiments toward consistent comparison of
theoretical results with experimental data.
\end{abstract}

\eject
\tableofcontents
\section{Introduction}

Heavy ion programs at Large Hadron Collider (LHC) in European
Organization for Nuclear Research (CERN) and at Relativistic Heavy Ion
Collider (RHIC) in Brookhaven National Laboratory (BNL) have focused
on the physics of strongly interacting matter of quarks and gluons
under extreme conditions, namely, on the physics of the quark gluon
plasma (QGP)~\cite{Yagi:2005yb}.  By colliding two heavy nuclei at
relativistic energies, the matter formed in the collision is expected
to be in the state of QGP in temperatures up to 400 MeV 
$\sim 4 \times 10^{12}$ K. Such a high temperature was reached in the
early universe about ten micro seconds after the Big Bang.  One of the
main goals of the heavy ion programs is to extract bulk and transport
properties of the QGP from analyses of experimental data.

So far, a vast body of experimental data have been obtained at RHIC
and LHC. Among them, the azimuthal anisotropy of the emitted
particles, so-called elliptic flow~\cite{STARv2-1,STARv2-2,STARv2-3,
  PHENIXv2-1,PHENIXv2-2, PHENIXv2-3,PHOBOSv2-1,PHOBOSv2-2,
  Back:2004mh,Aamodt:2010pa,ATLAS:2011ah,Velkovska:2011zz} is one of
the main observables to provide information of the bulk properties of
the QGP. In particular, the large observed value of elliptic flow at
RHIC was one of the main reasons to conclude that the matter produced
in the collisions at RHIC does indeed thermalise and form
QGP~\cite{Heinz:2001xi,Gyulassy:2004vg}.

In a non-central collision of two spherical nuclei, the reaction zone
has an an almond-like shape in the plane perpendicular to the
collision axis, so called transverse plane. Because of this geometry,
pressure gradient within the reaction zone is not azimuthally
isotropic, but it is larger along the impact parameter than to the
direction orthogonal to it. This leads to anisotropic expansion of the
system with more particles being emitted along the impact parameter
than orthogonal to it. It is customary to call the plane spanned by
the beam and the impact parameter a reaction plane, and thus we say
that larger emission along the impact parameter means larger emission
``in-plane'' than ``out-of-plane''. This anisotropic particle emission
can be quantified by Fourier expanding the azimuthal distribution of
observed particles. In particular, this kind of anisotropy is
quantified by the second Fourier coefficient of the expansion,
$v_2$. Since a finite $v_2$ in a Fourier expansion corresponds to an
elliptic shape, this anisotropy is commonly called \emph{elliptic
  flow}~\cite{Ollitrault}. Quite surprisingly ideal hydrodynamics,
which neglects all the dissipative effects, gave a good description of
the elliptic flow observed at
RHIC~\cite{Kolb:2000fha,Huovinen:2001cy,Teaney:2000cw,
  Teaney:2001av,Hirano:2001eu,Hirano:2002ds}.  Given a fact that the
expansion rate of the system formed in a heavy ion collision is
tremendously large ($\sim$10$^{24}$/sec), it is far from obvious that
dissipative effects can be neglected.  Since in ideal hydrodynamics
perfect local equilibrium is assumed to hold at any instant, couplings
among constituents of the fluid must be so strong that relaxation of
the system against any thermodynamic forces happens extremely quickly.
Thus, from the agreement of hydrodynamic prediction with the elliptic
flow data, an announcement of the ``discovery of perfect fluidity" was
made \cite{BNL}, and a new paradigm of the strongly coupled QGP (sQGP)
was established
\cite{Heinz:2001xi,Gyulassy:2004vg,sQGP1,sQGP2,sQGP3,Hirano:2005wx}.

However, the described shape of the reaction zone is realistic only as
an average over many collisions. Since the nuclei are not uniform but
consist of separate nucleons, we may expect the reaction zone in a
single event to depict similar granular structure to the nuclei. For
example, according to the Monte-Carlo Glauber
model~\cite{Miller:2007ri,Broniowski:2007nz,Alver:2008aq} the reaction
zone consists of several peaks (\emph{a.k.a.} hot spots) and valleys
originating from the configuration of nucleons in the colliding
nuclei. This irregular structure means that even if the underlying
shape of the reaction zone in a non-central collision is almond-like,
it has been distorted and tilted. Thus, if one fits an ellipse to the
reaction zone, its minor axis no longer aligns with the impact
parameter. As described, elliptic flow is generated by the anisotropy
of pressure gradient. Now, if the thermodynamic language 
of pressure
gradients is applicable in an individual event, the largest gradient
and thus the largest emission of particles is not along the reaction
plane. It is along the participant plane\footnote{Also called event
  plane is some of the literature.}, which is spanned by the beam and
the minor axis of the reaction zone. Thus, the particle distribution
should not be Fourier expanded with respect to the reaction plane, but
to the participant plane.

The importance of these event-by-event fluctuations of the shape and
orientation of reaction zone was discovered when trying to understand
the behaviour of $v_2$ in Cu+Cu collisions. It had been found that
when experimentally measured $v_2$ was divided by the modelled
eccentricity of the reaction zone\footnote{Sometimes called standard
  eccentricity $\varepsilon_{\mathrm{std}}$. For definitions, see
  Section~\ref{sec:ic}.}, $\varepsilon$, the ratio $v_2/\varepsilon$
scales with transverse density $(1/S) dN/dy$, where $S$ is the
modelled transverse area of the overlap region and $dN/dy$ is the
measured final particle multiplicity at
midrapidity~\cite{STARv2-2,Alt:2003ab}. However, the measured $v_2$ in
Cu+Cu collisions did not obey such a scaling, but when the
eccentricity $\varepsilon$ was replaced by the participant
eccentricity $\varepsilon_{\mathrm{part}}$, the scaling was
restored~\cite{PHOBOSv2-3}. The main difference is that
$\varepsilon$ is always evaluated with respect to the reaction plane,
whereas the participant eccentricity $\varepsilon_{\mathrm{part}}$ is
evaluated with respect to the participant plane in each individual
event, and thus takes into account the fluctuations in the orientation
of the reaction zone \cite{Miller:2003kd}.

Since the reaction zone has a complicated azimuthal structure, and the
elliptic flow was explained as a result of the azimuthal variation of
the pressure gradient, it is natural to expect that the Fourier
coefficients beyond $v_2$ would be nonzero as well. The third
coefficient in the Fourier expansion, $v_3$, is called triangular
flow~\cite{Alver:2010gr}. It is generated by the triangular component
of the shape of the fluctuating reaction zone, and some puzzling
phenomena in intermediate transverse momentum regions can be
interpreted as manifestation of triangular flow.  For example,
Mach-cone like structure was discovered in the away-side region in
di-hadron correlation functions at
RHIC~\cite{Adams:2005ph,Adler:2005ee,Adare:2008ae} when one subtracts
background elliptic flow component from it. Recently, it was found
that these di-hadron correlation functions can be reproduced by a sum
of independently analysed higher harmonic
components~\cite{Aamodt:2011by,Aad:2012bu}, which indicates that
Mach-cone like structure would be caused simply by collective
triangular flow.

If one Fourier-decomposes the azimuthal particle distribution, one can
obtain information how the system responds to the initial fluctuating
profile and from this response one may deduce what the properties of
the system itself are
\cite{Shuryak:2009cy,Staig:2010pn,Staig:2011wj,Mocsy:2010um,Sorensen:2011hm}.
This reminds an analysis in observational cosmology: Through
decomposition of power spectrum of cosmic microwave background
radiation into spherical harmonics, one can constrain important
cosmological constants and even mass/energy budget of the universe
\cite{Komatsu:2008hk}.

In the observational cosmology, analysis tools~\cite{cmbtoolbox}
played important roles in extracting cosmological parameters.  The
situation in the physics of relativistic heavy ion collisions is quite
similar to this \cite{oscar}: One has to develop analysis tools to
extract the properties of the QGP from experimental data.  From this
point of view, let us overview the dynamics of heavy ion collisions.
High energy heavy ion collisions contain rich physics and exhibit many
aspects of dynamics according to relevant energy and time scales.  Two
energetic, Lorentz-contracted, heavy nuclei collide with each other.
These nuclei can be described by the colour glass condensate (CGC), a
universal form of hadrons and nuclei at extremely high energies
\cite{Iancu:2002xk,Iancu:2003xm,Gelis:2010nm}. These collisions can
be viewed as collisions of two bunches of highly coherent and dense
gluons.  Just after the collisions, longitudinal colour electric and
magnetic fields, which are also known as the colour flux
tubes~\cite{Gatoff:1987uf} are formed between two passing nuclei.
Subsequent non-equilibrium evolution of these colour fields towards
locally thermalised QGP is called ``glasma"~\cite{Lappi:2006fp}.  Once
local thermalisation is achieved, a QGP fluid expands
hydrodynamically, cools down and turns into a hadronic gas.  Hadrons
continue to rescatter until the system is so dilute that interactions
become very rare, and hadrons stream freely towards the detectors.

Since the final observables are the result of all these various stages
of the collision, it is important to describe the heavy ion collision
as a whole.  So far, we have developed the following integrated
dynamical model~\cite{Hirano:2010jg,Hirano:2010je} to describe the
dynamics of relativistic heavy ion collisions.  For the initial stage,
initial conditions are calculated using the CGC picture
\cite{Drescher:2006ca,Drescher:2007ax,Hirano:2009ah}.  Using these
initial conditions, we describe fully three dimensional ideal
hydrodynamic expansion of the QGP fluid
\cite{Hirano:2001eu,Hirano:2002ds} using a realistic equation of state
from lattice QCD simulations
\cite{Cheng:2007jq,Bazavov:2009zn,Huovinen:2009yb}.  The late stage
evolution of the hadron gas is described using microscopic kinetic
theory \cite{Nara:1999dz}.  Technical details about numerical
simulations of ideal hydrodynamics and hadronic cascades can be found
in Ref.~\cite{Hirano:2012yy}.

In this paper, 
we discuss experimental observables,
in particular anisotropic
flow, at RHIC and LHC energies using an integrated dynamical model.  A
special emphasis will be put on discussion about initial conditions
and final flow analysis methods from an event-by-event analysis point
of view.  In theoretical calculations both the reaction plane and the
participant plane are trivially known, but in experiments it is
impossible to measure the reaction plane, and it is quite hard to
precisely determine the participant plane from the finite number of
observed particles. Thus, several flow analysis methods have been
proposed \cite{Poskanzer:1998yz,Borghini:2000sa,Borghini:2001vi} to
experimentally measure anisotropic flow.  Hence, a \textit{consistent}
comparison of hydrodynamic results with experimental observables is
non-trivial. In this paper we demonstrate the differences of several
experimental methods of flow analysis by using them to analyse the
output of the integrated dynamical model.

The paper is organised as follows. In Sec.~\ref{sec:model}
we describe and review
the hybrid models, in which
hydrodynamics is combined with hadronic cascade,
and hydrodynamic simulations
on an event-by-event basis.
In particular we describe each module and the interfaces between
  them of our integrated dynamical model
In Sec.~\ref{sec:result1}, we first
summarise the results obtained using smooth initial profiles, which
are the conventional initial conditions employed in hydrodynamic
simulations.  We next show results from event-by-event hydrodynamic
simulations in Sec.~\ref{sec:result2} emphasising the importance of
employing the same flow analysis method as in experiment.
Section~\ref{sec:conclusion} is devoted to the conclusion and outlook.

\section{Model \label{sec:model}}

Integrated dynamical models, in general, consist of three separate stages:
Initial conditions, hydrodynamics and hadronic cascade.
In our version, the
initial particle production in the collision of the nuclei is either
described by the MC-KLN version of the colour glass condensate, or
parametrised using the MC-Glauber model. These models provide the
initial state for the subsequent expansion of the matter, which we
describe by relativistic ideal hydrodynamics.
As for the equation of state, we employ results from the
state-of-the-art lattice QCD
simulations~\cite{Cheng:2007jq,Bazavov:2009zn}.
Once the matter is
dilute enough to form hadrons, we switch the description of the system
from fluid dynamics to microscopic hadron cascade. In this
section we
describe all these stages of the integrated model, and how we connect
hydrodynamics to cascade.
We also review current status of the equation of state and its 
application,
hybrid models, and event-by-event hydrodynamic simulations.

\subsection{\it Hydrodynamic Equations \label{sec:hydroeq}}

To describe a system in length scales much larger than a typical
microscopic length scale, like the mean free path, it is sufficient to
characterise it in terms of a few macroscopic fields: The
energy-momentum tensor $T^{\mu\nu}$, and conserved charge currents
$j_i^\mu$ (if any). In relativistic fluid dynamics, the equations of
motion are given by the conservation laws for these fields
\begin{equation}
 \label{eq:conservation}
   \partial_{\mu}T^{\mu\nu} = 0 {\hspace{1cm}} {\rm and} {\hspace{1cm}}
   \partial_{\mu}j^{\mu}_i = 0.
\end{equation}
Without any additional constraints these $4+n$ ($n$ is the number of
conserved currents) equations contain $10+4n$ unknown variables. To
close the system of equations, one either has to provide further
equations in the form of constituent equations for dissipative
currents (shear stress tensor $\pi^{\mu\nu}$, bulk pressure $\Pi$ and
energy flow/particle number diffusion $q_i$), or to eliminate some of
the variables by further approximations. In the following we apply the
latter approach and reduce the number of unknowns by assuming that the
fluid is in exact local thermodynamical equilibrium.

In a local thermodynamical equilibrium, the single particle phase-space
distribution for noninteracting fermions or bosons is
\begin{equation}
 \label{eq:thermal}
 f_0(p,x) = \frac{g}{(2\pi)^3}\frac{1}{\exp (p\cdot u(x) - \mu(x))/T(x) \pm 1}.
\end{equation}
When one applies this to the kinetic theory definitions of the
energy-momentum tensor and charged currents, one obtains
\begin{equation}
 \label{eq:idealTmunu}
    T^{\mu\nu}(x) = [\epsilon(x) + P(x)]\, u^\mu(x) u^\nu(x) - P(x)g^{\mu\nu},
    {\hspace{1cm}} {\rm and} {\hspace{1cm}}
    j^{\mu}_i(x) = n_i(x) u^\mu(x),
\end{equation}
respectively, where $\epsilon$ is the energy, and $n_i$ are the  charge
densities in the local rest frame of the fluid, $P$ is
the thermodynamic pressure and $u^\mu$ is the fluid flow
four-velocity. The Eqs.~(\ref{eq:idealTmunu}) imply that for a fluid
in local thermodynamical equilibrium the dissipative currents are
zero.  This consideration of a non-interacting gas in local
equilibrium is the starting point of the ideal fluid approximation:
One postulates that the energy-momentum tensor and charge currents are
of the form of Eqs.~(\ref{eq:idealTmunu}), and thus the dissipative
currents are zero by definition.

Such an approximation reduces the number of unknown variables in
Eqs.~(\ref{eq:conservation}) to $5+n$: the above mentioned densities,
pressure and three components of the flow four-velocity (note that the
usual normalisation $u_\mu u^\mu = 1$ reduces the number of unknowns
by one). To finally close the system of equations, an additional
equation is usually provided in the form of the equilibrium equation
of state (EoS) of the matter, which expresses the pressure and
densities in terms of thermodynamical parameters temperature $T$
and chemical potentials $\{\mu_i\}$, $P = P(T,\{\mu_i\})$. However, to
solve the Eqs.~(\ref{eq:conservation}), it is often practical to
provide the EoS in the form $P=P(\epsilon,\{n_i\})$ connecting the
pressure directly to the densities. The knowledge of temperature
and chemical potentials is not necessarily required to calculate the
evolution of the fluid itself.

Note that once the EoS, and boundary conditions (usually referred to as
initial conditions) for the set of differential equations in
Eqs.~(\ref{eq:conservation}) are fixed, the evolution is determined by
Eqs.~(\ref{eq:conservation}). In the ideal fluid approximation, the
only place where information about the nature of the constituents of
the fluid and their microscopic interactions enters, is the EoS.

In the implementation of our model, we solve
Eqs.~(\ref{eq:conservation}) numerically in all three spatial
dimensions.
We employ the Milne coordinates ($\tau$, $x$, $y$, $\eta_{s}$), 
where $\tau = \sqrt{t^2 - z^2}$ is proper time and 
$\eta_{s} = \frac{1}{2}\log\frac{t+z}{t-z}$ is space-time rapidity.

Since we are mostly interested in the observables at midrapidity in
the collider energies\footnote{$\sqrt{s_{\mathrm{NN}}}=200$ GeV
    at RHIC and $\sqrt{s_{\mathrm{NN}}}=2.76$ TeV at LHC.}, we can
ignore the baryon current
\cite{Adler:2001bp,Adler:2003cb,Abelev:2008ab,Floris:2011ru}~\footnote{The
  other conserved currents relevant for heavy ion collisions; electric
  charge, isospin and strangeness are in general either tiny or
  zero.}.  As usual in hydrodynamical models, we take the spatial
boundary condition to be vacuum at infinity~\cite{Rischke:1998fq},
\textit{i.e.}\ the hydrodynamical evolution proceeds independently
without any feedback from the cascade. The temporal boundary,
\textit{i.e.}\ the initial value(s) for the differential equations are
described in Section~\ref{sec:ic}.  We employ the Piecewise Parabolic
Method (PPM) \cite{Colella:1982ee} as an algorithm to solve the
equations of ideal hydrodynamics (Eqs.~(\ref{eq:conservation})). PPM
is known to be robust against strong shocks, therefore it is suitable
to apply for bumpy initial conditions in event-by-event hydrodynamic
simulations.  For details on PPM, see Ref.~\cite{Hirano:2012yy}.

\subsection{\it Equation of State \label{sec:eos}}

The equation of state (EoS) of strongly interacting matter can be
obtained either by using various models or by lattice QCD
calculations~\cite{Petreczky:2012rq}. Even if the recent lattice QCD
calculations of the EoS have provided continuum extrapolated
results~\cite{Borsanyi:2010cj}, there is a practical reason to use the
hadron resonance gas (HRG) model for the EoS at low temperatures. When
converting fluid to particles using the Cooper-Frye procedure as
described in section~\ref{sec:F2G}, the conservation laws are obeyed
without any further considerations if the degrees of freedom are the
same before and after particlisation. In other words, if the emitted
particles are the same particles the fluid consists of. If the fluid
is described as hadron resonance gas, the degrees of freedom and their
properties are well known, and these degrees of freedom are the
experimentally observable hadrons which distributions we eventually
want to calculate. HRG has also been shown to provide a reasonably
good approximation of the EoS of interacting hadron gas in
temperatures slightly below the pion mass~\cite{Venugopalan:1992hy},
and its use is therefore justified.

To show how an EoS combining a hadron resonance gas at low
temperatures, and lattice QCD results at high temperatures can be
made, we briefly review the construction of the $s95p$-v1.1
parametrization~\cite{Hirano:2010jg,Huovinen:2009yb}. The high
temperature part of this EoS is based on the lattice QCD results of
the hotQCD collaboration~\cite{Cheng:2007jq,Bazavov:2009zn}, and its
low temperature part contains the same hadrons and resonances as the
JAM hadron cascade~\cite{Nara:1999dz}. We have also used this EoS to
calculate the results discussed in Sections~\ref{sec:result1}
and~\ref{sec:result2}.

\begin{figure}[tb]
\begin{center}
\begin{minipage}[t]{85mm}
 \epsfig{file=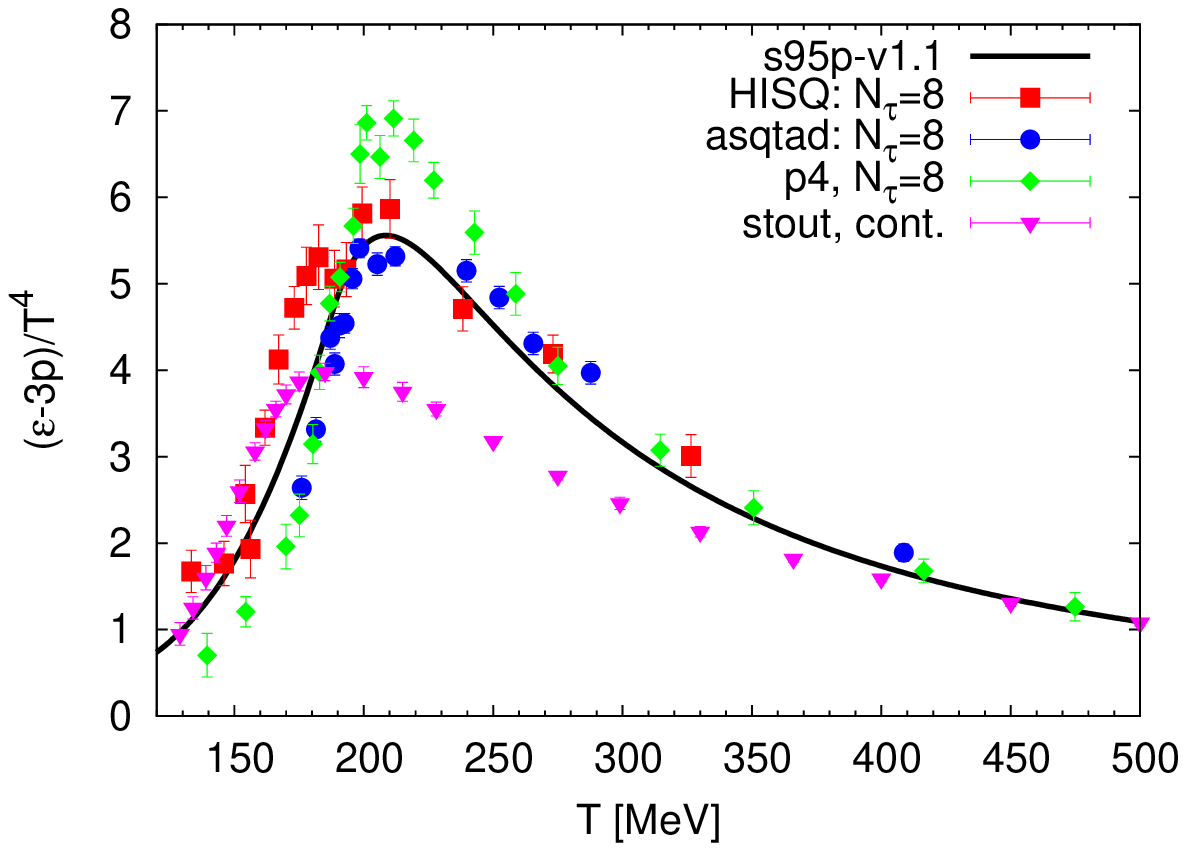,scale=0.7}
 \caption{The parametrised trace anomaly compared with lattice
   results calculated with \emph{p4}~\cite{Cheng:2009zi},
   \emph{asqtad} and \emph{HISQ/tree}~\cite{Bazavov:2010pg} actions as
   well as the continuum extrapolated result obtained using
   \emph{stout} action~\cite{Borsanyi:2010cj}.
   \label{fig:trace}}
\end{minipage}
 \hfill
\begin{minipage}[t]{85mm}
 \epsfig{file=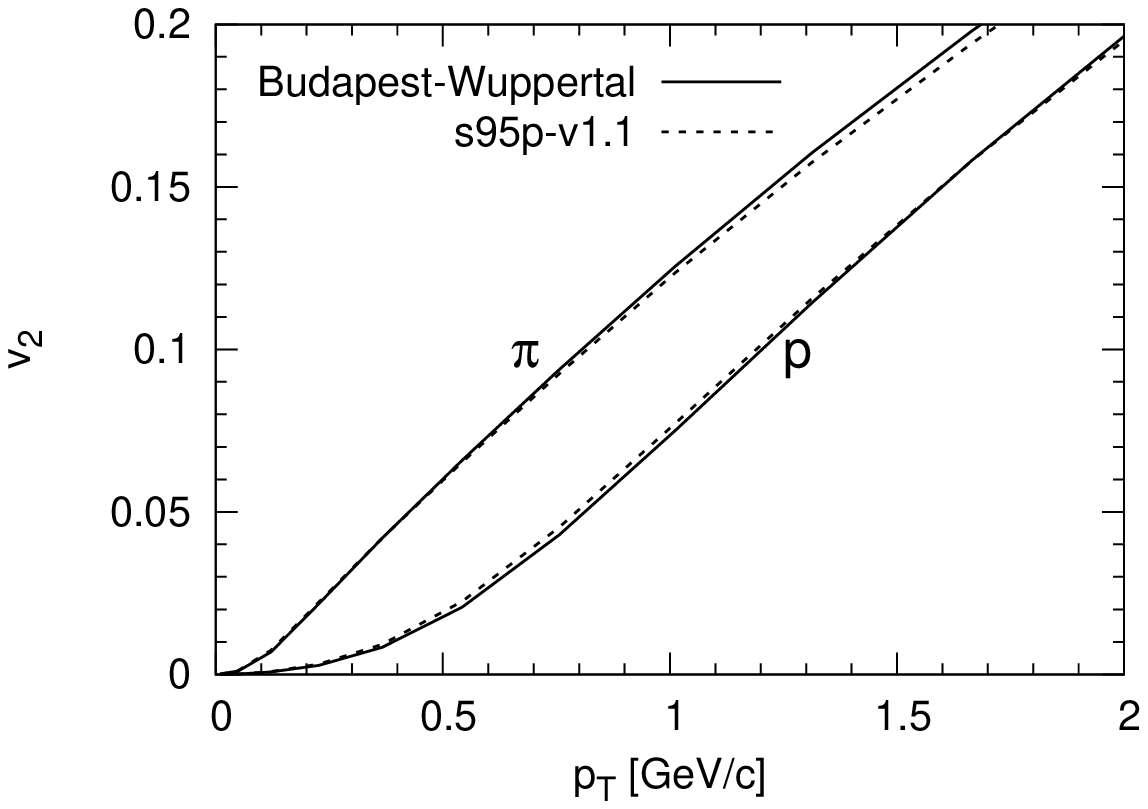,scale=0.7}
 \caption{The $p_T$-differential elliptic flow of pions and protons in
   $b=7$ fm Au+Au collisions at $\sqrt{s_{NN}} = 200$ GeV
   evaluated using ideal fluid hydrodynamical model using EoS
   $s95p$-v1~\cite{Huovinen:2009yb} and the parametrised
   Budapest-Wuppertal EoS~\cite{Borsanyi:2010cj}.
   \label{fig:v2EoS}}
\end{minipage}
\end{center}
\end{figure}

The starting point of the construction of the $s95p$ EoS is the trace
anomaly $\Theta(T)=\epsilon(T)-3P(T)$ evaluated on lattice, see
Fig.~\ref{fig:trace}. The lattice results are parametrised and
connected to the trace anomaly of HRG preserving the smooth crossover
nature of the transition. The trace anomaly is then converted to
pressure via
\begin{equation}
  \frac{P(T)}{T^4}-\frac{P(T_{\rm low})}{T_{\rm low}}
 =\int_{T_{\rm low}}^{T} \frac{d T'}{{T'}^5} \Theta(T),
 \label{eq:P-integral}
\end{equation}
where pressure at the lower integration limit, $T_{\rm low}$, is given
by HRG. Energy and entropy densities are subsequently obtained via
laws of thermodynamics. By construction such an EoS is limited to zero
net baryon density and zero net strangeness, but as mentioned zero
charge is a good approximation when describing the system at
midrapidity in collisions at RHIC and LHC.

The trace anomaly of the $s95p$-v1.1 EoS is shown in
Fig.~\ref{fig:trace}, and compared to recent lattice QCD results. It
differs considerably from the continuum extrapolated result by the
Budapest-Wuppertal collaboration~\cite{Borsanyi:2010cj}. However, the
Budapest-Wuppertal EoS deviates from HRG already at $T\approx
130$--140 MeV, which necessitates switching from fluid to cascade
below this temperature leading to much worse reproduction of data in
our calculations. The $s95p$ parametrisation follows HRG up to $T=183$
MeV temperature providing much more freedom in choosing the switching
temperature, and therefore we prefer to use it. To estimate the
uncertainty our choice of EoS produces in the particle anisotropies,
we have calculated the elliptic flow anisotropy $v_2$ in impact
parameter $b=7$ fm Au+Au collisions at the full RHIC energy
($\sqrt{s_{NN}}=200$ GeV), see Fig.~\ref{fig:v2EoS}.  In this
calculation we follow the procedure used in
Ref.~\cite{Huovinen:2009yb} to test various parametrizations of the
EoS. We use ideal fluid hydrodynamical model and assume chemical
equilibrium until kinetic freeze-out. The model is initialized using
an optical Glauber model with components proportional to the number of
participants and binary collisions (see Sec.~\ref{sec:ic} and
Refs.~\cite{Heinz:2001xi,Kolb:2001qz}) The parameters are chosen to
reproduce the centrality dependence of charged particle
multiplicity. The usual procedure requires choosing the freeze-out
temperature to reproduce the particle spectra in most central
collisions, but that would require the use of temperature
$T_{\mathrm{fo}}\approx 140$ MeV. As mentioned, the Budapest-Wuppertal
EoS deviates from HRG below that temperature, and thus converting
fluid to free particles in such a temperature violates conservation of
energy. Therefore we used freeze-out temperature $T_{\mathrm{fo}} =
125$ MeV for both EoSs even if it leads to slightly flatter $p_T$
distributions of pions and protons than experimentally observed. Both
EoSs, Budapest-Wuppertal and $s95p$ lead to very similar $p_T$
distributions at that temperature.  As can be seen in
Fig.~\ref{fig:v2EoS}, the difference in $p_T$-differential elliptic
flow is tiny as well, and smaller than the experimental errors. Thus
we consider the use of $s95p$-v1.1 EoS a reasonable approximation.

Similar insensitivity to the details of the EoS was seen in
Ref.~\cite{Huovinen:2009yb} where different parametrizations of
lattice EoS were tested. Even if the EoS governs the expansion of the
fluid and buildup of collective motion, the details of the EoS have
tiny observable consequences. The rule of thumb is that
the stiffer the EoS, \emph{i.e.}, the larger the speed of sound in the
fluid, the larger the flow velocity generated during the expansion.
But, even if larger flow velocity mean flatter $p_T$ spectrum (see
f.ex.~Ref.~\cite{Huovinen:2006jp}), this effect can be negated by
choosing the fluid to freeze-out earlier at larger temperature, by
assuming later thermalisation time and thus later start of the
hydrodynamical evolution, by changing the initial shape of the density
distribution (if the model allows) or by any combination of these
three. The integrated dynamical models discussed in this paper have no
freeze-out temperature, but the final particle distributions are
sensitive to the switching temperature from hydro to cascade, and all
these problems hamper these models as well.

\begin{figure}[tb]
 \begin{center}
   \epsfig{file=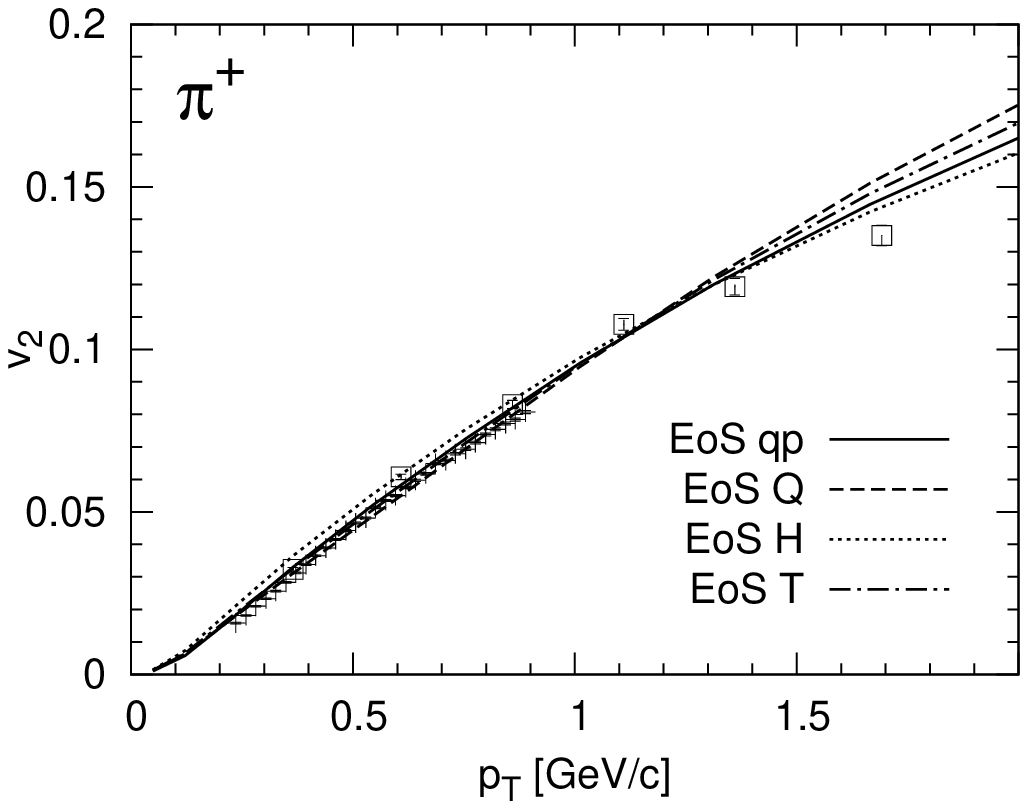,scale=0.7}\hspace*{-1mm}
   \epsfig{file=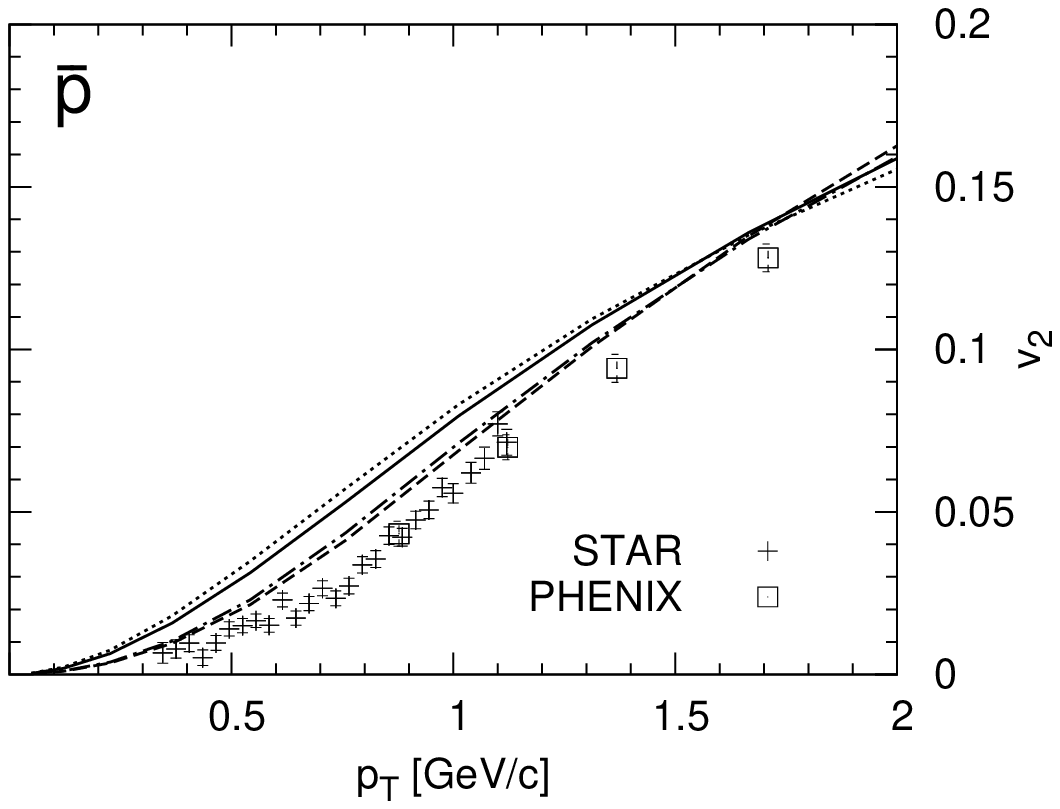,scale=0.7}
  \caption{Elliptic flow of pions and antiprotons vs.~transverse
      momentum in minimum bias Au+Au collisions at
      $\sqrt{s_{\mathrm{NN}}}=200$ GeV calculated using four different
      EoSs~\cite{Huovinen:2005gy} and compared with the data by the
      STAR~\cite{STARv2-4} and PHENIX~\cite{PHENIXv2-2}
      collaborations. The labels stand for a lattice QCD inspired
      quasiparticle model (qp), EoS with a first order phase
      transition (Q), a parametrized smooth but rapid crossover (T)
      and pure hadron resonance gas with no phase transition (H).}
 \label{fig:eoscomparisonI}
 \end{center}
\end{figure}

Based on the $p_T$ distributions on particles alone, we can basically
only say that the EoS must contain a large number of degrees of
freedom. Otherwise the $p_T$ distributions come too flat, see
 discussion and Fig.~5 in
Ref.~\cite{Sollfrank:1996hd}, nor does the EoS contain all the
observed particles and resonances. To say anything beyond the large
number of degrees of freedom requires the use of more sophisticated
observables like the azimuthal anisotropy $v_2$ or the HBT-radii,
a.k.a.\ femtoscopy. In general the changes in $v_2$ due to different
EoSs can be compensated by changing the freeze-out temperature. There
is, however, an exception. The $p_T$ differential anisotropy of
protons, $v_2(p_T)$, is to some extent sensitive to the order of phase
transition~\cite{Huovinen:2005gy}. This is demonstrated in
Fig.~\ref{fig:eoscomparisonI} where pion and proton $v_2(p_T)$ in
minimum bias Au+Au collisions at $\sqrt{s_{\mathrm{NN}}}=200$ GeV
are calculated using ideal fluid hydrodynamical
model, and four different EoSs. The EoSs are the bag model based
EoS with a first order phase transition between HRG and ideal parton gas (EoS~Q), pure hadron resonance
gas with no phase transition (EoS~H), a lattice inspired quasiparticle
model EoS (EoS qp, which is quite close to the present lattice QCD
EoSs) by Schneider and Weise~\cite{Schneider}, and an EoS with a
parametrised smooth crossover from 
HRG to ideal parton gas (EoS~T).  As seen in the figure, the
sensitivity of pion $v_2(p_T)$ on the EoS is tiny, but proton
$v_2(p_T)$ shows clear sensitivity on EoS. Surprisingly it is the EoS
Q with the first order phase transition which provides the best fit to
the data---a construction ruled out by the lattice QCD
calculations. The lattice inspired EoS qp leads to proton
$v_2(p_T)$ which is clearly above the data, and almost as large as the
$v_2(p_T)$ obtained using the purely hadronic EoS~H. The EoS~T, which
has a very rapid crossover leads to proton $v_2(p_T)$ which is almost
as close to the data than the one obtained using EoS~Q. This means
that the order of the phase transition does not affect the build up of
anisotropy, but how rapid the transition is does have an
effect. Nevertheless, EoS~T is ruled out by the present lattice data.

On the other hand, it can be argued that a soft EoS leads to a long
lifetime of the system, which is excluded by the HBT
measurements~\cite{Chojnacki:2007jc}. It has also been shown that if
one assumes a relatively hard, lattice inspired EoS, it is possible to
reproduce the measured HBT radii~\cite{Chojnacki:2007rq}, although in
that case the proton $v_2(p_T)$ is not reproduced. This apparent
contradiction appears when using ideal fluid dynamics. When
dissipative corrections are applied, the proton $v_2(p_T)$ can be
reproduced also when a lattice based hard EoS is used, see
Fig.~\ref{fig:v2ptPIDPHENIX} in Sect.~\ref{sec:RHICresult}, or
Ref.~\cite{Song:2011hk}. Thus there is no contradiction between
lattice QCD EoS and the observed particle anisotropies.

The previous discussion about the sensitivity of the particle
distributions to EoS is based on ideal fluid dynamics, since so far
there has been no systematic study addressing what we can learn about
the EoS of strongly interacting matter using either dissipative
hydrodynamics or integrated models. But since dissipative corrections
are supposed to be small in the fluid dynamical stage, it is highly
unlikely that adding dissipation would make the fluid evolution more
sensitive to the details of the EoS. Thus we may expect that what we
have learned about the effects of EoS on flow using ideal fluid
dynamics, holds for dissipative fluids as well.

\subsection{\it Fluid to Particles\label{sec:F2G}}

The QGP produced in relativistic heavy ion collisions expands, cools
down and goes through a transition to a hadronic gas.\footnote{It is
  not necessarily a phase transition. As mentioned in the previous
  section, the recent lattice QCD simulations predict a crossover
  rather than a phase transition between the QGP phase and the hadron
  phase.} At this late stage, the hadronic system is so diluted that
it would be hard to maintain equilibrium during the evolution.  Thus,
we switch from a macroscopic fluid dynamical picture to a microscopic
kinetic picture at a switching temperature $T_{\mathrm{sw}}$. In this
subsection, we discuss how to change from hydrodynamic description to
kinetic description. We will discuss dynamics of hadron cascade in the
next subsection.

We employ the Cooper-Frye formula \cite{Cooper:1974mv} to calculate
the single particle phase-space distribution for all hadrons in the
hadronic equation of state. For a hadron of species $i$, the
contribution to the distribution from a single fluid element located
at $x$ is
\begin{eqnarray}
\label{eq:CFlocal}
f_i(p,x) d^{3}x& = & \frac{d \Delta N_{i} }{d^{3}p}(x) \nonumber\\
& = &\frac{g_{i}}{(2\pi)^{3}E}
\frac{p \cdot \Delta \sigma(x)}{\exp[(p \cdot u(x) -\mu_{i}(x))/T(x)]\pm 1}.
\end{eqnarray}
Here $\Delta \sigma$ is a normal vector of a 
surface element of a constant temperature hypersurface
  $T(x) = T_{\mathrm{sw}}$\footnote{Note that this is not a
      freezeout hypersurface since in the subsequent stage hadrons
      still interact with each other.}.

In the actual calculations,
we Monte-Carlo sample the distributions of hadrons emitted from
  individual surface elements \cite{Hirano:2002ds}.
Regarding this, some comments on the conservation laws
are in order here:
\begin{enumerate}
\item The Cooper-Frye formula
 counts the \emph{net} number of 
emitted particles \cite{Cooper:1974mv}.
Here the net number means 
the number of particles moving outwards through the surface minus
  the number of particles moving inwards.
Thus the number obtained from Eq.~(\ref{eq:CFlocal})
could be negative in a case of 
a fluid element with either a space-like normal vector $(\Delta \sigma)^2 < 0$
or a time-like normal vector having a negative time component.
This so-called negative contribution problem of Cooper-Frye, 
is a long-standing problem in 
hydrodynamical modelling of relativistic heavy ion collisions.
Note that the \textit{total} number of particles
obtained by summing the  
contribution given by Eq.~(\ref{eq:CFlocal})
 over \textit{all} surface elements
is positive for all initial states relevant for heavy ion collisions.
Although the in-coming particles are necessary to ensure
the energy momentum conservation,
it is conceptually difficult to treat the negative number
in the subsequent kinetic approach.\footnote{
One possible solution would be to 
provide the cascade with the information of the location of the
  particlisation hypersurface, and remove from the cascade the hadrons
  which enter the space-time region within the hypersurface.}
However, at the time of particlisation at $T=T_\mathrm{sw}$ the
  collective flow is large, and thus the particle distributions are
  boosted to the direction of the flow, i.e. outwards. Therefore the
  yield of the in-coming particles is much smaller than the yield of
  out-going particles, and they form only a small correction 
 of the order of 5\%
  to the total multiplicity and the energy of the 
system\footnote{Somewhat larger corrections of the order of 10\%
    were reported in Ref.~\cite{Huovinen:2012is}. We believe that the
    reason is the different longitudinal structure of the system in
    these approaches.}.
As a first order approximation, we may thus ignore these
in-coming particles without violating the conservation laws significantly.

\item 
By sampling we create an integer number of particles
from a tiny fluid element
whose three dimensional volume is typically $\sim 0.1$ fm$^{3}$.
An expectation
value of the number of particles from this fluid element
in the grand canonical ensemble 
is, of course, not an integer and in our calculations 
 less than unity.
Therefore energy, momentum and charges are not conserved in each
  individual sampling, but only in average --- as is customary for a
  grand canonical ensemble.
This issue could be resolved
by an oversampling method:
At each fluid element,
$N$ times larger number of particles is sampled
with $N$ being large and, the subsequent dynamics of hadrons is simulated
with the cross section divided by $N$
to ensure the original Boltzmann
equation~\cite{Welke:1989dr,Molnar:2000jh}.
This procedure can maintain energy momentum conservation
of the order of $\mathcal{O}(1/N)$
at the particlisation.
However,
this would be numerically expensive and
neglect the effect of fluctuations
because different events are averaged over in the oversampling method.
A faster method which is called
``local-ensemble" method has been proposed
in Refs.~\cite{localensemble,Danielewicz:1991dh,Xu:2004mz}.
An alternative approach is to impose the requirement of the
  conservation of energy and charges on the sampling procedure as done
  in Refs.~\cite{Huovinen:2012is,Petersen:2008dd}. This approach
  maintains the effect of fluctuations, but requires generating
  several ensembles of particles, where only parts are kept to avoid
  bias.

\item In principle one  should switch from a
  macroscopic fluid picture to a microscopic particle picture
in a temperature region where both descriptions 
give similar results.
However, the single particle distribution
under \textit{local} equilibrium
never becomes
a solution of the Boltzmann equation:
One needs  viscous corrections to the local equilibrium
distribution to match the two solutions.
Since we use ideal fluid hydrodynamics without any dissipative
  corrections, the hydrodynamical evolution always differs from the
  cascade. Thus we cannot expect to find a region where the solutions
  of both models would agree, and simply regard the single particle
  phase-space distribution obtained using hydrodynamics as an initial
  condition for the hadron cascade.
\end{enumerate}

Keeping these issues and assumptions in mind,
we calculate a discrete single particle phase-space
distribution on an event-by-event basis.
First, we need the information of the particlisation hypersurface
  to apply the Cooper-Frye formula (Eq.~(\ref{eq:CFlocal})). We
  approximate the surface normal $\Delta\sigma^\mu$ in the following way:

\noindent
\textit{Bulk emission}: At each time step $\tau_i$ in a hydrodynamic
simulation,
we scan all the fluid elements to check whether
particlisation condition for bulk emission 
(a) $T(\tau_{i-1}) > T_{\mathrm{sw}} > T(\tau_{i})$
or (b) $T(\tau_{i-1}) < T_{\mathrm{sw}} < T(\tau_{i})$
is satisfied.
When this condition is satisfied, information about
a surface element vector (a) $\Delta\sigma^{\tau} = 
\tau_{\mathrm{sw}} \Delta x \Delta y \Delta \eta_{s}$
or  (b) $\Delta \sigma^{\tau} = -\tau_{\mathrm{sw}} \Delta x \Delta y \Delta \eta_{s}$
together with flow velocity $u_{\mathrm{sw}}^{\mu}$ at this point
are stored.
Here particlisation time $\tau$, flow velocities $v^{x}$ and $v^{y}$ and flow rapidity 
$Y_{f} = \tanh^{-1} v_{z}$ are linearly
interpolated between $\tau_{i}$ and $\tau_{i-1}$ such that
\begin{eqnarray}
w_{i}^{0} & = & \left| \frac{T(\tau_{i})-T_{\mathrm{sw}}}{T(\tau_{i}) - T(\tau_{i-1})} \right|,\enskip
w_{i-1}^{0} \ = \ \left| \frac{T_{\mathrm{sw}}-T(\tau_{i-1})}{T(\tau_{i}) - T(\tau_{i-1})} \right|,\\
\tau_{\mathrm{sw}} & = & \tau_{i-1}w_{i}^{0} + \tau_{i}w_{i-1}^{0}, \\
v_{\mathrm{sw}}^{x} & = & v^{x}(\tau_{i-1})w_{i}^{0} + v^{x}(\tau_{i})w_{i-1}^{0}, \\
v_{\mathrm{sw}}^{y} & = & v^{y}(\tau_{i-1})w_{i}^{0} + v^{y}(\tau_{i})w_{i-1}^{0}, \\
Y_{f, \mathrm{sw}} & = & Y_{f}(\tau_{i-1})w_{i}^{0} + Y_{f}(\tau_{i})w_{i-1}^{0}.
\end{eqnarray}

\noindent
\textit{Surface emission}: At each time step $\tau_{i}$,
we scan all the fluid elements in each direction to check whether
particlisation condition for surface emission
(a) $T(x_{i-1}) > T_{\mathrm{sw}} > T(x_{i})$
or (b) $T(x_{i-1}) < T_{\mathrm{sw}} < T(x_{i})$
is satisfied for an adjacent pair of surface elements. Here, for simplicity, we denote
only one dimensional dependence, say $x$ coordinate.
When this condition is satisfied, information about
surface vector (a) $\Delta \sigma^{x} = \tau_{i} \Delta \tau  \Delta y \Delta \eta_{s}$
or  (b) $\Delta \sigma^{x} = - \tau_{i} \Delta \tau  \Delta y \Delta \eta_{s}$
together with flow velocity $u_{\mathrm{sw}}^{\mu}$ at this point are stored.
Here  flow velocities $v^{x}$ and $v^{y}$ and flow rapidity 
$Y_{f}$ are linearly
interpolated  between $x_{i}$ and $x_{i-1}$ at $\tau_{i}$ such that
\begin{eqnarray}
w_{i}^{x} & = & \left| \frac{T(x_{i})-T_{\mathrm{sw}}}{T(x_{i}) - T(x_{i-1})} \right|, \enskip
w_{i-1}^{x}  =  \left| \frac{T_{\mathrm{sw}}-T(x_{i-1})}{T(x_{i}) - T(x_{i-1})} \right|, \\
v_{\mathrm{sw}}^{x} & = & v^{x}(x_{i-1})w_{i}^{x} + v^{x}(x_{i})w_{i-1}^{x}, \\
v_{\mathrm{sw}}^{y} & = & v^{y}(x_{i-1})w_{i}^{x} + v^{y}(x_{i})w_{i-1}^{x}, \\
Y_{f, \mathrm{sw}} & = & Y_{f}(x_{i-1})w_{i}^{x} + Y_{f}(x_{i})w_{i-1}^{x}. 
\end{eqnarray}
Unlike more sophisticated algorithms (see
Ref.~\cite{Huovinen:2012is}), which give relatively smooth surfaces,
this simple algorithm constructs a granular surface consisting of
``cubes''. At the limit of infinitely small elements, however, the
surfaces are equal. As well, for Cooper-Frye procedure the components
of the normal vectors of the surface elements are needed, and those
come out similarly in this approach and in the more sophisticated
approaches. The main difference is in the positions where the velocity
and densities are interpolated on the surface. This causes differences
proportional to the grid spacing, which defines the accuracy of the
numerics in general as well.

Using  this information, we generate a 
 hadron from the surface element.
We first calculate an 
expectation value of the number of hadrons of species $i$
out-going or in-coming through a hypersurface element
\begin{eqnarray}
\label{eq:CFpm}
\Delta N_{\pm}^{i} &=& g^{i}\int \frac{d^3p}{(2\pi)^{3}E}
   \frac{\Theta(\pm p\cdot \Delta \sigma)|p\cdot \Delta \sigma|}
        {\exp[(p\cdot u-\mu_{i})/T_{\mathrm{sw}}]
-\epsilon },
\end{eqnarray}
where $\epsilon = 1$ for bosons and $-1$ for fermions. $\Delta N_{\pm}$
are always positive by construction.
In the following, we neglect $\Delta N_{-}$ for simplicity as we
mentioned before.  We next create  a
hadron of species $i$ only when a randomly generated number $r_{1}$
($0<r_{1}<1$) is less than $\Delta N_{+}^{i}$.  Note that $\Delta
N_{+}^{i}$ is always less than unity in the usual setting of
simulations and its typical values are $\sim 0.01$. Such a low value
allows us to interpret $\Delta N_{+}^{i}$ as a probability to create a
particle, instead of sampling a Poisson distribution to decide whether
and how many particles are created.

If we create a hadron, we choose a momentum for it by sampling
the Lorentz invariant distribution 
\begin{eqnarray}
\frac{d^{3}p'}{E'}\frac{1}{\exp[(E'-\mu_{i})/T_{\mathrm{sw}}]-\epsilon}\Blue{.}
\end{eqnarray}
This is a momentum in the local rest frame of the fluid. We next
  Lorentz-boost it by flow velocity $u^{\mu}$ to obtain momentum in
  the centre-of-mass frame of the system. By construction the boosted
  momentum $p$ obeys the distribution
\begin{eqnarray}
\frac{d^{3}p}{E}\frac{1}{\exp[(p \cdot u-\mu_{i})/T_{\mathrm{sw}}]-\epsilon}.
\end{eqnarray}
We repeat this procedure until the obtained momentum
satisfies $p\cdot \Delta \sigma>0$.
Next we consider an weight $p\cdot \Delta \sigma$ in Eq.~(\ref{eq:CFpm}).
Suppose $r_{\mathrm{max}}$ is the maximum  
value of
$p\cdot \Delta\sigma$, which varies 
from hypersurface element to element. We generate another random
  number $r_2$, $0<r_2<r_{\mathrm{max}}$, and require that the
  momentum of the hadron fulfils $r_{2} < p\cdot \Delta\sigma$. If
  that is not the case, we discard the momentum, and start the process
  again by sampling the thermal momentum distribution.
Finally, 
we choose a position for this particle 
from a
uniform distribution inside the surface element.
The emission time is
either $\tau_{\mathrm{sw}}$ for bulk emission or $\tau_{i}$ for
surface emission.

We go through all the elements of the particlisation surface, and
  generate in this way an ensemble of hadrons and resonances with well
  defined positions $x_\mu^{i}$ and momenta 
  $p_\mu^{i}$~\footnote{Note that JAM allows different initial times
  $t_i$ for each hadron. Thus hadrons enter the cascade and begin interacting
  at the time when they are emitted from the fluid.}.
  We use this ensemble as the initial condition for the hadron cascade JAM,
  which we use to model the rest of the hadronic rescattering stage.
This will be discussed in the next section.

The switching procedure, namely
calculating the contribution to the particle distributions of
  \emph{all} hadrons in the EoS from \emph{all} hypersurface elements
  of the particlisation surface
is numerically expensive.
Among all the constituents of hybrid calculations---initial
  conditions, hydrodynamic simulation, switching process and hadronic
  cascade---it has been the bottleneck.
In event-by-event hybrid simulations,
this rather practical issue  must be resolved to gain
high statistics.
In Appendix, we show in detail how to integrate
the Cooper-Frye formula at less numerical costs \cite{murase}.

\subsection{\it Hadronic Cascade \label{sec:cascade}}

Hadronic transport models can be used to describe the system in the
low density hadronic phase of the
evolution. In this work we use the
microscopic transport model JAM~\cite{Nara:1999dz,Isse:2005nk} for
that purpose.  In JAM, the trajectories of all hadrons and resonances,
including those produced in resonance or string decays, are propagated
along their classical trajectories like in other microscopic hadronic
transport models such as RQMD~\cite{rqmd1,rqmd2,rqmd4,rqmd5}, and
UrQMD~\cite{urqmd1,urqmd2,urqmd3}. To achieve a more sophisticated
hadronic EoS, a mean field can be included within a framework of
either Boltzmann-Uehling-Uhlenberck (BUU) model~\cite{Bertsch:1988ik},
or Quantum Molecular Dynamics (QMD)~\cite{Aichelin:1991xy} approach.
However, all results in this work are obtained without any mean field.

In the hadronic transport models, time evolution of system is
described by a sum of incoherent binary hadron-hadron ($hh$)
collisions.  Two body collisions are realised by the closest distance
approach: Two particles collide if their minimum distance $b$ in the
centre-of-mass (c.m.) flame of two colliding particles is smaller than
the distance given by the geometrical interpretation of cross section:
\begin{equation}
  b \leq \sqrt{\frac{\sigma_\mathrm{tot}}{\pi}}\ ,
\end{equation}
where $\sigma_\mathrm{tot}$ denotes the total cross section at the
energy $\sqrt{s}$. For two particles with their positions $x_1$ and
$x_2$, and four momenta $p_1$ and $p_2$, the Lorentz invariant
expression for impact parameter $b$ is given by
\begin{equation}
  b^2 = -(x_1-x_2)^2 + \frac{[P\cdot (x_1-x_2)]^2}{P^2}
           + \frac{[q\cdot (x_1-x_2)]^2}{q^2}\Blue{,}
\end{equation}
where $P=p_1+p_2$,
and $q= (p_1 - p_2) - \frac{[P\cdot (p_1-p_2)]^2}{P^2}P$.

Inelastic $hh$ collisions are modelled by resonance formation at low
energies and by formation of colour strings at high energies.
Threshold between resonance and string formation is set to about 4 GeV
for baryon-baryon (BB), 3 GeV for meson-baryon (MB) and 2 GeV for
meson-meson (MM) collisions.  In the string formation process, we use
the same distribution for the light-cone momentum transfer as in the
HIJING model~\cite{Wang:1991hta,hijing2,hijing3,hijing4}. Quark
content of a string is assumed to be the same as the quark content of
a corresponding hadron before excitation, as in the Fritiof
model~\cite{fritiof,fritiof2}.

The string decays are performed by the Lund string
model~\cite{lund,lund2,Sjostrand:2006za}. Formation points and times
for newly produced particles are determined from string decay by yo-yo
formation point~\cite{bialas}.  Formation time is about 1 fm/$c$ with
the string tension $\kappa=1$ GeV/fm.  In a baryon-like string,
hadrons are produced by the quark-antiquark pair creation in the
colour flux-tube between the quark and diquark. The antiquark from
the pair creation is combined with the constituent quark in the in the
string to form a first rank hadron. This hadron has an original
constituent quark. We assign a formation time for the quarks from the
quark-antiquark pair creation, but not for the original constituent
quark. Thus, for example, the original constituent quark inside a
newly formed meson can scatter, but with a reduced cross section
$1/2\sigma_{MM}$. In general, leading hadrons which contain original
constituent quarks can scatter during their formation time with cross
sections reduced according to the additive quark model. The
importance of this quark(diquark)-hadron interaction for the
description of baryon stopping at CERN/SPS energies has been reported
by Frankfurt group~\cite{rqmd2,rqmd4,rqmd5,urqmd1,urqmd2}.

Experimentally well-known total and elastic cross sections, such as
$pp$, $pn$, $\pi^{+} p$, $K^{-} p$ and $\bar{p}p$, are parametrised in
JAM.  Cross sections involving the resonances are assumed to be the
same as the corresponding stable hadron cross sections with the same
quark content.  For example, $\rho p$ cross section is the same as
$\pi p$ cross section.

In nucleon-nucleon scattering, nonstrange baryonic resonance
excitation channels
\begin{equation}
 NN \to NR, \qquad NN \to RR \ ,
\end{equation}
where $R$ means a nucleon resonance ($N(1440)$-$N(1990))$ or a
$\Delta$ resonance ($\Delta(1232)$-$\Delta(1950))$ up to 2 GeV, are
implemented.  These resonance formation cross sections are fixed by
pion production cross sections.  Inverse processes such as $NR \to NN$
are computed employing the detailed balance formula where the finite
width of the resonance is taken into
account~\cite{Danielewicz:1991dh,Wolf2,detbal2}.  The lifetime of
resonance, $t$, is randomly chosen according to an exponential decay
law $\exp(-t\gamma \Gamma(M))$, where $\Gamma(M)$ is the
energy-dependent width of the resonance and $\gamma=E/M$ is the
Lorentz factor.

As an example of a meson-baryon scattering, the total cross section
for the $\pi N$ incoming channel is decomposed to
\begin{equation}
 \sigma_{\mathrm{tot}}(s) = \sigma_{\mathrm{el}} + \sigma_{\mathrm{ch}}(s) 
 + \sigma_{\mathrm{BW}}(s) + \sigma_{s-S}(s) + \sigma_{t-S}(s),
\end{equation}
where $\sigma_{\mathrm{el}}$, $\sigma_{\mathrm{ch}}(s)$,
$\sigma_{\mathrm{BW}}(s)$, $\sigma_{s-S}(s)$, and $\sigma_{t-S}(s)$
denote the elastic, charge exchange, $s$-channel resonance formation,
$s$-channel string formation and $t$-channel string formation cross
section, respectively.  Resonance formation cross section
$\sigma_{BW}(s)$ is computed using the Breit-Wigner
formula~\cite{rqmd4} by summing up cross sections to form resonances
   $R=N(1440)$-$N(1990)$, $\Delta(1232)$-$\Delta(1950)$,
   $\Lambda(1405)$-$\Lambda(2110)$,
   $\Sigma(1385)$-$\Sigma(2030)$
   and $\Xi(1535)$-$\Xi(2030)$.

In the case of $\bar{K}N$ incoming channel, we add $t$-channel hyperon
production cross sections such as $K^{-} p \to \pi^{0} \Lambda$.  The
cross section of the inverse process $\pi Y \to \bar{K} N$,
$Y=\Lambda, \Sigma$ is calculated using the detailed balance formula.
The kaon-nucleon ($KN$) incoming channel does not have $s$-channel
resonance formation, but $t$-channel resonance production processes
$KN \leftrightarrow K\Delta$, $KN\leftrightarrow K(892)N$,
$KN\leftrightarrow K(892)\Delta$ are included.  The Breit-Wigner
formula is used to evaluate the cross section for resonance production
in meson-meson scatterings as well.  Meson resonance states are
included up to about 1800 MeV.

Additive quark model~\cite{rqmd2,rqmd4,urqmd2,urqmd3} is used for the
experimentally unknown cross sections such as an incoming channel
involving multistrange hadrons, \emph{e.g.} $\phi$ meson-pion
scattering.  Strangeness suppression factor is correctly included in
the additive quark model: We have $\sigma_{\pi N} \approx 26$ mb and
$\sigma_{K N} \approx 21$ mb consistent with the experimental data
above resonance region.

\subsection{\it Brief overview of hydro + hadronic cascade models \label{sec:afterburner}}

\begin{figure}[tb]
\begin{center}
\begin{minipage}[t]{85mm}
 \epsfig{file=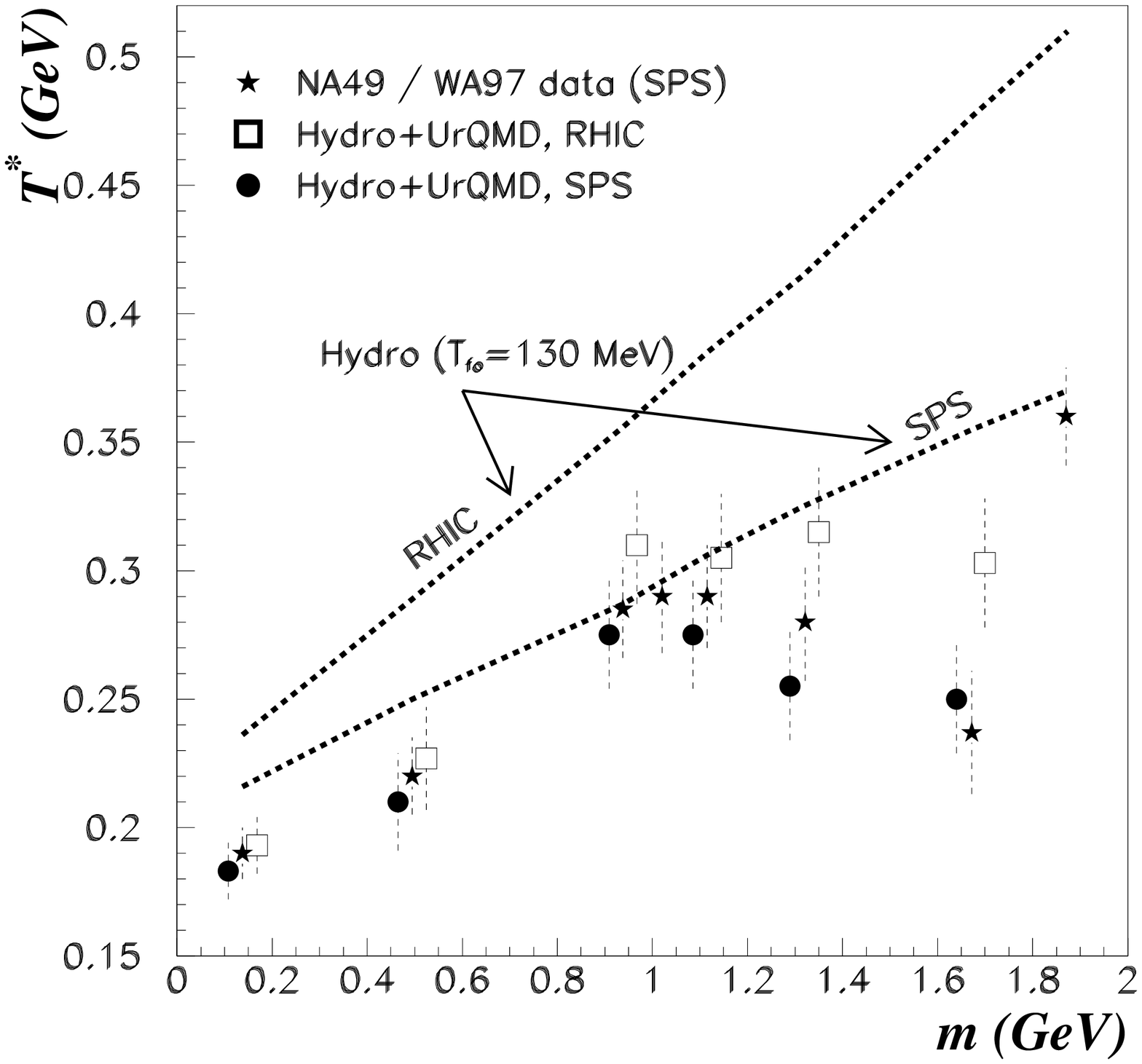,scale=0.4}
 \caption{
Inverse slopes of the $m_{T}$ spectra for various 
strange and non-strange hadrons at midrapidity
from the (1+1)-D hybrid model
are compared with SPS data (star symbols) \cite{WA97,NA49}.
lines are results from pure hydrodynamic simulations
with freezeout temperature $T_{\mathrm{fo}}=130$ MeV.
Open squares and closed circles are
results from the hybrid model at RHIC and SPS,
respectively.
Figure is  taken from Ref.~\cite{Dumitru:1999sf}.
\label{fig:hybrid1}}
\end{minipage}
 \hfill
\begin{minipage}[t]{85mm}
 \epsfig{file=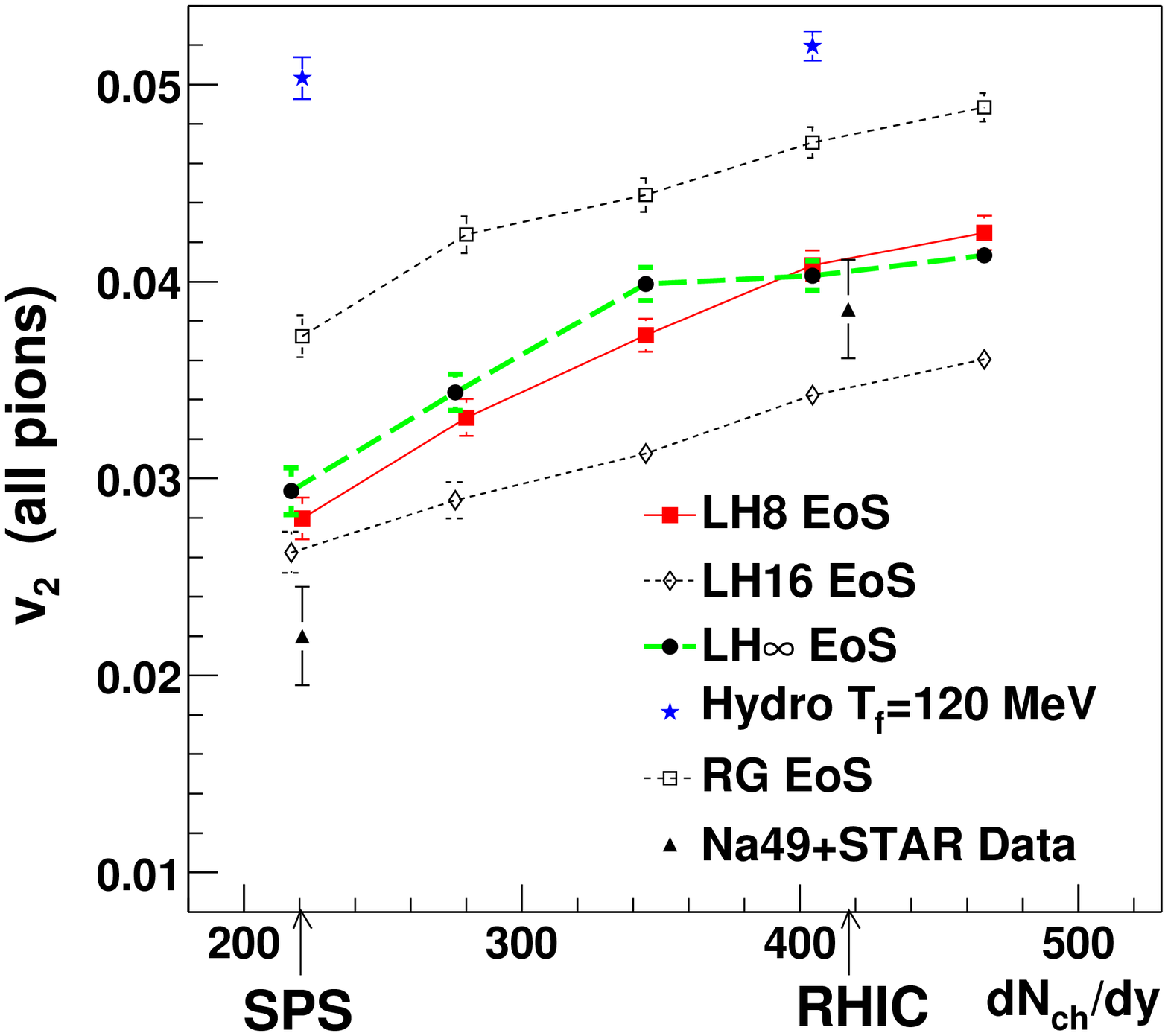,scale=0.4}
 \caption{
Elliptic flow parameters $v_{2}$ from four different
equations of state
as functions of the charged particle multiplicity
from the (2+1)-D hybrid model
is compared with SPS \cite{NA49_v2} and RHIC \cite{Ackermann:2000tr} data (triangle
symbols).
Impact parameter in the simulations is taken to be $b=6$ fm.
Star symbols are results from pure hydrodynamic simulations
with freezeout temperature $T_{f} = 120$ MeV. 
Figure is taken from Ref.~\cite{Teaney:2000cw}.
   \label{fig:hybrid2}}
\end{minipage}
\end{center}
\end{figure}

In this subsection, we briefly overview the
current status of hydro + hadronic cascade model
(sometimes called the ``hybrid" model).

\begin{figure}[tb]
\begin{center}
\begin{minipage}[t]{85mm}
 \epsfig{file=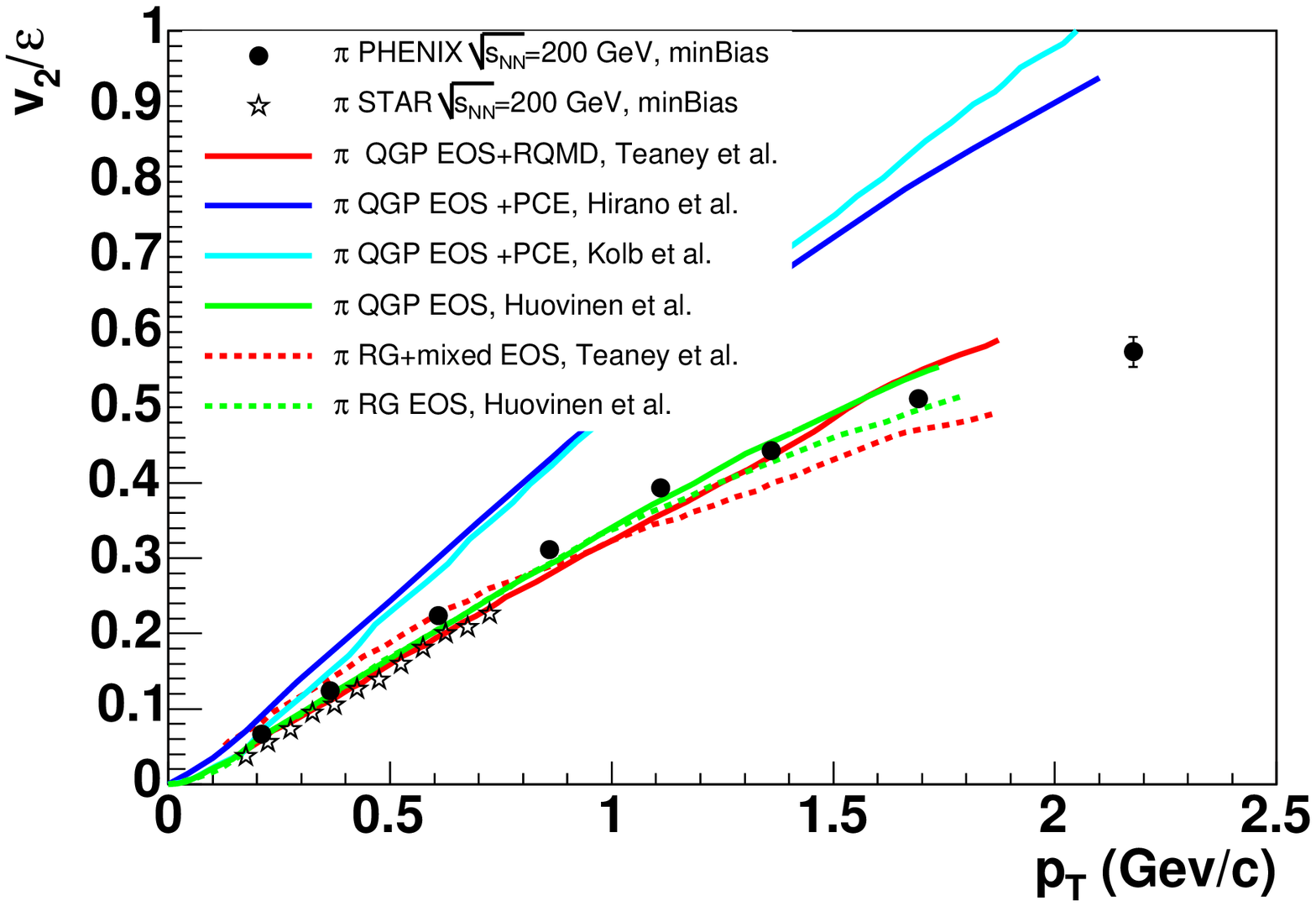,scale=0.42}
\end{minipage}
 \hfill
\begin{minipage}[t]{85mm}
 \epsfig{file=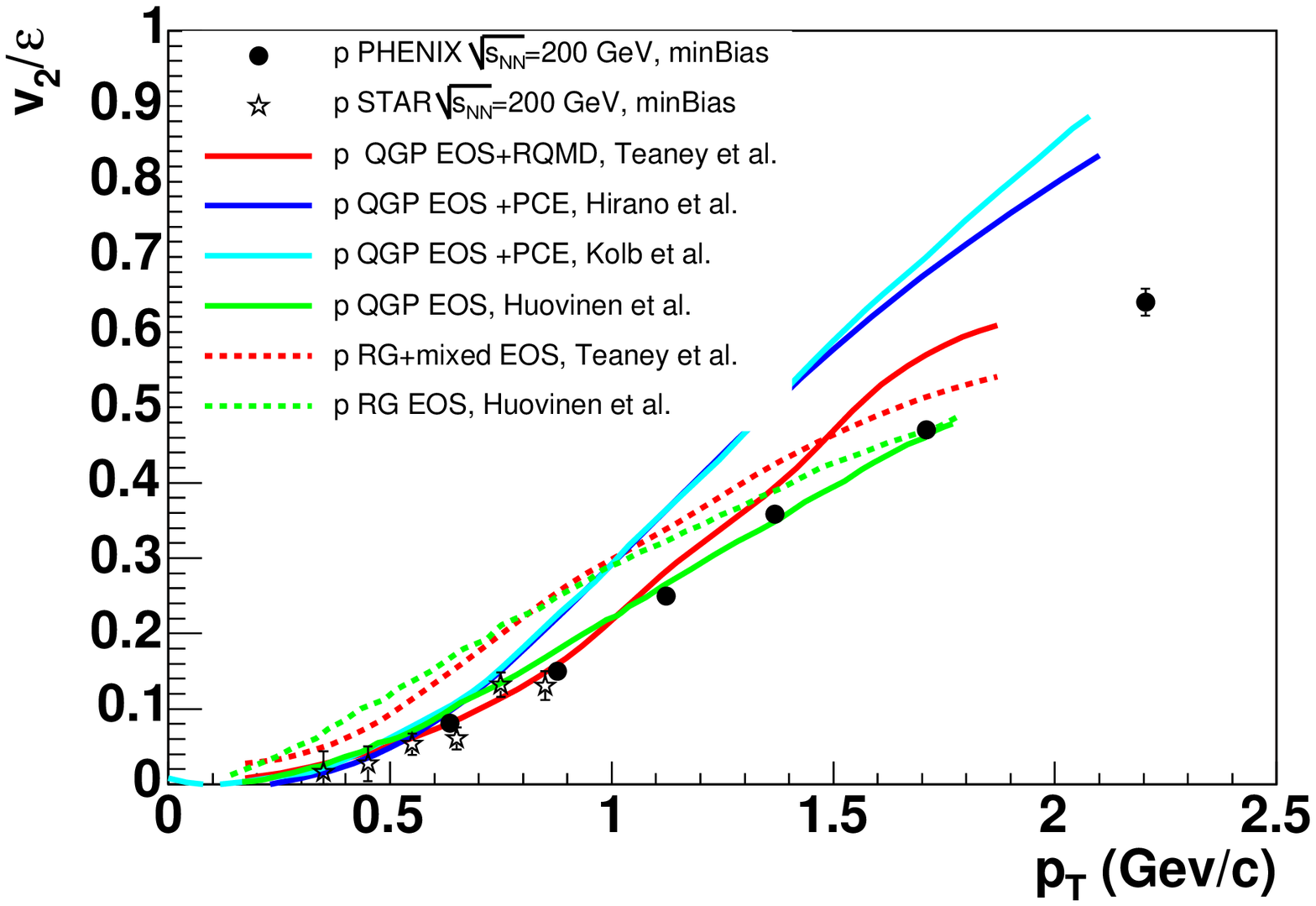,scale=0.42}
\end{minipage}
 \caption{
A compilation of hydrodynamic results as of year 2004.
$v_{2}/\varepsilon$
as a function of transverse momentum for pions (left) and for protons (right)
from hydrodynamic models are compared with STAR and PHENIX 
minimum bias data.
For details, see text and Ref.~\cite{Adcox:2004mh}.
   \label{fig:phenix_white}}
\end{center}
\end{figure}

The very first work of hydro + cascade approaches was done by Dumitru
{\it et al.} \cite{Dumitru:1999sf}.  Motivation in this first study
was to describe particle species dependence of
freezeout process in a consistent manner.  They solved hydrodynamic
equations by assuming boost invariant longitudinal flow and
cylindrical symmetry.  As for the equation of state, a bag model which
exhibits the first order phase transition was employed with 
$T_{c} = 160$ MeV.  They switched description of dynamics from
hydrodynamics to hadronic cascade at hadronisation.  The $m_{T}$
inverse slope parameters for various hadrons including multi-strange
ones were calculated both in pure hydrodynamics (with
$T_{\mathrm{fo}}=130$ MeV) and in the hydro + cascade approach as
shown in Fig.~\ref{fig:hybrid1}.  In pure hydrodynamics, it is known
that the inverse slope parameter increases monotonically with the
hadron mass.  On the other hand, the slope parameter of multi strange
hadrons from the hadron cascade approach are almost identical
regardless of the mass difference \cite{Dumitru:1999sf}, which is
clearly different from a tendency of the pure hydrodynamic result
mentioned above.  They made a further analysis of kinetic freezeout in
the subsequent papers \cite{Bass:1999tu,Bass:2000ib} within this
hybrid approach.

A systematic analysis of SPS and RHIC data
was done by Teaney {\it et al.}~within a (2+1) dimensional hydro + cascade model \cite{Teaney:2000cw,Teaney:2001av},
where an event generator, RQMD, was employed for the hadronic cascade model.
The importance of hadronic dissipation 
in interpreting the elliptic flow data
was first demonstrated in this study.
In the SPS and RHIC energy regions, pure ideal hydrodynamics
predicts $v_{2}/\varepsilon \sim 0.2$-$0.25$ 
depending on
the equation of state employed in the simulations.
This is sometimes called ``hydrodynamic limit".
Experimental data of $v_{2}/\varepsilon$ increase
with the transverse particle density
$(1/S) dN_{\mathrm{ch}}/d\eta$ \cite{Alt:2003ab},
where $S$ is the transverse area,
and reach 
the ``hydrodynamic limit'' of $\sim 0.2$
in central collisions at the top RHIC energy.
As mentioned, ideal hydrodynamics predicts roughly constant
  $v_2/\epsilon$, and does not reproduce this data.
On the other hand, as shown in Fig.~\ref{fig:hybrid2},
this monotonic increase is described by the hydro + RQMD model
\cite{Teaney:2000cw,Teaney:2001av}
in which finite cross sections of hadronic interactions
lead to dissipation and, consequently,
integrated $v_{2}$ is considerably reduced in comparison with
pure ideal hydrodynamic calculations.

\begin{figure}[tb]
\begin{center}
\begin{minipage}[t]{9 cm}
\epsfig{file=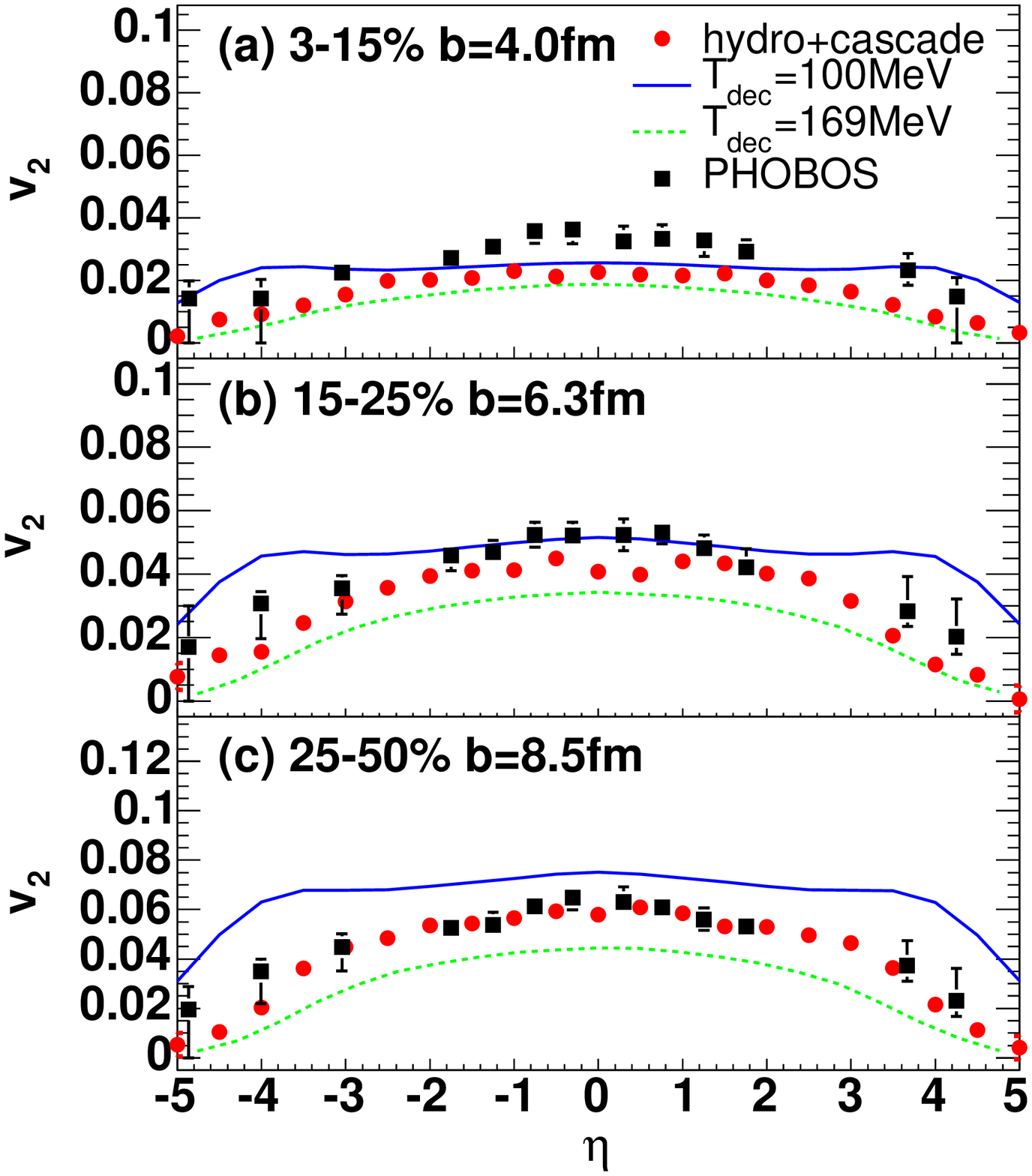,scale=0.45}
\end{minipage}
\begin{minipage}[t]{9 cm}
\epsfig{file=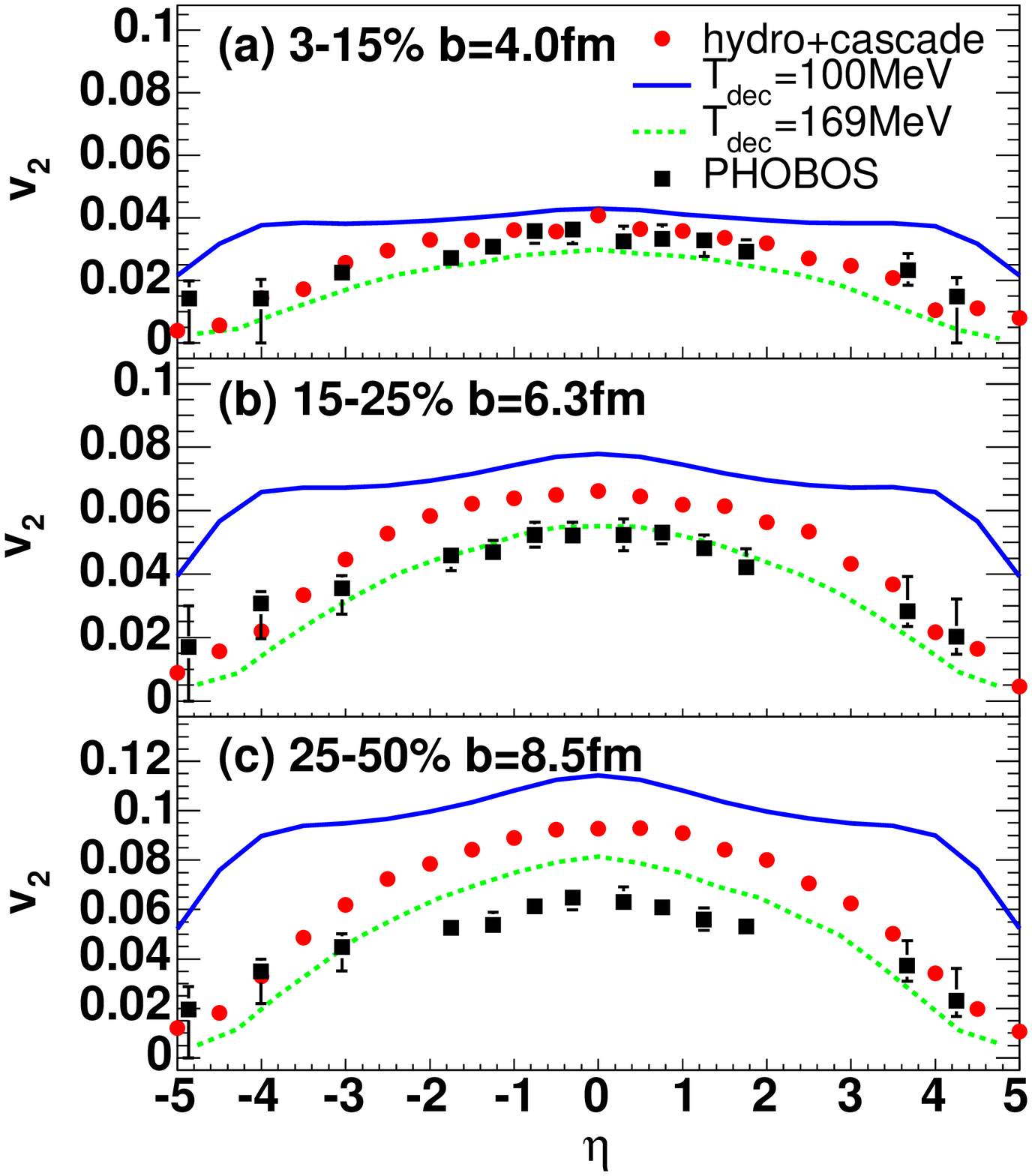,scale=0.45}
\end{minipage}
\caption{
Pseudorapidity dependence of $v_{2}$
in 3-15\%, 15-25\% and 25-50\% centralities
from Glauber-BGK (left)
and KLN (right) initial conditions
are compared with the PHOBOS data \cite{Back:2004mh}.
Plots are results from the full 3D hybrid model.
Solid and dashed curves are results from
pure hydrodynamic simulations with
freezeout (decoupling) temperature $T_{\mathrm{dec}}=100$ MeV and 169
MeV, respectively.
Since the temperature of the boundary between the QGP and hadron phases
in these calculations
is $T=170$ MeV,
results with $T_{\mathrm{dec}}=169$ MeV correspond
to the ones just after complete hadronisation.
Figures are taken from Ref.~\cite{Hirano:2005xf}.
\label{fig:v2eta}}
\end{center}
\end{figure}

Figure \ref{fig:phenix_white} shows 
a compilation of hydrodynamic results
made by the PHENIX Collaboration \cite{Adcox:2004mh} as of year 2004.
Among several hydrodynamic approaches, 
it was only the hydro + RQMD model  that
reproduced particle identified spectra and $v_{2}(p_{T})$ data at the same time.
$p_{T}$ spectra for pions and $v_{2}(p_{T})$ for pions and protons are
reproduced 
using ideal hydrodynamics where the fluid is in chemical equilibrium.
However, such a model fails to
reproduce the yield of protons since 
the freeze-out temperature required to fit the slopes of the
  $p_T$ distributions is well below the temperature required to fit
  the observed particle ratios.
To solve this issue, a partial chemical equilibrium (PCE) 
model \cite{Hirano:2002ds} is employed.
In this case, chemical freezeout 
is incorporated
in the equation of state and the system
is in kinetic but not in chemical equilibrium below a chemical
  freeze-out temperature.
Such a model leads to successful 
reproduction of yields and spectra for pions and protons, but a slope of pion $v_{2}(p_{T})$ 
is steeper than that of the data \cite{Hirano:2002ds} as shown in
Fig.~\ref{fig:phenix_white} (left).
The importance of the hadronic dissipative effects 
in simultaneous reproduction of yields, spectra and differential $v_{2}$ 
was  discussed in detail later in Ref.~\cite{Hirano:2005wx}.

Hirano \textit{et al.}  \cite{Hirano:2005xf,Hirano:2007ei} combined a
fully (3+1) dimensional ideal hydrodynamics with a hadronic cascade
model, JAM.  Our integrated dynamical approach presented in this paper
is based on this model.  One of the
advantages of the fully three dimensional simulations without assuming
Bjorken scaling solution is to be able to
obtain elliptic flow parameter as a function of pseudorapidity,
$v_{2}(\eta)$.  The charged hadron
$v_{2}(\eta)$ has been measured by the PHOBOS
Collaboration and it has a maximum at midrapidity and decreases as
moving away from midrapidity \cite{PHOBOSv2-1,PHOBOSv2-2}. 
In a full three dimensional ideal hydrodynamic simulation with
  $T_{\mathrm{fo}} = 100$ MeV~\cite{Hirano:2002ds}, $v_2$ does not
  depend strongly on rapidity. Thus the main dependence on
  pseudorapidity comes from the Jacobian between rapidity and
  pseudorapidity \cite{Kolb:2001yi}, and the calculated $v_2(\eta)$ is
  flatter than the measured, whereas the three dimensional hybrid
  approach reproduces the observed $v_2(\eta)$ in the whole rapidity
  region in non-central collisions when the Glauber initial conditions
  are used~\cite{Hirano:2005xf}, see Fig.~\ref{fig:v2eta}. \label{discuss:v2eta}
The space-time rapidity dependence
of life time of the QGP fluid
plays an important role in understanding
the (pseudo-)rapidity dependence of $v_{2}$ since it takes time for 
the system to fully develop elliptic flow.
Since $dN_{\mathrm{ch}}/d\eta$ decreases with
increasing $\eta$,
initial entropy and, in turn, initial temperature
decreases with increasing $\eta_{s}$.
Consequently, the lifetime of the QGP also decreases with increasing
$\eta_{s}$, and we may expect lower values of $v_2$ at large $\eta_s$ if we
  evaluate $v_2$ immediately after hadronisation. In these
  calculations phase transition took place at $T_c = 170$ MeV, and the
  evaluation of $v_2(\eta)$ on a $T = 169$ MeV surface leads to a
  shape peaking at midrapidity as expected, see Fig.~\ref{fig:v2eta},
although the overall magnitude is less than the PHOBOS data
\cite{PHOBOSv2-1,PHOBOSv2-2}.
Additional generation of elliptic flow during the hadronic cascading stage
fills this gap to reproduce the PHOBOS data.
Compared with a purely ideal hydrodynamic treatment of the hadronic gas,
the hadron cascade contains dissipative effects through finite
mean free path.
This indicates the importance of hadronic dissipative effects
in particular in forward/backward regions.

It is interesting to note the deviations of the 3D hybrid model
  results from the PHOBOS data as well~\cite{Hirano:2005xf}. First,
 $v_{2}(\eta)$ from the full 3D hybrid model
is smaller than the data at 3-15\% centrality when the Glauber type
initial conditions are employed (see top panel of Fig.~\ref{fig:v2eta}
(left)), which implies the necessity of 
eccentricity fluctuations in the initial conditions.
Second, the hybrid model with the KLN initialisation
leads to larger $v_{2}$ than the data in semi-central 
to peripheral collisions (see Fig.~\ref{fig:v2eta} (right)),
which  
leaves room for finite although very small QGP viscosity.

\begin{figure}[tb]
\begin{center}
\epsfig{file=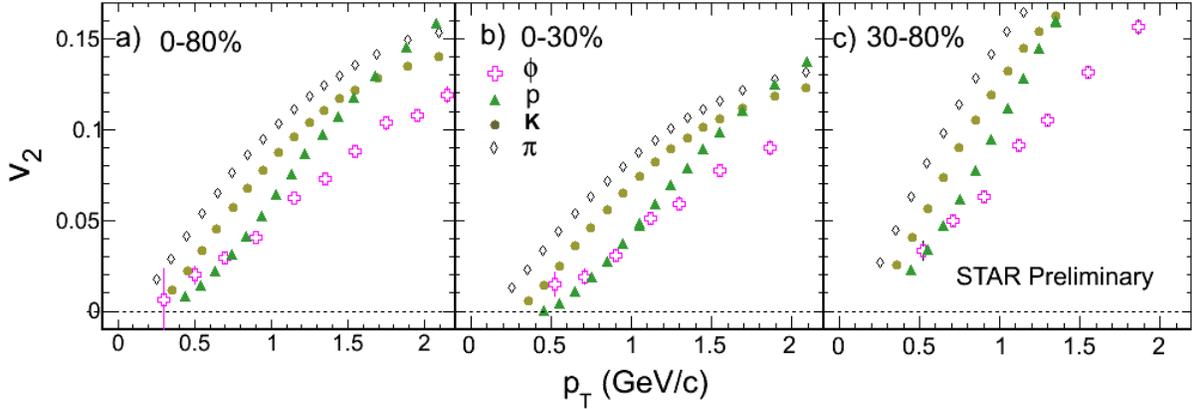,scale=0.45}
\caption{
Transverse momentum dependence of particle identified elliptic flow parameters
in 0-80\%, 0-30\% and 30-80\% centrality
from STAR Collaboration.
Violation of mass ordering 
is clearly seen below $\sim$0.8 GeV/$c$ in 0-30\%
centrality.
Figures are taken from Ref.~\cite{Nasim:2012gz}.
\label{fig:STARphi}}
\end{center}
\end{figure}

Another important finding from
this full 3D hybrid model is a violation of mass ordering
in differential elliptic flow parameter $v_{2}(p_{T})$ \cite{Hirano:2007ei}. 
Because of the interplay of thermal and collective motion, we
  expect that 
  $m_1 < m_2 \Rightarrow v_2(p_T,m_1) > v_2(p_T,m_2)$~\cite{Huovinen:2001cy}. 
  This mass ordering, however, holds only for particles frozen out in
  the same temperature having the same collective flow velocity. We
  expect that particles like $\phi$ mesons, which have very small
  cross sections and thus hardly rescatter, freeze-out earlier. Such a
  particle's $v_2(p_T)$ would be typical for larger temperature and
  smaller flow velocity. For particles with mass around 1 GeV mass,
  like $\phi$ mesons, freezing out earlier would mean larger
  $v_2(p_T)$ at small $p_T$. The hybrid model calculations predicted
  this kind of behaviour for the $\phi$ meson \cite{Hirano:2007ei},
  and this violation of mass ordering was recently confirmed by the
  STAR collaboration~\cite{Nasim:2012gz} (See also
  Fig.~\ref{fig:STARphi}).

\begin{figure}[tb]
\begin{center}
\begin{minipage}[t]{85mm}
 \epsfig{file=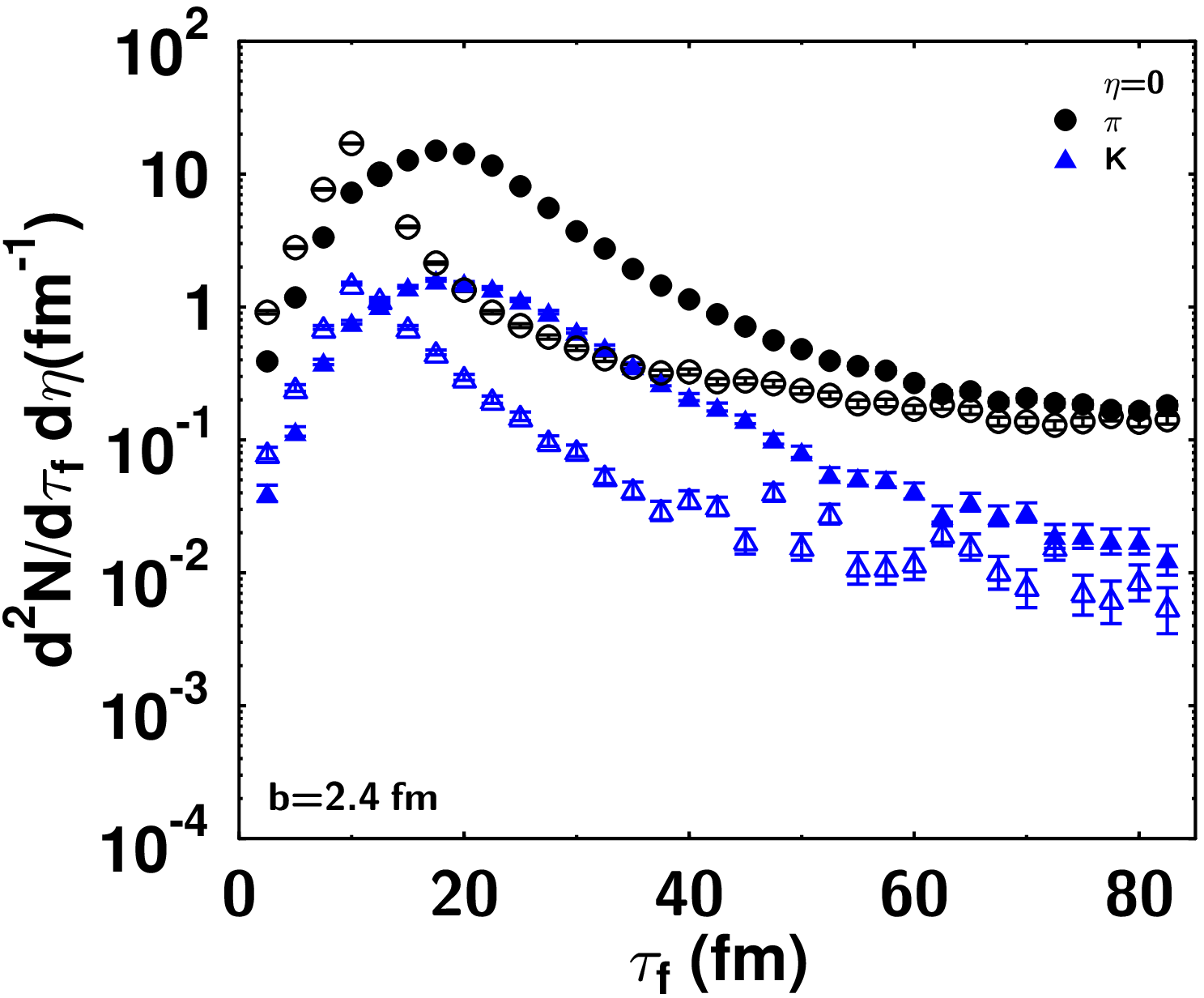,scale=0.5}
 \caption{
Freezeout-time distribution for pions (circles)
and kaons (triangles)
at an impact parameter $b=2.4$ fm
at the RHIC energy.
Open (Closed) symbols correspond to results without (with)
hadronic rescatterings.
Figure is  taken from Ref.~\cite{Nonaka:2006yn}.
\label{fig:hybrid3}}
\end{minipage}
 \hfill
\begin{minipage}[t]{85mm}
\epsfig{file=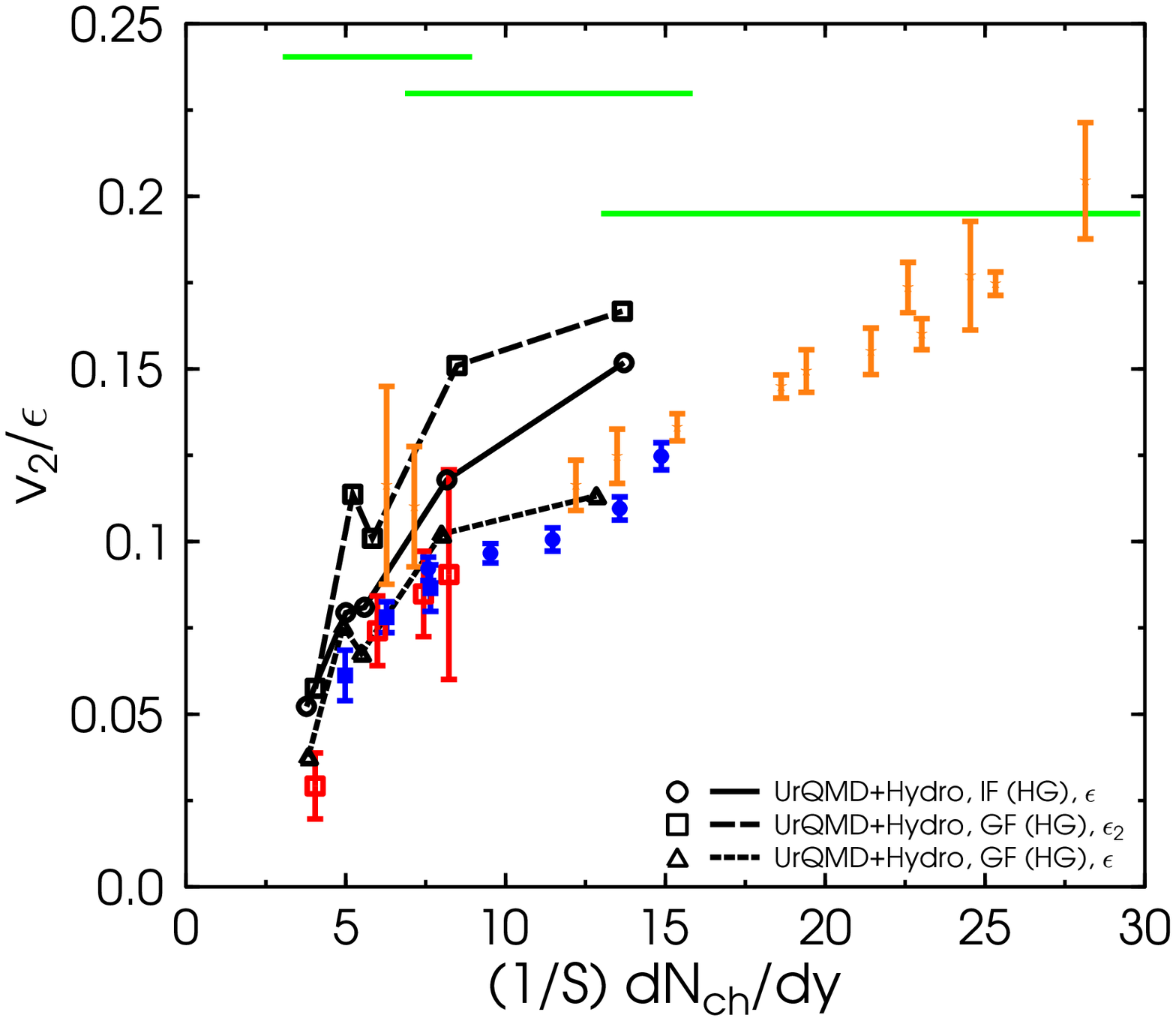,scale=0.45}
 \caption{
$v_{2}/\varepsilon$ as a function of transverse 
density $(1/S)dN_{ch}/dy$ from
a full 3D hybrid model is compared with
experimental data at AGS, SPS and RHIC energies.
Results with isochronous freezeout and
gradual freezeout are shown as open circles and triangles,
respectively.
Results with gradual freezeout
using event-by-event $\varepsilon$ averaged over many events
is shown as open squares.
Figure is taken from Ref.~\cite{Petersen:2009vx}.
   \label{fig:hybrid4}}
\end{minipage}
\end{center}
\end{figure}

The ideas of the Frankfurt group's first work 
were taken over
by Nonaka and Bass \cite{Nonaka:2006yn} who used
a fully three dimensional
ideal hydrodynamic model coupled to UrQMD.
Similar reduction of $v_{2}$ in forward/backward rapidity regions
as shown in Fig.~\ref{fig:v2eta}
was found in their results as well.
Hadronic rescattering effects are
seen in Fig.~\ref{fig:hybrid3}
as shifts of the peaks of freeze-out time distributions from
  $\sim$ 10 fm/$c$ to $\sim$ 18 fm/$c$. The effect is similar for
  pions and kaons.

Petersen {\it et al.}\  combined full three dimensional
ideal hydrodynamics in the Cartesian coordinate with UrQMD
 \cite{Petersen:2008dd,Petersen:2009vx,
  Petersen:2010md,Petersen:2010cw,Qin:2010pf,Petersen:2011fp,Petersen:2011sb}.
This is the first 
hybrid simulation on an event-by-event basis, which will be
discussed later.
One of the distinct features
of this model is to employ
isochronous particlisation
when energy density of all fluid elements drops below
5 times ground state energy density ($\sim 730$ MeV/fm$^3$).
They also discussed all fluid elements
in one transverse slice rather than the whole region
dropping below this value as an alternative criterion \cite{Petersen:2009vx}.
By using this (called gradual freezeout),
one can take account of time dilatation in the forward rapidity region
where the fluid moves faster in the Cartesian coordinate.
An advantage of this method is to be able to avoid
negative contribution in the Cooper-Frye formula
since the hypersurface element is always time-like
vector $d\sigma^{\mu} = (d^3 x, {\bm 0})$ and $p^{\mu} d\sigma_{\mu}$
term in this formula is positive definite.
On the other hand,
it may happen that freezeout occurs in some fluid elements
only when the temperature becomes very small ($T<100$ MeV)
and the system is already diluted sufficiently.
In the context of hadronic rescattering effects,
they mainly focused on collisions at SPS energies and lower
and found that 
their hybrid model can nicely describe elliptic flow at those
  energies
\cite{Petersen:2008dd,Petersen:2009vx,Petersen:2010md}, see
  Fig.~\ref{fig:hybrid4}. Recently this model has been extended to
  allow isothermal or iso density particlisation as
  well~\cite{Huovinen:2012is}.

\begin{figure}[tb]
\begin{center}
\begin{minipage}[t]{70mm}
 \epsfig{file=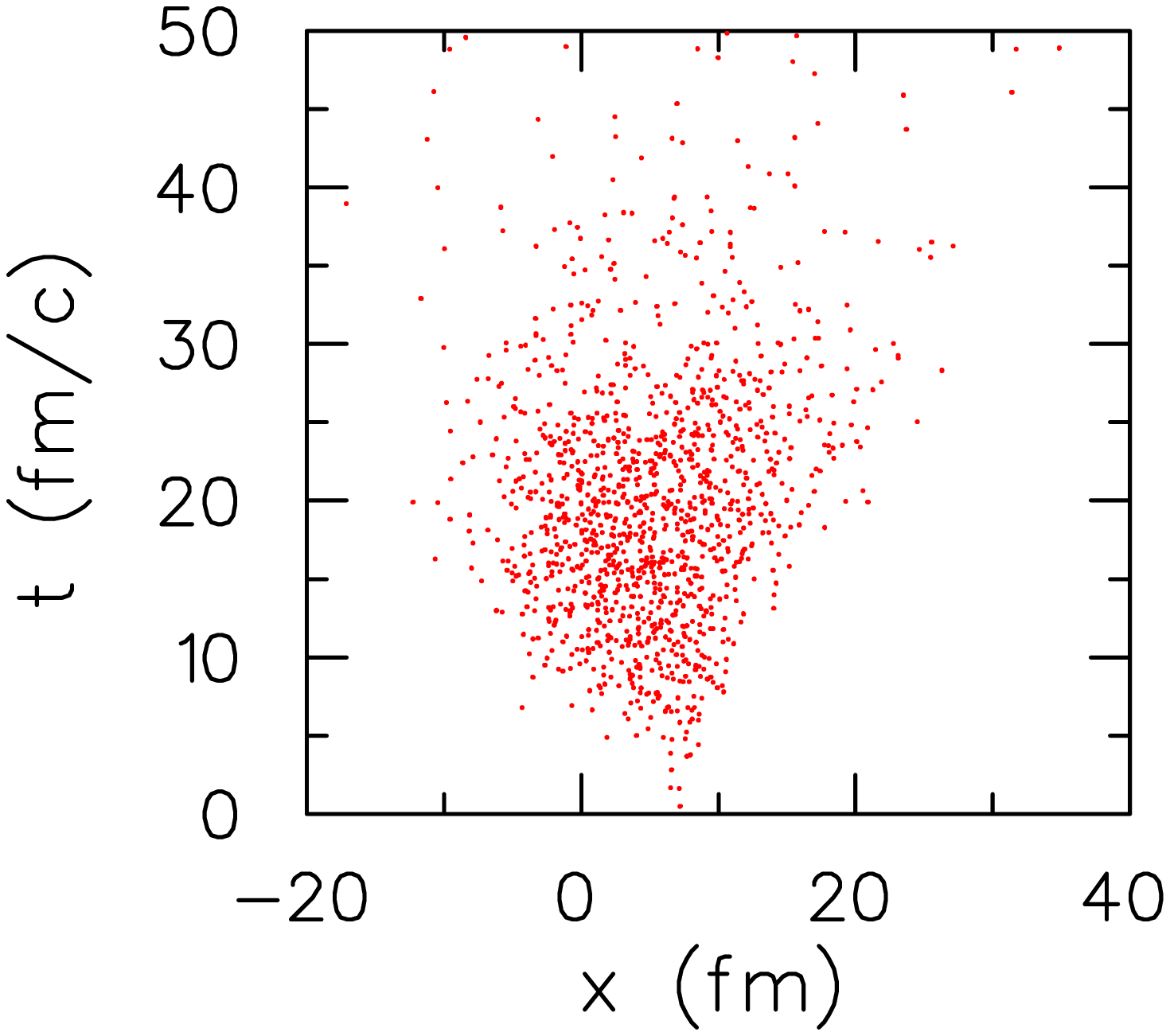,scale=0.45}
 \caption{
Positions of last interaction points for pions with $p_{x} = 300$, $p_{y} = 0$ MeV/$c$.
Figure is  taken from Ref.~\cite{Pratt:2008sz}.
\label{fig:hybrid5}}
\end{minipage}
 \hfill
\begin{minipage}[t]{105mm}
\epsfig{file=figures1/Fig1,scale=0.45}
\caption{
Comparison of $v_{2}/\varepsilon$ vs. $(1/S)(dN_{\mathrm{ch}}/dy) $
curves with experimental
data from the STAR Collaboration. 
(a) the MC-KLN model and (b) the MC-Glauber model. 
Figures are taken from Ref.~\cite{Song:2010mg}.
\label{fig:hybrid8}}
\end{minipage}
\end{center}
\end{figure}

Pratt and Vredevoogd~\cite{Pratt:2008sz}
were the first to develop
a hybrid model based on viscous hydrodynamics. Their goal was to 
understand femtoscopic observables at RHIC, and they assumed radial
symmetry and boost invariance both in hydrodynamical calculation and
hadron cascade. In their model
 particlisation happens at a switching energy
density $e_{\mathrm{sw}}=400$ MeV/fm$^{3}$ instead of
a constant temperature.
They kept information about particlisation hypersurface
and a hadron which returns to the interior of the surface ($e>e_{\mathrm{sw}}$)
during the cascade description is deleted from simulations.
This would correspond to negative contribution
to the Cooper-Frye formula. 
According to their estimation, a percentage of the absorbed particles
is only about one percent.
Fig.~\ref{fig:hybrid5} shows
positions of last interaction points for pions
with $p_{x} = 300$, $p_{y} = 0$ MeV/$c$.
Modestly positive correlation between the outward position
and the emission time is seen, which would result in $R_{\mathrm{out}}/R_{\mathrm{side}}$
to be close to unity.

Werner {\it et al.}
combined an event generator, EPOS, full (3+1) dimensional ideal
hydrodynamics and an hadronic cascade, 
UrQMD~\cite{Werner:2010aa,Werner:2010ny,Werner:2010ss,Werner:2012xh}.
In addition to nuclear collisions,
they also applied their hybrid model to proton-proton collisions
at the LHC energies.
They concluded there exists collective flow
even in pp collisions. Their results on ridge phenomenon
will be discussed later in Sec.~\ref{sec:e-by-e_hydro}.

Detailed analyses of elliptic flow parameters based on a (2+1) dimensional hybrid model
combining viscous hydrodynamics with UrQMD
have been made by Song {\it et al.}~\cite{Song:2011hk,Song:2010mg,Song:2010aq,Song:2011qa}.
They mainly focused on extraction of $\eta/s$ from a comparison of
$v_{2}/\varepsilon$ results with data and found $\eta/s$ is not larger than
twice the conjectured minimum bound, $1/4\pi$ \cite{Kovtun:2004de} 
(See Fig.~\ref{fig:hybrid8}).
From a hybrid model viewpoint,
a systematic analysis of switching temperature dependence
was made.
They concluded that
there exists no safe window of temperature below $T_{\mathrm{ch}}=165$ MeV
to switch from viscous hydrodynamics to UrQMD~\cite{Song:2011hk}.
This means that below $165$ MeV
UrQMD describes expanding matter far from equilibrium.
On the other hand they did not test switching temperatures larger
  than 165 MeV, and thus it is not possible to say whether a
  temperature range exists in their model where the exact value of
  the switching temperature does not affect the results.

The $\chi^{2}$-fitting to spectra, $v_{2}$ and HBT radii
in 0-10\% central collisions at the RHIC energy
was done by Soltz {\it et al.}\ using
(2+1) dimensional viscous hydrodynamic simulations
combined with the UrQMD cascade code \cite{Soltz:2012rk}.
Although the fit was done to
a relatively small number of data sets,
they were able to exclude two sets of initial conditions,
namely $N_{\mathrm{part}}$
density without pre-equilibrium flow and $N_{\mathrm{coll}}$ density with
pre-equilibrium flow, and to constrain initial temperature
and the ratio
of shear viscosity to entropy density for the other two sets of initial conditions,
$N_{\mathrm{part}}$
density with pre-equilibrium flow and $N_{\mathrm{coll}}$ density without
pre-equilibrium flow.

The simultaneous implementation of full three dimensionality, viscosity,
hadronic  cascade
and event-by-event initialisation
was first made by Ryu {\it et al.} \cite{Ryu:2012at}.
Their results are preliminary at the moment, since they do not
  take into account the dissipative corrections to particle
  distributions at particlisation, and the statistics in their
  calculations is so far limited leading to large statistical error bars.
Nevertheless, this is one of the promising approaches
to investigate the transport properties of the QGP.

Table \ref{table:h2c}
summarises the current  hydro + cascade models
by focusing on the cascade model and the switching temperature.

\begin{table}[htb]
\begin{center}
\caption{
Current status of the hydro + cascade models.
\label{table:h2c}
}
\begin{tabular}{|l|l|l|l|}
\hline
Authors and References & Hadronic Cascade & Hydrodynamics  & $T_{\mathrm{sw}}$ [MeV] \\
\hline
\hline
Bass {\it et al.} \cite{Dumitru:1999sf,Bass:1999tu,Bass:2000ib} & UrQMD & (1+1)-D ideal & 160 \\
\hline
Teaney {\it et el.} \cite{Teaney:2000cw,Teaney:2001av} & RQMD & (2+1)-D ideal & 160 \\
\hline
Hirano {\it et al.} \cite{Hirano:2005xf,Hirano:2007ei} & JAM & (3+1)-D ideal & 169, 155 \\
\hline
Nonaka and Bass \cite{Nonaka:2006yn} & UrQMD & (3+1)-D ideal & 150 \\
\hline
Petersen {\it et al.}~\cite{Petersen:2008dd,Petersen:2009vx,
  Petersen:2010md,Petersen:2010cw,Qin:2010pf,Petersen:2011fp,Petersen:2011sb}
& UrQMD & (3+1)-D ideal & $^{*1}$\\
\hline
Pratt and Vredevoogd~\cite{Pratt:2008sz} & $^{*2}$ & (1+1)-D viscous & $^{*3}$ \\
\hline
Werner {\it et al.}~\cite{Werner:2010aa,Werner:2010ny,Werner:2010ss,Werner:2012xh} & UrQMD & (3+1)-D ideal & 166 \\
\hline
Song {\it et al.} \cite{Song:2010mg,Song:2010aq,Song:2011hk,Song:2011qa}
 & UrQMD & (2+1)-D viscous & 165$^{*4}$ \\
\hline
Soltz {\it et al.}~\cite{Soltz:2012rk}  & UrQMD & (2+1)-D viscous & 165 \\
\hline
Ryu {\it et al.}~\cite{Ryu:2012at}& UrQMD & (3+1)-D viscous & 170\\
\hline
\end{tabular}
\end{center}
*$^{1}$ Switching energy density is taken to be $\sim$ 730 MeV/fm$^{3}$.

*$^{2}$ A rather simple hadronic cascade model is employed here~\cite{Pratt:2008sz}.

*$^{3}$ Switching energy density is taken to be 400 MeV/fm$^{3}$.

*$^{4}$ Sensitivity of the results to the value of $T_{\mathrm{sw}}$ is also 
      investigated.
\end{table}

\subsection{\it Initial Conditions \label{sec:ic}}

The results of hydrodynamic calculations depend strongly on initial
conditions since hydrodynamics requires solving initial value problems
of partial differential equations. In principle one should obtain the
initial conditions for hydrodynamic evolution by solving the
non-equilibrium evolution of the matter created in the primary
collision of nuclei, but unfortunately this is one of the outstanding
problems in heavy-ion physics. Therefore we skip the description of
how the matter equilibrates, and 
rely on models assuming that the matter has equilibrated, and
  that the densities are given by the density of produced gluons
  immediately after the primary collision (MC-KLN) or by the density
  of participating nucleons or binary collisions (MC-Glauber). Due to
  the lack of models, it is very difficult to quantify how
  pre-equilibrium dynamics would affect the interpretation of data and
  our understanding of the initial state of hydrodynamical
  evolution. It has been argued that flow built up during
  thermalisation would strongly affect the femtoscopic
  data~\cite{Pratt:2008sz}, but so far we do not know the mechanism
  creating pre-equilibrium flow, and thus do not know how large it
  could be. As well, it has been argued that the pre-equilibrium
  processes affect the granularity of the initial state in
  event-by-event calculations~\cite{Song:2011hk}, but we cannot
  calculate how large this smearing effect should be.

The Glauber model~\cite{Miller:2007ri,Broniowski:2007nz,Alver:2008aq}
has been widely used to fix the initial conditions of hydrodynamic
simulations. In the various implementations of Glauber model, one
either initialises the energy or entropy density, and assumes it to be
proportional to the number density of participants, binary collisions,
or some combination of those two~\cite{Heinz:2001xi,Kolb:2001qz}.  On
the other hand, one expects highly coherent dense gluon system, called
colour glass condensate
(CGC)~\cite{Iancu:2002xk,Iancu:2003xm,Gelis:2010nm}, to appear in high
energy hadronic and nuclear collisions.  One may describe the dynamics
of gluon fields before local equilibration by solving classical
Yang-Mills equation~\cite{KV1,KV2,KV3,KNV1,KNV2,KNV3,Lappi1,Lappi2}.
The $k_T$-factorisation formulation is widely used to compute the
inclusive 
cross section for produced gluons~\cite{Gribov,KLN2,KLN3,KLN4,fKLN,
  Gelis:2006tb,Albacete:2007sm,Albacete:2010bs,
  Levin:2011hr,Levin1,Levin2,Tribedy:2010ab,Tribedy,
  Dumitru:2011wq,mckt,Albacete:2011fw}
in hadronic collisions.  Here we shall employ the Monte-Carlo
implementations of $k_T$-factorisation formulation
(MC-KLN)~\cite{MCKLN1,Albacete:2012xq,MCKLN2} and Glauber model
(MC-Glauber)~\cite{Miller:2007ri,Broniowski:2007nz,Alver:2008aq} to
obtain initial conditions of hydrodynamical simulations. These
Monte-Carlo approaches include fluctuations of the positions of
nucleons inside colliding nuclei, which allows us to generate a set of
initial conditions which fluctuate event-by-event.
We do not include fluctuations of
energy deposition/entropy generation per collisions~\cite{Muller:2011bb},
which results in negative binomial distribution of
 multiplicity~\cite{Qin:2010pf,Tribedy:2010ab,Tribedy,Dumitru:2012yr},
since rapidity dependence of this kind of fluctuation 
is not known well.

In MC-Glauber model, for each event the positions of nucleons
  inside the two colliding nuclei
  are randomly sampled according to a nuclear density
distribution (\textit{e.g.}, Woods-Saxon function).  
One of the nuclei, and nucleons within, is shifted by
a randomly-chosen impact parameter $b$ with probability
$b\,db$.  A nucleon-nucleon collision is assumed to take place if
their distance $d$ in the transverse plane orthogonal to the beam axis
fulfils the condition
\begin{equation}
  \label{eq:collisioncriterion}
  d \leq \sqrt{\frac{\sigma_\text{in}(\sqrt{s})}{\pi}}\ ,
\end{equation}
where $\sigma_\text{in}(\sqrt{s})$ denotes
the inelastic nucleon-nucleon cross section at the c.m.~energy $\sqrt{s}$.
Incident energy dependent total $pp$ cross section is
parametrised based on Regge theory, which parameters have been determined
by the Particle Data Group~\cite{pdg1996}:
\begin{equation}
 \sigma_{\mathrm{tot}}(\sqrt{s}) = X s^\epsilon + Ys^{-\eta}
\end{equation}
with $X=22$, $Y=56.1$, $\epsilon=0.079$ and $\eta=0.46$  for $pp$ collision.
Elastic cross section is computed using PYTHIA
parametrisation~\cite{Sjostrand:2006za,Schuler:1993td,Schuler:1993wr}:
\begin{equation}
 \sigma_{\mathrm{el}} = \frac{\sigma_{\mathrm{tot}}^2}
                             {16\pi B_{\mathrm{el}}(s)}, \quad
 B_{\mathrm{el}}(s) = 4b_p + 4s^{0.0808} - 4.2 \ (\text{GeV}^{-2})\ ,
\end{equation}
where $b_p= 2.3$. This parametrisation leads to the following values
for the inelastic cross section:
$\sigma_\text{in} = \sigma_{\mathrm{tot}} - \sigma_{\mathrm{el}}
=39.53, 41.94$ and 61.36 mb at 
$\sqrt{s}=130, 200$ and 2760 GeV, respectively.

It should be noted that the standard Woods-Saxon parameters
shown in, \textit{e.g.}, Ref.~\cite{DeJager:1987qc}
cannot be directly used to distribute nucleons inside a nucleus
because of the finite interaction range in our approach.
We need to modify nuclear density parameters
so that a convolution of nucleon profiles leads to the measured 
Woods-Saxon profile~\cite{Hirano:2009ah}:
\begin{eqnarray}
\rho(\vec{x}) & = &\int \Delta(\vec{x}-\vec{x}_0)\rho_{\mathrm{WS}}(\vec{x}_0)d^{3}x_0 , \\
\label{eq:finiteprofile}
\Delta(\vec{x}-\vec{x}_0) & = & \frac{\theta(r_{N} -\mid \vec{x}-\vec{x}_0 \mid)}{V_{N}},\\
V_{N} & = & \frac{4 \pi r_{N}^3}{3}, \quad 
r_{N}  =  \sqrt{\frac{\sigma_{NN}^{\mathrm{in}}}{\pi}}, \quad
\rho_{\mathrm{WS}}  =  \dfrac{\rho_0}{\exp\left(\dfrac{r-r_0}{\delta r} \right)+1}.
\label{eq:WS}
\end{eqnarray}
\begin{figure}[htb]
\begin{center}
\includegraphics[width=3.4in]{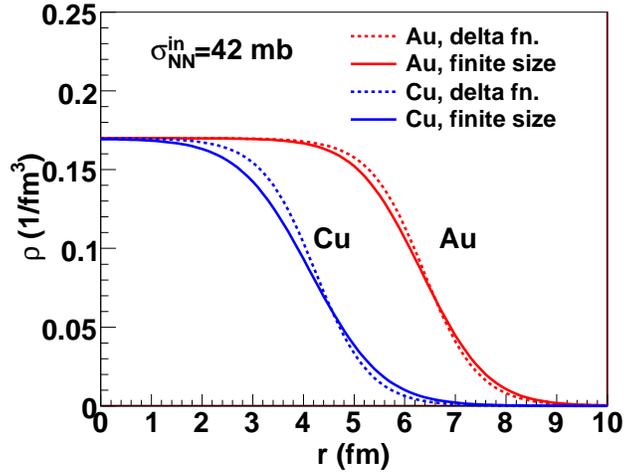}
\caption{
Nuclear density as a function of nuclear radius.
Solid lines show nuclear density distribution for gold and copper
nuclei in which a finite nucleon profile 
is implemented and positions of nucleons 
are sampled according to the Woods-Saxon distribution
with default parameter sets.
Dashed lines show the Woods-Saxon distribution with default parameter sets. 
}
\label{fig:WS}
\end{center}
\end{figure}
%
Figure \ref{fig:WS} shows the standard Woods-Saxon profile (dashed)
and the nuclear density profile taking into account finite interaction
ranges above but keeping the standard Woods-Saxon parameters (solid).
In both gold and copper nuclei, the finite nucleon profile
in Eq.~(\ref{eq:finiteprofile}) makes the nuclear surface
more diffused if one uses the standard Woods-Saxon parameters
to distribute nucleons in a nucleus.
Without adjustment of the Woods-Saxon parameters for the finite nucleon
profile, eccentricity becomes smaller  by $\sim 10\%$ \cite{Drescher:2006ca}.
So we re-parametrise the distribution of nucleon position
to reproduce the Woods-Saxon distribution 
with default parameters in Eq.~(\ref{eq:WS}) \cite{Broniowski:2007nz}.
The adjustment of the Woods-Saxon parameters
has not been considered in
most of Monte Carlo
approaches of the collisions including event generators
If one wants to discuss eccentricity and elliptic flow coefficient
$v_2$ within $\sim$10\% accuracy, this adjustment should be
taken into account.
If the Gaussian form is used as a nucleon profile,
one obtains better agreement with the measured nucleus charged
distribution~\cite{Heinz:2011mh}.
However, hard sphere form for the nucleon profile
is sufficient for the discussion of $v_2$ in this work.
In our MC-Glauber model,
we find the default Woods-Saxon distribution
is reproduced by 
larger radius parameter and 
smaller diffuseness parameter (\textit{i.e.}, sharper
boundary of a nucleus)
than the default parameters:
$\rho_0 =$ 0.1695 (0.1686) fm$^{-3}$, $r_0  = $ 6.42 (4.28) fm and
$\delta r  = $ 0.44 (0.50) fm for a gold (copper) nucleus
at $\sqrt{s_{NN}}=200$ GeV
and $\rho_0 = 0.161$ fm$^{-3}$, $r_0  = $ 6.68 fm and $\delta r  = $ 0.38 for a lead
nucleus at $\sqrt{s_{NN}}=$2.76 TeV.
In Ref.~\cite{Heinz:2011mh}, one finds the parametrisation 
in the case of Gaussian shape for nucleons.
This kind of effect exists in almost all Monte Carlo approaches
to the collisions, including event generators.
If one wants to discuss eccentricity and elliptic flow coefficient $v_2$
within $\sim 10\%$ accuracy, this effect should be taken into account.
We also calculate initial entropy distribution in U+U collisions
at $\sqrt{s_{NN}}=200$ GeV
by changing the nuclear density from gold to uranium.
To take account of the prolate deformation of
uranium nuclei, we parametrise the
radius parameter in the Woods-Saxon distribution
as 
\begin{eqnarray}
R(\theta, \phi) & = & r_{0}\left( 1+\beta_{2}Y_{20}(\theta, \phi)
 + \beta_{4}Y_{40}(\theta, \phi) \right),
\end{eqnarray}
where $Y_{lm}$ is the spherical harmonic function,
$r_{0} = 6.86$ fm,
$\beta_{2} = 0.28$ and $\beta_{4} = 0.093$ \cite{Filip:2009zz}.
Note that to account of
the finite interaction range of nucleons in the Monte Carlo approach,
we have again adjusted $R_{0}$ above 
and the diffuseness parameter $\delta r = 0.44$ fm
and the saturation density $\rho_{0} = 0.166$ fm$^{-3}$
to retain the nuclear density as in the original Woods-Saxon distribution
\cite{Hirano:2009ah}.
We also take into account
that colliding uranium nuclei are randomly
oriented in each event.

Next, we compute particle production at each grid point
in the transverse plane.
In the MC-Glauber approach, we assume that the initial entropy profile
in the transverse plane is proportional to a linear combination of
the number density of participants and that of binary collisions:
\begin{equation}
\label{eq:Glauber}
s_0(\bm{r}_\perp) 
 \equiv 
 \left. \frac{dS}{\tau_0 d^{2}r_{\perp} d\eta_s} \right|_{\eta_s = 0}
 =  \frac{C}{\tau_0}\left(\frac{1-\alpha}{2}
    \rho_{\mathrm{part}}(\bm{r}_\perp)
  + \alpha\, \rho_{\mathrm{coll}}(\bm{r}_{\perp})
   \right),
\end{equation}
where $\tau_0=0.6$ fm/$c$ is a typical initial time for the hydrodynamical
simulation.
Parameters $C=15.0$ and $\alpha=0.18$ have been fixed through
comparison with the centrality dependence of $p_{T}$ spectra
in Au+Au collisions at RHIC \cite{Adler:2003cb}.
At the LHC energy, $C=41.4$ and $\alpha = 0.08$ are chosen \cite{Hirano:2010je}
so that we reproduce the ALICE data on centrality dependence 
of multiplicity in Pb+Pb collisions at $\sqrt{s_{NN}}=2.76$ TeV \cite{ALICEdNdeta,ALICEdNdeta2,Aamodt:2010ft}.

The participant density, $\rho_\text{part}(\bm{r}_\perp)$, and the number
density of binary collisions, $\rho_{\mathrm{coll}}(\bm{r}_{\perp})$, in
Eq.~(\ref{eq:Glauber}) are obtained from the previously described positions of
nucleons, and the criterion for their interaction,
Eq.~(\ref{eq:collisioncriterion}). At each grid point, the participant density
is the sum of the number of those nucleons in both nuclei,
which scatter within the
radius $r_0=\sqrt{\sigma_\text{in}/\pi}$ around the gridpoint, divided by the
area $\sigma_\text{in}$: \begin{equation} \rho_\text{part}(\bm{r}_\perp)
=\rho_\text{A}(\bm{r}_\perp) + \rho_\text{B}(\bm{r}_\perp) =
\frac{N_{A,w}(\bm{r}_\perp) + N_{A,w}(\bm{r}_\perp)}{\sigma_\text{in}}
\end{equation}

Similarly, the density of the number of binary collisions at each grid point
is obtained by counting the number of binary collisions $N_{\mathrm{coll}}$ 
within the area $\sigma_\text{in}$, where the position of a binary collision
is assumed to be the average position of the two colliding nucleons:
\begin{equation}
   \rho_{\mathrm{coll}}(\bm{r}_{\perp})
   =\frac{N_{\mathrm{coll}}(\bm{r}_\perp) }{\sigma_\text{in}}\ .
\end{equation}
which may be also obtained by the expression
$\rho_A(\bm{r}_\perp)\rho_B(\bm{r}_\perp)\sigma_\text{in}$.

Note that we calculate the eccentricity of the initial state
  based on the densities defined above, whereas in the PHOBOS
  MC-Glauber
  model~\cite{Miller:2007ri,Broniowski:2007nz,Alver:2008aq},
  eccentricity is computed based on the actual positions of point-like
  particles. The conversion of positions to densities causes an
  additional smearing over the region
  $\sigma_\text{in}$~\cite{Sorensen:2011hm,Nara:2011ee}. Therefore
  the eccentricities in our calculations are smaller than
  eccentricities in the PHOBOS MC-Glauber model. 
In the literature the conversion of positions to densities is
  often done by replacing each position by a Gaussian density
  profile~\cite{Petersen:2008dd,Holopainen:2010gz,Schenke:2010rr}. Such
  a procedure again leads to slightly different eccentricities.

In the Monte-Carlo KLN (MC-KLN)
 model~\cite{MCKLN1,Albacete:2012xq,MCKLN2},
the number distribution of gluon production
at each transverse grid is given by
the $k_T$-factorisation formula~\cite{KLN2,KLN3,KLN4}
\begin{eqnarray}
  \frac{dN_g}{d^2 r_{\perp}dy} & = &\kappa
   \frac{4N_c}{N_c^2-1}
    \int
    \frac{d^2p_\perp}{p^2_\perp}
      \int \frac{d^2k_\perp}{4} \;\alpha_s(Q^2)\nonumber \\
       &\times &   \phi_A(x_1,(\bm{p}_\perp + \bm{k}_\perp)^2/4)\;
       \phi_B(x_2,(\bm{p}_\perp{-}\bm{k}_\perp)^2/4)~,
      \label{eq:ktfac}
\end{eqnarray}
with $N_c=3$ the number of colours.
Here, $p_\perp$ and $y$ denote the
transverse momentum and the rapidity of the produced gluons, respectively. 
The light-cone momentum fractions of the colliding gluon
ladders are then given by $x_{1,2} = p_\perp\exp(\pm y)/\sqrt{s_{NN}}$,
where $\sqrt{s_{NN}}$ denotes the centre of mass energy.
Running coupling $\alpha_s(Q^2)$ is evaluated at the scale
$Q^2=\max( (\bm{p}_\perp -\bm{k}_\perp)^2/4,(\bm{p}_\perp
+\bm{k}_\perp)^2/4)$.
An overall normalisation factor $\kappa$ is chosen to fit
the multiplicity data in most central Au+Au collisions at RHIC.
In the MC-KLN model, saturation momentum
is parametrised by assuming that the saturation momentum squared
is 2 GeV$^2$ at $x=0.01$
in Au+Au collisions at $b=0$ fm at RHIC where $\rho_\text{part}=3.06$
fm$^{-2}$ \cite{KLN2,KLN3,KLN4,KLN1}:
\begin{equation}
Q_{s,A}^2 (x; \bm{r}_\perp)  =  2\ \text{GeV}^2
\frac{\rho_{A}(\bm{r}_\perp)}{1.53\ \text{fm}^{-2}}
\left(\frac{0.01}{x}\right)^{\lambda} \ .
\label{eq:qs2}
\end{equation}
$\lambda$ is a free parameter which is expected
to be in the range of $0.2<\lambda<0.3$ from
Hadron Electron Ring Accelerator (HERA) global analysis for
$x<0.01$~\cite{HERA,HERA2}.
In MC-KLN, we assume the gluon distribution
function as
\begin{equation}
\label{eq:uninteg}
  \phi_{A}(x,k_\perp^2;\bm{r}_\perp)\sim
    \frac{1}{\alpha_s(Q^2_{s,A})}\frac{Q_{s,A}^2}
       {{\rm max}(Q_{s,A}^2,k_\perp^2)}~.
\end{equation}

We assume that initial conditions of hydrodynamical simulations
are obtained by identifying the gluons' momentum rapidity $y$ with
space-time rapidity $\eta_s$
\begin{equation}
s_0(\bm{r}_\perp) 
\propto \frac{dN}{\tau_0 d^{2}r_\perp d\eta_s}\ .
\end{equation}
We note that gluon production itself also
  fluctuates~\cite{Qin:2010pf,Tribedy:2010ab,Tribedy,Dumitru:2012yr},
  but we do not take those fluctuations into account in our model.

To quantify the anisotropy of the initial distributions, we define
  the anisotropies $\varepsilon_n$, and the corresponding orientation
  angles $\Phi_n$ \cite{Alver:2010gr,Sorensen:2011hm}:
\begin{eqnarray}
\label{eq:epsilonn}
\varepsilon_{n}\{\mathrm{PP}\} & = & \frac{| \langle \bm{r}_{\perp}^{2} e^{in\varphi} \rangle_{x}|}{\langle \bm{r}_{\perp}^{2} \rangle_{x}}\\
\label{eq:Phin}
n \Phi_{n} & = & \arg \langle \bm{r}_{\perp}^{2} e^{in\varphi} \rangle_{x}
\end{eqnarray}
where $\langle \cdots \rangle_{x}$ represents
a weighted average over the transverse plane at a fixed space-time
  rapidity, with the initial density distribution as a weight.
Although one may take other definitions
 \cite{Petersen:2010cw,Teaney:2010vd,Alver:2010dn,Bhalerao:2011yg,Bhalerao:2011bp,Gardim:2011xv},
we restrict our discussion to Eqs.~(\ref{eq:epsilonn}) and (\ref{eq:Phin}) throughout this paper.
As for the initial density, we use the entropy density  throughout this work.
Here $\bm{r}_{\perp}$ is the transverse two dimensional vector
measured from the centre of mass defined by 
     $\langle \bm{r}_{\perp} \rangle_{x} = \bm{0}$
and $\varphi$ is its coordinate angle.
For example, the  anisotropy $\varepsilon_2$ becomes
\begin{eqnarray}
\varepsilon_{2}\{\mathrm{PP}\} & = & \frac{ \sqrt{\langle x_{\perp}^{2}-y_{\perp}^{2}\rangle_{x}^{2} +4 \langle x_{\perp} y_{\perp}  \rangle_{x}^{2}}}
{\langle x_{\perp}^{2} + y_{\perp}^{2} \rangle_{x}},\\
x_{\perp} & = & x- \langle x \rangle_x,\\
y_{\perp} & = & y- \langle y \rangle_x,
\label{eq:epsilon2}
\end{eqnarray}
which is also known as the participant eccentricity $\varepsilon_{\mathrm{part}}$
or eccentricity with respect to the participant plane. To keep
  our terminology consistent with the terminology in literature,
  we define participant plane as the plane spanned by a unit vector
  pointing to direction $\Phi_2-\pi/2$, and the beam axis.
Taking a real part of Eq.~(\ref{eq:epsilonn}) instead of its absolute value,
one obtains the anisotropy with respect to the reaction plane, also known as
the standard eccentricity $\varepsilon_{\mathrm{std}}$ (although with an opposite sign),
\begin{eqnarray}
\label{eq:e2rp}
\varepsilon_{2}\{\mathrm{RP}\} = -\varepsilon_{\mathrm{std}} & = & \frac{\Re \langle \bm{r}_{\perp}^{2} e^{i 2\varphi} \rangle_{x}}{\langle \bm{r}_{\perp}^{2} \rangle_{x}} 
 =  \frac{ \langle x_{\perp}^{2}-y_{\perp}^{2} \rangle_{x}}{\langle x_{\perp}^{2}+y_{\perp}^{2} \rangle_{x}}.
\end{eqnarray}
 
MC-KLN and MC-Glauber models create an ensemble of initial
  distributions for event-by-event calculations, but it is also
  possible to construct an averaged density profile which includes
  some effects of fluctuations. This is done by rotating each
  distribution by its orientation angle $\Phi_n$ around its centre of
  mass, shifting the distributions so that the origin of the
  coordinates is at its centre of mass, $(\langle x \rangle_{x},
  \langle y \rangle_{x})$, and averaging over these shifted and
  rotated distributions. In the literature the required angle is often
  defined as in Ref.~\cite{Alver:2008zza}:
\begin{eqnarray}
\tan 2 \psi_{2} & = & \frac{2\sigma_{xy}}{\sigma_y^2-\sigma_x^2}, \\
\sigma_x^2 & = & \langle x^2 \rangle_{x} - \langle x \rangle_{x}^2,\\
\sigma_y^2 & = & \langle y^2 \rangle_{x} - \langle y \rangle_{x}^2,\\
\sigma_{xy} & = & \langle xy \rangle_{x} - \langle x \rangle_{x}\langle y \rangle_{x},
\end{eqnarray}
where the angle $\psi_2$ is related to the orientation angle $\Phi_2$
defined in Eq.~(\ref{eq:Phin}) as $\Phi_{2} = -\psi_{2}$. 
As was described in the introduction,
  elliptic flow arises when the system expands preferentially along
  its participant plane. In this procedure the participant planes of
  various events are aligned and set equal to the reaction plane.
That such an initial state 
contains some effects of eccentricity fluctuation
even though the profile is smooth is
manifested in the finite eccentricity even in most
  central collisions where the impact parameter is zero.
We call this initialisation model ``B".
Compared with this,  the procedure
averaging over many initial distributions
without shift or rotation
corresponds to a conventional initialisation
without the effect of
eccentricity fluctuation and is called  model ``A".
When we use the averaged initial conditions of models ``A'' and
  ``B'', called later the smooth initial profiles,
 we assume longitudinal boost invariance
and calculate observables only at midrapidity. In particular, in the case of the MC-KLN model,
we evaluate gluon production at midrapidity using
  Eq.~(\ref{eq:ktfac}), and assume it to be the same at all
  rapidities.
In actual hydrodynamic simulations, we prepare
the lattice in the longitudinal direction up to $\eta_{s} = 6$
and solve hydrodynamic equations with boost invariant
initial conditions. We have checked that the boundary of the lattice
does not affect the boost invariant solutions.

\begin{table}[h]
\begin{center}
\caption{Centrality definition using $N_{\mathrm{part}}$ in Au+Au, U+U and Cu+Cu collisions
 at $\sqrt{s_{NN}} =200$ GeV and in Pb+Pb collisions $\sqrt{s_{NN}} =2.76$ TeV .}
\label{table:cent}
\begin{tabular}{rrrrrrrrrrr}
Centrality(\%) & 0-5 & 5-10 & 10-15 & 15-20 & 20-30 & 30-40 & 40-50 & 50-60 & 60-70 & 70-80\\
\hline
Au+Au $N_{\mathrm{part, max}}$ &  394& 327 & 279 &  237& 202 & 144 & 99 & 65 & 39 &21\\
Au+Au $N_{\mathrm{part, min}}$ &  327 & 279 &  237&  202& 144 & 99 & 65 & 39 & 21 &10\\
U+U $N_{\mathrm{part, max}}$ & 476 & 389 & 330 & 281  & 239  &  170 & 117 & 77 & 46& 25\\
U+U $N_{\mathrm{part, min}}$ & 389 & 330 & 281 & 239 & 170 & 117 & 77 & 46 &25 & 12\\
Cu+Cu $N_{\mathrm{part, max}}$ & 126 & 102 & 88 & 75 & 65 & 47 & 33 & 22 & 14& 9\\
Cu+Cu $N_{\mathrm{part, min}}$  & 102 & 88  & 75 & 65 & 47 & 33 & 22 & 14 & 9& 5 \\
Pb+Pb $N_{\mathrm{part, max}}$ & 416 & 356 & 305 & 261 & 223 & 161 & 112 & 74 & 46& 25\\
Pb+Pb $N_{\mathrm{part, min}}$  & 356 & 305  & 261 & 223 & 161 & 112 & 74 & 46 & 25& 12 \end{tabular}
\end{center}
\end{table}

In models A and B centrality is defined using the number of events 
as a function of $N_{\mathrm{part}}$.
One can categorise the whole events into sub events from top 5\%, 5-10\%
and so on
according to $N_{\mathrm{part}}$.
In Table \ref{table:cent},
we show the maximum and minimum numbers of participants
for each centrality bin in Au+Au, U+U, Cu+Cu collisions
at the RHIC energy and in Pb+Pb collisions at the LHC
energy.

\begin{figure}[tb]
\begin{center}
\begin{minipage}[t]{9 cm}
\epsfig{file=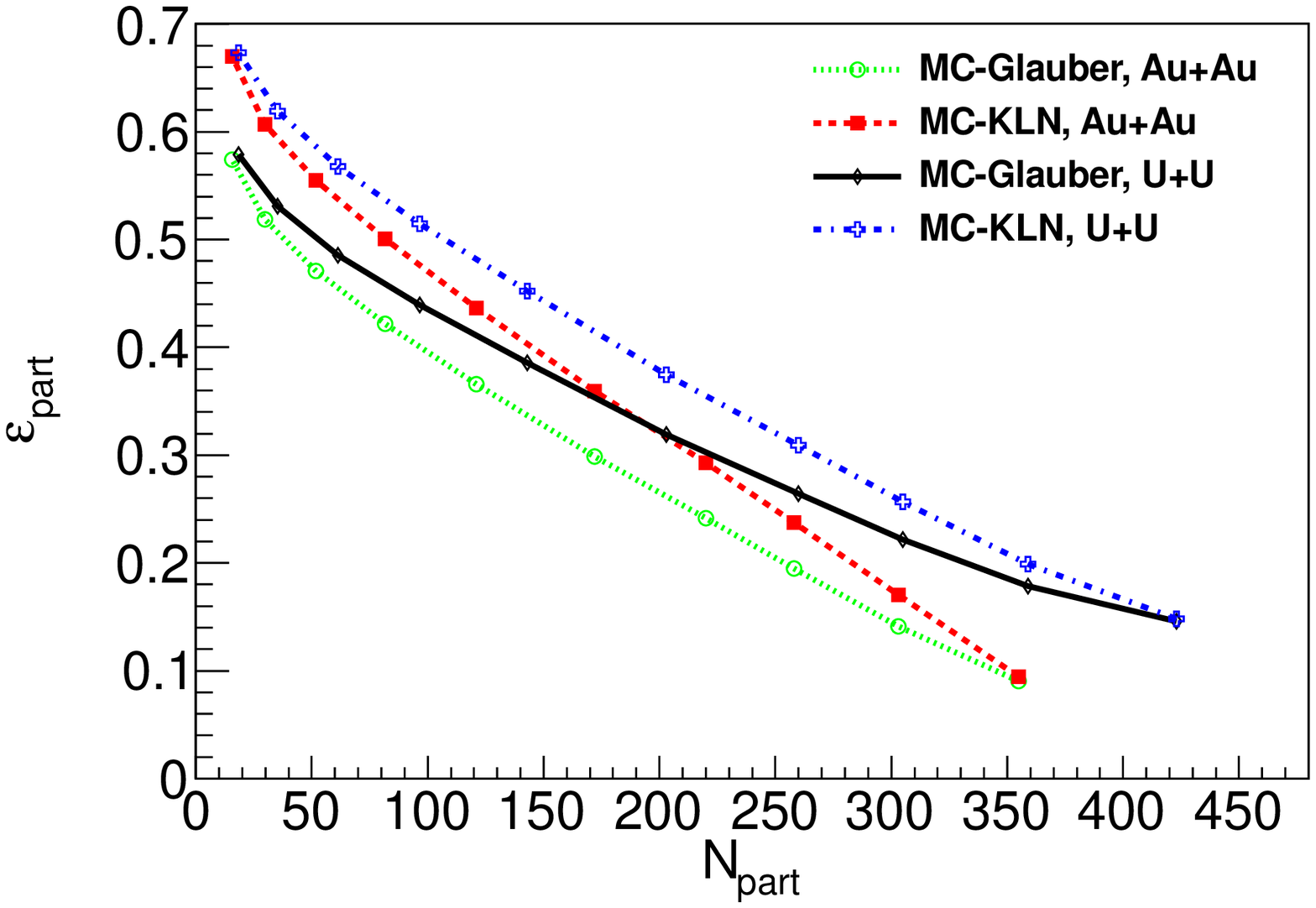,scale=0.45}
\end{minipage}
\begin{minipage}[t]{9 cm}
\epsfig{file=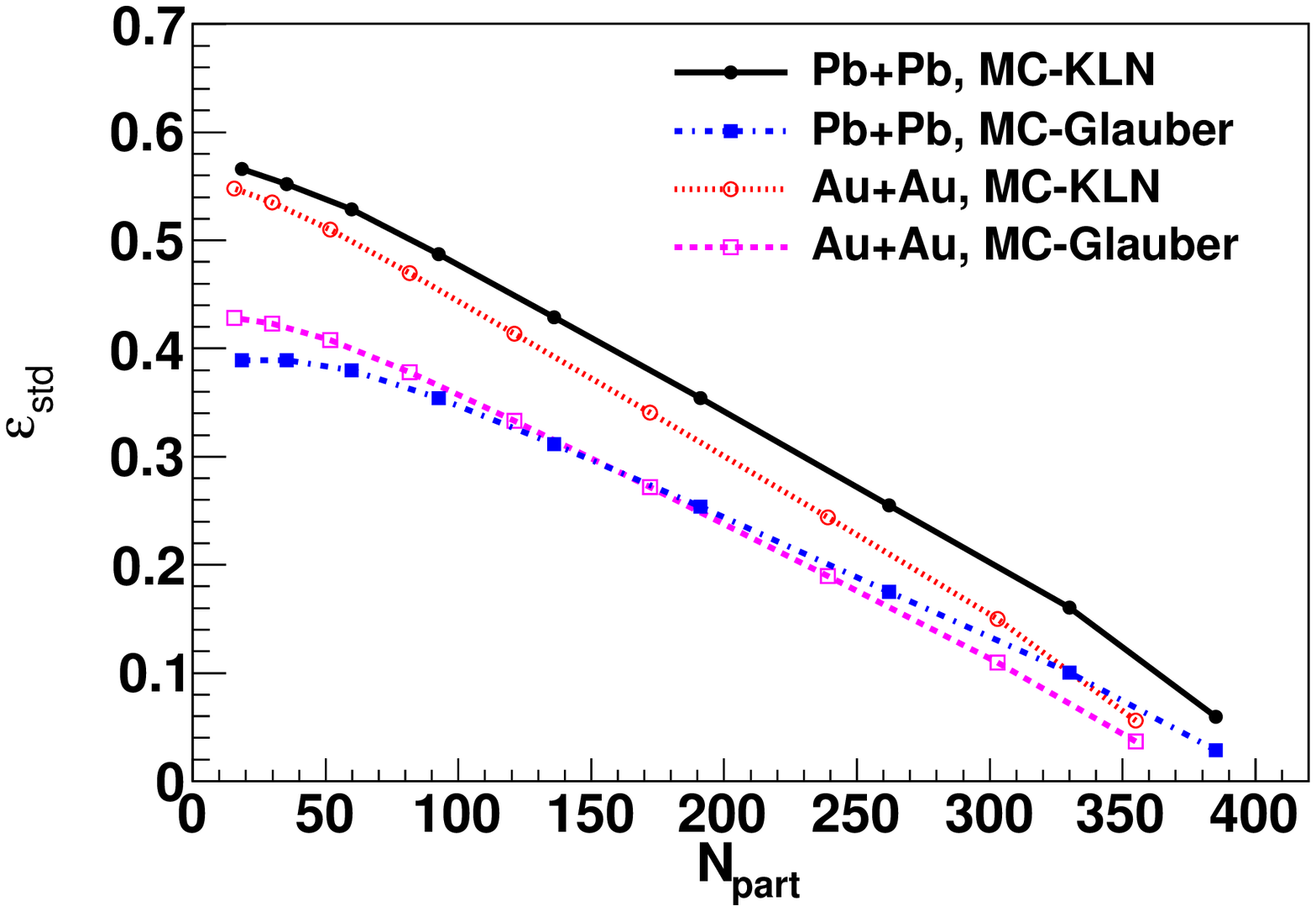,scale=0.45}
\end{minipage}
\caption{The second order anisotropy 
with respect to participant plane (model ``B'')
in Au+Au and U+U collisions at 
$\sqrt{s_{NN}}$=200 GeV (left) and
with respect to reaction plane (model ``A'')
in Au+Au collisions at 
$\sqrt{s_{NN}}$=200 GeV
and in Pb+Pb collisions at 
$\sqrt{s_{NN}}$=2.76 TeV (right).
Results from the MC-KLN and MC-Glauber models are 
compared with each other.
Figures are taken from Refs.~\cite{Hirano:2010jg, Hirano:2010je}
\label{fig:epspart}
}
\end{center}
\end{figure}

Figure \ref{fig:epspart} (left) shows
the initial eccentricity with respect to participant plane 
(model ``B'')
in Au+Au and U+U collisions at $\sqrt{s_{NN}} = 200$ GeV
as a function of the number of participants.
At each of the ten centrality bins the average eccentricity and
  the average number of participants $\langle N_{\mathrm{part}} \rangle$ 
  were calculated using both the MC-Glauber and the MC-KLN models.
Since the eccentricity is measured in
the participant plane, it is finite
even in the very central (0-5\%) Au+Au collisions.
As previously known,
the MC-KLN model leads to $\sim$20-30\% larger eccentricity than 
the MC-Glauber model except in the most central events 
 \cite{Hirano:2005xf,fKLN}.
In most central 5\% of U+U collisions eccentricity reaches
0.146 in the MC-Glauber model and 0.148 in the MC-KLN model.
The eccentricity is larger in U+U than in Au+Au collisions. 
Due to the deformed shape of uranium nucleus, this holds not only 
at fixed number of participants, but also at fixed centrality.
The difference, however, 
decreases with decreasing centrality, and there is almost 
no difference in the very peripheral events (70-80\%).

Shown in Fig.~\ref{fig:epspart} (right)
is the initial eccentricity 
with respect to reaction plane (model ``A'')
as a function of $N_{\mathrm{part}}$
in Pb+Pb collisions at $\sqrt{s_{NN}}=$ 2.76 TeV
and in Au+Au collisions at $\sqrt{s_{NN}}=$ 200 GeV.
Again, the $k_t$-factorised formula of KLN model generates
larger eccentricity than the Glauber model does
\cite{Hirano:2005xf,fKLN}.
In the MC-KLN model,
eccentricity in Pb+Pb collisions
at $\sqrt{s_{NN}} = 2.76$ TeV
is apparently larger than that in Au+Au collisions
at $\sqrt{s_{NN}} = 200$ GeV 
when $N_{\mathrm{part}}$ is fixed.
However, at a fixed centrality, the difference between them is very small \cite{Hirano:2010jg}.
On the other hand, in the MC-Glauber model,
eccentricity in Pb+Pb collisions
at $\sqrt{s_{NN}} = 2.76$ TeV
is slightly smaller than
that in Au+Au collisions
at $\sqrt{s_{NN}} = 200$ GeV 
for a fixed centrality.

This is due to the smearing process we use to obtain a smooth
initial profile for hydrodynamic evolution \cite{Song:2011hk}. As mentioned, we use the
inelastic cross section in $p+p$ collisions, $\sigma_{\mathrm{in}}$,
when converting the positions of collision points to densities,
  and this effectively smears the distributions.
 This cross section 
is $\sim 1.5$ times larger at LHC than at RHIC, and thus the smearing
area, $S = \sigma_{\mathrm{in}}$ \cite{Drescher:2006ca,Drescher:2007ax},
is also larger at LHC, and the eccentricity is reduced. Our smearing procedure also
leads to a smaller eccentricity than the PHOBOS
MC-Glauber model\footnote{In the MC-Glauber model in the
literature~\cite{Miller:2007ri}, one assumes $\delta$ function
profile for each collision point in $\rho_{\mathrm{part}}$
distribution rather
than a box-like profile in the present work.}. 
The effect of
smearing is smaller in the MC-KLN initialisation, and we have
checked that the eccentricity at LHC turns out to be essentially the
same as at RHIC when the smearing area is the same.

Instead of shifting, rotating or averaging transverse profiles,
we directly use each individual initial density given by
these Monte-Carlo approaches
to perform event-by-event hydrodynamic simulations.
In these event-by-event calculations, we perform full three dimensional hydrodynamic simulations
without assuming boost invariance.
In the case of MC-KLN Eq.~(\ref{eq:ktfac}) 
provides the rapidity distribution of density as well.
On the other hand, the Glauber model does not tell  the longitudinal structure of the 
density distribution.
Motivated by analyses in Refs.~\cite{BGK,AG05},
we parametrise initial entropy density distribution as \cite{Hirano:2005xf}
\begin{eqnarray}
\label{eq:modifiedBGK}
 s_{0}(\tau_{0}, \eta_{s}, \bm{r}_\perp) & = &\frac{dS}{\tau_{0} d\eta_s d^{2}r_{\perp}} \nonumber \\
 &=&  \frac{C}{\tau_{0}} \theta\bigl(Y_b{-}|\eta_s|\bigr)\, f^{pp}(\eta_s) 
  \Biggl[\frac{1-\alpha}{2} 
  \left(\frac{Y_b{-}\eta_s}{Y_b}\,\rho_{A}(\bm{r}_\perp) 
      + \frac{Y_b{+}\eta_s}{Y_b}\,\rho_{B}(\bm{r}_\perp) \right)
 + \alpha\, \rho_{\mathrm{coll}}(\bm{r}_\perp)\Biggr],
\end{eqnarray}
where 
$Y_b$ is the beam rapidity and $f^{pp}$ is a parametrisation 
of the shape of rapidity distribution in $pp$ collisions,
\begin{eqnarray}
\label{eq:pp}
\frac{dS^{pp}}{d\eta_{s}} & = & \int d^{2}r_{\perp}\frac{dS^{pp}}{d\eta_s d^{2}r_{\perp}}
=C\theta(Y_{b} -|\eta_{s}|)  f^{pp}(\eta_s) 
\nonumber \\
& = & C\theta(Y_{b} -|\eta_{s}|)\exp\left[-\theta(|\eta_s|{-}\Delta\eta)\,
  \frac{(|\eta_s|{-}\Delta\eta)^2}{\sigma_\eta^2}\right],
\end{eqnarray}
where $\Delta\eta$ and $\sigma_\eta$ are adjustable parameters.
Equation (\ref{eq:modifiedBGK})
is designed so that the density is independent of the space-time
  rapidity $\eta_s$ only around midrapidity when the densities of
  participating nucleons are the same in both nuclei,
  $\rho_{A}(\bm{r}_\perp)=\rho_{B}(\bm{r}_\perp)$.
Equation (\ref{eq:modifiedBGK}) reduces to Eq.~(\ref{eq:Glauber}) when
one plugs in $\eta_{s} = 0$.
We call this parametrisation as the modified BGK model \cite{Hirano:2005xf}.
As mentioned, we do not consider fluctuation of
particle production processes and, consequently,  
longitudinal profile becomes a smooth function.
Therefore, there exists some correlation of 
particle production in the rapidity direction.

Initial parameters in the modified BGK model
 in Au+Au collisions
 at $\sqrt{s_{NN}} =200$ GeV 
are chosen to reproduce
$dN_{\mathrm{ch}}/d\eta_{s}$ measured by
PHOBOS Collaboration \cite{PHOBOS_dNdeta}.
The resultant parameters are $\Delta \eta = 1.3$ and $\sigma_{\eta} = 2.1$.
At the time of this writing, the measured pseudorapidity 
dependence of multiplicity
in Pb+Pb collisions at $\sqrt{s_{NN}} =2.76$ TeV is still preliminary.
The parameters $\Delta \eta = 1.9$ and $\sigma_{\eta} = 3.2$ are
  chosen to give in central $0 < b < 5$ fm events an average
  $dN_{\mathrm{ch}}/d\eta$ similar to the value obtained using the
  MC-KLN initialisation.
Once the experimental data is finalised, 
we can adjust these parameters again.

Throughout this work, initial flow velocity is chosen as
the Bjorken flow $u^{\tau}=1$ and $u^{x}=u^{y}=u^{\eta_{s}}=0$ \cite{Bjorken:1982qr}.
In actual simulations, initial energy density is also needed. One calculates
it from the initial entropy density utilising
numerical table of EoS, $\epsilon=\epsilon(s)$.
Note that we have neglected baryon density in our calculations.

\begin{figure}[tb]
\begin{center}
\epsfig{file=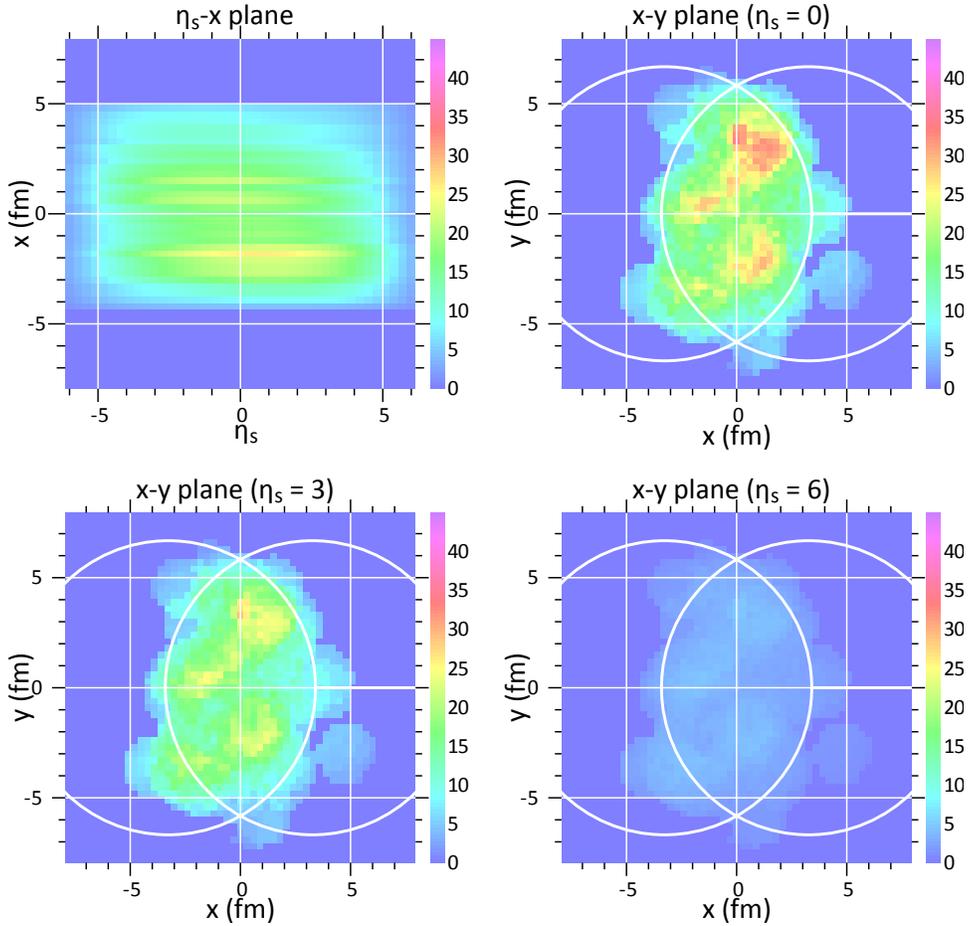,scale=0.55}
\caption{
An example of initial condition in Pb+Pb collisions at 
$\sqrt{s_{NN}}$=2.76 TeV from the MC-KLN model.
Initial entropy distribution (scaled down by 8.5) is shown in a plane
at 
$y=\langle y \rangle_{x}$ (top-left), $\eta_{s}=0$ (top-right),
 $\eta_{s}=3$ (bottom-left) and
$\eta_{s}=6$ (bottom-right).
Both circles with a radius $r_{0}=6.68$ fm
represent a colliding nucleus. 
\label{fig:iniKLN}
}
\end{center}
\end{figure}

\begin{figure}[tb]
\begin{center}
\epsfig{file=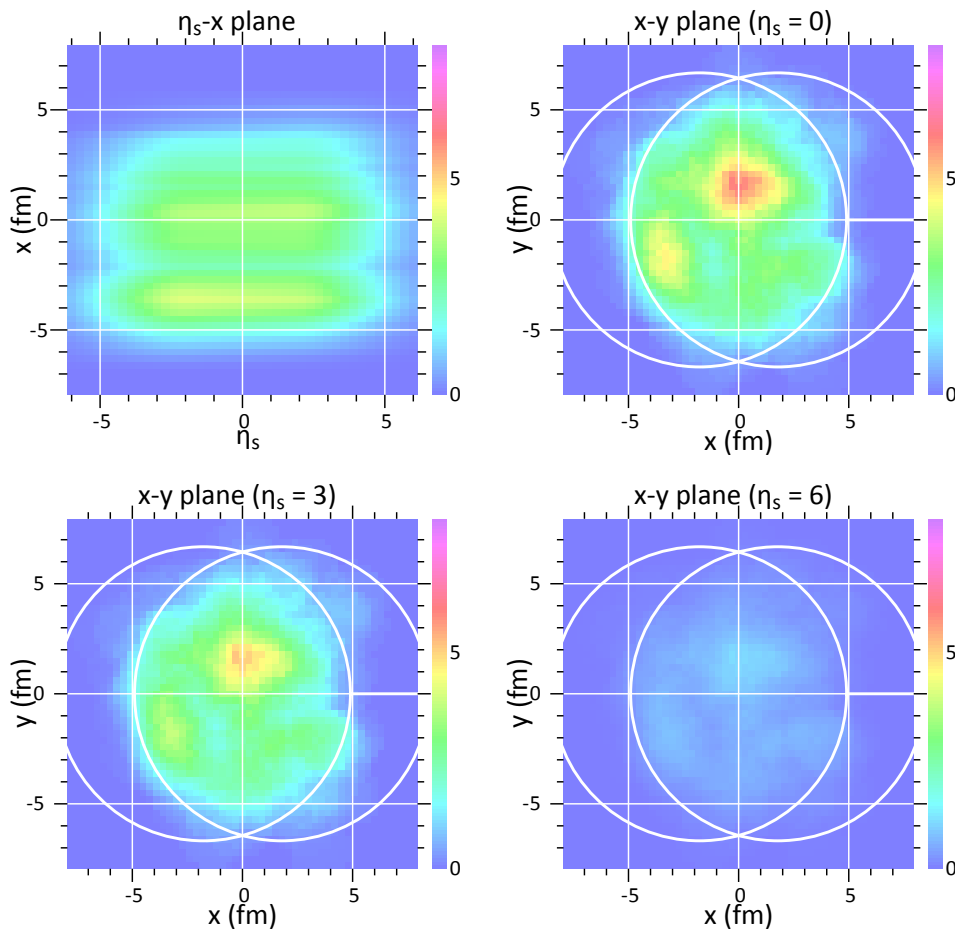,scale=0.55}
\caption{
An example of initial condition in Pb+Pb collisions at 
$\sqrt{s_{NN}}$=2.76 TeV from the MC-Glauber model
with an extension using the BGK model.
Initial entropy distribution (scaled down by 25)
is shown in a plane at 
$y=\langle y \rangle_{x}$ (top-left), $\eta_{s}=0$ (top-right),
 $\eta_{s}=3$ (bottom-left) and
$\eta_{s}=6$ (bottom-right).
Both circles with a radius $r_{0}=$6.68 fm 
represent a colliding nucleus.
\label{fig:iniGlauber}
}
\end{center}
\end{figure}

Figures \ref{fig:iniKLN} and \ref{fig:iniGlauber} show
a sample of the initial profile
in Pb+Pb collisions at $\sqrt{s_{NN}}$ = 2.76 TeV
from the MC-KLN model and the MC-Glauber model, respectively.
Longitudinal streak-like structures are seen 
in $y=\langle y \rangle_{x}$ fm (top-left panel in both figures),
where $\langle y \rangle_{x}$ has been averaged over space-time
rapidity.
This structure comes simply from smooth longitudinal profiles
at the collision point in the transverse plane
described in Eqs.~(\ref{eq:ktfac}) or (\ref{eq:modifiedBGK}).
Due to this, similar transverse profiles are seen
at all space-time rapidities: 
Hot spots are always located in the same position 
in the transverse plane.

\begin{figure}[tb]
\begin{center}
\begin{minipage}[t]{9 cm}
\epsfig{file=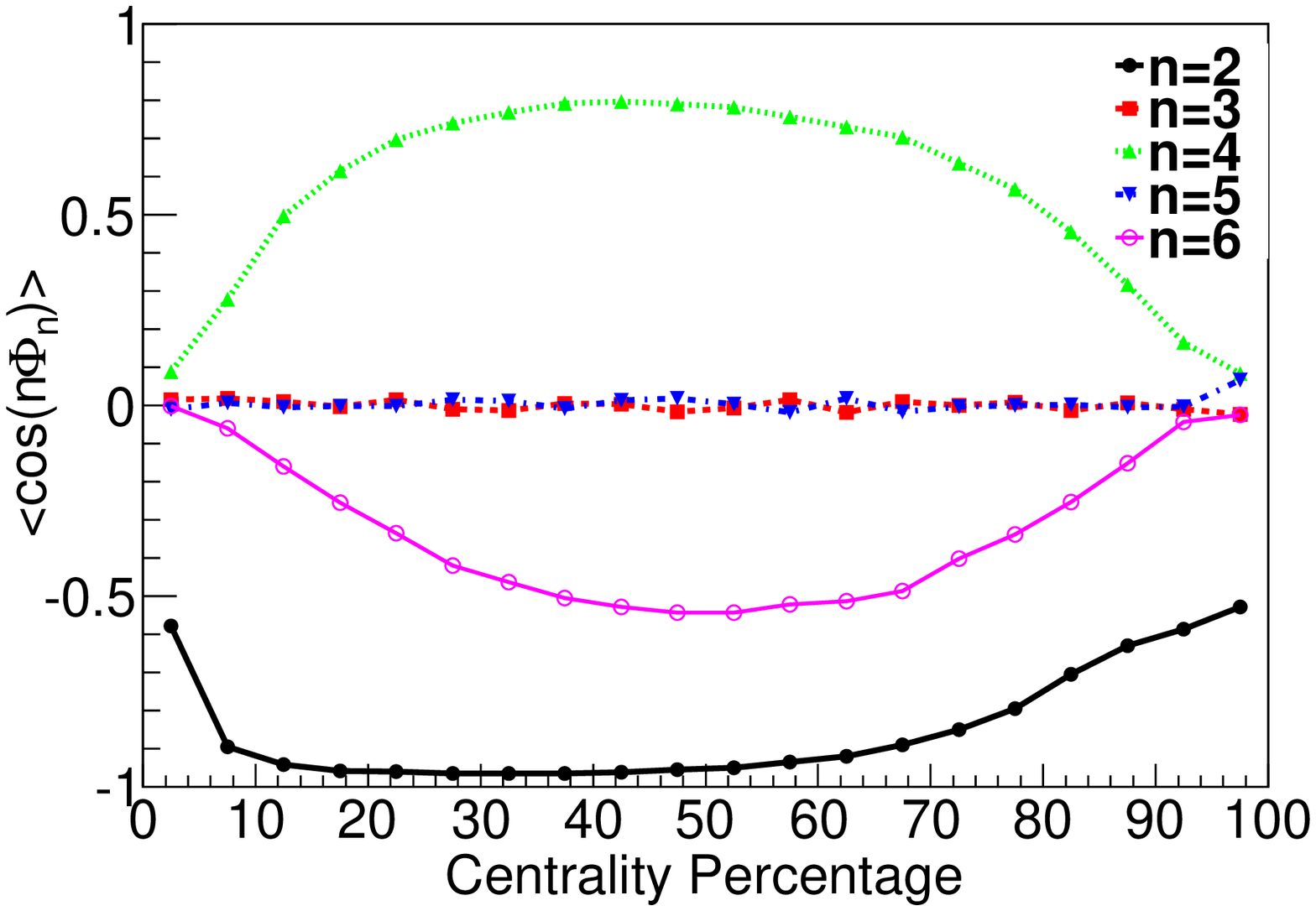,scale=0.45}
\end{minipage}
\begin{minipage}[t]{9 cm}
\epsfig{file=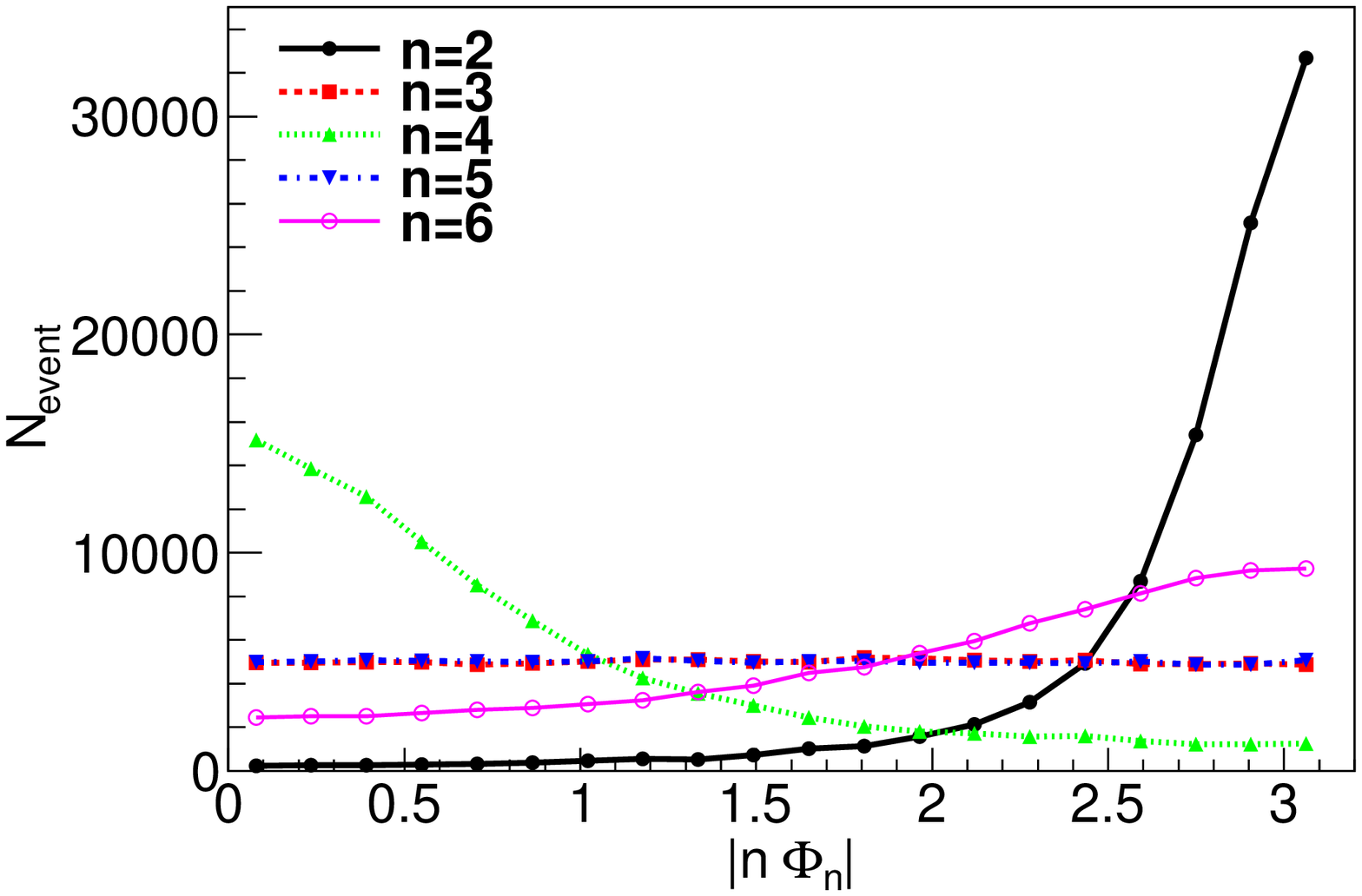,scale=0.45}
\end{minipage}
\caption{
Centrality dependence of average orientation
 angle (left)
and event distribution of orientation 
angle (right).
The total number  of events is 10$^5$ and the number of bins
is 20. 
\label{fig:Phin}}
\end{center}
\end{figure}

Figure \ref{fig:Phin} (left) shows the
centrality dependence of the average orientation 
angle $ \cos (n \Phi_{n}) $. 
Harmonics with even $n$  do not vanish,
which is expected from an almond-like
geometry of non-central collisions.
The second and the sixth orientation angles negatively correlate with the reaction plane,
whereas the fourth 
orientation angle shows positive correlation.
On the other hand, $\langle \cos n \Phi_{n} \rangle $ for the odd $n$
vanishes, which results from a fact that there is no correlation between $\Phi_{n} (n=3, 5)$
and the reaction plane shown in Fig.~\ref{fig:Phin} (right).
In Fig.~\ref{fig:Phin} (right) the number of events as a function
  of the absolute value of the orientation angle $|n\Phi_{n}|$ is
  shown for a sample of 10$^{5}$ minimum bias events. These events are
  binned according to the orientation angle of the initial state
  measured from the reaction plane.
As clearly seen, $|2\Phi_{2}|$ has a prominent peak at $\pi$,
which comes from a fact that initial profile
looks like an almond shape elongated in the $y$-direction on average.
Note that the angle $\Phi_2$ thus gives the angle between the
  major axis of the almond and the reaction plane.
Other even harmonics, $4\Phi_4$
  and $6\Phi_6$, have broad peaks at 0 and $\pi$, respectively.
Orientation angles with odd $n$ are
randomly distributed due to absence of any correlation
with respect to the reaction plane.
The width of event distribution might be important
in understanding the fluctuation of 
the anisotropies of the particle distribution, $\delta v_{n}$,
although we postpone this study to a future work.

\begin{figure}[tb]
\begin{center}
\begin{minipage}[t]{6 cm}
\epsfig{file=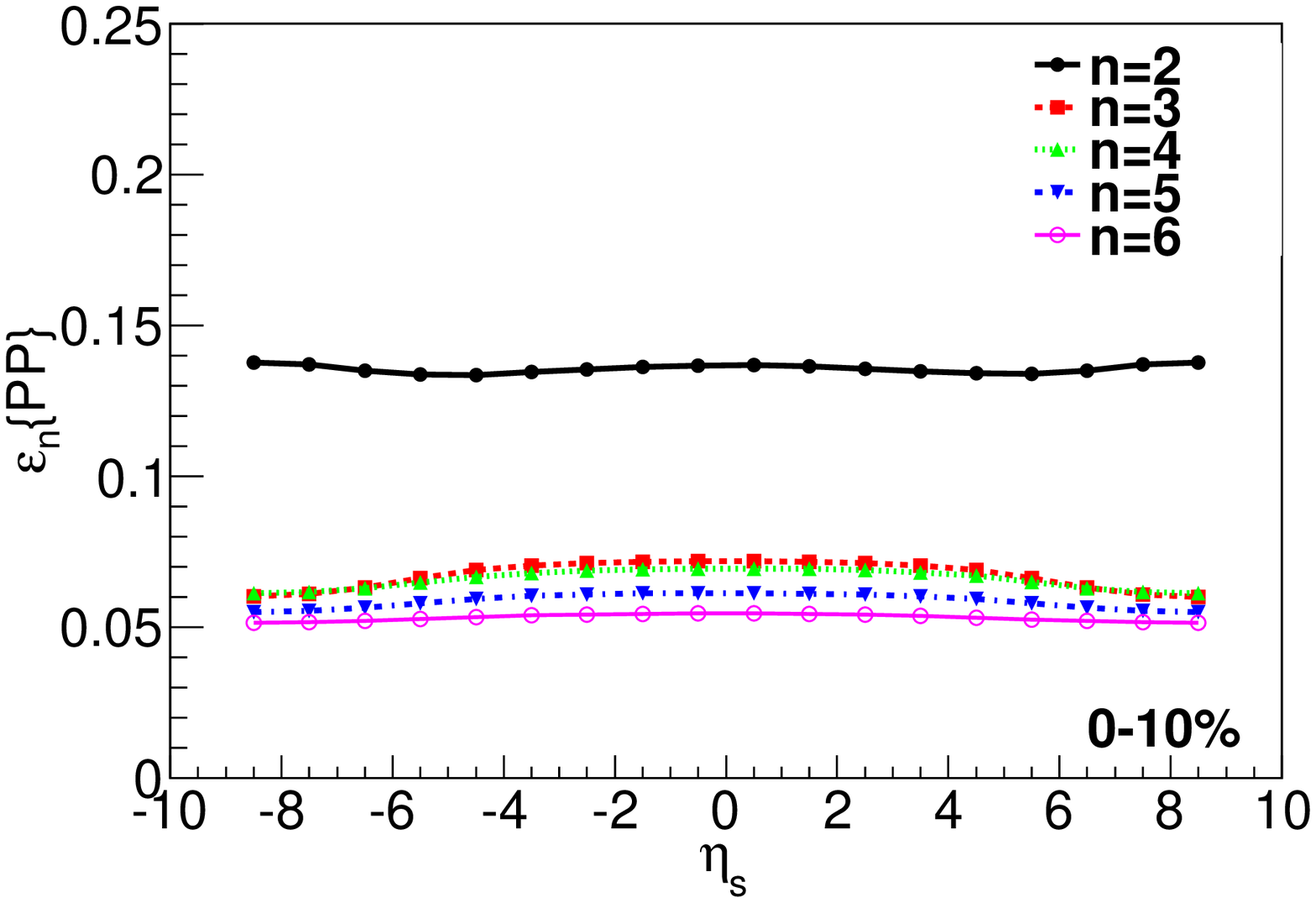,scale=0.3}
\end{minipage}
\begin{minipage}[t]{6 cm}
\epsfig{file=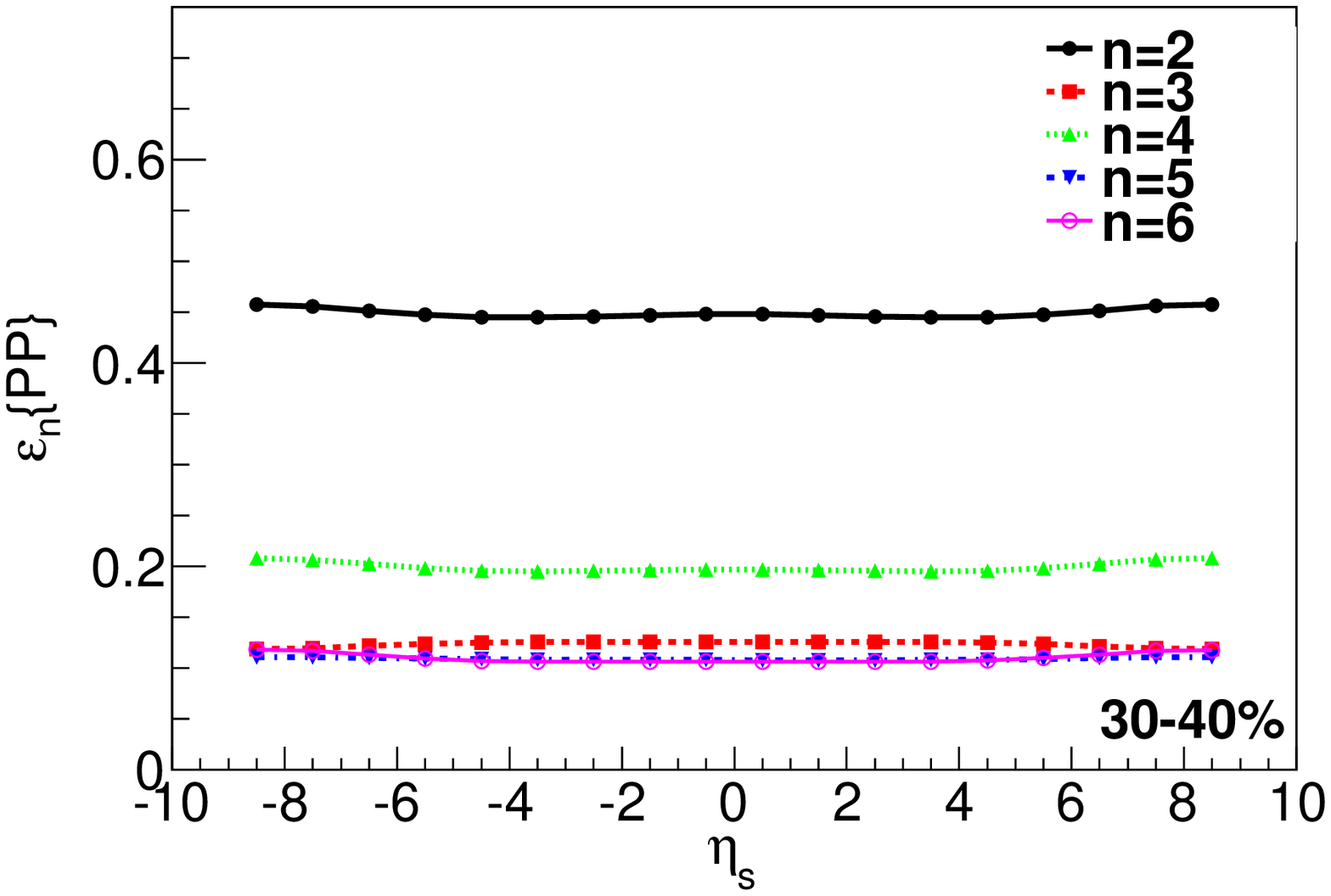,scale=0.3}
\end{minipage}
\begin{minipage}[t]{6 cm}
\epsfig{file=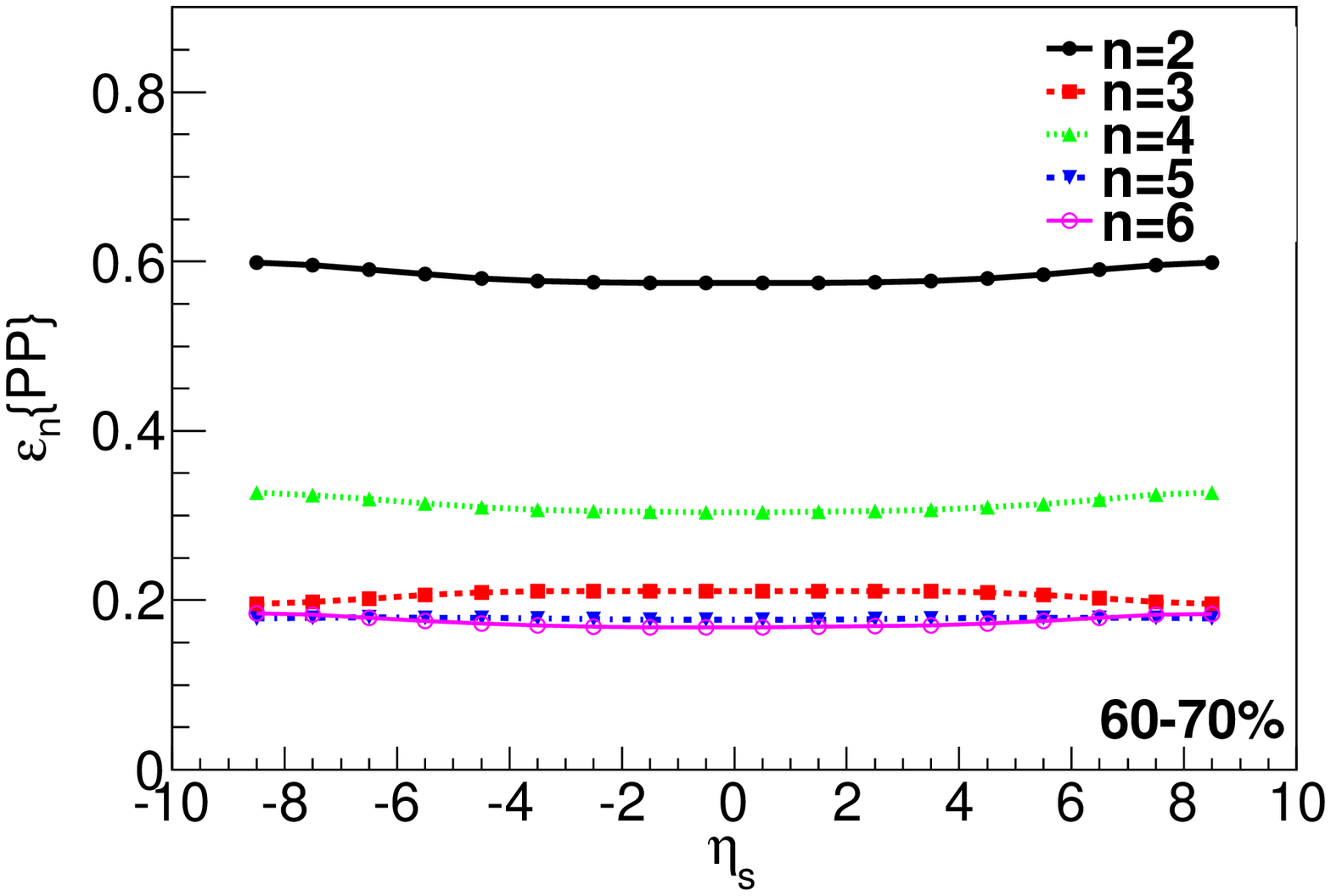,scale=0.3}
\end{minipage}
\caption{
Space-time rapidity dependences of
$\varepsilon_{n}\{\mathrm{PP}\}$ in 0-10\% (left), 30-40\% (middle) and
60-70\% (right) centrality at LHC energy
using the MC-KLN initialisation.
\label{fig:epsilonpp-eta}
}
\end{center}
\end{figure}

\begin{figure}[tb]
\begin{center}
\begin{minipage}[t]{6 cm}
\epsfig{file=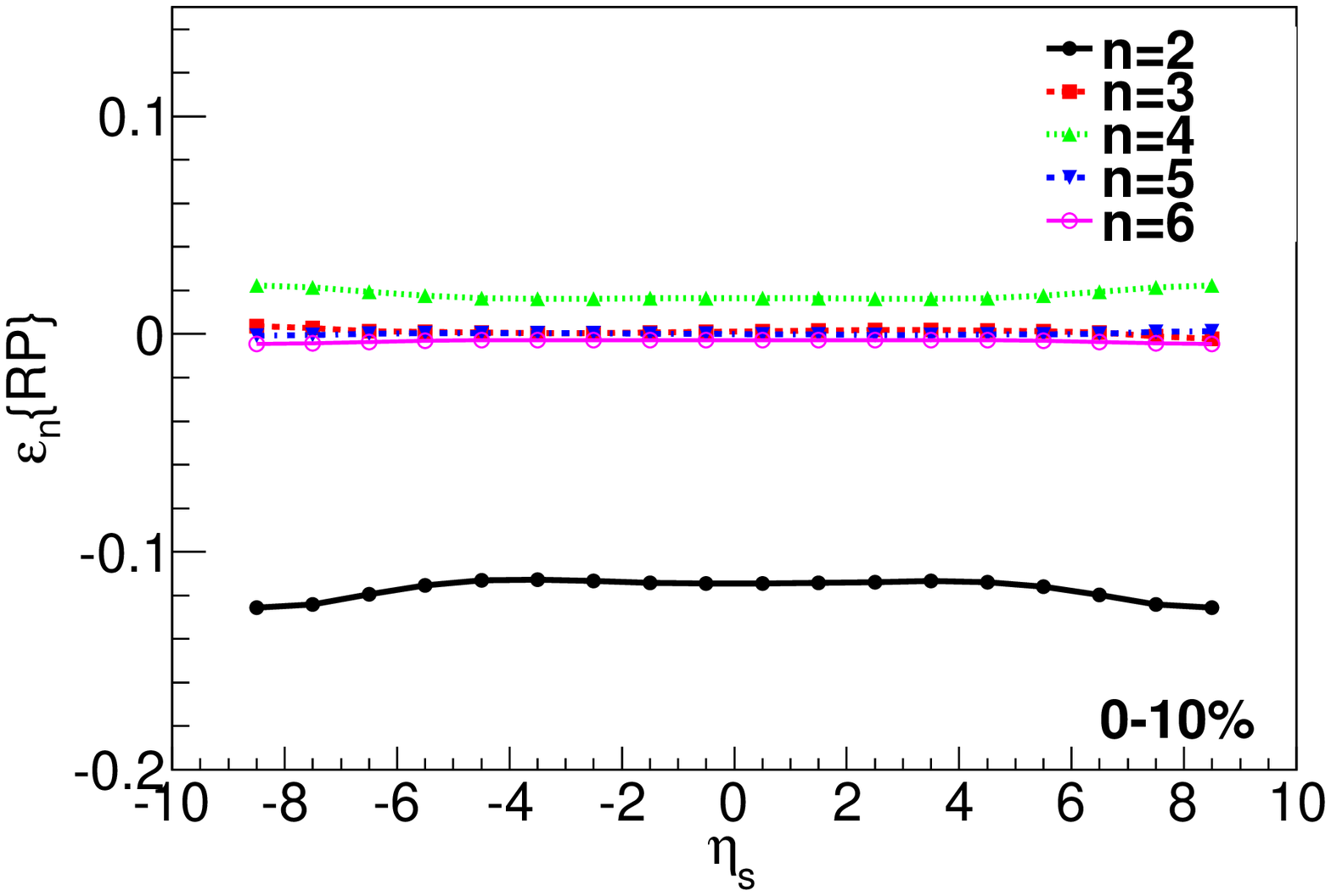,scale=0.3}
\end{minipage}
\begin{minipage}[t]{6 cm}
\epsfig{file=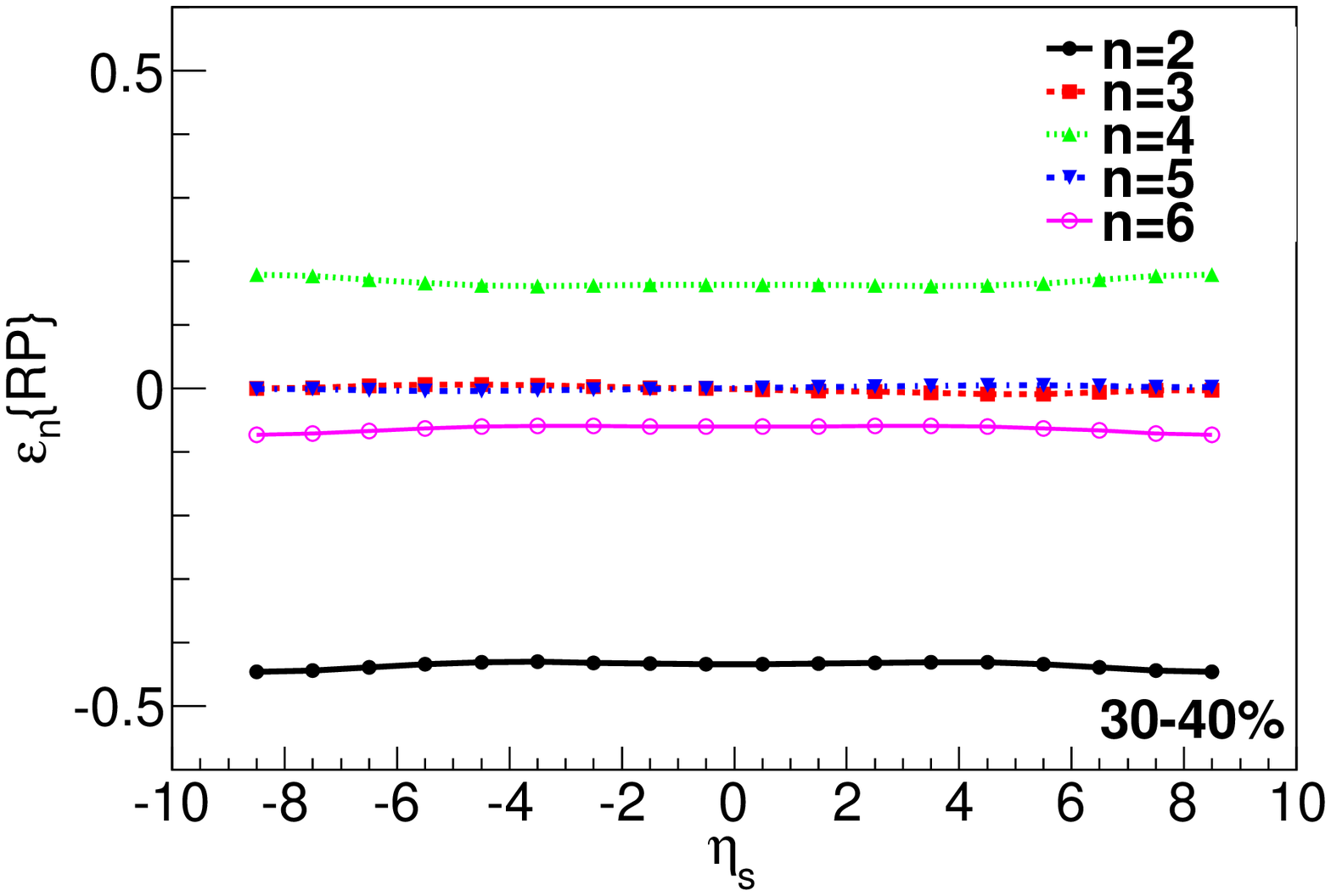,scale=0.3}
\end{minipage}
\begin{minipage}[t]{6 cm}
\epsfig{file=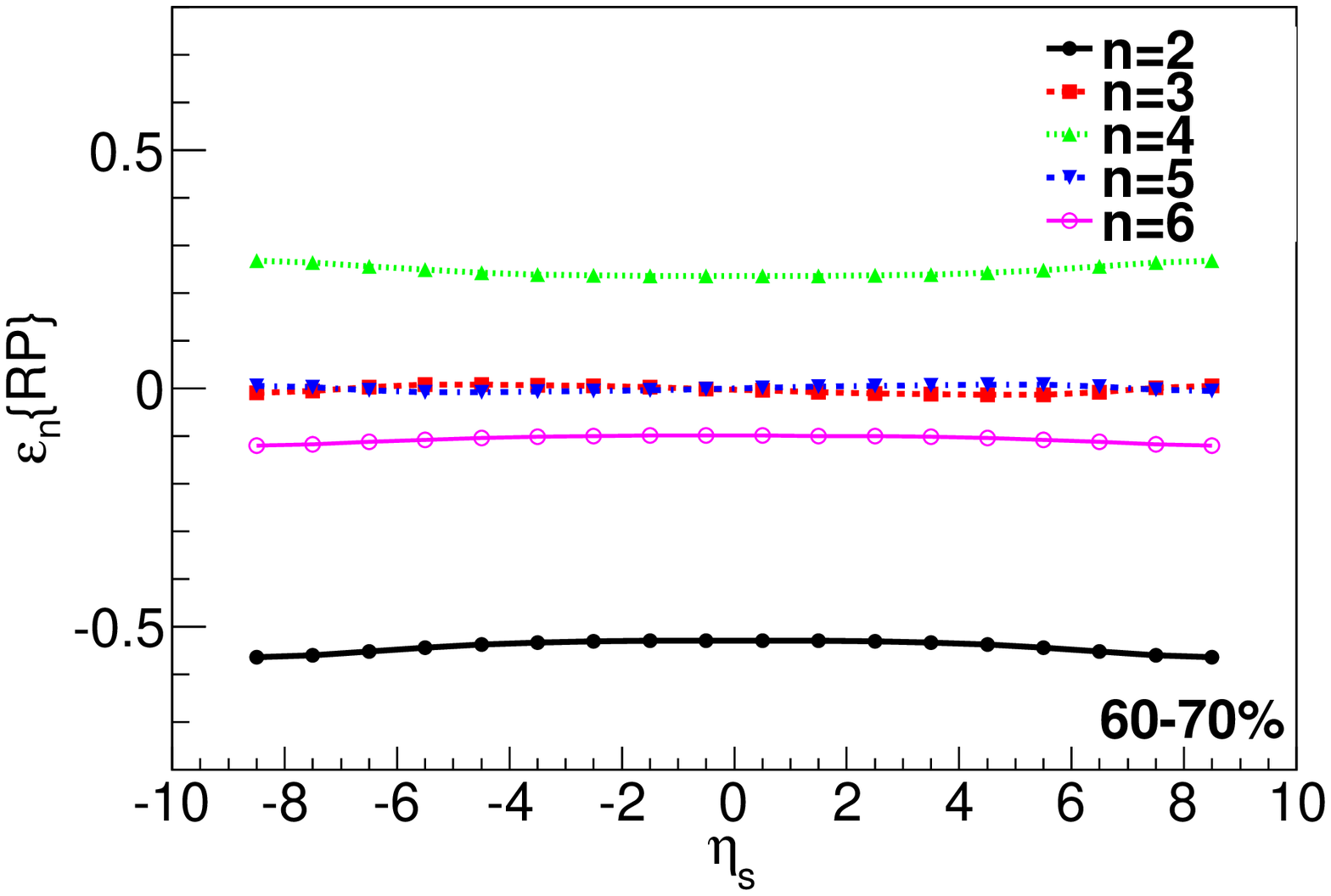,scale=0.3}
\end{minipage}
\caption{
The same as Fig.~\ref{fig:epsilonpp-eta}
but for $\varepsilon_{n}\{\mathrm{RP}\}$.
\label{fig:epsilonrp-eta}
}
\end{center}
\end{figure}

\begin{figure}[tb]
\begin{center}
\begin{minipage}[t]{6 cm}
\epsfig{file=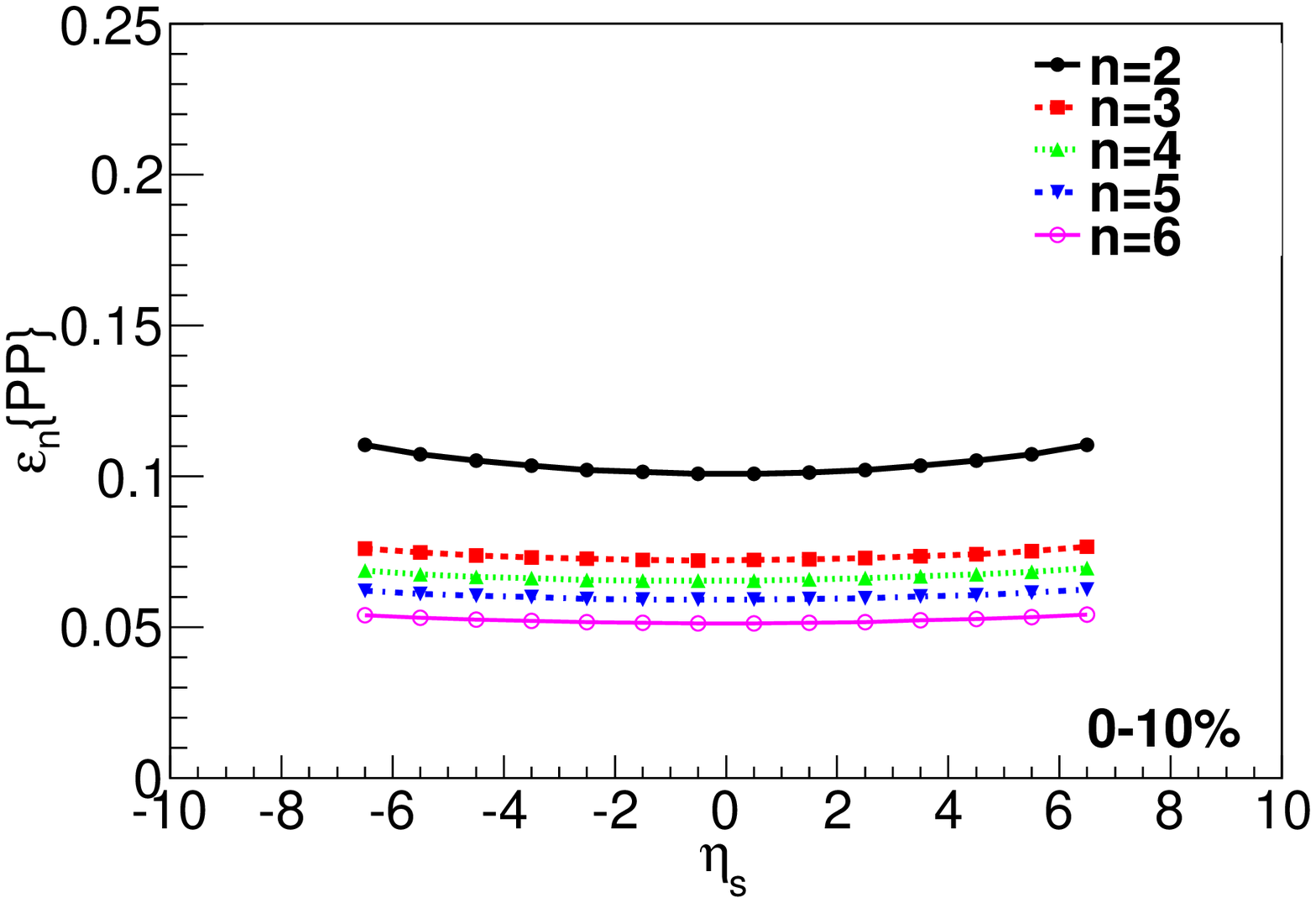,scale=0.3}
\end{minipage}
\begin{minipage}[t]{6 cm}
\epsfig{file=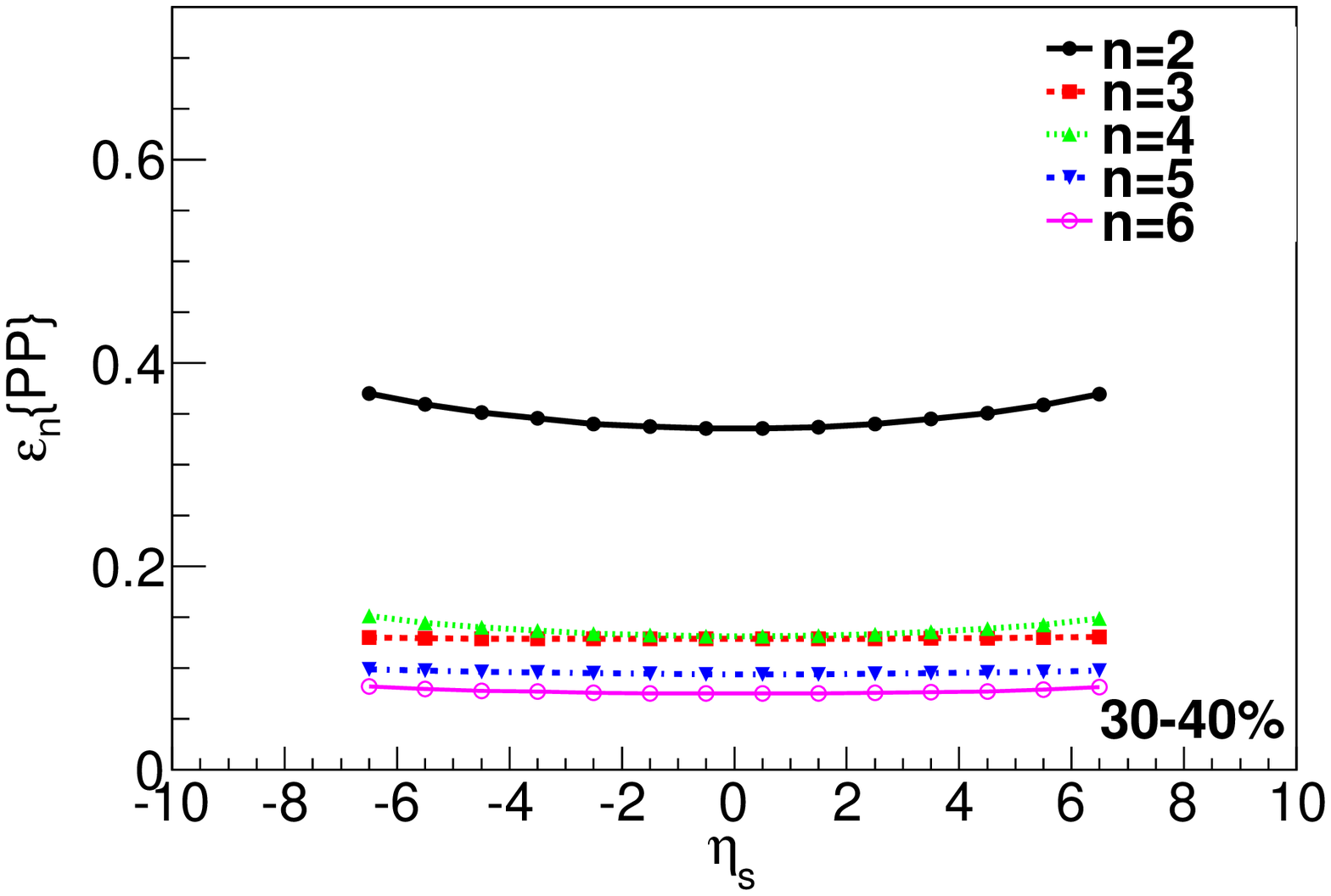,scale=0.3}
\end{minipage}
\begin{minipage}[t]{6 cm}
\epsfig{file=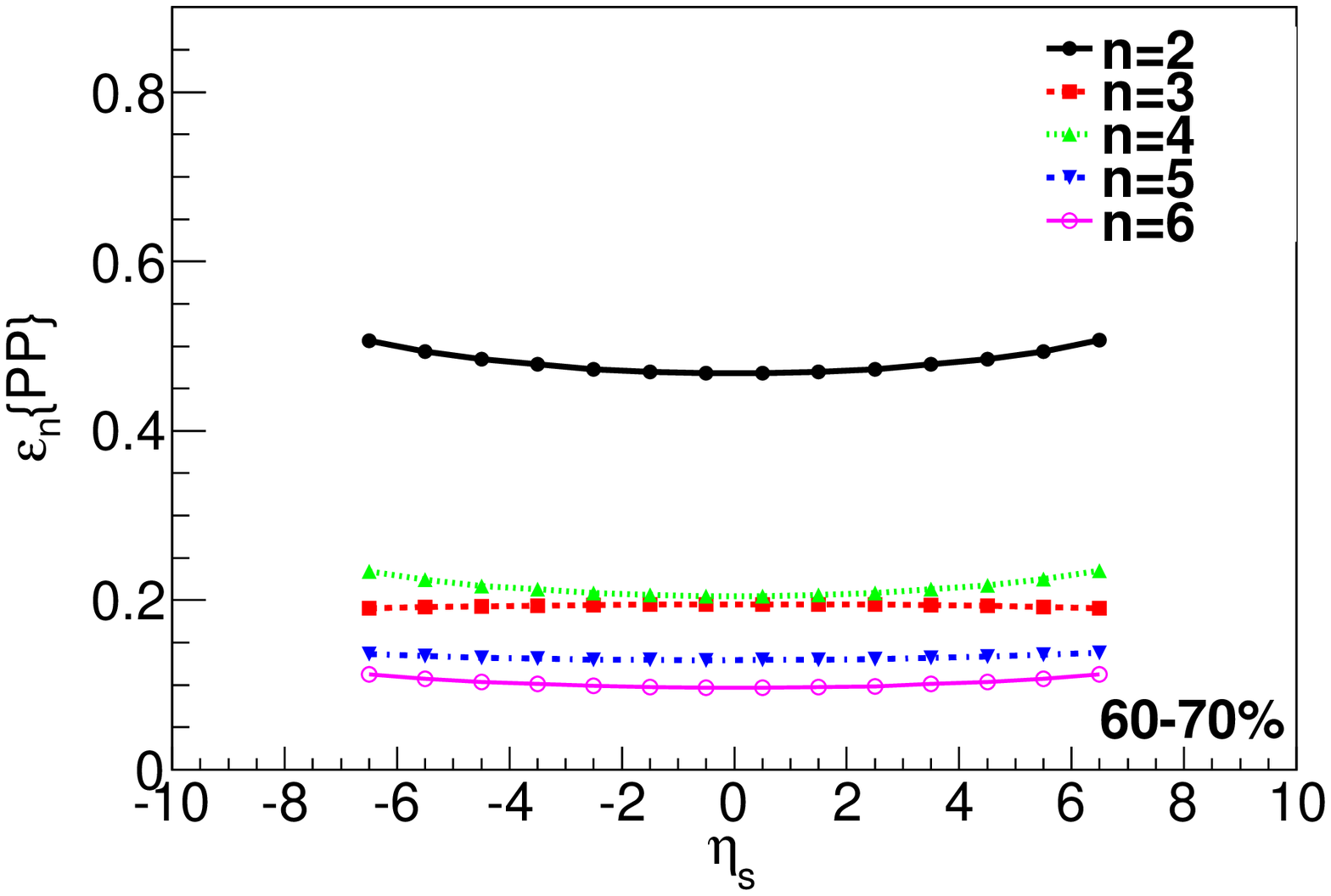,scale=0.3}
\end{minipage}
\caption{
The same as Fig.~\ref{fig:epsilonpp-eta}
but using the MC-Glauber initialisation.
\label{fig:epsilonpp-eta-glbgk}
}
\end{center}
\end{figure}

\begin{figure}[tb]
\begin{center}
\begin{minipage}[t]{6 cm}
\epsfig{file=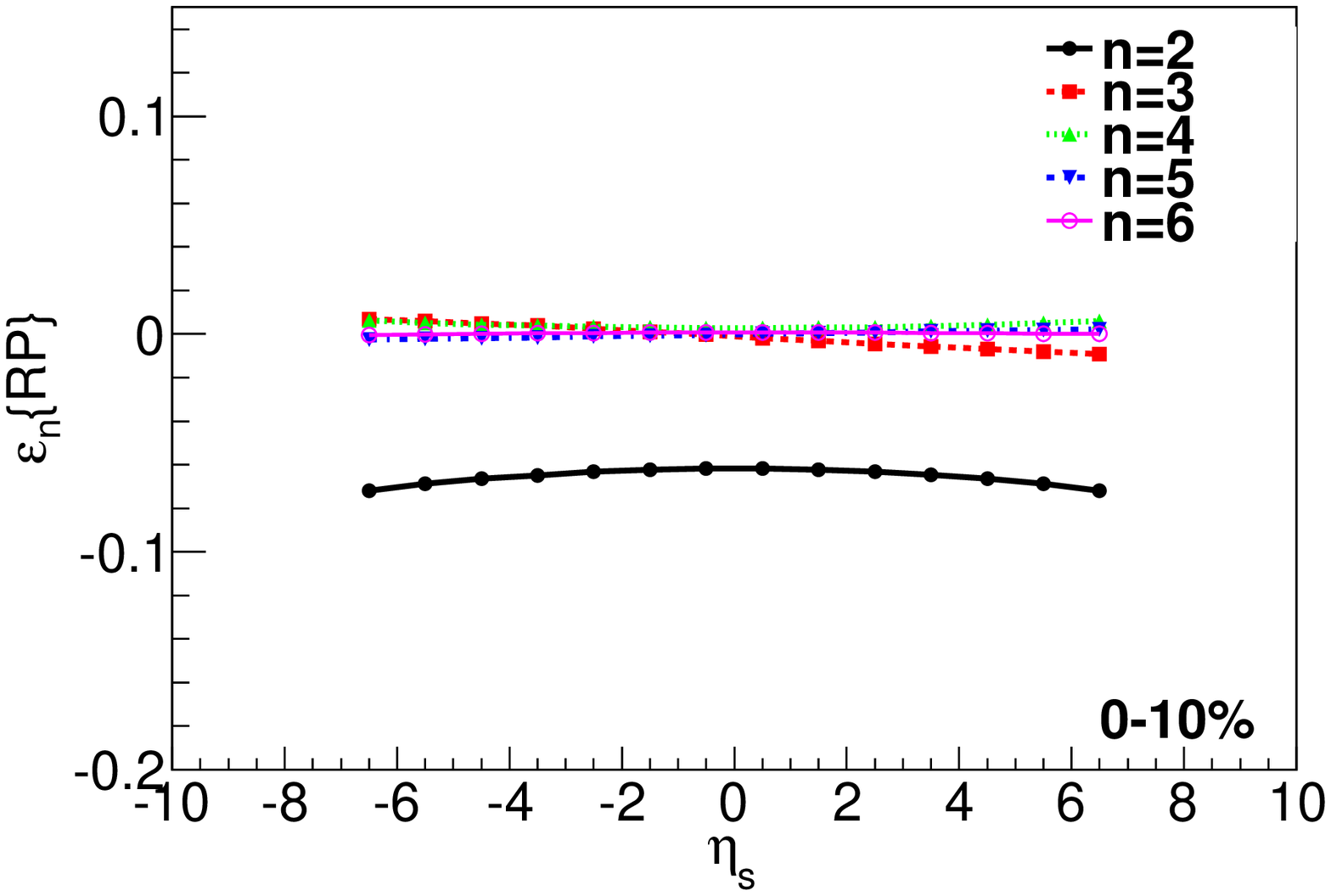,scale=0.3}
\end{minipage}
\begin{minipage}[t]{6 cm}
\epsfig{file=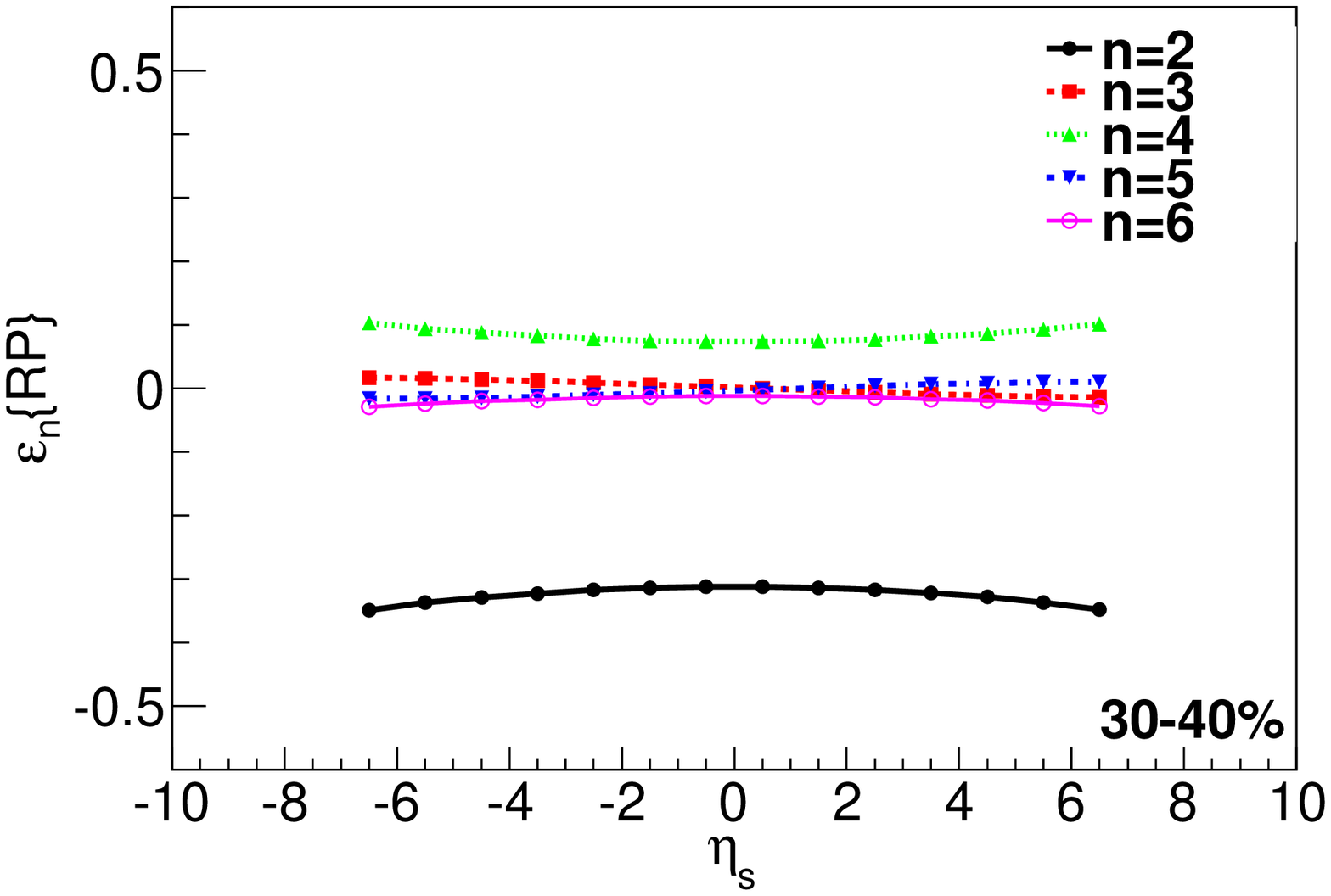,scale=0.3}
\end{minipage}
\begin{minipage}[t]{6 cm}
\epsfig{file=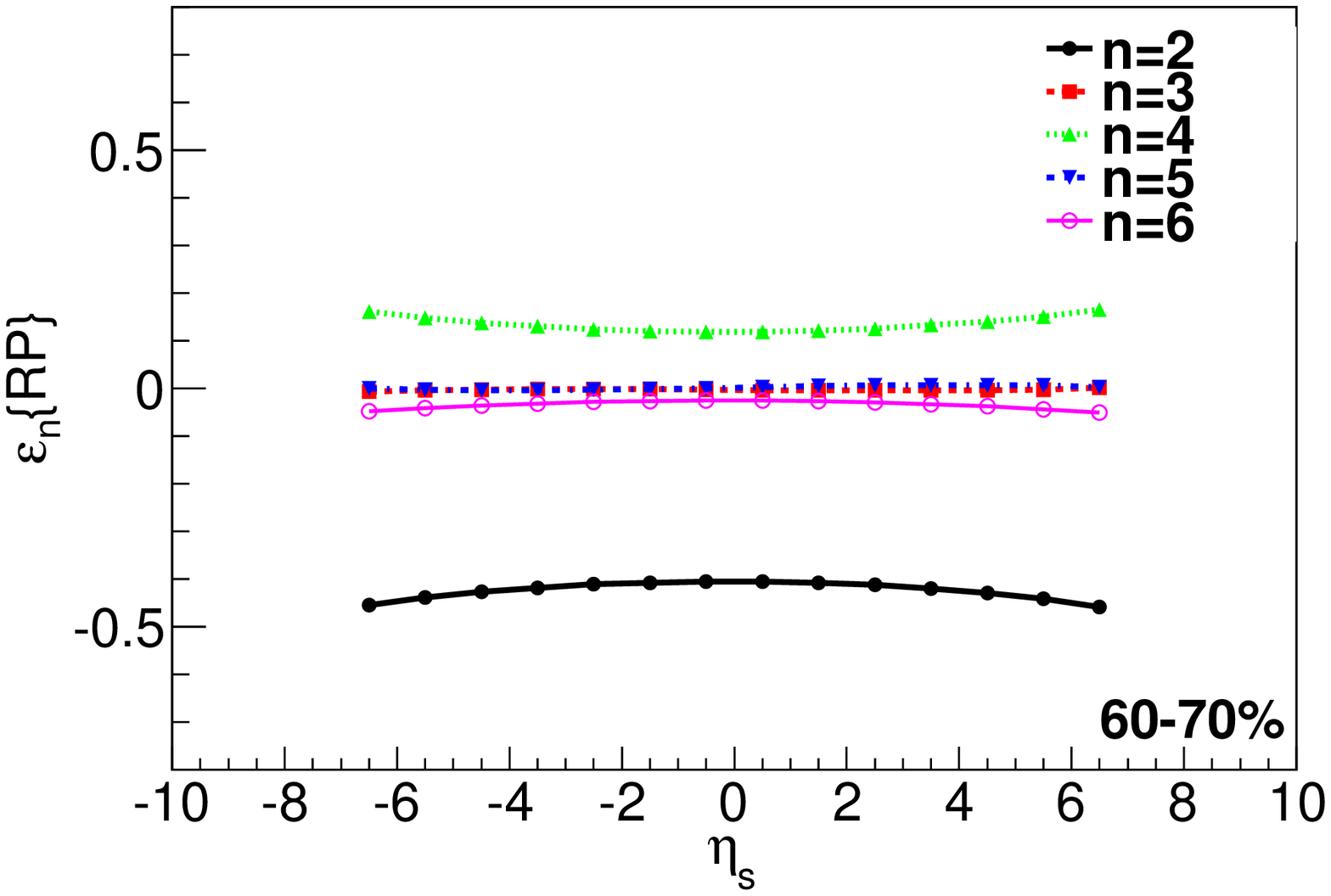,scale=0.3}
\end{minipage}
\caption{
The same as Fig.~\ref{fig:epsilonrp-eta}
but using the MC-Glauber initialisation.
\label{fig:epsilonrp-eta-glbgk}
}
\end{center}
\end{figure}

\begin{figure}[tb]
\begin{center}
\begin{minipage}[t]{9 cm}
\epsfig{file=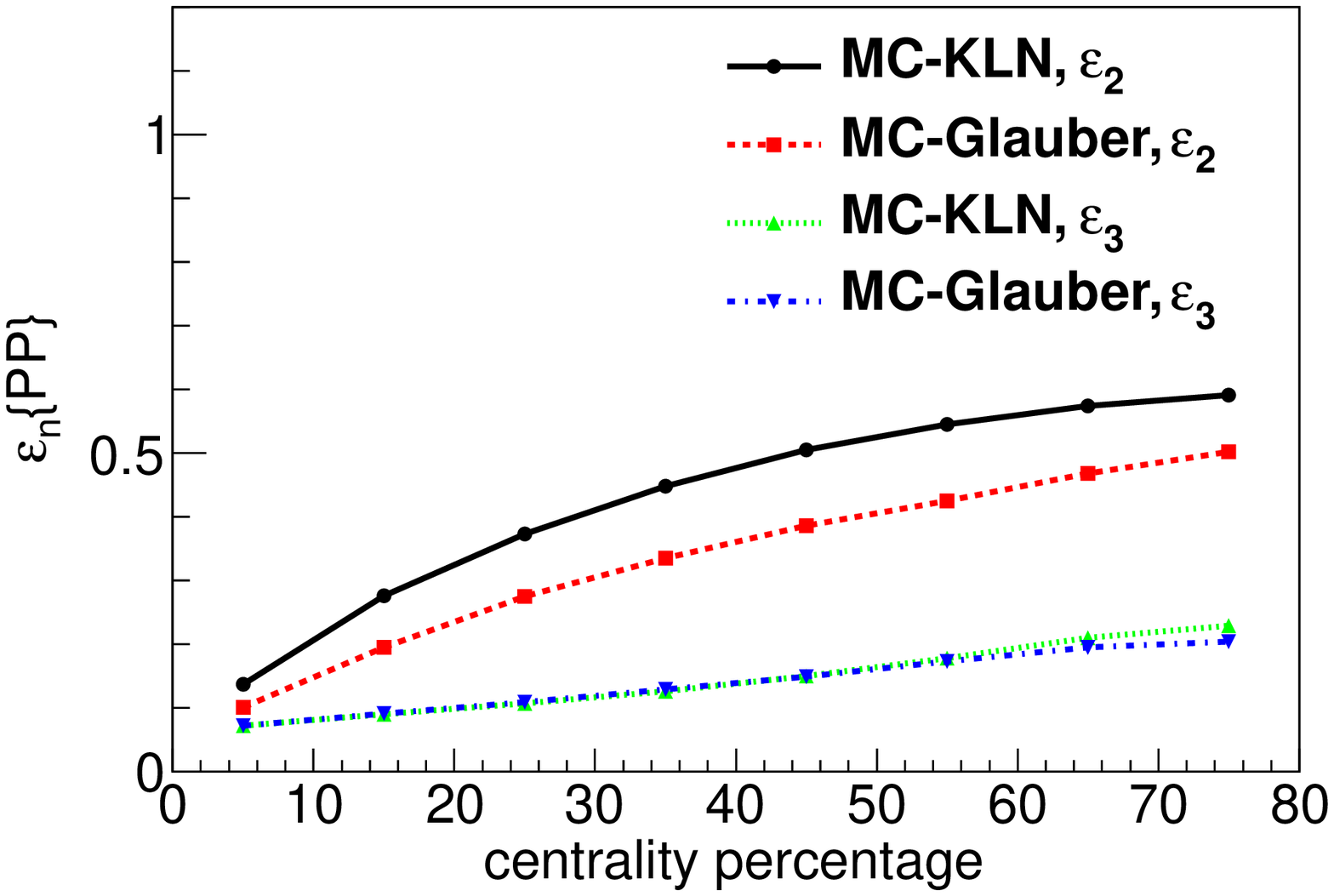,scale=0.45}
\end{minipage}
\begin{minipage}[t]{9 cm}
\epsfig{file=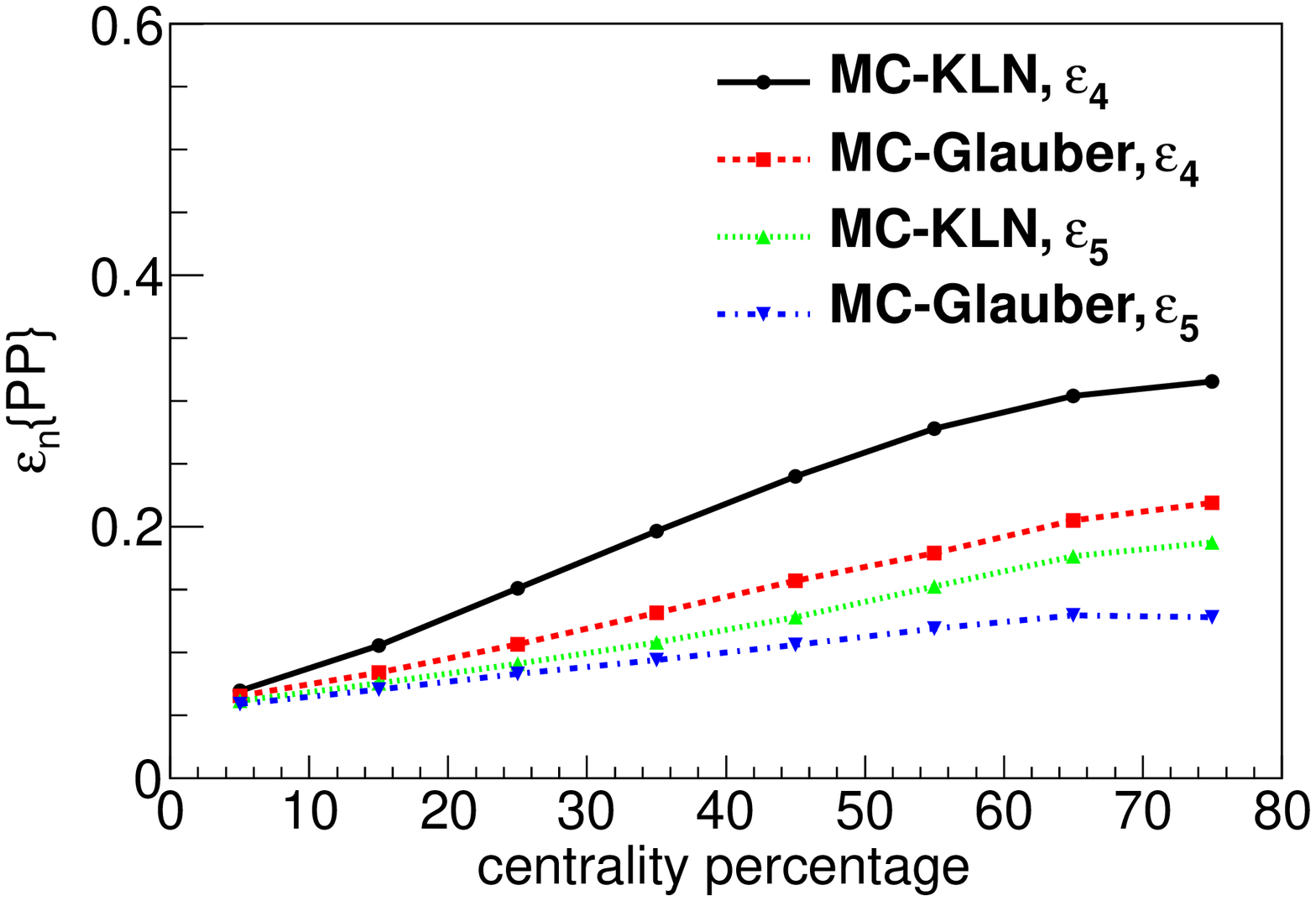,scale=0.45}
\end{minipage}
\caption{
Centrality dependence of $\varepsilon_{2}$ and $\varepsilon_{3}$ (left)
and $\varepsilon_{4}$ and $\varepsilon_{5}$ (left) for charged hadrons
at midrapidity ($0<\eta_{s}<1$)
in Pb+Pb collisions at $\sqrt{s_{NN}}=2.76$ TeV.
Results from the MC-KLN model are compared
with the ones from the MC-Glauber model.
\label{fig:KlnVsGlauberEps}}
\end{center}
\end{figure}

Figures \ref{fig:epsilonpp-eta} and \ref{fig:epsilonrp-eta} show
space-time rapidity dependences
of $\varepsilon_{n}\{\mathrm{PP}\}$ and
$\varepsilon_{n}\{\mathrm{RP}\}$, respectively, 
using the MC-KLN model, and Figures \ref{fig:epsilonpp-eta-glbgk} and
  \ref{fig:epsilonrp-eta-glbgk} using the MC-Glauber model (extended
  in the longitudinal direction using the BGK model). The anisotropies
  $\varepsilon_n$ are evaluated
in 0-10\% (left), 30-40\% (middle) and 60-70\% (right)
centrality classes in Pb+Pb collisions at $\sqrt{s_{NN}}=$ 2.76 TeV.
Since the density profiles are smooth and  
have a streak-like structure
in the longitudinal direction
as shown in Figs.~\ref{fig:iniKLN} and~\ref{fig:iniGlauber},
$\varepsilon_{n}\{\mathrm{PP}\}$ are almost
independent of $\eta_{s}$.
For MC-KLN initialization
$\varepsilon_{n}\{\mathrm{PP}\}$ for $n=3$, 5, and 6 are
close to each other
at all centralities, whereas the values differ for MC-Glauber
  initialization. This indicates different origin of fluctuations in
  these two models.
If $v_{n}$ was roughly proportional to
$\varepsilon_{n}\{\mathrm{PP}\}$,
$v_{n}$ would be independent of
rapidity.
However, it is not the case at least for $v_{2}(\eta)$ at RHIC.
Even though $\varepsilon_{2}$ is almost independent of
space-time rapidity \cite{Hirano:2001eu},
final $v_{2}$ has a broad peak at midrapidity
due to relatively larger hadronic dissipative effects
in forward/backward rapidity regions \cite{Hirano:2005xf}.
 This can be also interpreted
as follows.
$v_{2}$ increases during  the QGP evolution and does not so much
in the hadronic evolution.
So rapidity dependence of $v_{2}$ is a key
to understand longitudinal structure of the QGP. 
Since $\varepsilon_{n}\{\mathrm{PP}\}$
does not depend on space-time rapidity as shown in Fig.~\ref{fig:epsilonpp-eta},
$v_{n}(\eta)$ should contain the direct information about the longitudinal
structure of the QGP.
We will discuss (pseudo-)rapidity dependence of $v_{n}$ later.
The negative $\varepsilon_{2}\{\mathrm{RP}\}$ 
in Figs.~\ref{fig:epsilonrp-eta} 
and~\ref{fig:epsilonrp-eta-glbgk} is due to
our definition of $\varepsilon_{2}\{\mathrm{RP}\}$, 
Eq.~(\ref{eq:e2rp}), which has a different sign than the usual 
definition of eccentricity.
Compared with finite $\varepsilon_{n}\{\mathrm{PP}\}$,
$\varepsilon_{n}\{\mathrm{RP}\}$ vanishes for odd $n$ since
odd harmonics result solely from initial fluctuations and
do not correlate with reaction plane. 
Although longitudinal profiles, 
Eqs.~(\ref{eq:modifiedBGK}) and (\ref{eq:pp}), in the MC-Glauber model
gives similar rapidity and centrality dependence to the MC-KLN model,
absolute values of $\varepsilon_{n}\{\mathrm{PP}\}$
and $\varepsilon_{n}\{\mathrm{RP}\}$ are different
except for $n=3$ 
as shown  in Fig.~\ref{fig:KlnVsGlauberEps}.

\subsection{\it Brief overview of event-by-event initial conditions \label{sec:e-by-e_hydro}}

In this subsection, we review 
hydrodynamic modelling 
of relativistic heavy ion collisions
by focusing particularly on
initial conditions on an event-by-event basis.

One of the first works along these lines were
the boost-invariant calculations by Gyulassy \emph{et al.}~\cite{Gyulassy:1996br} where
HIJING~\cite{Wang:1991hta,hijing2,hijing3,hijing4} event generator was
used to evaluate the initial conditions.
At collider energies, mini-jets were expected to become
one of the dominant sources of fluctuations.
In HIJING, particle production is modelled by string excitation
and its decay into hadrons by using Lund jet fragmentation scheme
for soft processes, and hard pQCD processes
are generated based on an eikonal multiple collision formalism
and PYTHIA event generator \cite{Sjostrand:2006za}.

\begin{figure}[htb]
\begin{center}
\epsfig{file=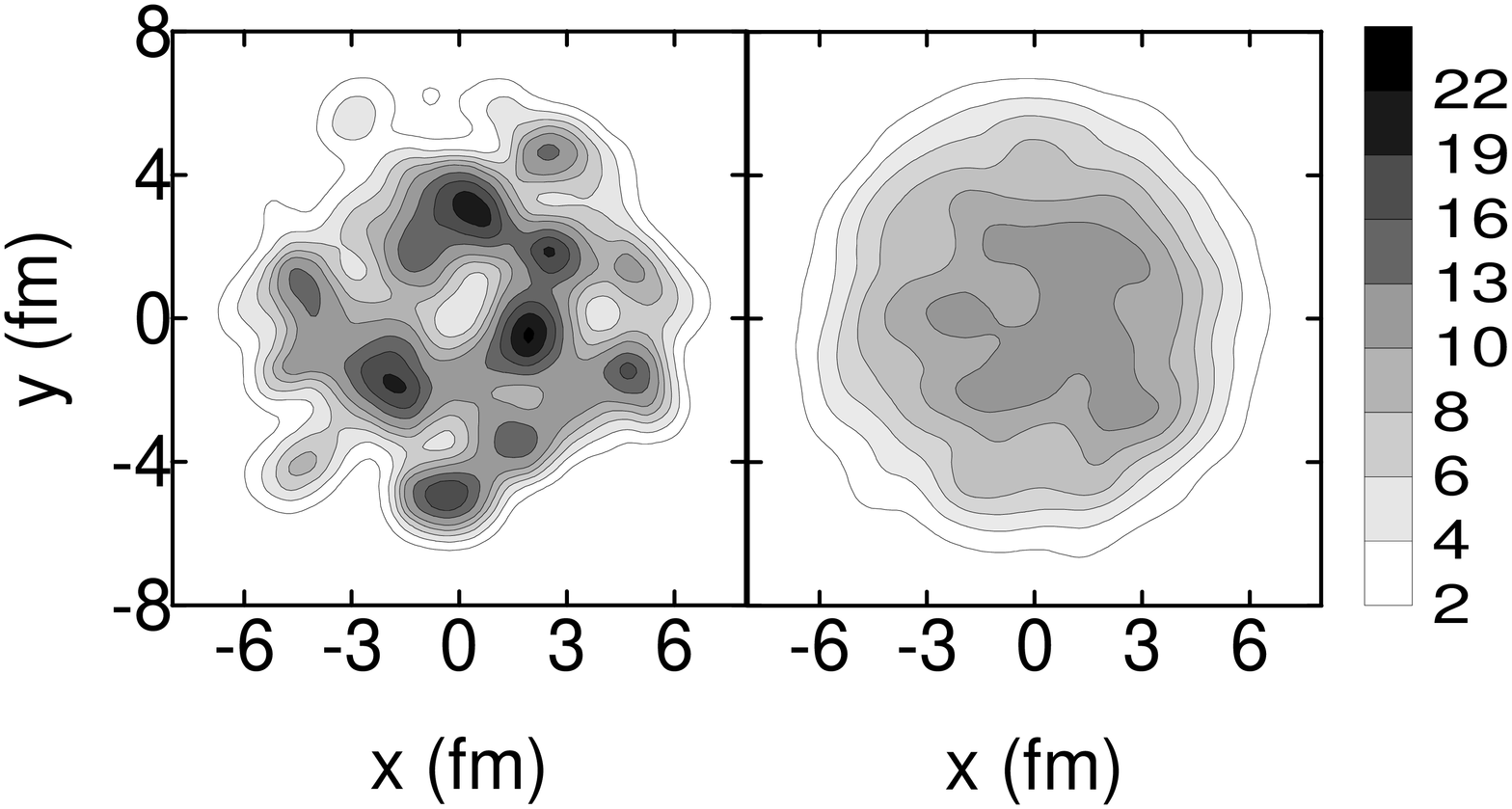,scale=0.5, angle=270}
\caption{
Examples of initial conditions in central Au+Au
collisions given by NeXus at midrapidity. The energy
density is plotted in units of GeV/fm$^3$. (Left) 
Example of an
event. (Right) Average over 30 events.
Figure taken from Ref.~\cite{Hama:2004rr}.
}
\label{fig:Nexus}
\end{center}
\end{figure}

Another pioneering work was done by the Rio and Sao Paolo groups, who used
NeXus event generator~\cite{Drescher:2000ha} to calculate initial
condition of fully (3+1) dimensional ideal hydrodynamic
simulations~\cite{Osada:2001hw,Aguiar:2001ac,Hama:2004rr,Andrade:2006yh,Andrade:2008xh,Takahashi:2009na,
  Gardim:2011qn}. 
They were the first to point out the importance of initial state
  fluctuations when interpreting the elliptic flow data. Namely, the
  calculated $v_2$ is different depending on whether one first
  evaluates an average initial state, evolves it hydrodynamically, and
  calculates the $v_2$, or whether one evolves the initial states
  event-by-event, calculates $v_2$ in every event, and averages these
  calculated values~\cite{Andrade:2006yh,Andrade:2008xh}.
NeXus is a Monte-Carlo event generator based on
the Gribov-Regge theory and the  pQCD parton model.
To convert the output of NeXus to the initial state of
  hydrodynamic evolution, they calculate the energy momentum tensor
  and conserved currents using the kinetic theory definitions.
Obviously, energy momentum tensor obtained in this way
is far from the one in equilibrium.
Nevertheless, the energy momentum tensor can be decomposed and
  energy density and velocity obtained
{\it a.~la.~}Landau
\begin{equation}
T^{\mu}_{\enskip \nu} u^{\nu} = \epsilon u^{\mu}.
\end{equation}
Using $u^{\mu}$,
one can also obtain baryon density from baryon current
$n_{B} = u_{\mu}N^{\mu}_{B}$.
Once $\epsilon$ and $n_{B}$ are obtained,
pressure is calculated by using the equation of state 
$P=P(\epsilon, n_{B})$.
Energy momentum tensor for perfect fluids
is then based on this energy density, pressure and flow velocity,
  and the non-ideal terms in the original energy-momentum tensor are
  ignored. A drawback of this procedure is that energy and momentum
  are not strictly conserved~\cite{Cheng:2010mm} because the non-ideal
  terms are ignored and the EoS of the fluid may be different from the
  EoS of NeXus.
Figure \ref{fig:Nexus} shows
initial energy density in a single event (left)
and that averaged over 30 events (right).
The bumpy structure in a single event is smeared by taking event averages
in the initial condition.

The NeXus event generator and hydrodynamic approach
was further utilised to evaluate two-pion correlation functions on
event-by-event basis by Ren \emph{et al.}~\cite{Ren:2008sc}.
The main motivation to consider initial fluctuations
was to understand RHIC HBT data 
$R_{\mathrm{out}}/R_{\mathrm{side}} \sim 1$~\cite{Adler:2001zd,Adcox:2002uc}.

\begin{figure}[htb]
\begin{center}
\begin{minipage}[t]{9 cm}
\epsfig{file=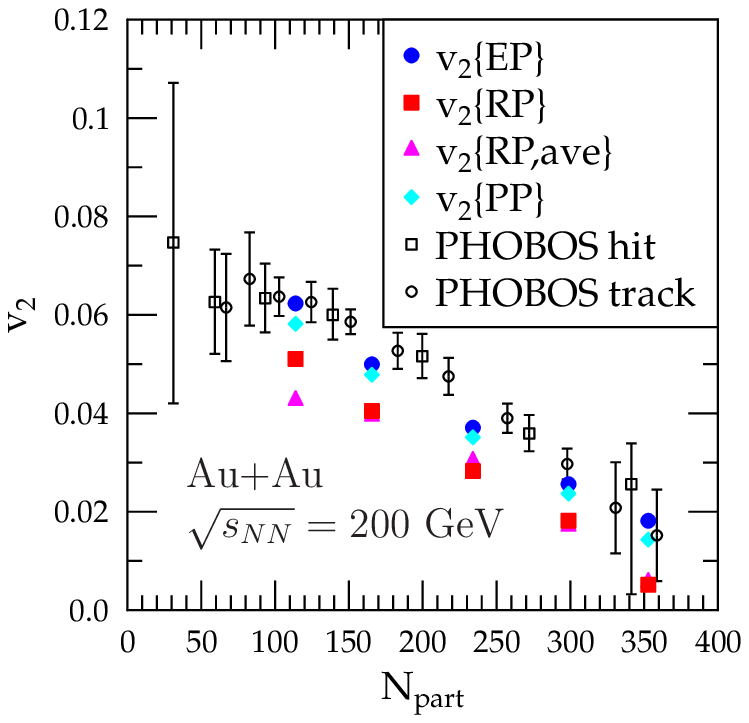,scale=1.0}
\end{minipage}
\begin{minipage}[t]{9 cm}
\epsfig{file=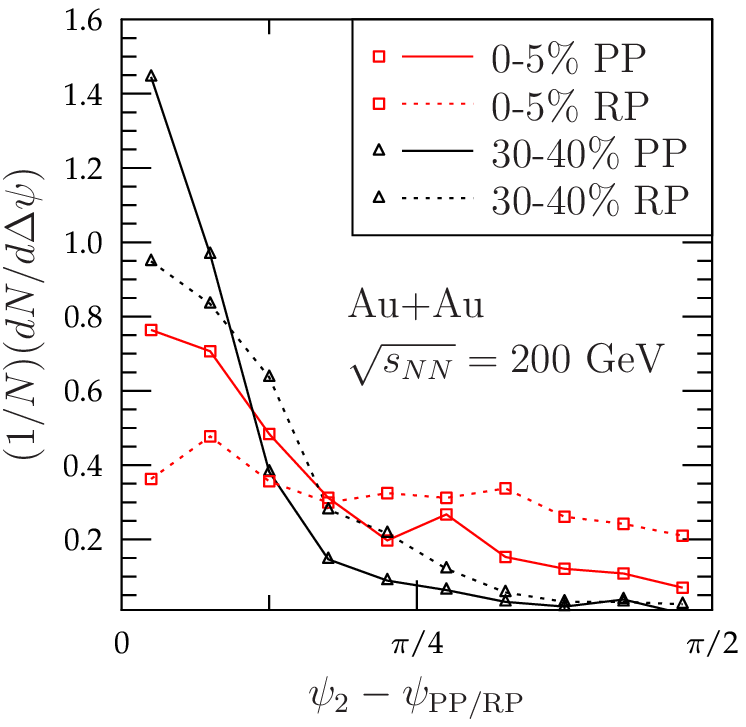,scale=1.0}
\end{minipage}
\caption{
(Left) 
Elliptic flow parameter $v_{2}$ 
with $\sigma = 0.4$ fm in  Au+Au collisions at $\sqrt{s_{NN}} = 200$ GeV  are compared
with the PHOBOS data.
(Right) 
Distribution of difference of angles between
event and reaction planes
and event and participant planes in central (0-5\%) and
semicentral (30-40\%) collisions.
Figures are taken from Ref.~\cite{Holopainen:2010gz}.
}
\label{fig:Finland}
\end{center}
\end{figure}

Later, the Jyv\"askyl\"a group performed boost-invariant event-by-event ideal
hydrodynamic simulations~\cite{Holopainen:2010gz} using Monte-Carlo
Glauber initialisation~\cite{Miller:2007ri}, and applied their results
to analyse thermal photon spectra \cite{Chatterjee:2011dw} and jet
quenching \cite{Renk:2011qi}. 
With an option of eWN (energy density using wounded nucleons),
initial energy density is calculated as
\begin{equation}
\label{eq:ene_trans}
\epsilon(x, y) = \frac{K}{2 \pi \sigma^2} \sum_{i=1}^{N_{\mathrm{part}}}
\exp \left[-\frac{(x-x_{i})^2+(y-y_{i})^2}{2\sigma^2} \right].
\end{equation} 
Here $(x_{i}, y_{i})$ is the transverse position of a participant from
Monte-Carlo Glauber simulations. $\sigma$
is a smearing parameter of the collision point being either 0.4 fm or 0.8 fm
in this model. A parameter $K$ controls the absolute value of particle yields.
In most hydrodynamical calculations the elliptic flow parameter $v_2$ has
been evaluated with respect to either reaction plane
or participant plane.
The Jyv\"askyl\"a group 
were the first to use the event plane method to evaluate the
  hydrodynamically calculated $v_2$, and thus to follow the
  experimental procedure as closely as possible.
See, \textit{e.g.}, Fig.~\ref{fig:Finland} (left).
They also analyse distribution of difference of angle at the second
order (elliptic flow) between event and reaction/participant planes as
shown in Fig.~\ref{fig:Finland} (right).

\begin{figure}[htb]
\begin{center}
\epsfig{file=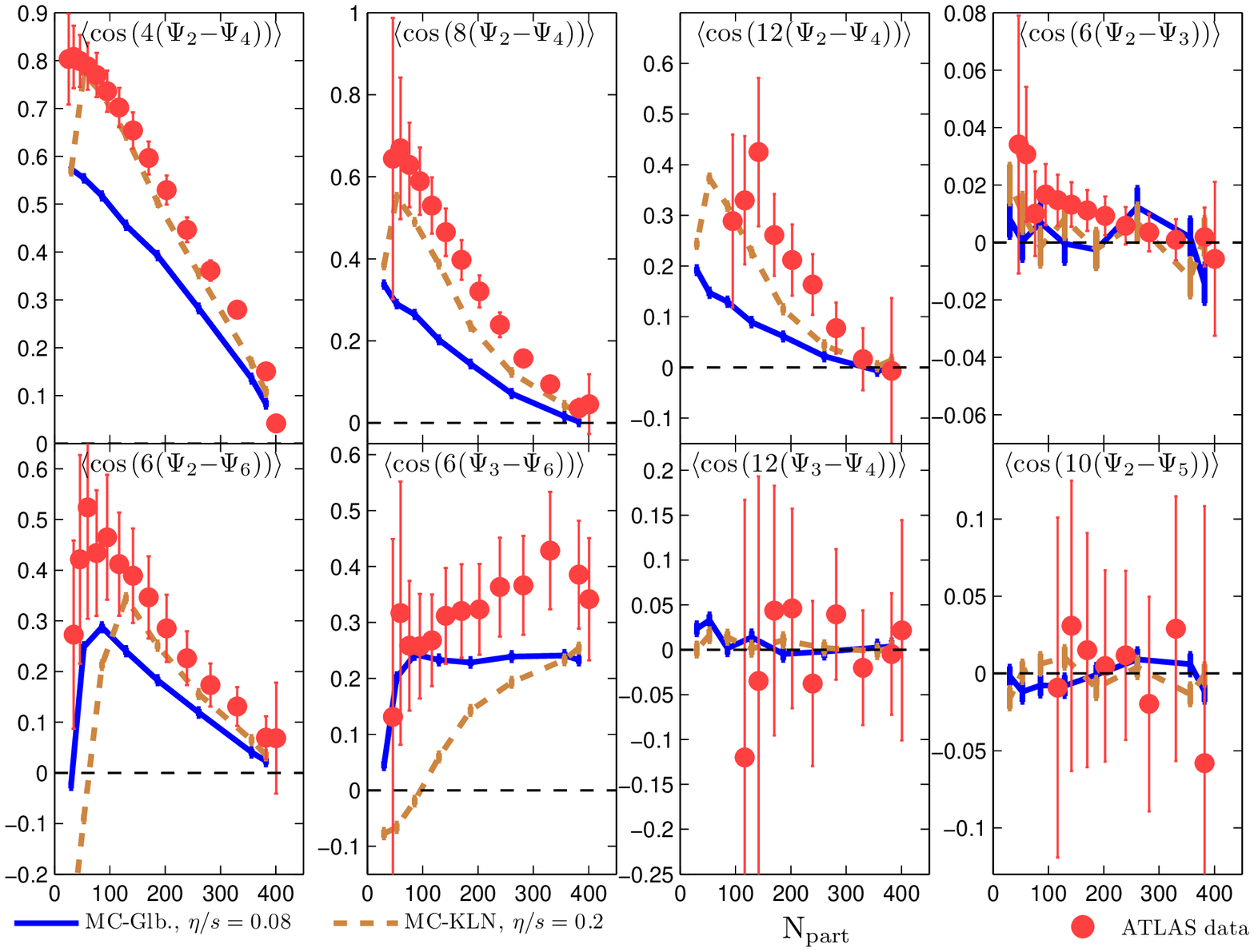,scale=0.6}
\caption{
Centrality dependences of correlation between different order of event plane angles
are compared with the ATLAS data~\cite{Jia:2012sa}.
The MC-Glauber (solid) and MC-KLN (dashed) initial profiles 
are propagated  
using viscous hydrodynamics with $\eta/s=0.08$ and 0.2, respectively.
Figure is taken from Ref.~\cite{Qiu:2012uy}.
\label{fig:QiuHeinz}
}
\end{center}
\end{figure}

The Monte-Carlo Glauber and KLN models were employed to
initialise ideal \cite{Qiu:2011iv} and viscous \cite{Qiu:2012uy}
fluid evolution also by the Ohio State group.
In their MC-Glauber initialization they assumed that 
initial entropy density profile, rather than 
energy density profile like in Eq.~\ref{eq:ene_trans}, 
is proportional to
linear combination of soft wounded nucleon distribution and
hard binary collision distribution.
Correlations between the participant  and the event plane angles \cite{Qiu:2011iv},
between the two participant plane angles and between the two event plane
angles \cite{Qiu:2012uy}
were extensively studied by using event-by-event hydrodynamic simulations
in (2+1) dimension assuming boost invariance in the longitudinal direction.
Relations between
different orders of harmonics
were first discussed in Ref.~\cite{Qiu:2012uy} and  will be
 discussed  in detail in Sec.~\ref{sec:result2}. 
Figure \ref{fig:QiuHeinz} shows
centrality dependences of correlation between
different order of event plane angles
at the LHC energy
using event-by-event viscous hydrodynamic
simulations.
Although there are small deviations from the ATLAS data~\cite{Jia:2012sa},
overall tendency is reproduced in this approach.

\begin{figure}[htb]
\begin{center}
\begin{minipage}[t]{8.5 cm}
\epsfig{file=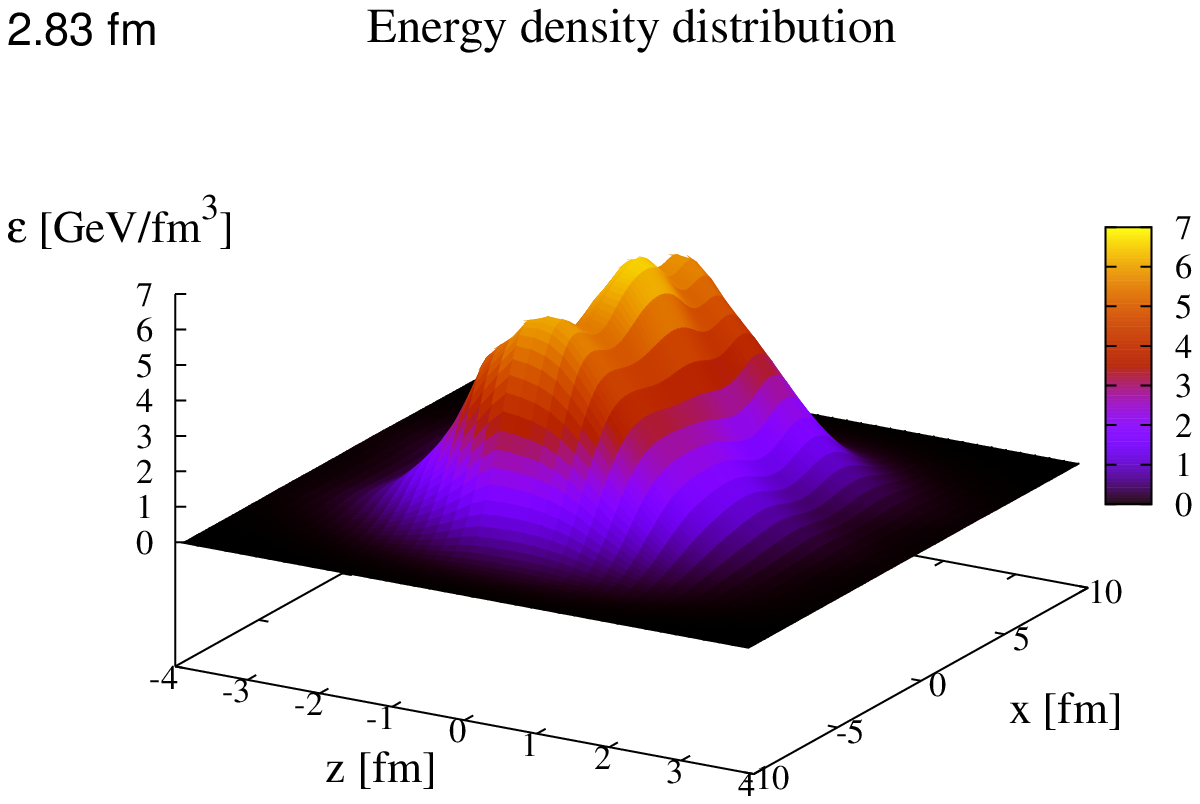,scale=0.7}
\caption{
(Left) Initial energy density
distribution
in the reaction plane of single central event
in Pb+Pb collisions at $E_{\mathrm{lab}}=40A$ GeV from
UrQMD.
Figure taken from Ref.~\cite{Petersen:2008dd}.
\label{fig:UrQMD}
}
\end{minipage}
\hfill
\begin{minipage}[t]{8.5 cm}
\epsfig{file=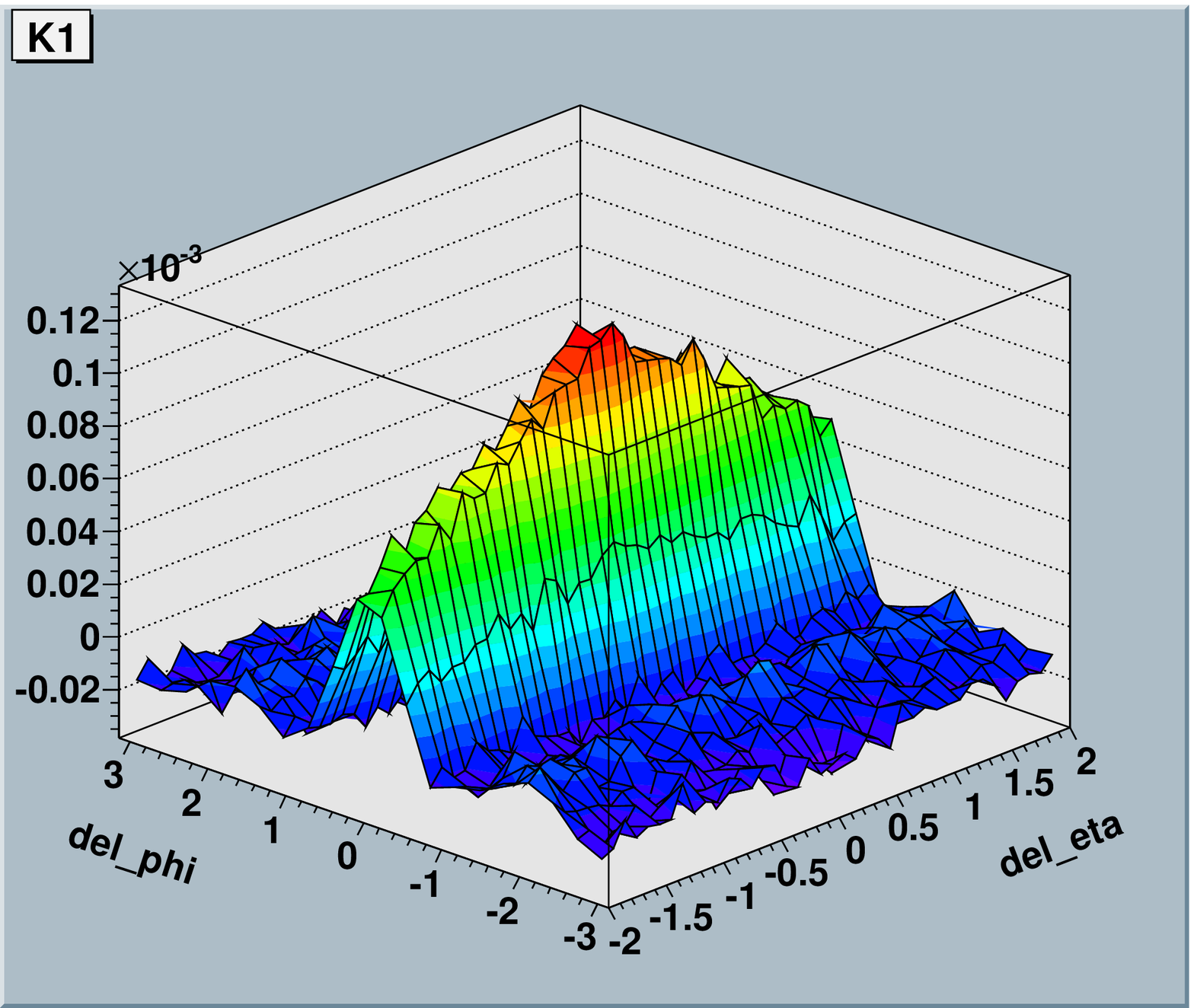,scale=0.4}
\caption{
(Right) Dihadron correlation in $\Delta \eta$-$\Delta \phi$ plane
in central Au+Au collisions at $\sqrt{s_{NN}}=200$ GeV from
event-by-event EPOS+ideal hydrodynamic simulations.
Figure taken from Ref.~\cite{Werner:2010aa}.
\label{fig:EPOS}
}
\end{minipage}
\end{center}
\end{figure}

As mentioned in Sec.~\ref{sec:afterburner}, the
Frankfurt group has developed an
integrated hybrid model based on fully (3+1) dimensional ideal
hydrodynamics and UrQMD. In their model UrQMD is utilised both for
generating the initial conditions and for describing the evolution in
the hadronic 
phase~\cite{Petersen:2008dd,Petersen:2009vx,Petersen:2010md,Petersen:2010cw,Qin:2010pf,Petersen:2011fp,Petersen:2011sb}.
Hydrodynamic simulations as well as UrQMD are
performed in the three-dimensional Cartesian coordinate.
At initial time $t_{\mathrm{start}}$
at which two colliding nuclei are maximally overlapped, $t_{\mathrm{start}} = 2R/\gamma v$,
where $R$ ($v$) is a radius (velocity) of a colliding nucleus,
initial energy density in the computational frame
is calculated as
\begin{equation}
\epsilon (x, y, z) = \sum_{p}\frac{\gamma_{z}}{(2\pi)^{3/2} \sigma^2} E_{p} \exp\left[-\frac{(x-x_{p})^2+(y-y_{p})^2+\gamma_{z}^2(z-z_{p})^2}{2 \sigma^2} \right],
\end{equation}  
where $E_{p}$ is the total energy of a particle (also in the computational frame)
from string fragmentations
and $\gamma_{z}$ Lorentz gamma factor in the beam direction.
The width of the Gaussian is chosen to be $\sigma = 1$ fm as a default value, which
is a little larger compared with other approaches.
Consequently the resultant initial energy density distribution
is smoother than in other Monte-Carlo
  approaches requiring smearing. See, \textit{e.g.}, Fig.~\ref{fig:UrQMD}.

Werner \emph{et al.}~\cite{Werner:2010aa,Werner:2010ny,Werner:2010ss,
Werner:2012xh} utilised
the EPOS event
generator \cite{Werner:2005jf} to generate the initial
conditions for (3+1) dimensional ideal fluid evolution.
EPOS is the successor of NeXus and 
is based on Pomerons and partons.
Nuclear effects such as Cronin effect,
parton saturation, and screening are introduced into EPOS.
Energy momentum tensor
is calculated from four-momenta of string segments $\delta p^{\mu}$
\begin{equation}
\label{eq:epostmunu}
T^{\mu \nu} (x) = \sum_{i} \frac{\delta p^{\mu}_{i}\delta p^{\nu}_{i}}{\delta p^{0}_{i}}g(x-x_{i}),
\end{equation}
where the summation is taken over for each $i$-th segment
and $g$ is a Gaussian type smearing function with a width 0.25 fm.
The energy momentum tensor is then converted to energy density
  and velocity using a similar procedure than what the Rio and Sao Paolo
  groups use.
Since these string segments decayed from flux tubes
correlate in the longitudinal direction,
transverse profiles look quite similar
at each space-time rapidity.
Consequently, this leads to the so-called ridge structure
in the di-hadron correlation function as shown in Fig.~\ref{fig:EPOS}. 

Parametrised initial conditions including higher order
deformation were used in viscous hydrodynamic simulations
to discuss triangular flow by Alver \emph{et al.}~\cite{Alver:2010dn}. 
This is not actually an event-by-event hydrodynamic simulation.
Nevertheless, it captures some features of higher order
deformation in the initial profiles.
An idea behind this is quite similar to the model ``B" 
explained in the previous subsection in our study.
Initial energy density in the transverse plane is parametrised as
\begin{equation}
\epsilon(x, y) = \epsilon_{0} \exp\left\{-\frac{r^2[1+\varepsilon_{n}\cos n(\phi-\psi_{n})]}{2\rho^2} \right\},
\end{equation}
where $r = \sqrt{x^2 +y^2}$ is radial and $\phi$ is azimuthal angle in
the polar coordinate.  $\varepsilon_{n}$ is the magnitude of the
deformation and $\psi_{n}$ is a reference angle. $\rho$ is roughly
root-mean-square radius of the produced matter and taken to be 3 fm.
The deformation $\varepsilon_{n}$ is estimated by either the
MC-Glauber or the MC-KLN model.  They tuned $\eta/s$ to reproduce
centrality dependence of $v_{2}$ at the RHIC energy.  Resultant values
are $\eta/s = 0.08$ and 0.16 for the MC-Glauber and the MC-KLN model,
respectively. Using this parameter, they could reproduce $v_{3}$ as a
function of $N_{\mathrm{part}}$ in the MC-Glauber model. Whereas,
their $v_{3}$ using the MC-KLN model with $\eta/s=0.16$ are
significantly smaller than the $v_{3}$ data.  Thus they concluded
$v_{3}$ data set a severe constraint to the initialisation model in
viscous hydrodynamic simulations.

\begin{figure}[htb]
\begin{center}
\begin{minipage}[t]{6 cm}
\epsfig{file=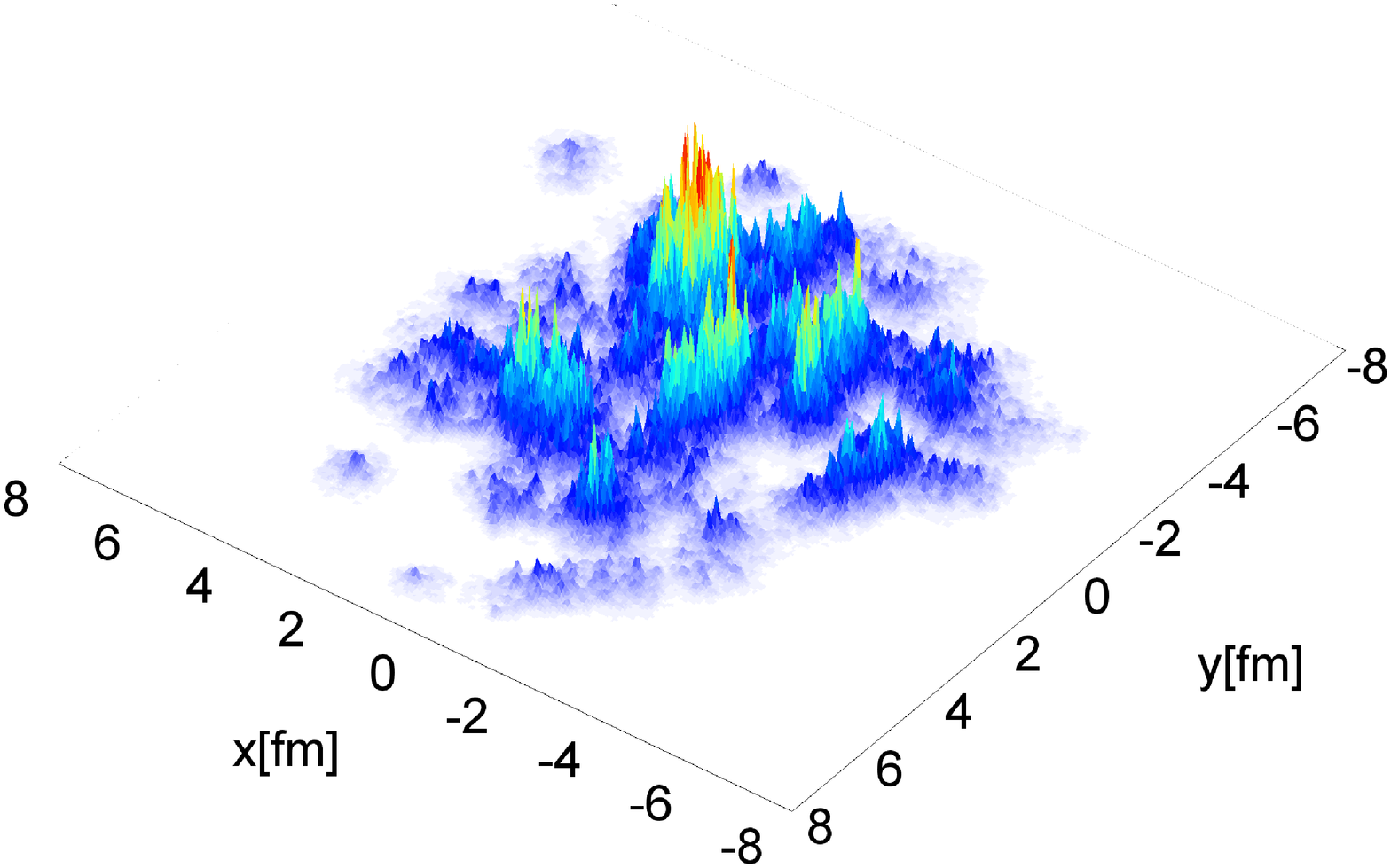,scale=0.2,angle=270}
\end{minipage}
\begin{minipage}[t]{6 cm}
\epsfig{file=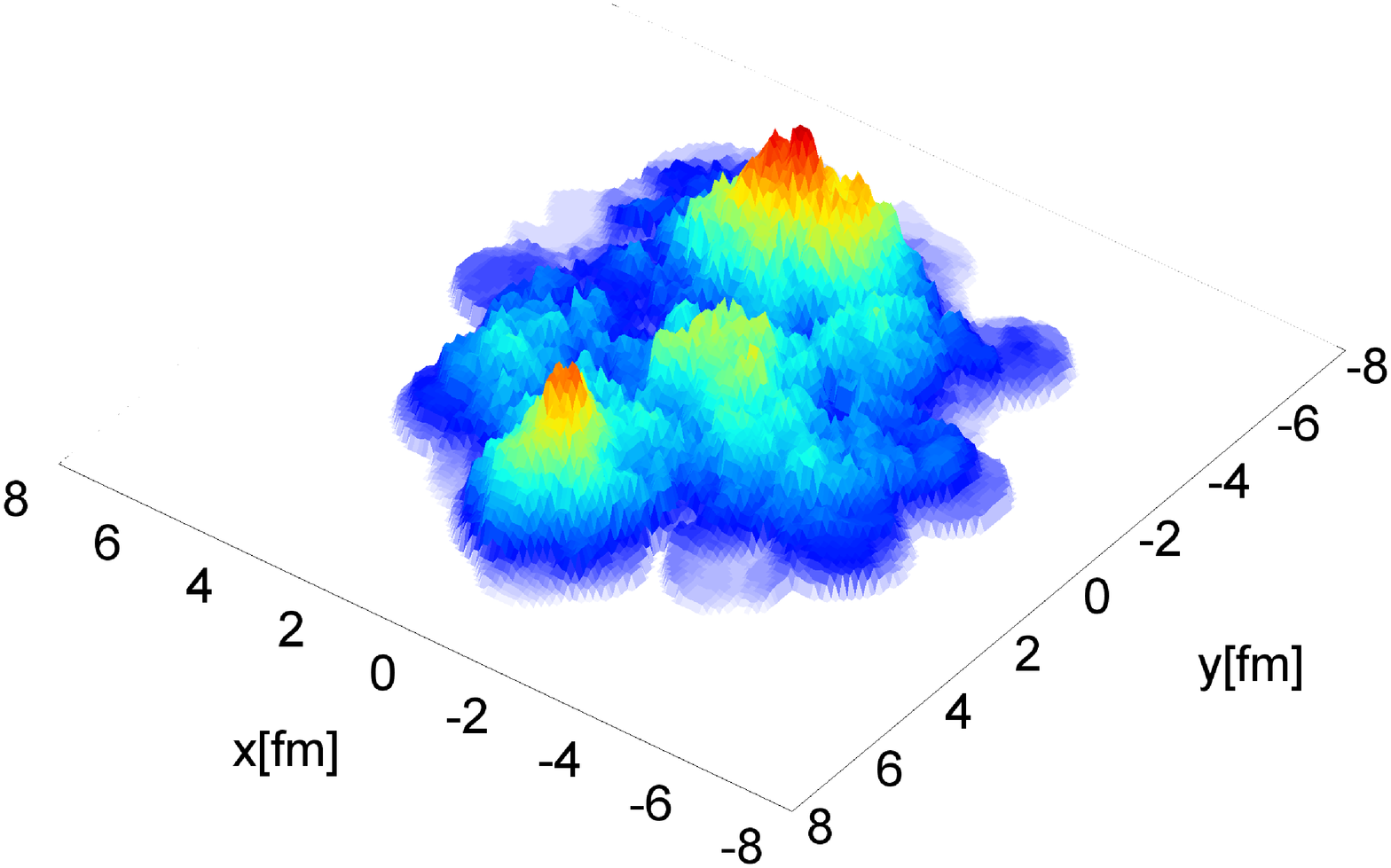,scale=0.2,angle=270}
\end{minipage}
\begin{minipage}[t]{6 cm}
\epsfig{file=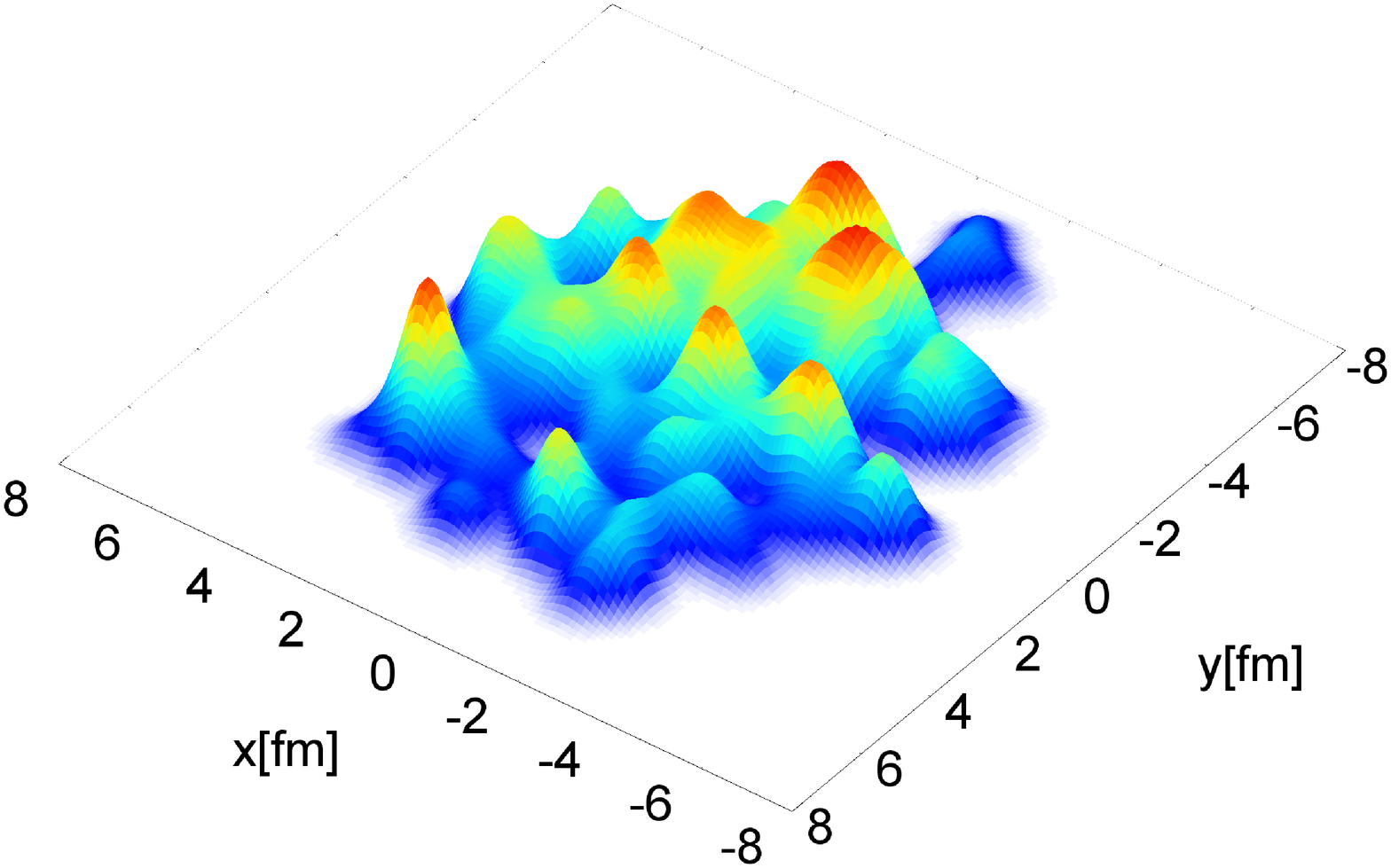,scale=0.2,angle=270}
\end{minipage}
\caption{
Examples of initial energy density distribution 
from the IP-Glasma model at $\tau = 0$ fm (left),
the MC-KLN model (middle) and the MC-Glauber model (right).
Figures are taken from Ref.~\cite{Schenke:2012wb}
}
\label{fig:IP-Glasma}
\end{center}
\end{figure}

Fully (3+1) dimensional viscous hydrodynamic simulations on an event-by-event
basis were performed by Schenke 
\emph{et al.}~\cite{Schenke:2010rr,Schenke:2011bn}.
They have also recently
evaluated fluctuating Glasma initial conditions by solving classical
Yang-Mills equations, and used them as the initial state for
hydrodynamical calculations~\cite{Schenke:2012wb,Gale:2012rq}.
In their first papers they utilise the MC-Glauber
model and initialise the energy density distribution
in the transverse plane as in Eq.~(\ref{eq:ene_trans}) \cite{Schenke:2010rr}.
In the longitudinal direction, 
they assume the energy density follows 
Bjorken scaling solution
near midrapidity and falls like a half Gaussian
near beam rapidity \cite{Schenke:2010rr}.
Recently, the IP-Glasma model
\cite{Schenke:2012wb,Gale:2012rq,Schenke:2012hg} was employed to initialise energy density in hydrodynamic simulations.
The IP-Glasma model solves the classical Yang-Mills equations
in which initial charge distributions of two colliding nuclei
are sampled from a Gaussian distribution with impact parameter
and Bjorken $x$ dependent color charge distributions.
Parametrization of $x$ and impact parameter dependence of 
saturation scale is taken from the IP-Sat (Impact Parameter Saturation) model
\cite{IPsat1,IPsat2}.
In this model event-by-event energy distribution exhibits the
  expected negative binomial distribution and it described the
  observed multiplicity distribution up to a constant scaling
  factor~\cite{Schenke:2012hg}.  Quite remarkably it correctly
  predicts the event-by-event distribution of $v_2$, $v_3$ and
  $v_4$~\cite{Gale:2012rq},
which should be important in understanding
initial fluctuations.
Fluctuations in the IP-Glasma
model have a length scale
of the order of the inverse of the saturation scale
$Q_{s}^{-1}(\bm{x}_{\perp})\sim 0.1$-$0.2$ fm
which is smaller than typical length scales $0.4 \lsim \sigma \lsim 1$ fm
in other calculations.
Figure \ref{fig:IP-Glasma}
shows comparison of initial energy density distribution
among the IP-Glasma, MC-KLN and MC-Glauber models.
Finer structure is seen in the result from the IP-Glasma model.

A simple parametrisation for initial energy density
was employed by Chaudhuri 
in his (2+1)-dimensional
viscous hydrodynamic simulations  
\cite{Chaudhuri:2011qm,Chaudhuri:2011pa}
\begin{equation}
\epsilon(x, y) = \epsilon_{0} \sum_{i=1}^{N_{\mathrm{part}}} \exp\left[-\frac{(\bm{r}-\bm{r}_{i})^2}{2\sigma^2}\right]. 
\end{equation}
The width is set to $\sigma = 1$ fm 
and $\epsilon_{0}$ is chosen to 
reproduce multiplicity in the experimental data. 
$N_{\mathrm{part}}$ is calculated using the optical Glauber
  model, and the positions of the hotspots, $\bm{r}_i$, are assumed to
  be Gaussian distributed.
It is noted that, since the hot-spot would originate from a pair of participating nucleons,
a scheme to randomly sample independent position of hot-spots
using averaged distribution of $N_{\mathrm{part}}$ in the transverse plane
does not capture the actual distribution of correlated hot-spots from
the MC-Glauber model.
Later, the effect of choice of smearing profile on final $v_{2}$ and $v_{3}$
in the context of the MC-Glauber model
was examined by choosing 
the Wood-Saxon profile instead of the Gaussian one \cite{RihanHaque:2012wp}.
They found that the anisotropy coefficients $v_2$ and $v_3$ were not
  affected by the form of the smearing profile nor by the value of the
  smearing parameter, whereas coefficients $v_4$ and $v_5$ showed some
  sensitivity to the smearing.

Bozek {\it et al.} employed Monte-Carlo Glauber model GLISSANDO \cite{Broniowski:2007nz}
for initialisation in their event-by-event 
full (3+1)-dimensional viscous hydrodynamic simulations \cite{Bozek:2012fw}.
Initialization of the entropy profile in the transverse plane is quite similar
to others:
\begin{eqnarray}
s(x,y) & = & \kappa \sum_{i} g_{i}(x,y) [(1-\alpha) + N_{i}^{\mathrm{coll}}\alpha],\\
g_{i}(x,y) & = & \frac{1}{2 \pi w^2}\exp \left[-\frac{(x-x_{i})^2+(y-y_{i})^2}{2w^2}\right].
\end{eqnarray}
Here the summation is taken over participants and 
$N_{i}^{\mathrm{coll}}$ is the number of collisions
of the $i$-th participant.
Similar to other groups, the transverse position of nucleons
is smeared from a point-like source to a Gaussian profile with a width $w = 0.4$ fm.
Soft-hard mixture $\alpha = 0.125$ leads to reproduction of
centrality dependence of $dN_{\mathrm{ch}}/d\eta$ at the top RHIC energy
within the full (3+1)-dimensional viscous hydrodynamic simulations.
Note that, 
since the Glauber approach tells us only profiles in the transverse plane,
longitudinal profiles of the produced matter have to be also modeled 
by taking account of momentum conservation among colliding
nucleons at certain transverse position \cite{Bozek:2012fw}.
\begin{eqnarray}
f_{\pm}(\eta_{\mathrm{s}}) & = & \left(1\pm \frac{\eta_{\mathrm
{s}}}{y_{\mathrm{beam}}}\right)f(\eta_{\mathrm{s}}),\\
f(\eta_{s}) & = & \exp \left[-\frac{(\eta_{\mathrm{s}}-\eta_{0})^2}{2\sigma_{\eta}^2}\theta(|\eta_{\mathrm{s}}|-\eta_{0}) \right],
\end{eqnarray}
where $y_{\mathrm{beam}}$ is the beam rapidity.
Using this model, they calculated transverse momentum fluctuations 
at the RHIC energy and 
found the experimental data can be explained by
the event-by-event fluctuations of initial profiles.

Recently, Pang {\it et al.} employed the AMPT event generator 
for initialisation in event-by-event (3+1)-dimensional 
ideal hydrodynamic simulations \cite{Pang:2012he}.
Information about phase space density from the AMPT simulation
at some initial time $\tau_{0}$
is used to calculate local energy-momentum tensor
\begin{equation}
\label{eq:PangEM}
T^{\mu \nu} (\tau_{0}, x, y, \eta_{s}) = K \sum_{i}\frac{p_{i}^{\mu}p_{i}^{\nu}}{p_{i}^{\tau}}
\frac{1}{2\pi\sigma_{r}^2}\exp\left[-\frac{(x-x_{i})^2+(y-y_{i})^2}{2\sigma_{r}}\right]\frac{1}{\tau_{0}\sqrt{2\pi \sigma_{\eta_{s}^2}}}\exp\left[-\frac{(\eta_{s}-\eta_{s,i})^2}{2\sigma_{\eta_{s}}^2}\right],
\end{equation}
where $p^{\tau}= m_{T}\cosh(Y-\eta_{s})$ for the $i$-th parton.
The widths in transverse and longitudinal directions
are chosen as being $\sigma_{r} = 0.6$ fm and $\sigma_{\eta_{s}} = 0.6$, respectively.
They allowed one parameter $K$ to reproduce
experimentally-measured multiplicity at midrapidity.
In actual calculations at the LHC energy,
$K = 1.6$ and $\tau_{0} = 0.2$ fm.
Through this approach, one can include
fluctuations of the density profile in the longitudinal direction
as well as in the transverse plane
and of flow velocities in the initial conditions.
They found that  initial fluctuations in rapidity
distributions lead to expanding hot spots in the longitudinal direction
and that fluctuations in the initial flow velocities lead to harder
transverse momentum spectra of final hadrons due to non-vanishing
initial radial flow velocities.

Zhang {\it et al.} investigated the effect of initial
fluctuation on jet energy loss
using (2+1) dimensional ideal hydrodynamic model~\cite{Zhang:2012ik}.
They found that, compared with smooth initial conditions,
a jet loses slightly more energy
in the expanding QGP with
fluctuating initial conditions.

Table \ref{table:e-by-e}
summarises current status of event-by-event hydrodynamic
simulations
by focusing on initialisation models and
dimension  of hydrodynamic simulations.
Note that there are many more models for initializing the
  event-by-event calculations than there are hadronic cascade models
  used in the hydro + cascade models (Table~\ref{table:h2c}).

\begin{table}[htb]
\caption{
Event-by-event hydrodynamic simulations.
\label{table:e-by-e}
}
\begin{tabular}{|l|l|l|l|}
\hline
Authors and References & Initialisation Model & Dimension & Ideal/Viscous \\
\hline
\hline
Gyulassy {\it et al.} \cite{Gyulassy:1996br} & HIJING & (2+1)-D & Ideal\\
\hline
Aguiar {\it et al.}\cite{Aguiar:2001ac,Hama:2004rr,Andrade:2006yh,Andrade:2008xh,Takahashi:2009na,
  Gardim:2011qn} & NeXus & (3+1)-D & Ideal\\
\hline
Ren {\it et al.} \cite{Ren:2008sc} & NeXus & (3+1)-D & Ideal\\
\hline
Holopainen {\it et al.} \cite{Holopainen:2010gz,Chatterjee:2011dw,Renk:2011qi} & MC-Glauber & (2+1)-D & Ideal\\
\hline
Qiu and Heinz \cite{Qiu:2011iv} & MC-Glauber, MC-KLN & (2+1)-D & Ideal\\
\hline
Petersen {\it et al.}~\cite{Petersen:2008dd,Petersen:2009vx,
  Petersen:2010md,Petersen:2010cw,Qin:2010pf,Petersen:2011fp,Petersen:2011sb}
& UrQMD & (3+1)-D & Ideal\\
\hline
Werner {\it et al.}~\cite{Werner:2010aa,Werner:2010ny,Werner:2010ss,Werner:2012xh} & EPOS & (3+1)-D & Ideal\\
\hline
Alver {\it et al.}~\cite{Alver:2010dn} & Parametrisation & (2+1)-D & Viscous\\
\hline
Schenke {\it et al.}~\cite{Schenke:2010rr,Schenke:2011bn,Schenke:2012wb} &
MC-Glauber, IP-Glasma & (3+1)-D & Viscous\\
\hline
Chaudhuri {\it et al.}~\cite{Chaudhuri:2011qm,RihanHaque:2012wp}
 & Parametrisation & (2+1)-D & Viscous \\ 
\hline 
Bozek and Broniowski~\cite{Bozek:2012fw} & MC-Glauber & (3+1)-D & Viscous\\
\hline
Pang {\it et al.}~\cite{Pang:2012he} & AMPT & (3+1)-D & Ideal \\
\hline
Zhang {\it et al.}~\cite{Zhang:2012ik} & MC-Glauber & (2+1)-D & Ideal \\
\hline
This study & MC-Glauber, MC-KLN & (3+1)-D & Ideal \\
\hline
\end{tabular}
\end{table}

\section{Results from smooth initial profile \label{sec:result1}}

In this section, we show results from the integrated dynamical model
starting from conventional smooth initial entropy 
density distributions
to describe a high-energy heavy ion reaction as a whole at LHC and RHIC.
We show $p_{T}$ spectra for pions, kaons and protons,
$v_{2}$ for charged hadrons
as functions of centrality and $p_{T}$
and $v_{2}(p_{T})$ for identified hadrons.
For the models of initialisations, MC-KLN and MC-Glauber models 
are employed and the results obtained using
them are compared with each other.
We will compare some of these results with those from event-by-event
hydrodynamic simulations later.

\subsection{\it Results at RHIC\label{sec:RHICresult}}

\begin{figure}[htb]
\begin{center}
\begin{minipage}[t]{6 cm}
\epsfig{file=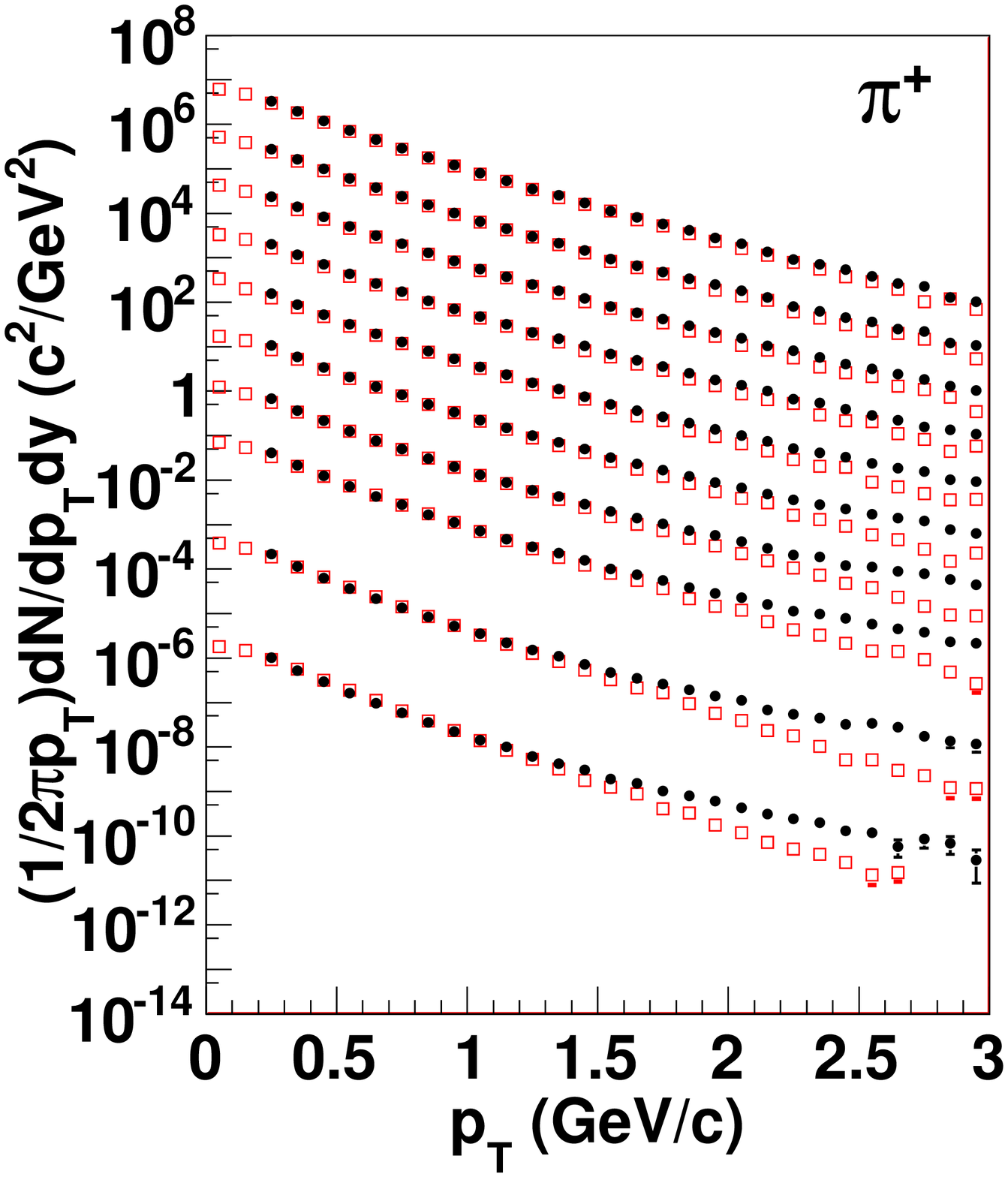,scale=0.3}
\end{minipage}
\begin{minipage}[t]{6 cm}
\epsfig{file=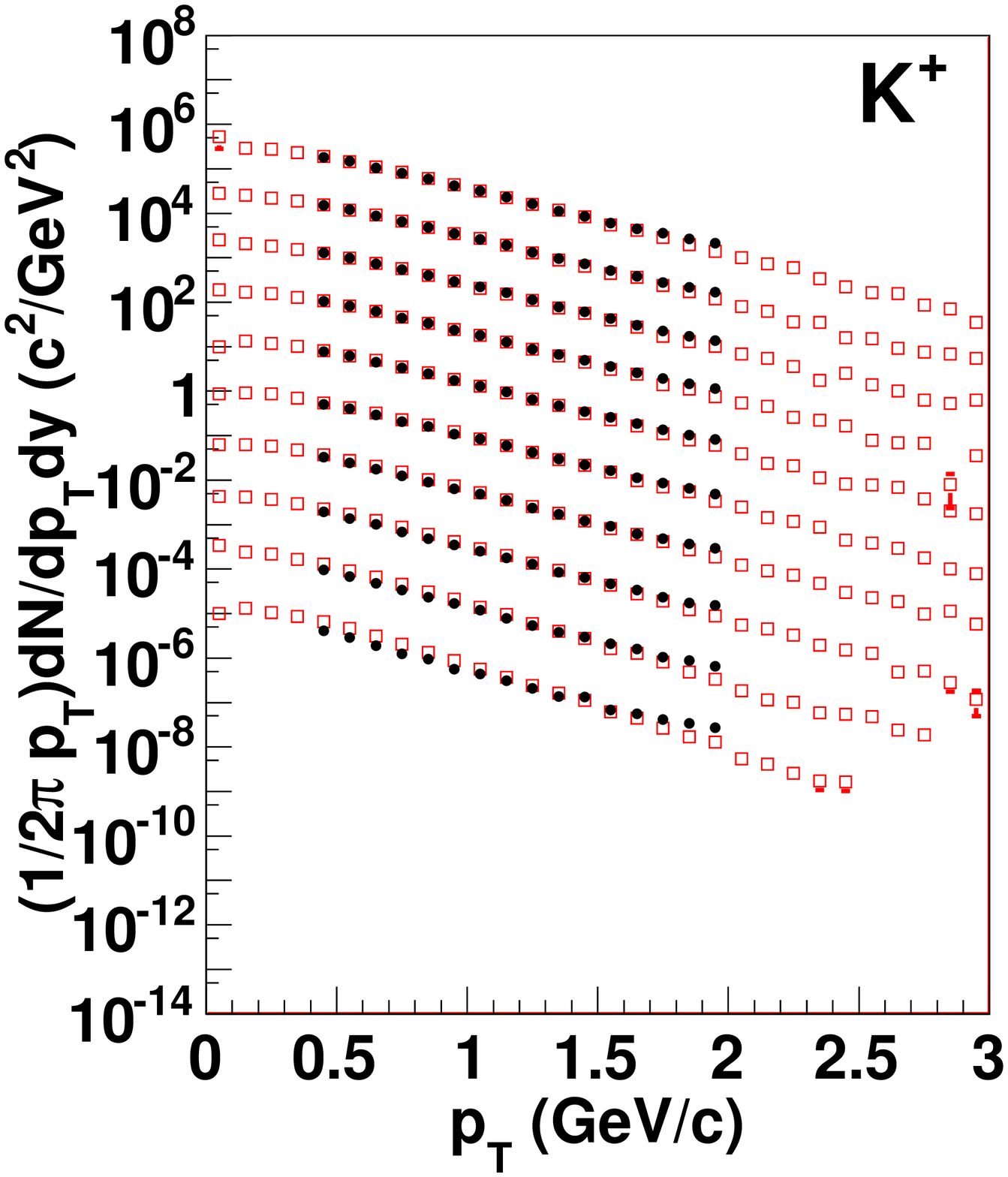,scale=0.3}
\end{minipage}
\begin{minipage}[t]{6 cm}
\epsfig{file=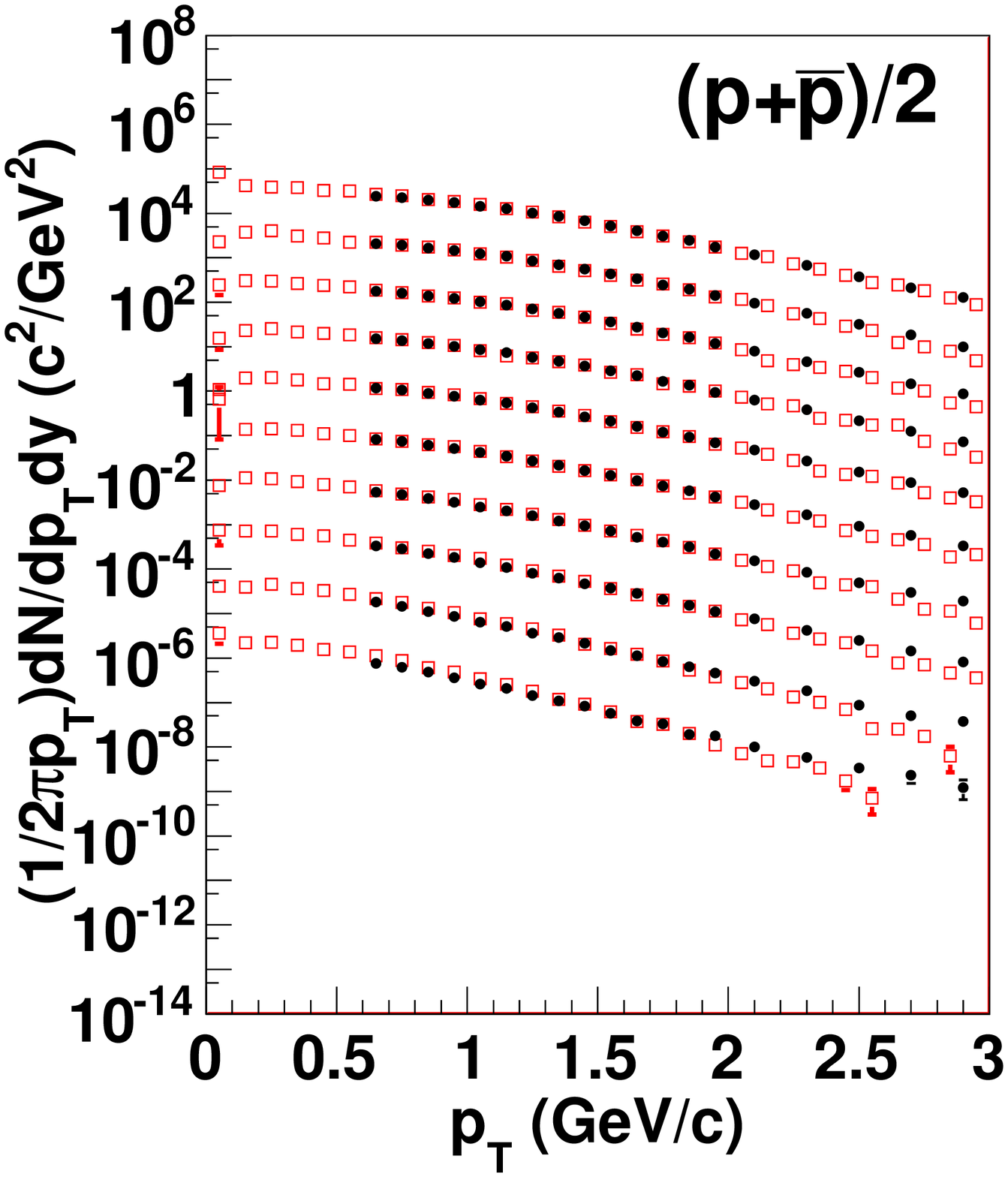,scale=0.3}
\end{minipage}
\caption{
Transverse momentum distributions of
 $\pi^{+}$ (left), $K^{+}$ (middle)
and ($p$+$\bar{p}$)/2 (right)
in $\sqrt{s_{NN}}$= 200 GeV Au+Au collisions 
at 0-5, 5-10, 10-15, 15-20, 20-30, 30-40,
40-50, 50-60, 60-70 and
70-80\% centralities from top to bottom.
Both the PHENIX data~\cite{Adler:2003cb} (filled circle) and results calculated
 using the KLN initial conditions (open square) are shown.
To show all these results, each spectrum is
multiplied by 10$^{n}$
with $n = 4, 3, 2, \cdots, -5$ from top to bottom for kaons and protons.
For pions,  $n = 4, 3, 2, \cdots, -3$, $-5$ and $-7$.
\label{fig:ptKLN}
}
\end{center}
\end{figure}

\begin{figure}[htb]
\begin{center}
\begin{minipage}[t]{6 cm}
\epsfig{file=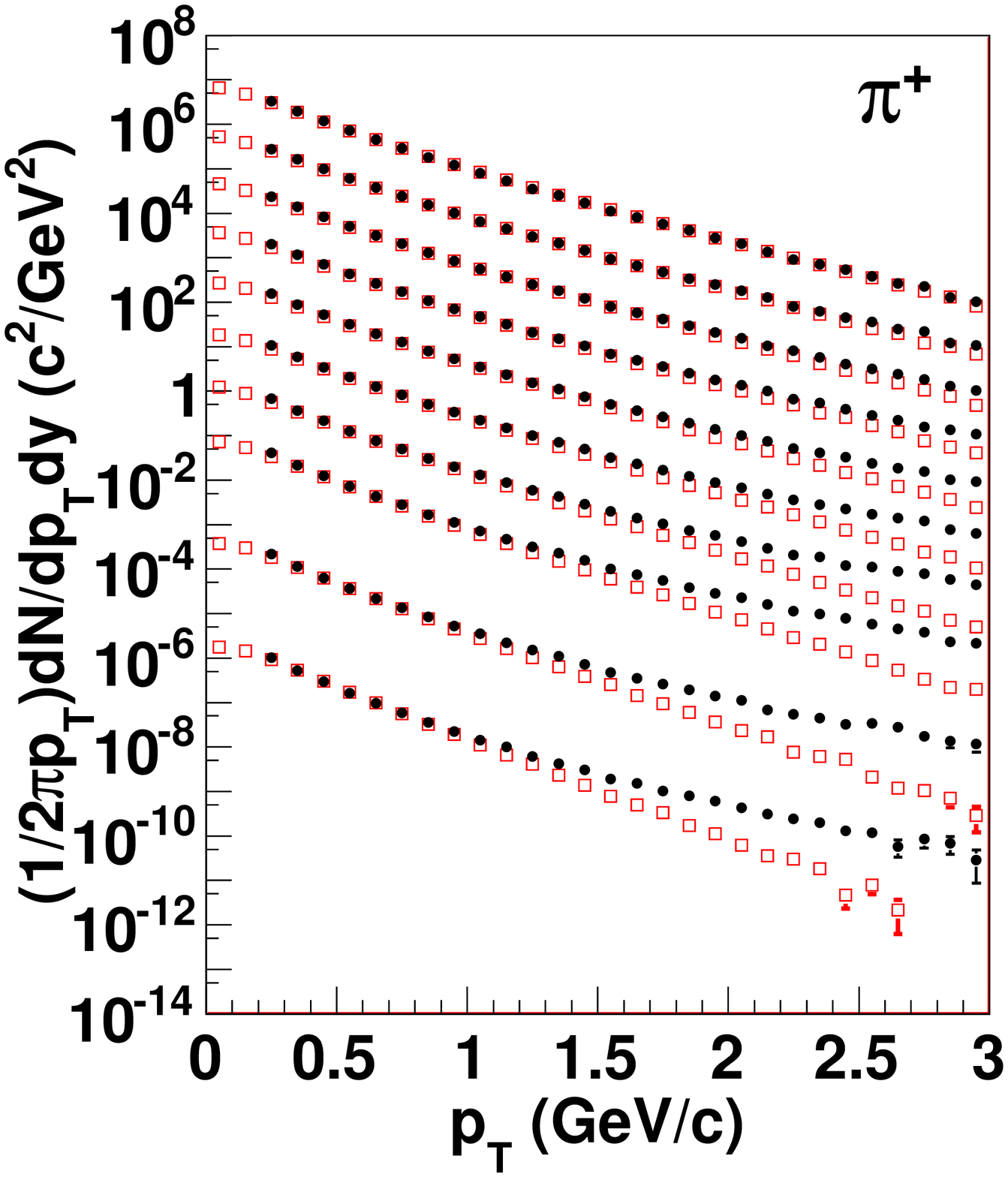,scale=0.3}
\end{minipage}
\begin{minipage}[t]{6 cm}
\epsfig{file=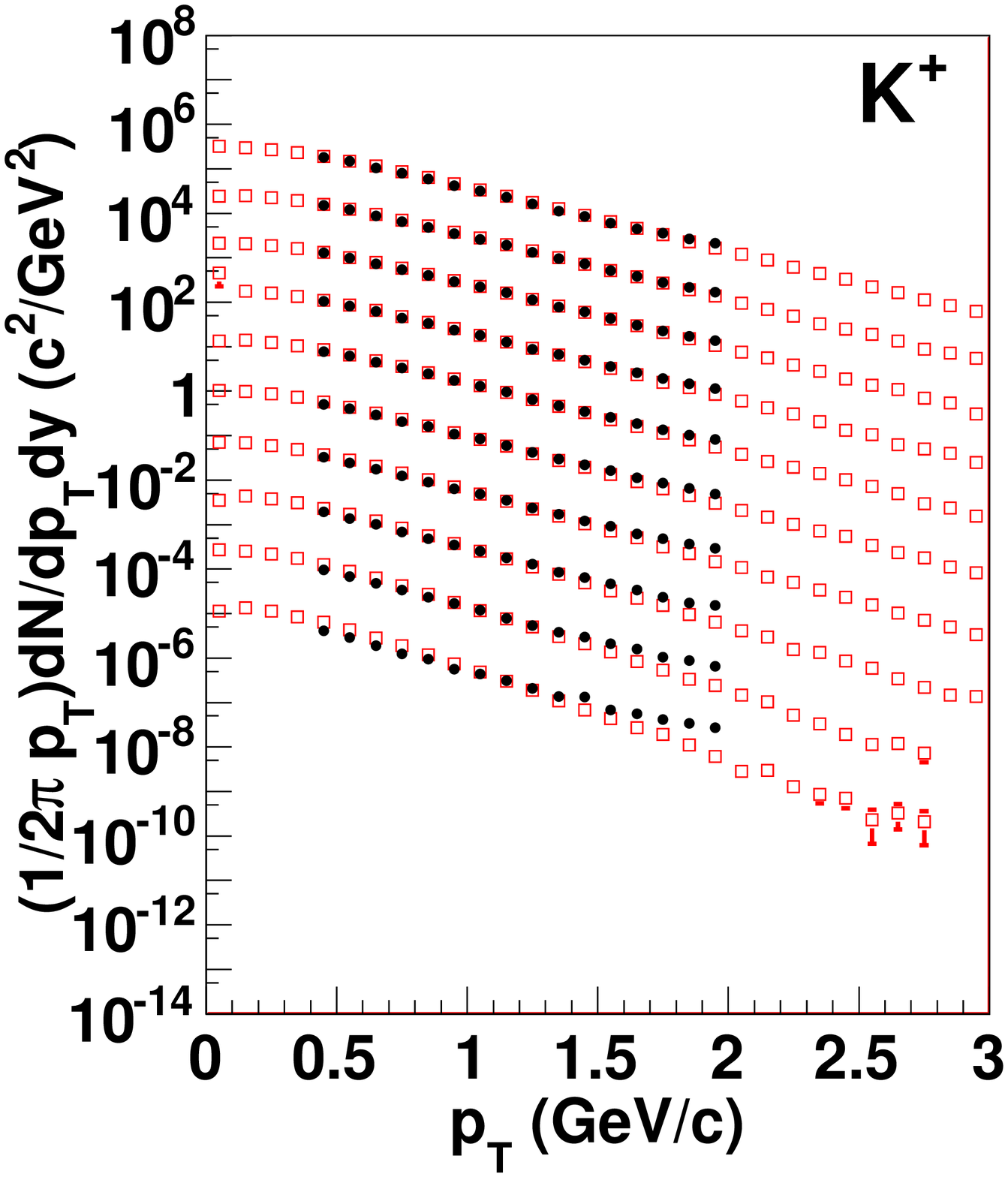,scale=0.3}
\end{minipage}
\begin{minipage}[t]{6 cm}
\epsfig{file=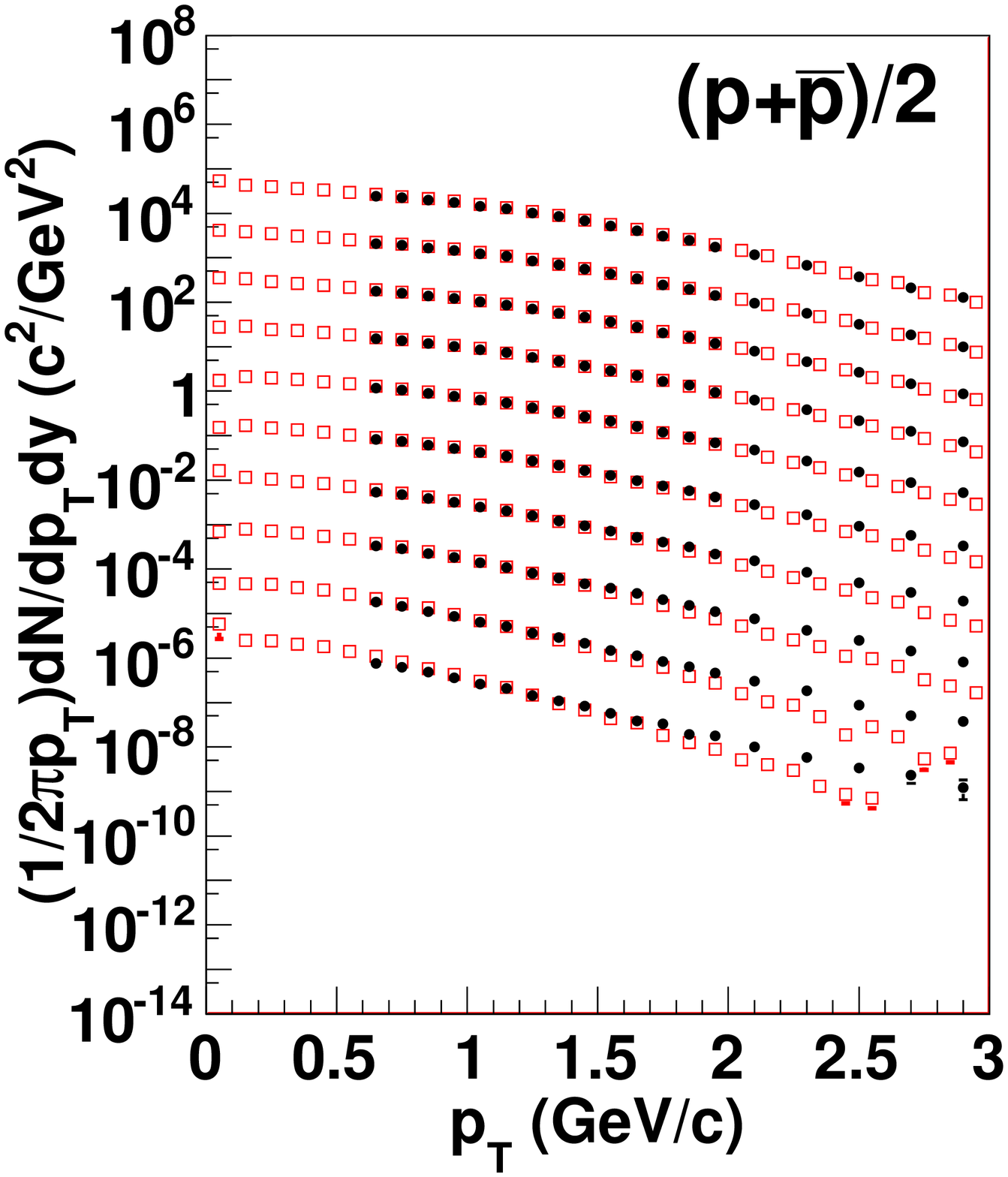,scale=0.3}
\end{minipage}
\caption{
The same as Fig.~\ref{fig:ptKLN} but 
 using the Glauber model initial conditions.
\label{fig:ptGlauber}
}
\end{center}
\end{figure}

In Figs.~\ref{fig:ptKLN} and \ref{fig:ptGlauber} we show
  transverse momentum distributions of positive pions and kaons, and
  the average of protons and antiprotons around midrapidity ($\mid
  \eta \mid < 0.35$) in $\sqrt{s_{NN}}$ = 200 GeV Au+Au
  collisions. The results calculated using the KLN model and Glauber
  model initial conditions (Figs.~\ref{fig:ptKLN} and
  \ref{fig:ptGlauber}, respectively) are compared with the PHENIX data
  \cite{Adler:2003cb}.
Although initial conditions are taken from ``Model B" in 
both cases,
results from ``Model A" (simple average over many samples)
 are almost identical to these.
In central collisions, we reproduce the PHENIX data \cite{Adler:2003cb}
well up to $p_{T}\sim$ 3 GeV/$c$.
The $p_{T}$ region where the model works well
becomes smaller as going to peripheral collisions:
For example, in 70-80\% centrality, we reproduce $p_{T}$ spectrum for pions
up to $p_{T} \sim 1$ GeV/$c$ and, beyond this, other components
such as recombination and/or jet fragmentation, which are
missing in the current integrated model, may be dominant.
$p_{T}$ slopes from the KLN model are a little harder than
those from the Glauber model.
It should be noted here that
we chose the switching temperature  $T_{\mathrm{sw}}$ = 155 MeV
to obtain the  observed particle ratios
of these identified hadrons, not to reproduce $p_{T}$ slope.
Final particle spectra are expected to be independent of a choice of the switching temperature.
However, this is not the case in the current 
calculations: The particle yields depend on the
switching temperature and thus it can be fixed
using the particle ratios.
Similar sensitivity to the value of $T_{\mathrm{sw}}$ was seen
  also in Ref.~\cite{Song:2011hk}, where a systematic analysis of
  switching temperature dependence was made.

\begin{figure}[tb]
\begin{center}
\begin{minipage}[t]{9 cm}
\epsfig{file=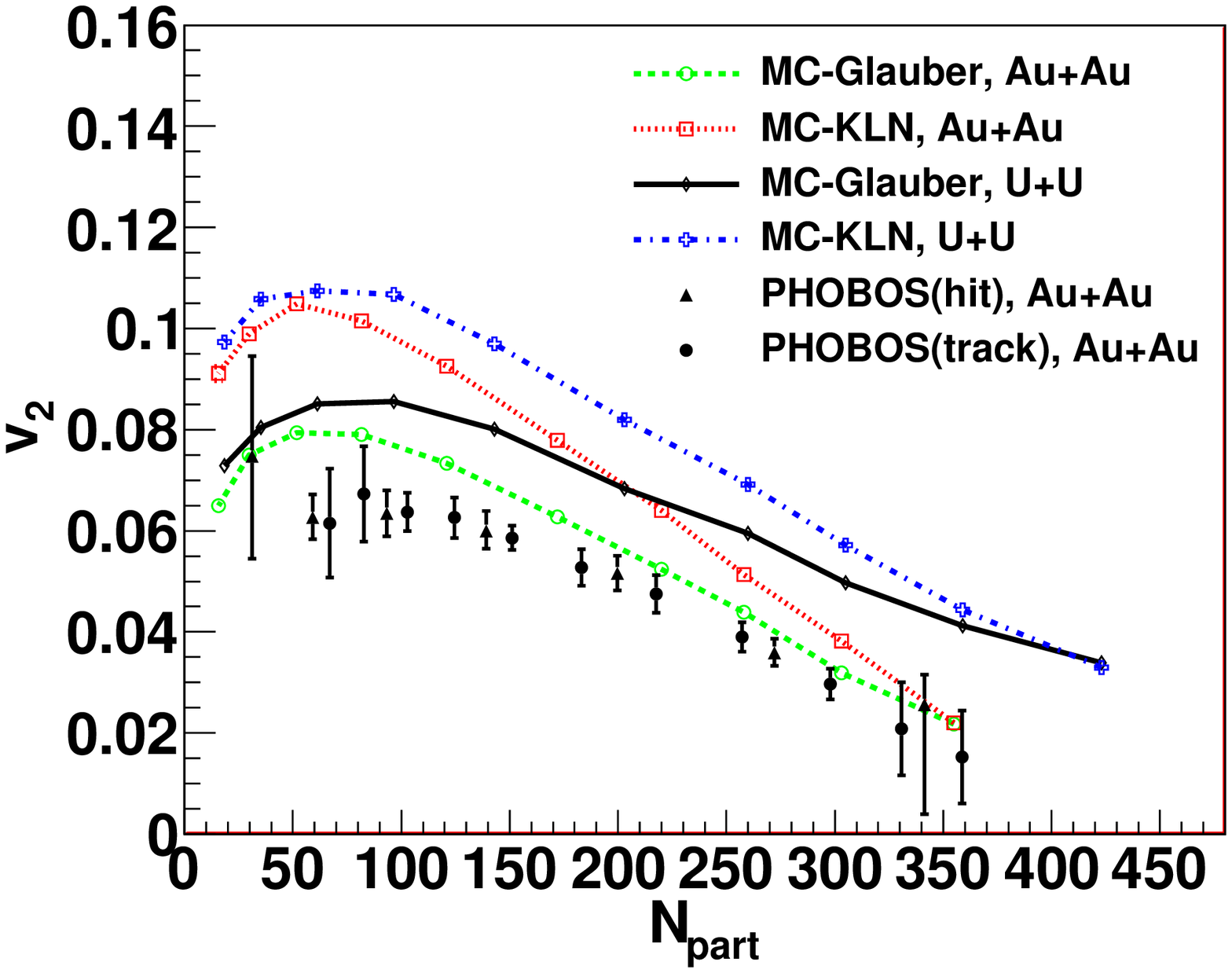,scale=0.45}
\end{minipage}
\begin{minipage}[t]{9 cm}
\epsfig{file=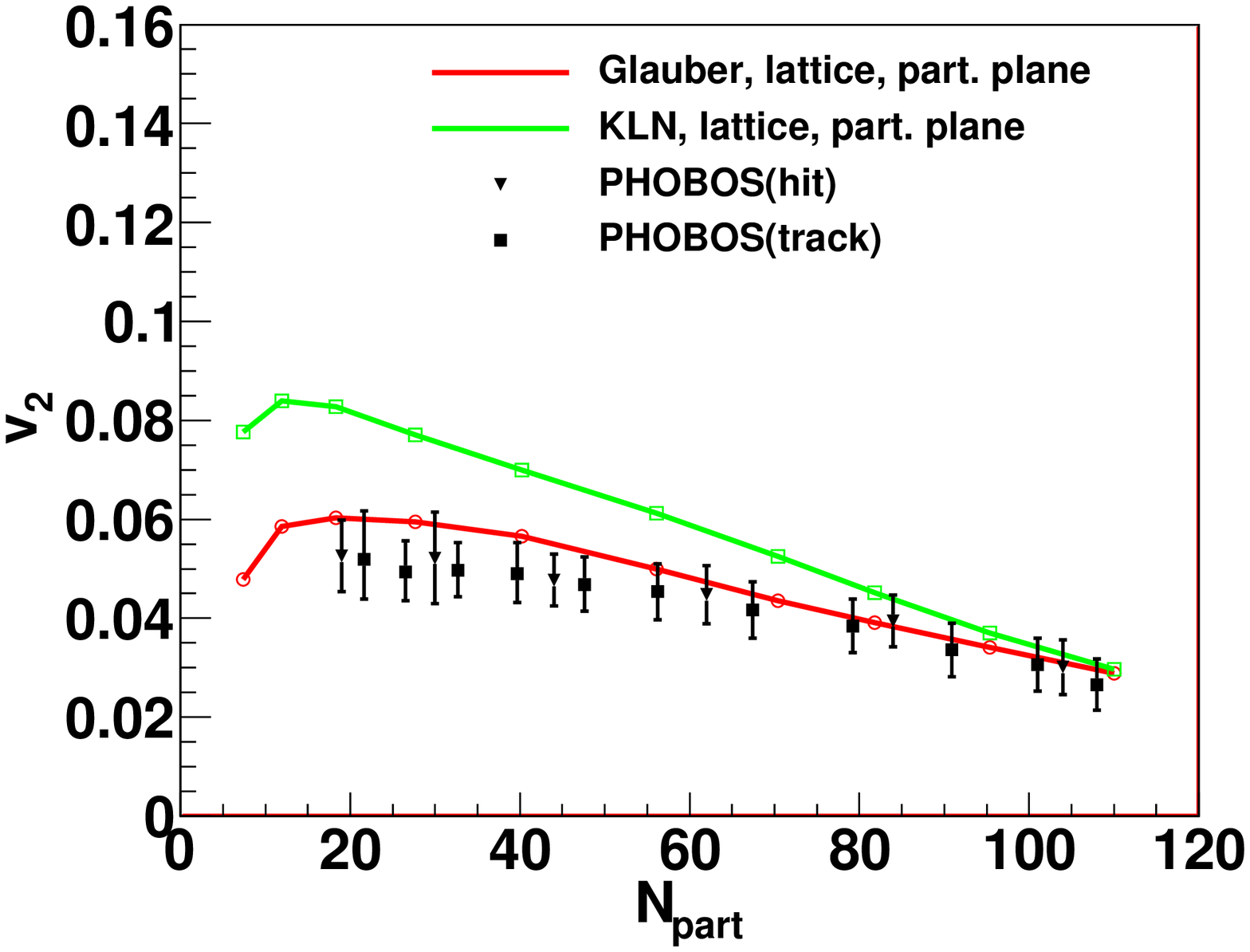,scale=0.45}
\end{minipage}
\caption{
$N_{\mathrm{part}}$ dependence of $v_{2}$ evaluated using model ``B'' (see the text) 
in Au+Au  (left) and Cu+Cu (right) collisions
at $\sqrt{s_{NN}}$ = 200 GeV
are compared with the PHOBOS data \cite{PHOBOSv2-2,PHOBOSv2-3}.
Predictions in U+U collisions at $\sqrt{s_{NN}}$ = 200 GeV
are also shown in the left figure (taken from Ref.~\cite{Hirano:2010jg}).
\label{fig:v2centPHOBOS}}
\end{center}
\end{figure}

In Fig.~\ref{fig:v2centPHOBOS} (left),
$v_{2}$ in Au+Au collisions is compared with
$v_{2}$ in U+U collisions at $\sqrt{s_{NN}}$ = 200 GeV.
Here initial conditions are taken from ``Model B"
and the momentum distribution is
integrated over $\mid \eta \mid < 1$ and the whole $p_{T}$ region.
Since larger eccentricity leads to
larger momentum anisotropy and $v_{2}$, the systematics of
$v_2(N_{\mathrm{part}})$ is similar to that of
$\varepsilon_{\mathrm{part}}(N_{\mathrm{part}})$ as shown in Fig.~\ref{fig:epspart} (left).
$v_{2}$ is larger in
U+U collisions than in Au+Au collisions, and KLN initialisation
leads to larger $v_{2}$ than Glauber initialisation.
$v_2$ first increases
with decreasing $N_{\mathrm{part}}$,
which reflects increasing initial eccentricity.
When $N_{\mathrm{part}}$ falls below  $\sim 50$, $v_{2}$, however, begins 
to decrease. This is due to the short lifetime of the system which 
does not allow the flow to fully build up, and to the large fraction 
of the lifetime spent in the hadronic phase where dissipative effects
are large.

In Au+Au collisions, results from the Glauber initialisation
almost reproduce the PHOBOS data \cite{PHOBOSv2-2}.
This indicates that there is little room for QGP viscosity
in the model calculations.
On the other hand, apparent discrepancy
between the results from the KLN initialisation
and the PHOBOS data means that 
viscous corrections during the fluid evolution are required.
Figure~\ref{fig:v2centPHOBOS} (right) shows
a comparison of results with the PHOBOS data in Cu+Cu collisions \cite{PHOBOSv2-3}.
Again, the Glauber model initialisation 
almost reproduces PHOBOS data, while
the KLN initialisation leads to larger $v_{2}$
than  the PHOBOS data.

As expected, the system in U+U
collisions at $\sqrt{s_{NN}}$ =200 GeV is denser than in Au+Au
collisions at the same energy. 
At initial time $\tau_{0}$ = 0.6
fm/$c$, the maximum temperature (energy density) in the most central
5\% of U+U collisions is $T_{0}=367$ MeV ($e_{0}=33.4$ GeV/fm$^{3}$)
and $T_{0}=361$ MeV ($e_{0}=31.4$ GeV/fm$^{3}$) in the Au+Au
collisions of the same centrality. This corresponds to charged
particle transverse densities of 25.4 and 24.1, respectively, which
means that the transverse density in U+U collisions is indeed larger,
but only by $\sim$ 6\%.\footnote{With sufficient statistics, one may
  make more severe centrality cut (e.g., 0-3\%) to obtain larger
  transverse particle density.  Multiplicity fluctuation in the
  centrality cut, which we do not take into account, could also
  enhance the transverse particle density.}

\begin{figure}[tb]
\begin{center}
\begin{minipage}[t]{9 cm}
\epsfig{file=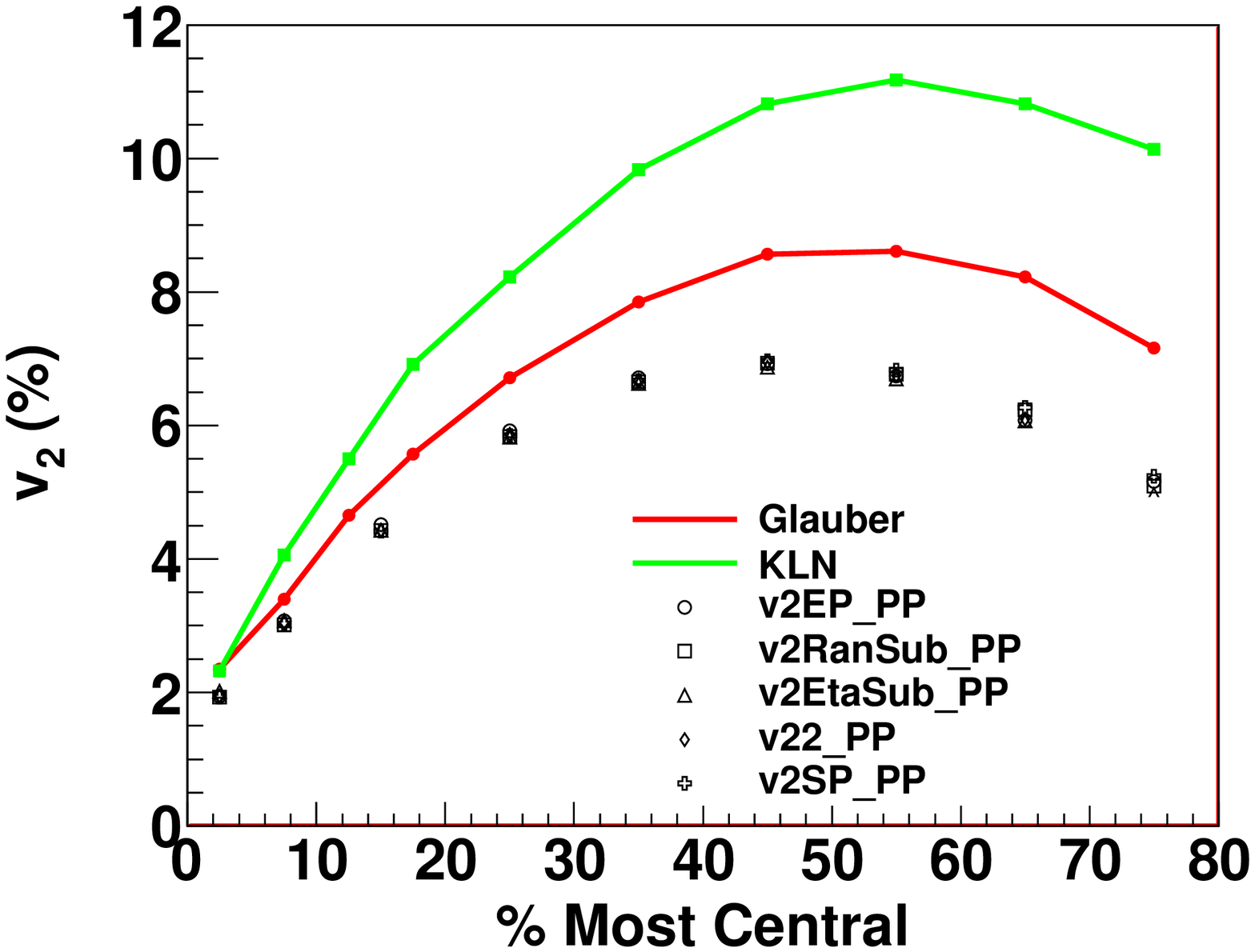,scale=0.45}
\end{minipage}
\begin{minipage}[t]{9 cm}
\epsfig{file=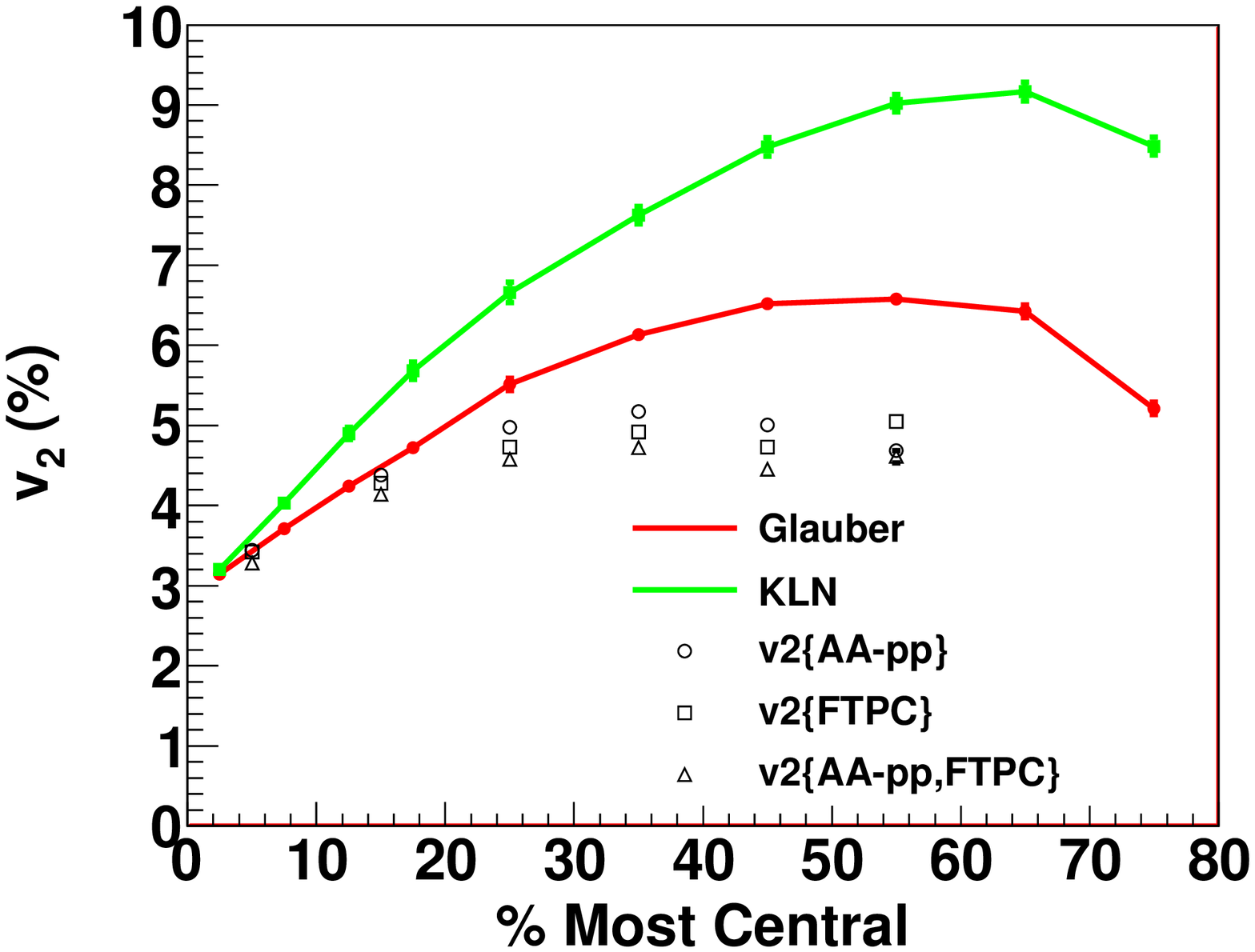,scale=0.45}
\end{minipage}
\caption{
Centrality dependence of $v_{2}$ with respect to
participant plane (model ``B'') in Au+Au  (left)
and Cu+Cu (right) collisions
at $\sqrt{s_{NN}}$ = 200 GeV
are compared with the STAR data \cite{STARv2-4,Abelev:2010tr}
($0.15 < p_{T} < 2$ GeV/$c$ and $\mid \eta \mid < 1$).
Experimental data in Au+Au collisions are as corrected
in Ref.~\cite{Ollitrault:2009ie}.
\label{fig:v2centSTAR}}
\end{center}
\end{figure}

Figure \ref{fig:v2centSTAR} shows again the centrality dependences
of $v_{2}$
in Au+Au (left) 
and Cu+Cu (right) collisions at $\sqrt{s_{NN}}$ = 200 GeV.
Here initial conditions are taken from Model ``B"
and the momentum distribution is
integrated over  $\mid \eta \mid < 1$ and $0.15 < p_{T} < 2$ GeV/$c$
according to experimental setup in STAR \cite{STARv2-4,Abelev:2010tr}.
Experimental data in Au+Au collisions \cite{STARv2-4}
have been corrected to subtract non-flow effects \cite{Ollitrault:2009ie} and, thus,
all data sets from various flow analysis methods coincide with each other.
Notice that the data have not been corrected yet in Cu+Cu collisions \cite{Abelev:2010tr}.
Similar to the results in Fig.~\ref{fig:v2centPHOBOS},
the Glauber initialisation little overshoots the STAR data, while
the KLN initialisation leads to larger $v_{2}$ than 
the Glauber 
initialisation clearly overshooting the STAR data.

\begin{figure}[tb]
\begin{center}
\begin{minipage}[t]{6 cm}
\epsfig{file=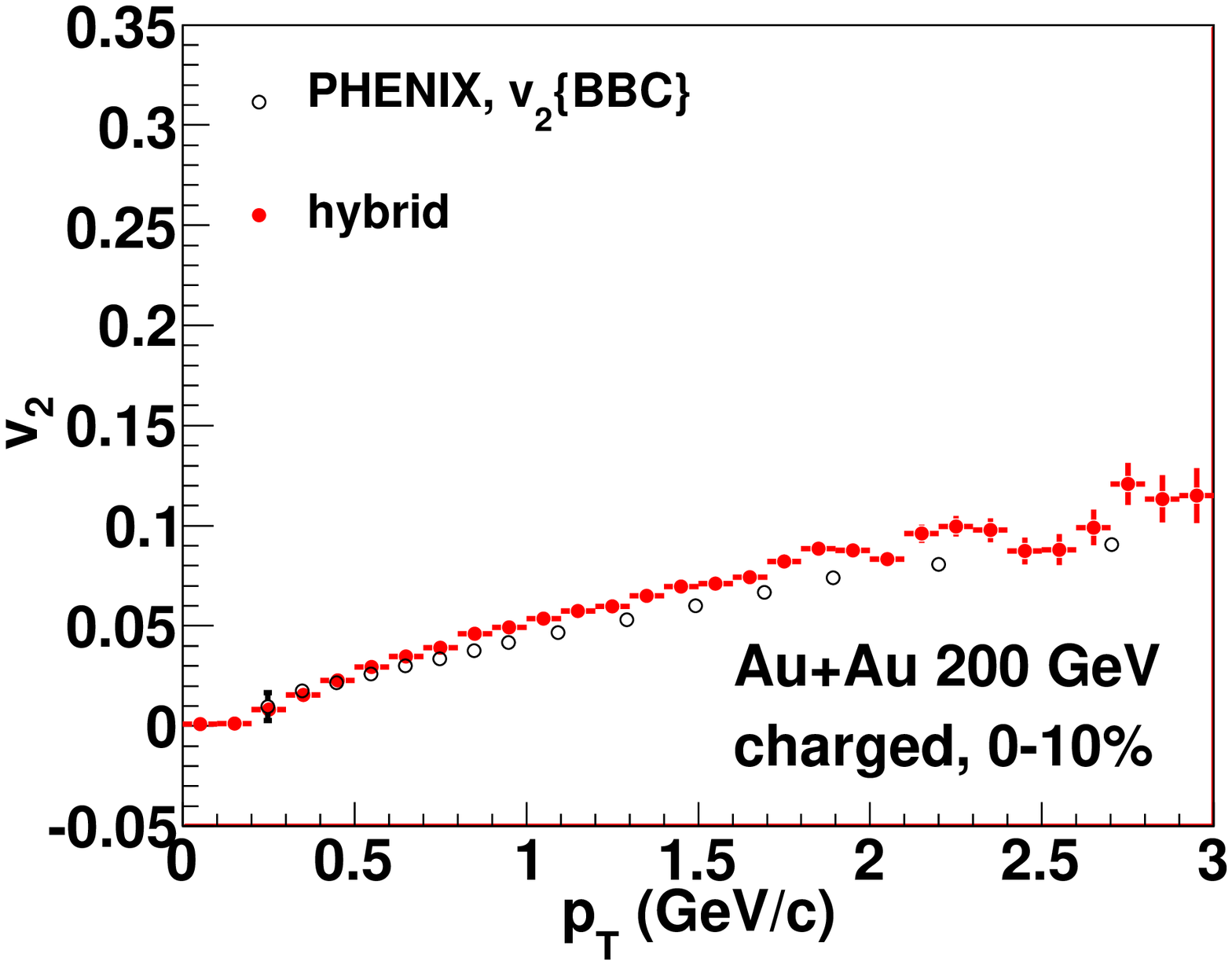,scale=0.3}
\end{minipage}
\begin{minipage}[t]{6 cm}
\epsfig{file=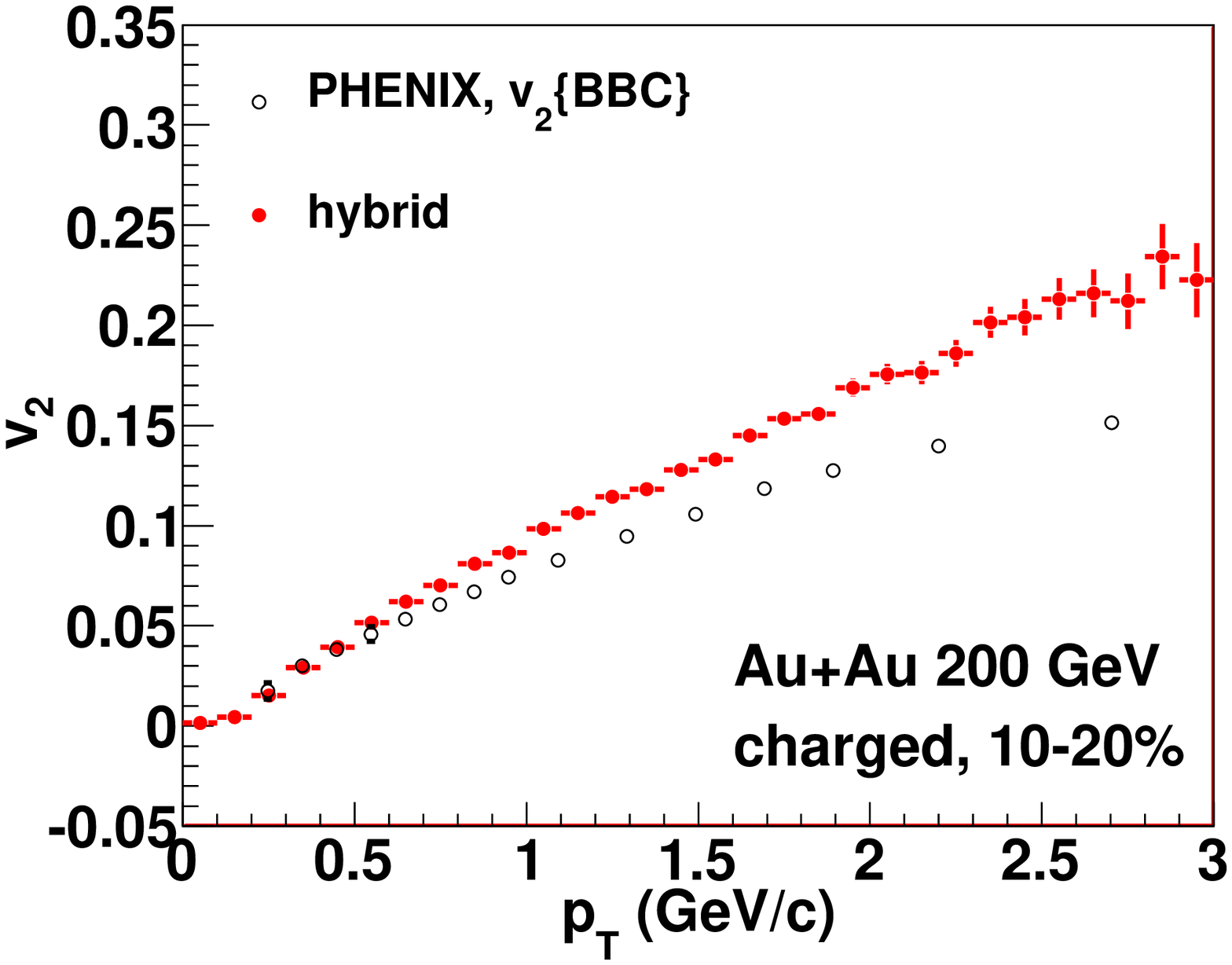,scale=0.3}
\end{minipage}
\begin{minipage}[t]{6 cm}
\epsfig{file=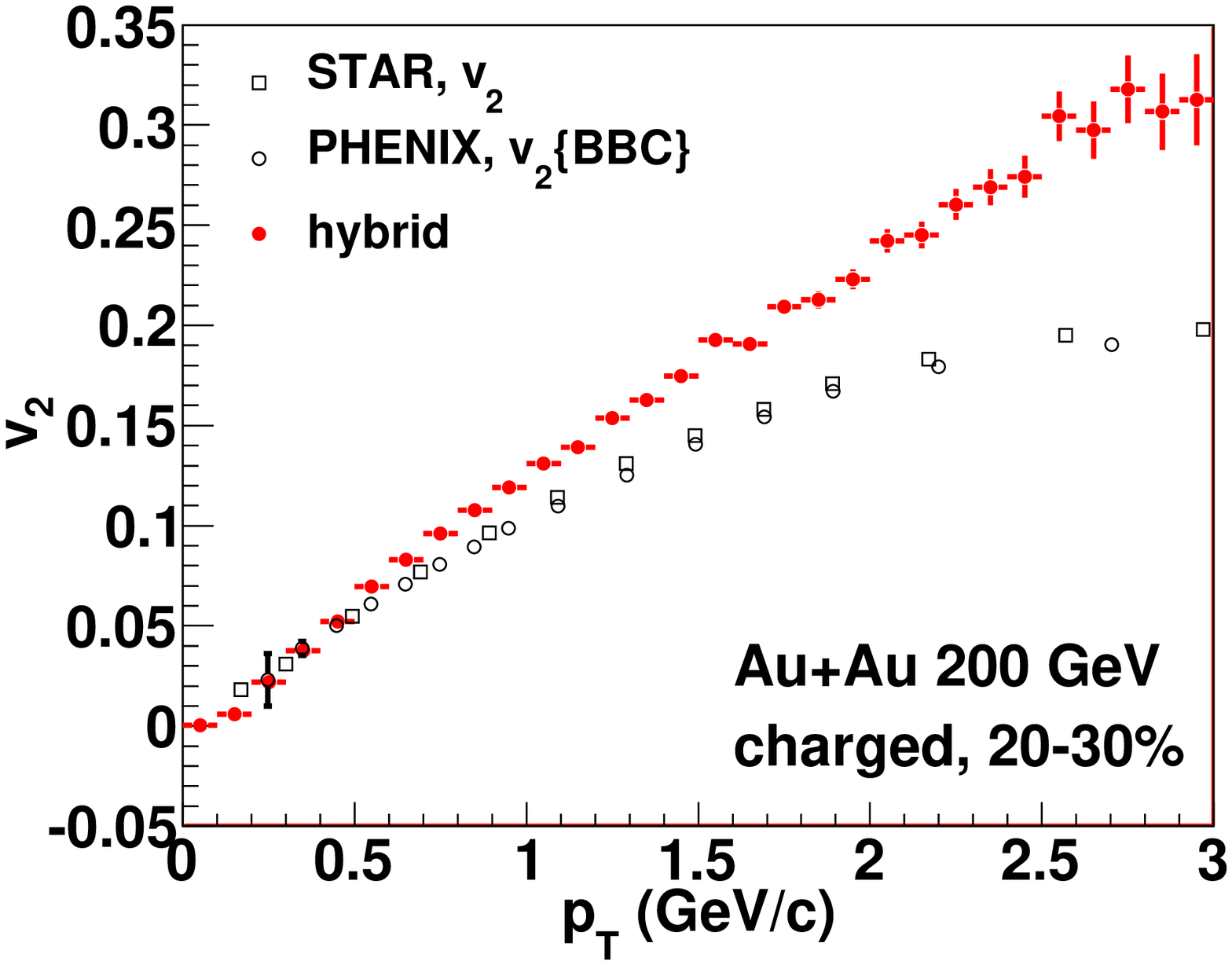,scale=0.3}
\end{minipage}
\begin{minipage}[t]{6 cm}
\epsfig{file=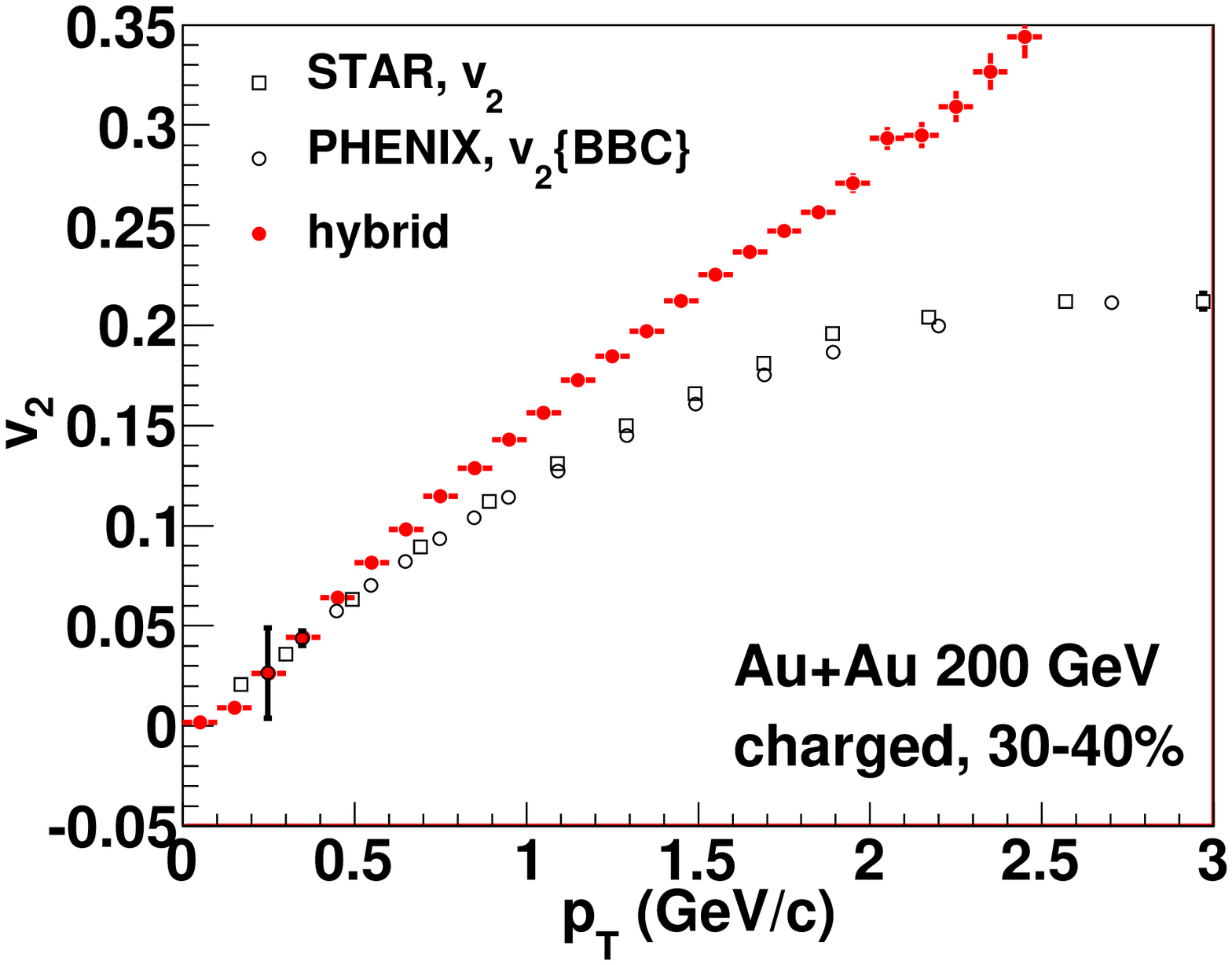,scale=0.3}
\end{minipage}
\begin{minipage}[t]{6 cm}
\epsfig{file=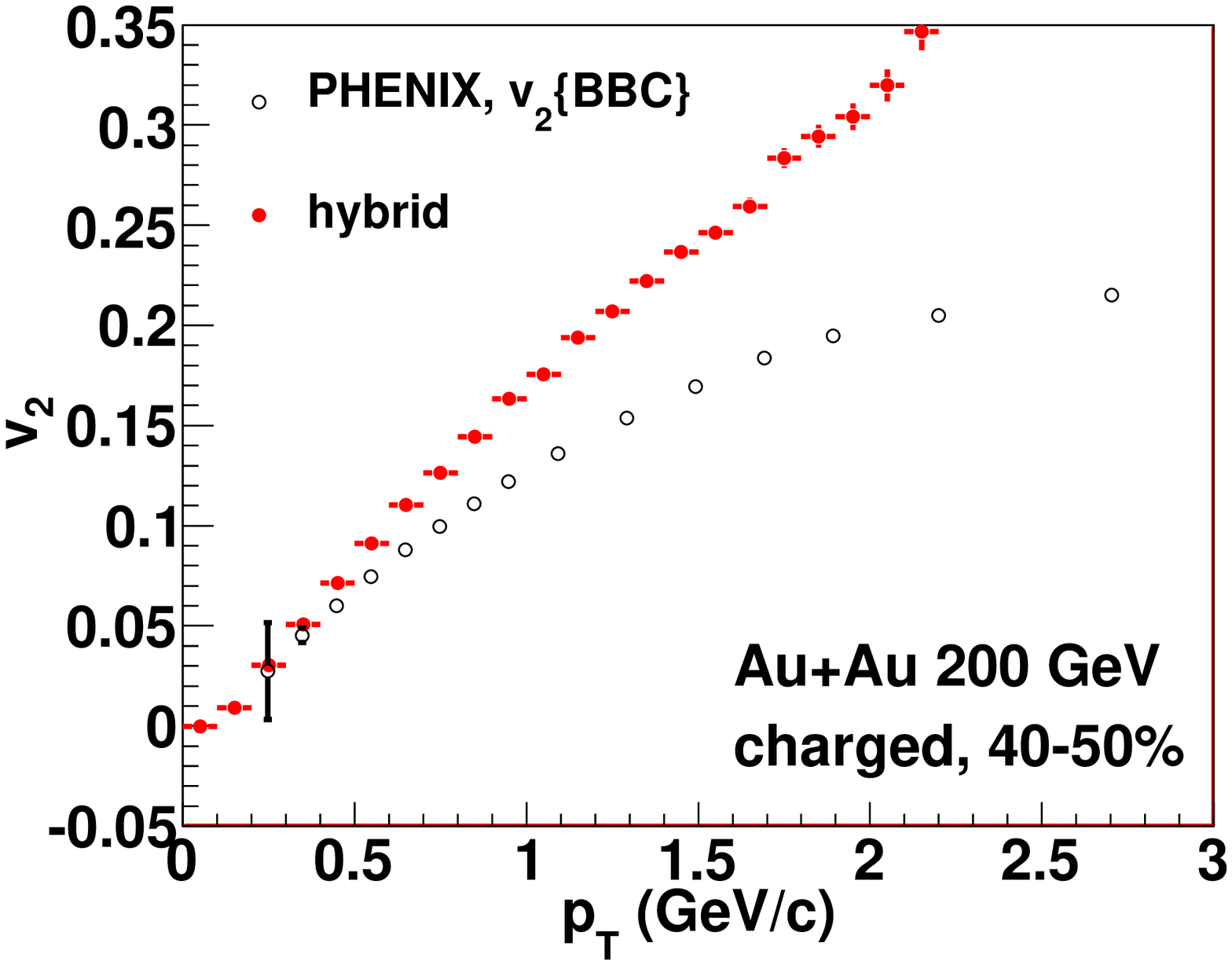,scale=0.3}
\end{minipage}
\begin{minipage}[t]{6 cm}
\epsfig{file=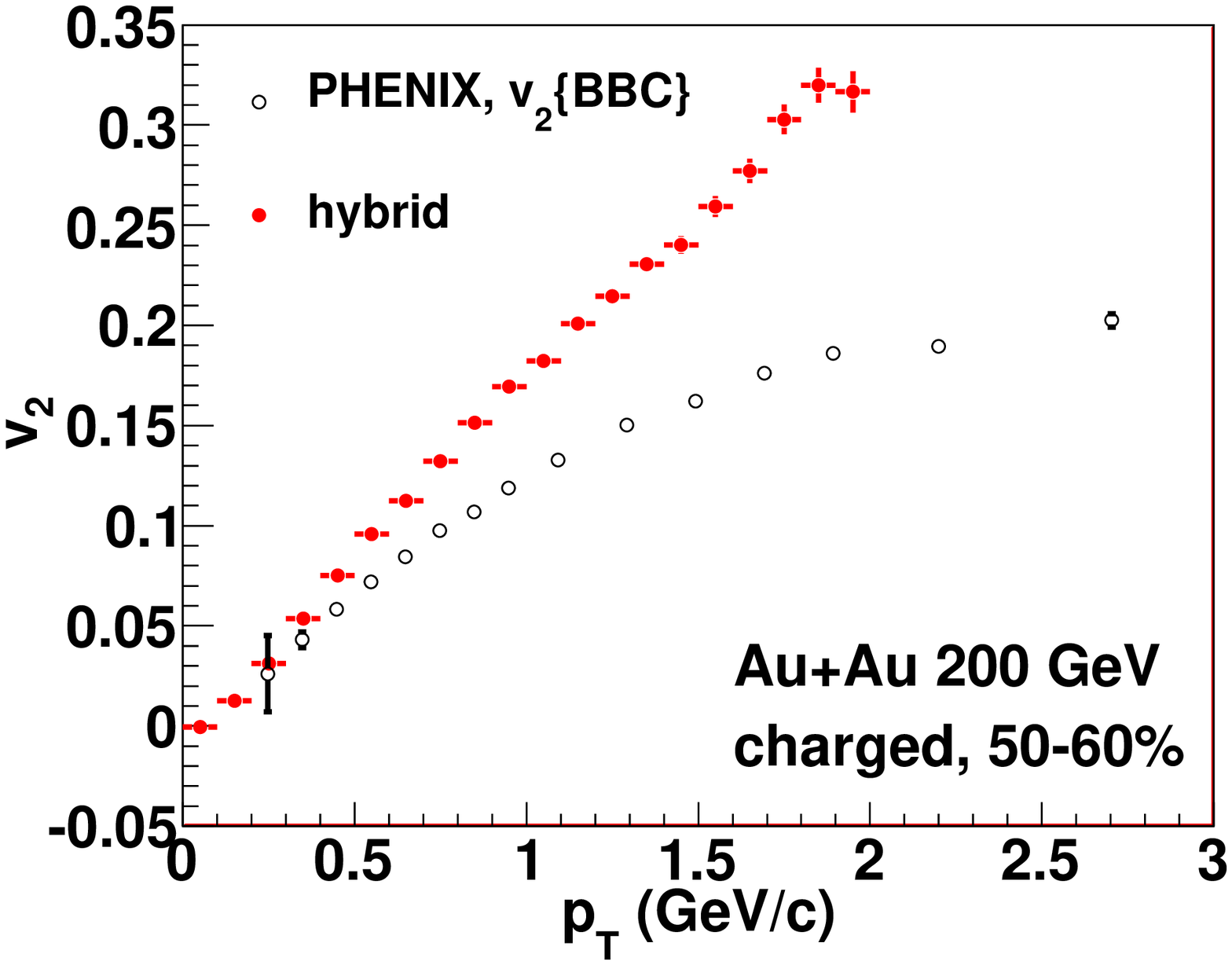,scale=0.3}
\end{minipage}
\caption{
Transverse momentum dependence of
$v_{2}$ of charged hadrons 
in $\sqrt{s_{NN}}$ = 200 GeV Au+Au collisions  in 0-10\% (top left),
10-20\% (top middle), 20-30\% (top right),
30-40\% (bottom left), 40-50\% (bottom middle) and 50-60\% (bottom right)
centralities. Results calculated with respect to the participant plane
(model ``B'') using the Glauber model initial conditions
are compared with the PHENIX \cite{Afanasiev:2009wq} and STAR data.
\label{fig:v2ptChPHENIX}
}
\end{center}
\end{figure}

In Figure \ref{fig:v2ptChPHENIX} the calculated $v_{2}(p_{T})$
  for charged hadrons in $\sqrt{s_{NN}}$ = 200 GeV Au+Au collisions is
  compared with the PHENIX \cite{Afanasiev:2009wq} and STAR
  \cite{STARv2-4} data in 0-10\% (top left), 10-20\% (top middle),
  20-30\% (top right), 30-40\% (bottom left), 40-50\% (bottom middle)
  and 50-60\% (bottom right) centralities. The calculation was done
  using the Glauber model initial state in the model ``B'' setting.
In 0-10\% central  collisions, results from the model is almost identical to the PHENIX
$v_{2}$\{BBC\} data. However, the data deviate from theoretical results
as going to peripheral collisions, which suggests the importance of viscous effects
in peripheral collisions where the transverse flow is more anisotropic and
the size of the system is smaller than in central collisions.

\begin{figure}[tb]
\begin{center}
\begin{minipage}[t]{6 cm}
\epsfig{file=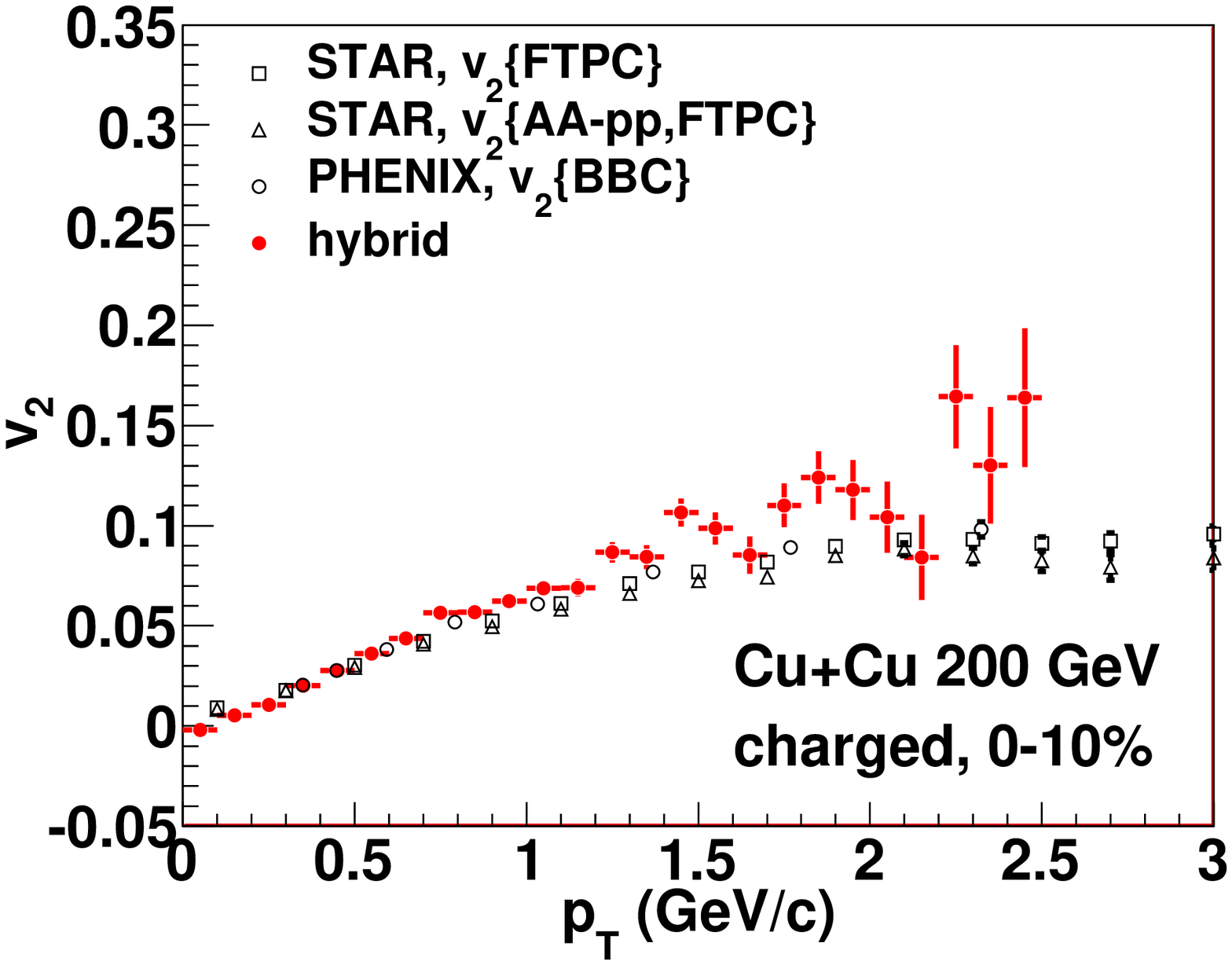,scale=0.3}
\end{minipage}
\begin{minipage}[t]{6 cm}
\epsfig{file=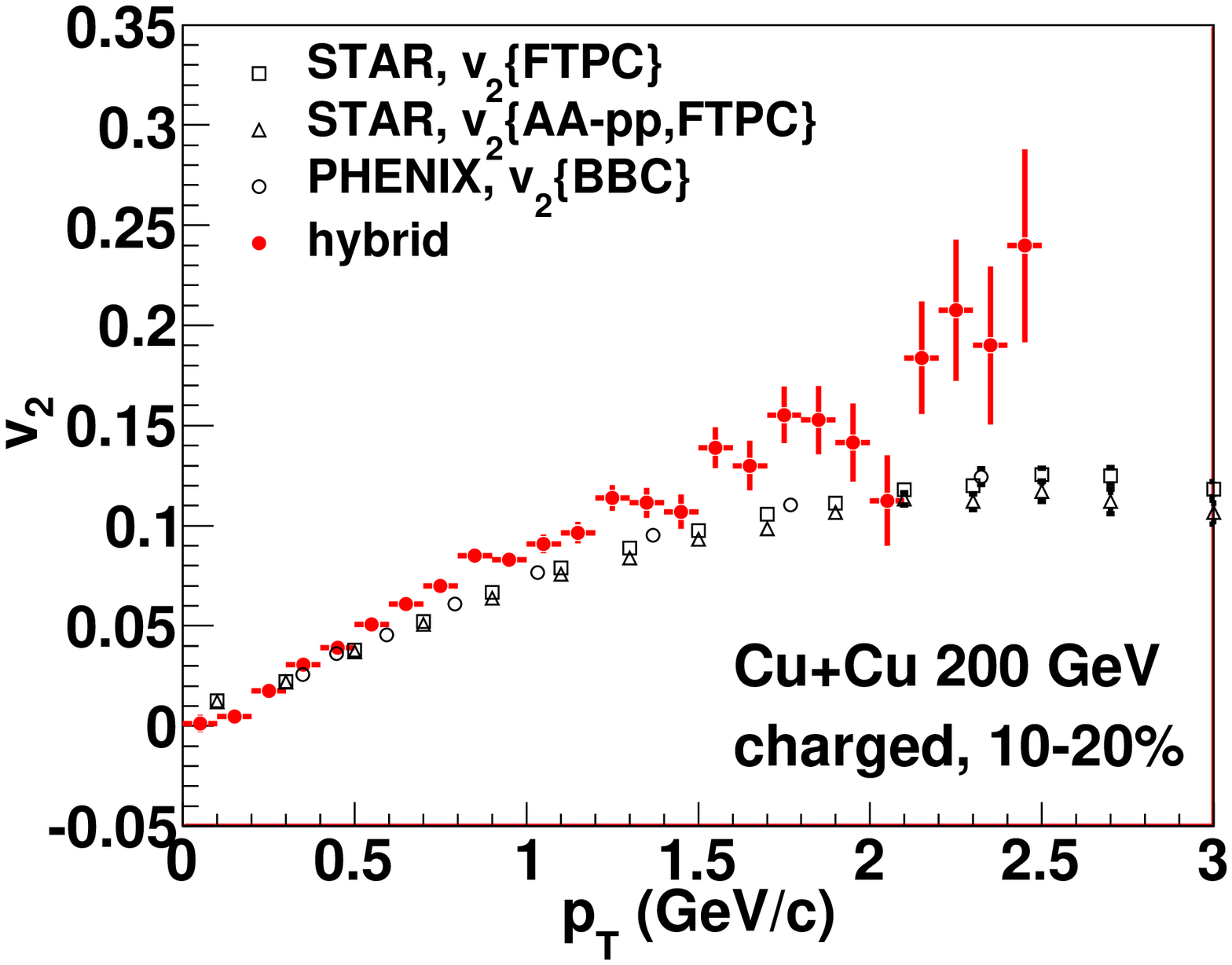,scale=0.3}
\end{minipage}
\begin{minipage}[t]{6 cm}
\epsfig{file=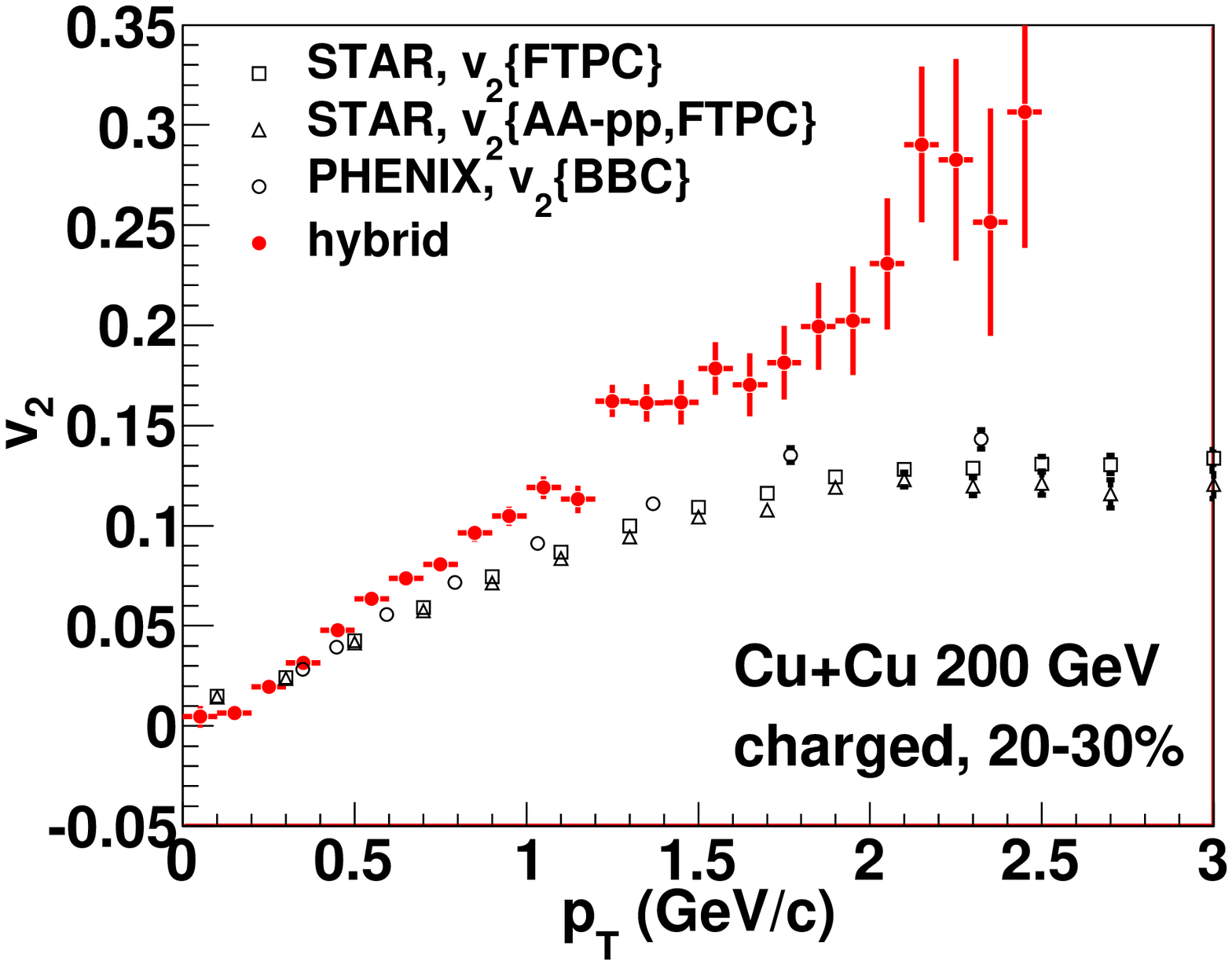,scale=0.3}
\end{minipage}
\begin{minipage}[t]{6 cm}
\epsfig{file=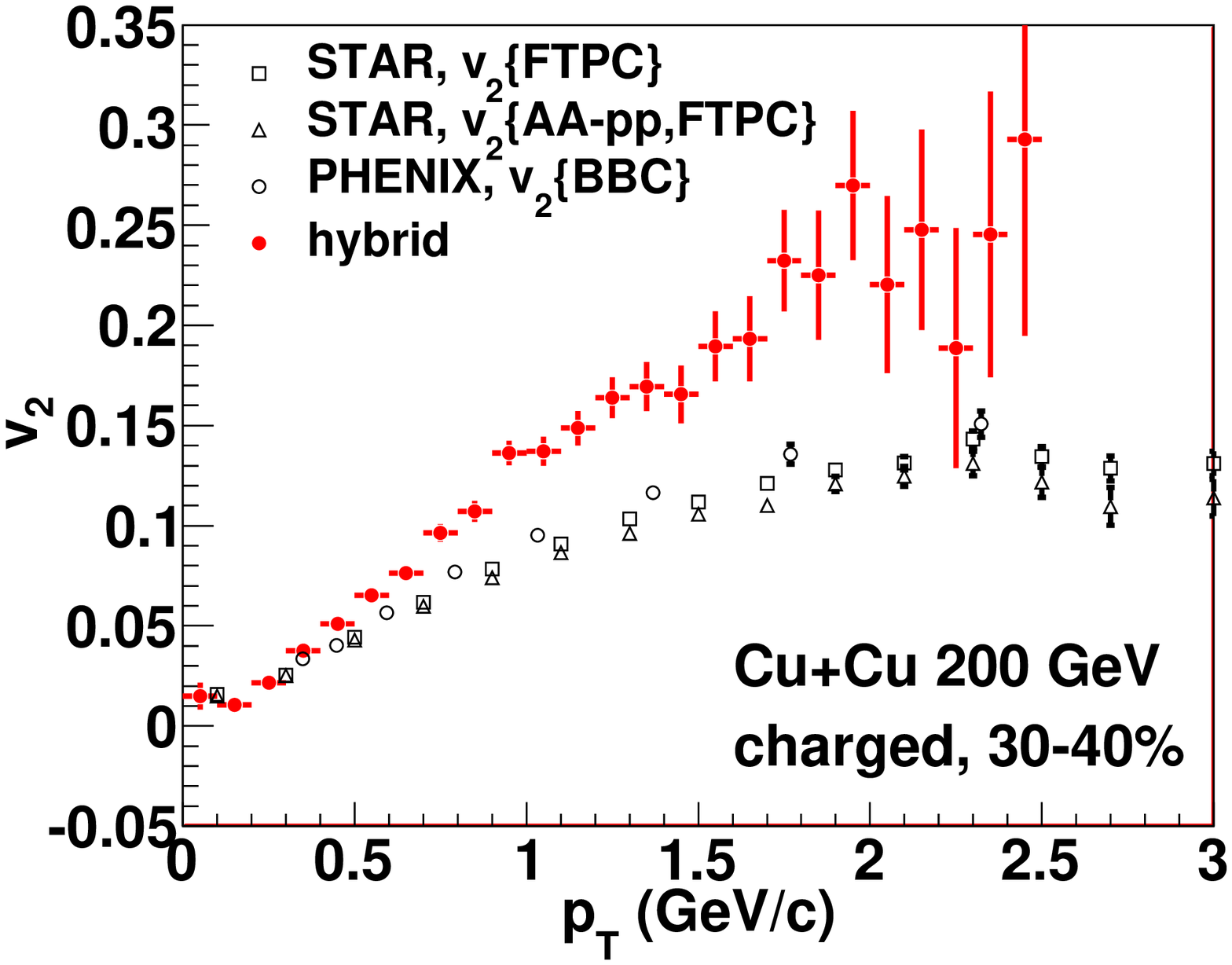,scale=0.3}
\end{minipage}
\begin{minipage}[t]{6 cm}
\epsfig{file=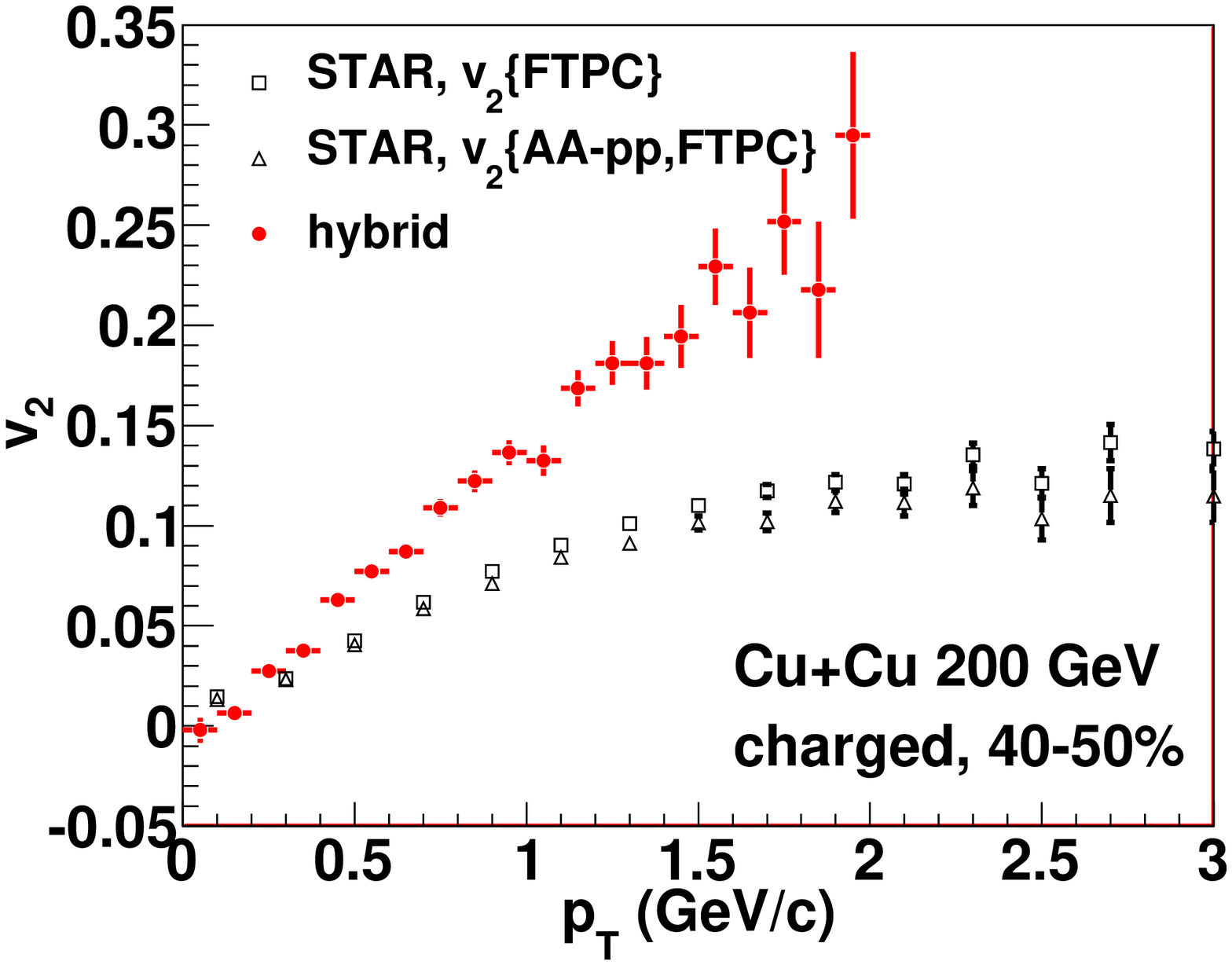,scale=0.3}
\end{minipage}
\begin{minipage}[t]{6 cm}
\epsfig{file=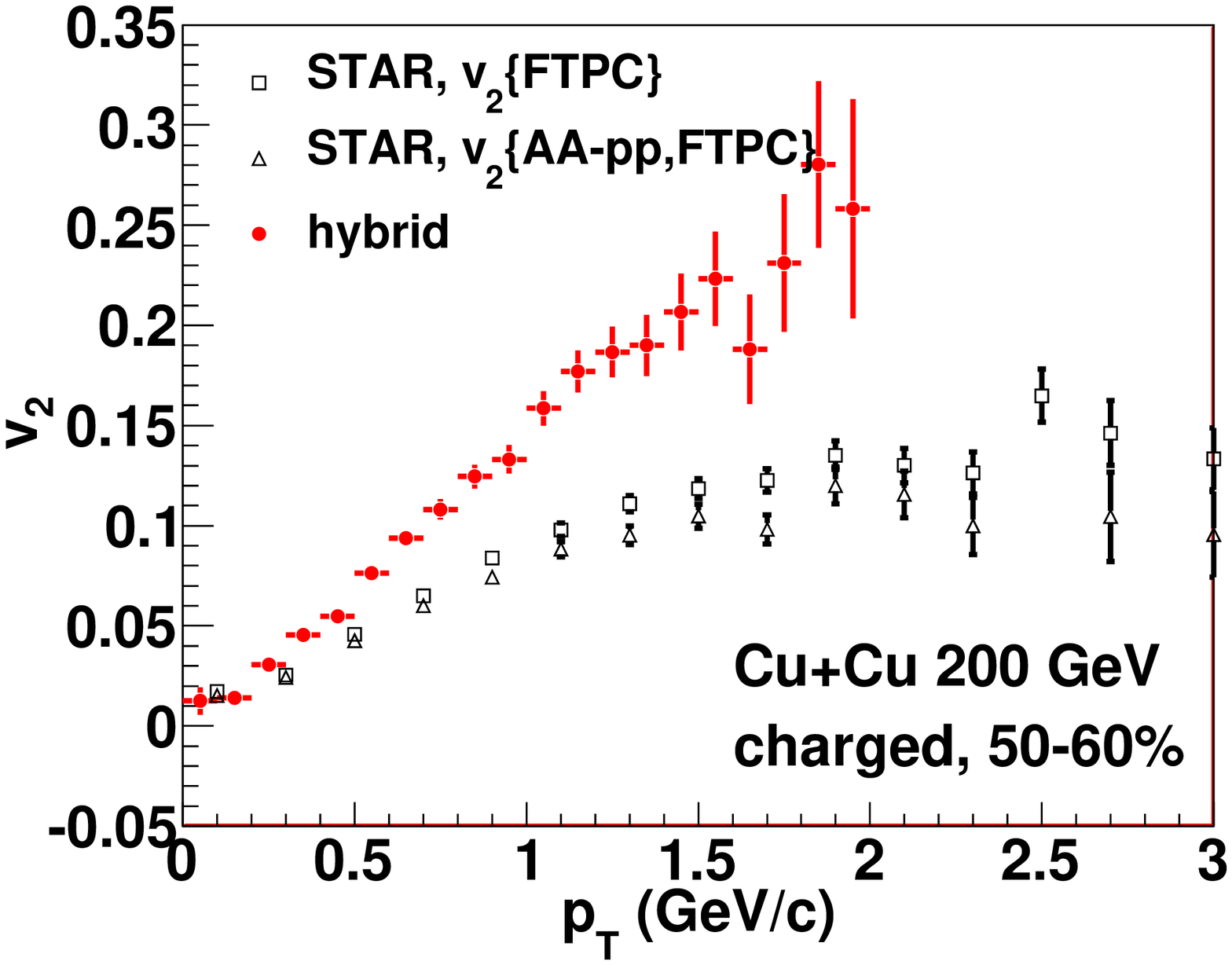,scale=0.3}
\end{minipage}
\caption{
The same as Fig.~\ref{fig:v2ptChPHENIX} but
in Cu+Cu collisions.
Data are from PHENIX \cite{Adare:2006ti} and STAR \cite{Abelev:2010tr}.
\label{fig:v2ptChCuPHENIX}
}
\end{center}
\end{figure}

As shown in Fig.~\ref{fig:v2ptChCuPHENIX}$, 
v_{2}(p_{T})$ pattern in Au+Au collisions is quite similar to that
in Cu+Cu collisions even though the size of the system
in the latter case is smaller than in the former case.
Experimental data are taken from PHENIX \cite{Adare:2006ti}
and STAR \cite{Abelev:2010tr}.

\begin{figure}[tb]
\begin{center}
\begin{minipage}[t]{6 cm}
\epsfig{file=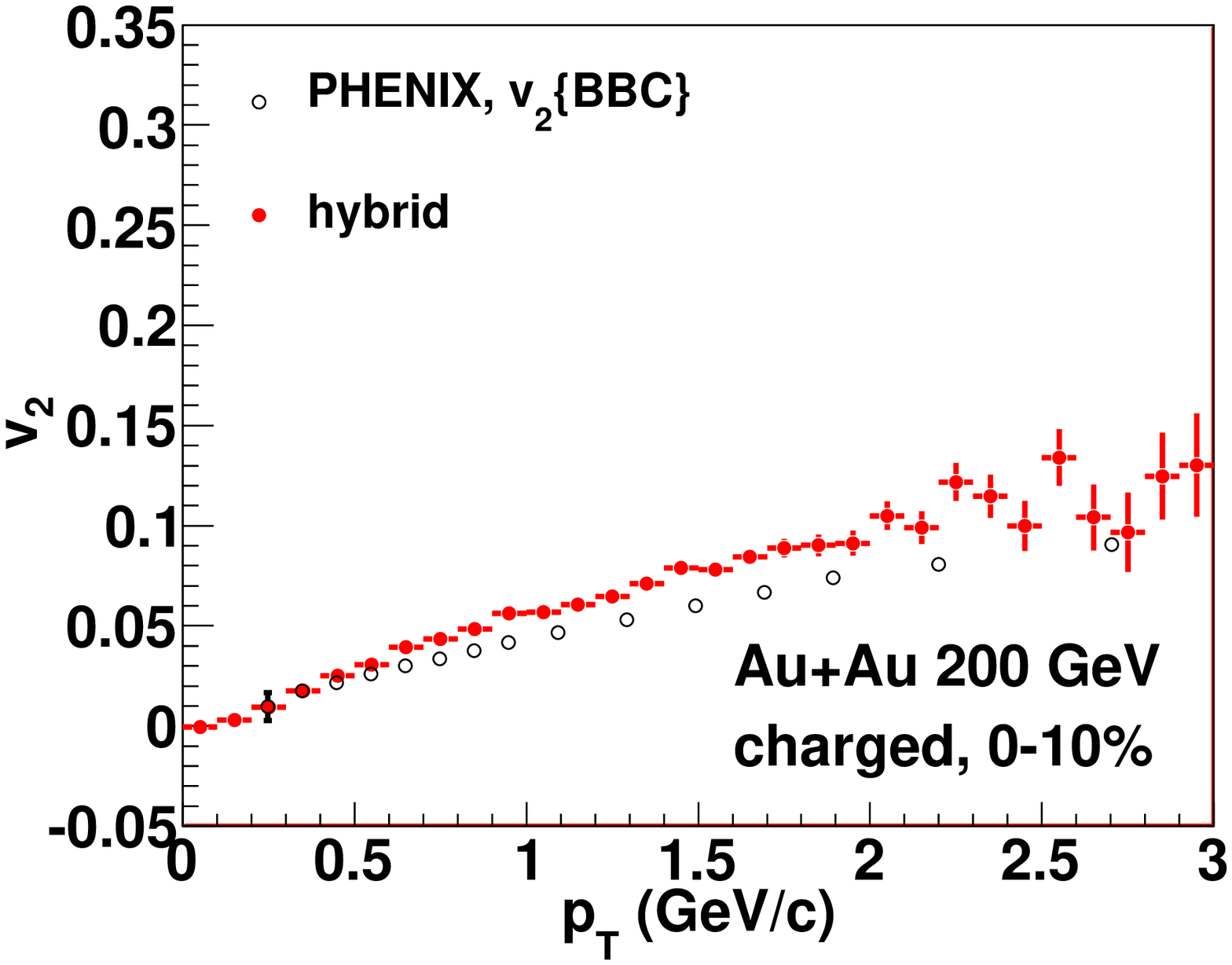,scale=0.3}
\end{minipage}
\begin{minipage}[t]{6 cm}
\epsfig{file=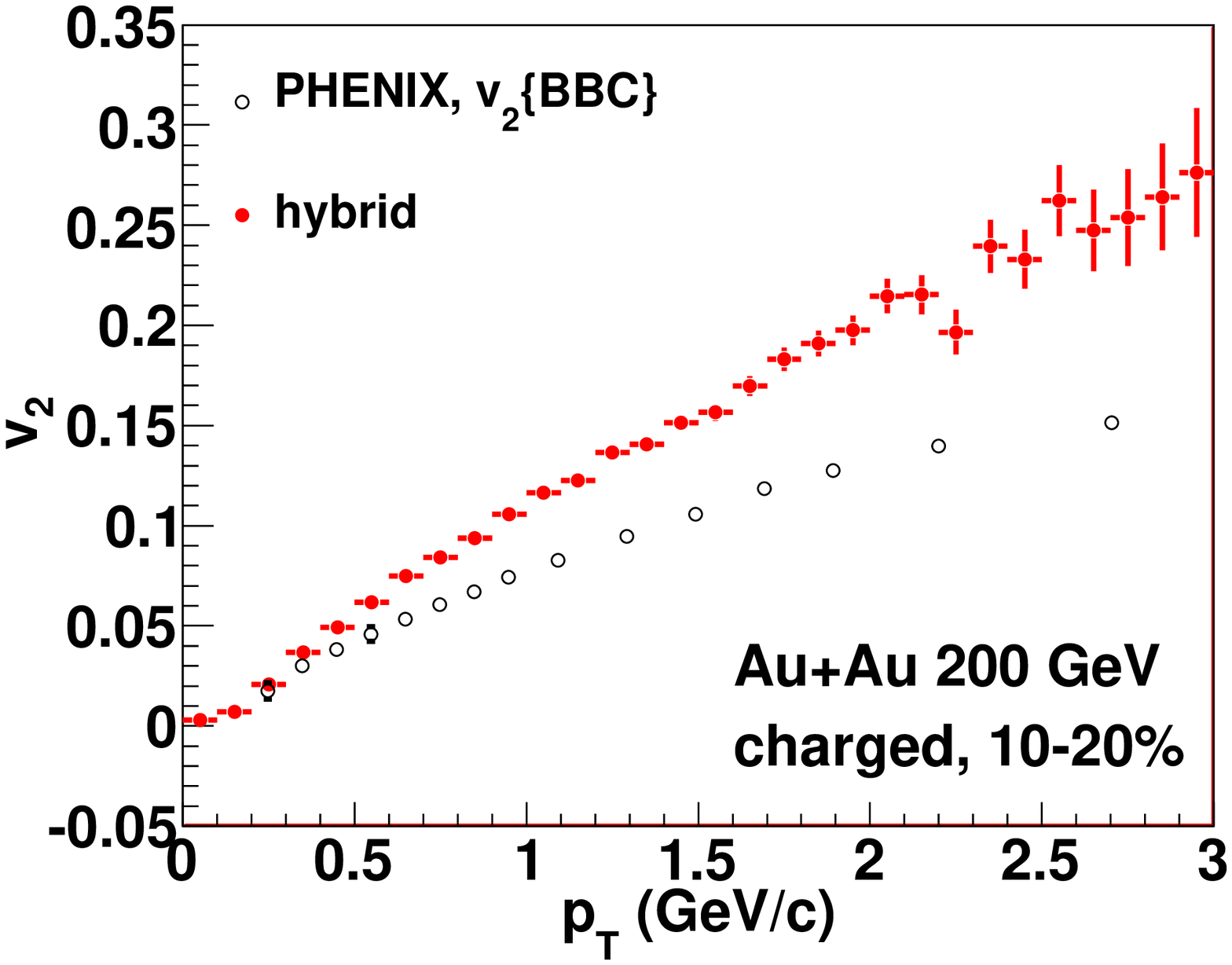,scale=0.3}
\end{minipage}
\begin{minipage}[t]{6 cm}
\epsfig{file=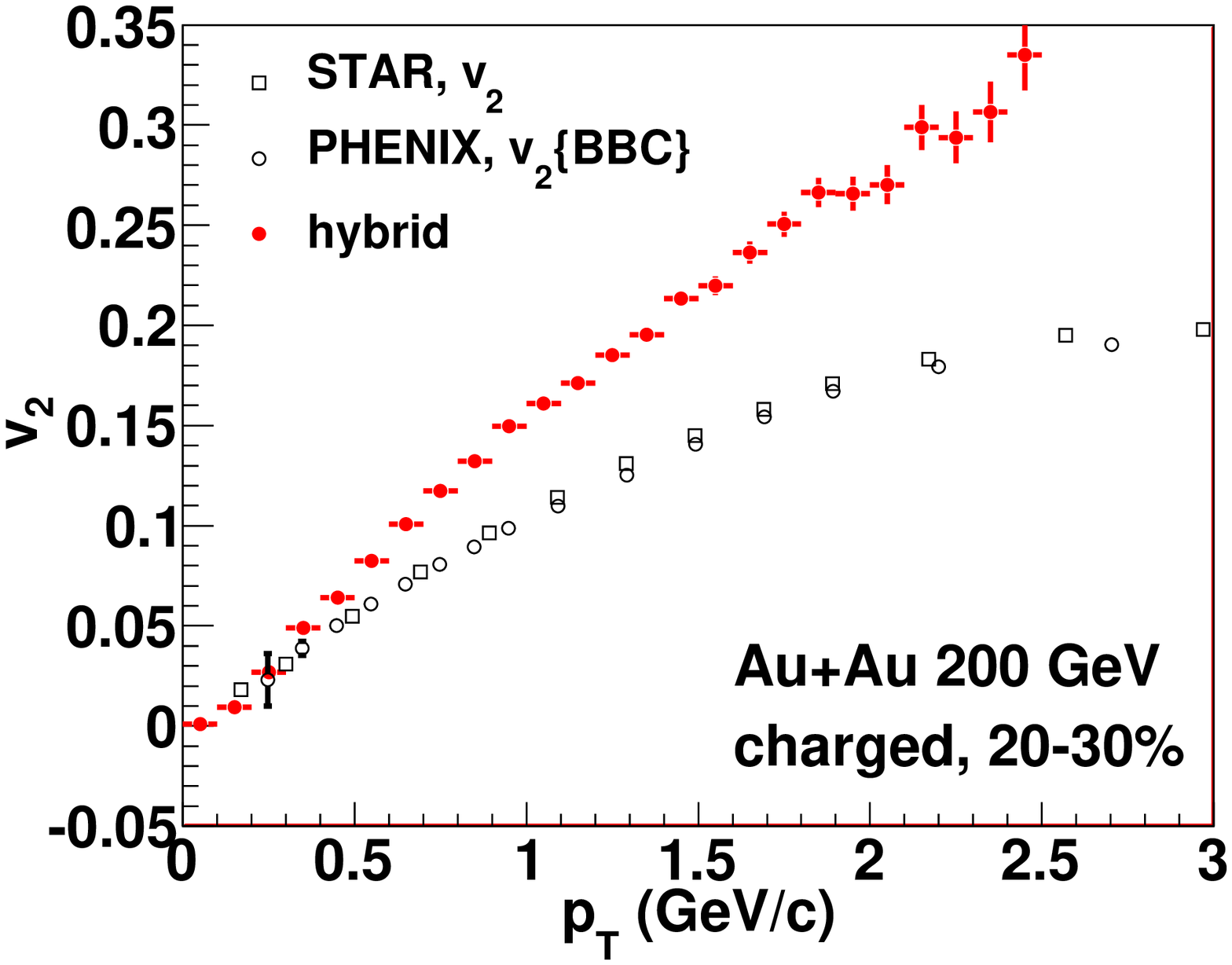,scale=0.3}
\end{minipage}
\begin{minipage}[t]{6 cm}
\epsfig{file=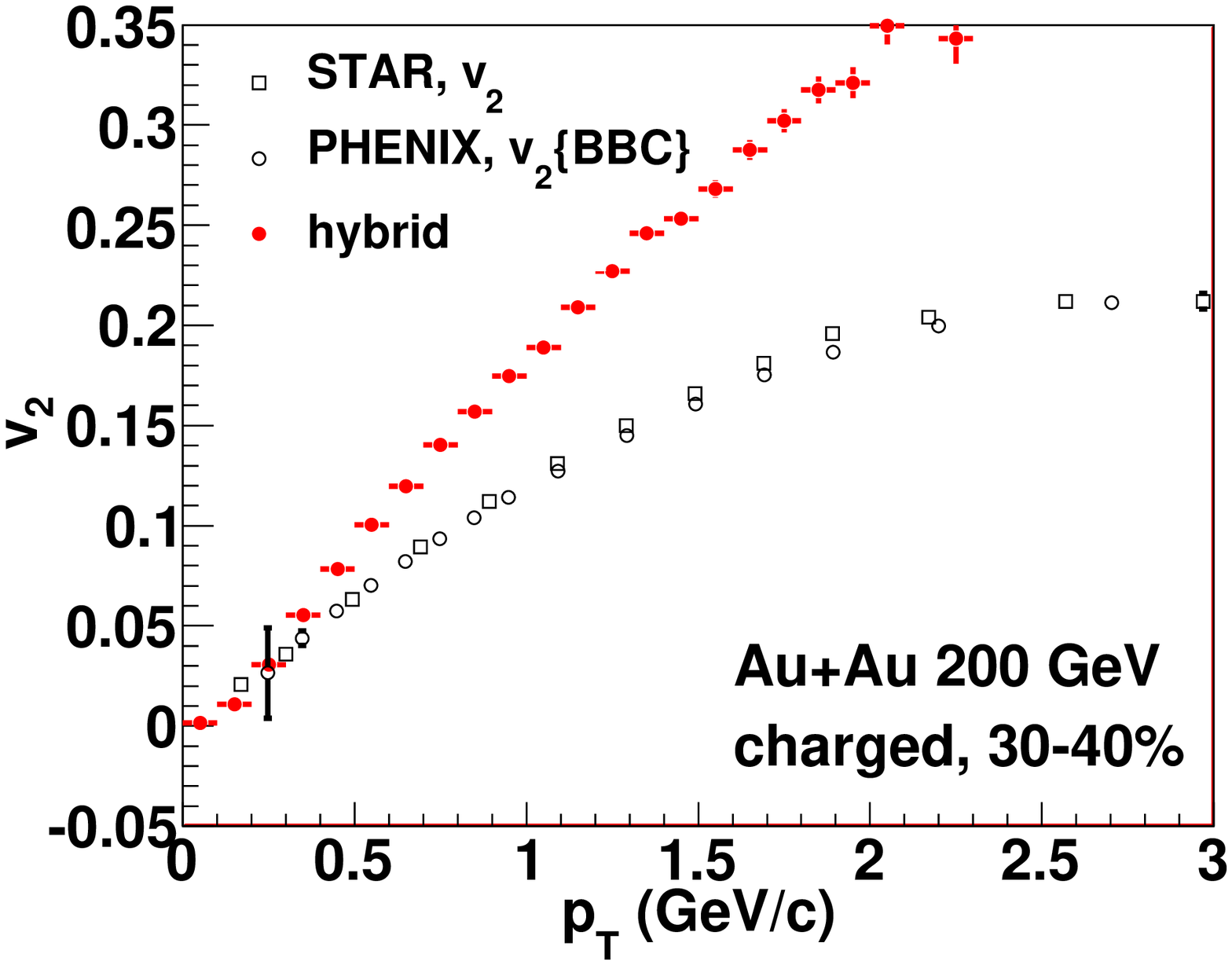,scale=0.3}
\end{minipage}
\begin{minipage}[t]{6 cm}
\epsfig{file=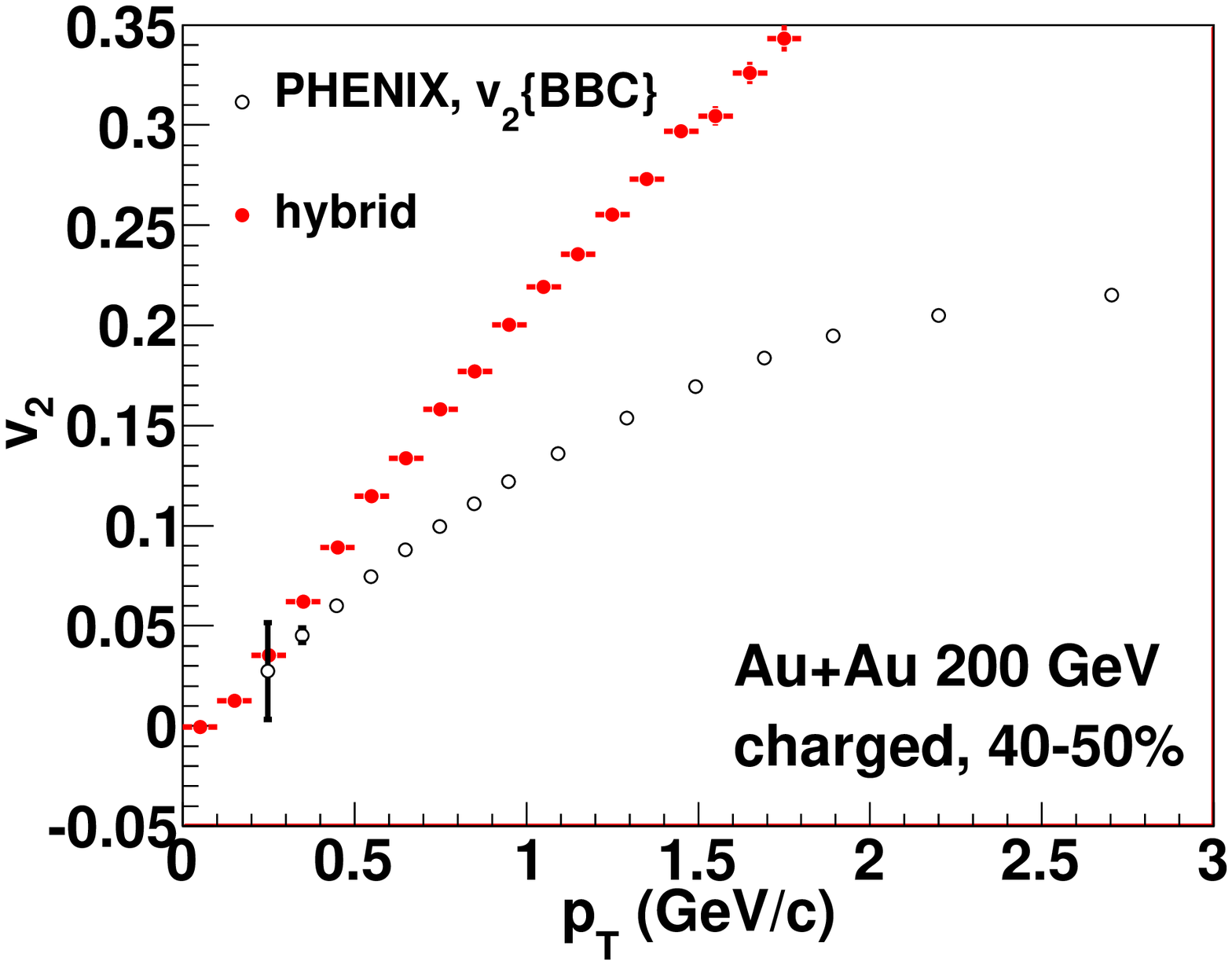,scale=0.3}
\end{minipage}
\begin{minipage}[t]{6 cm}
\epsfig{file=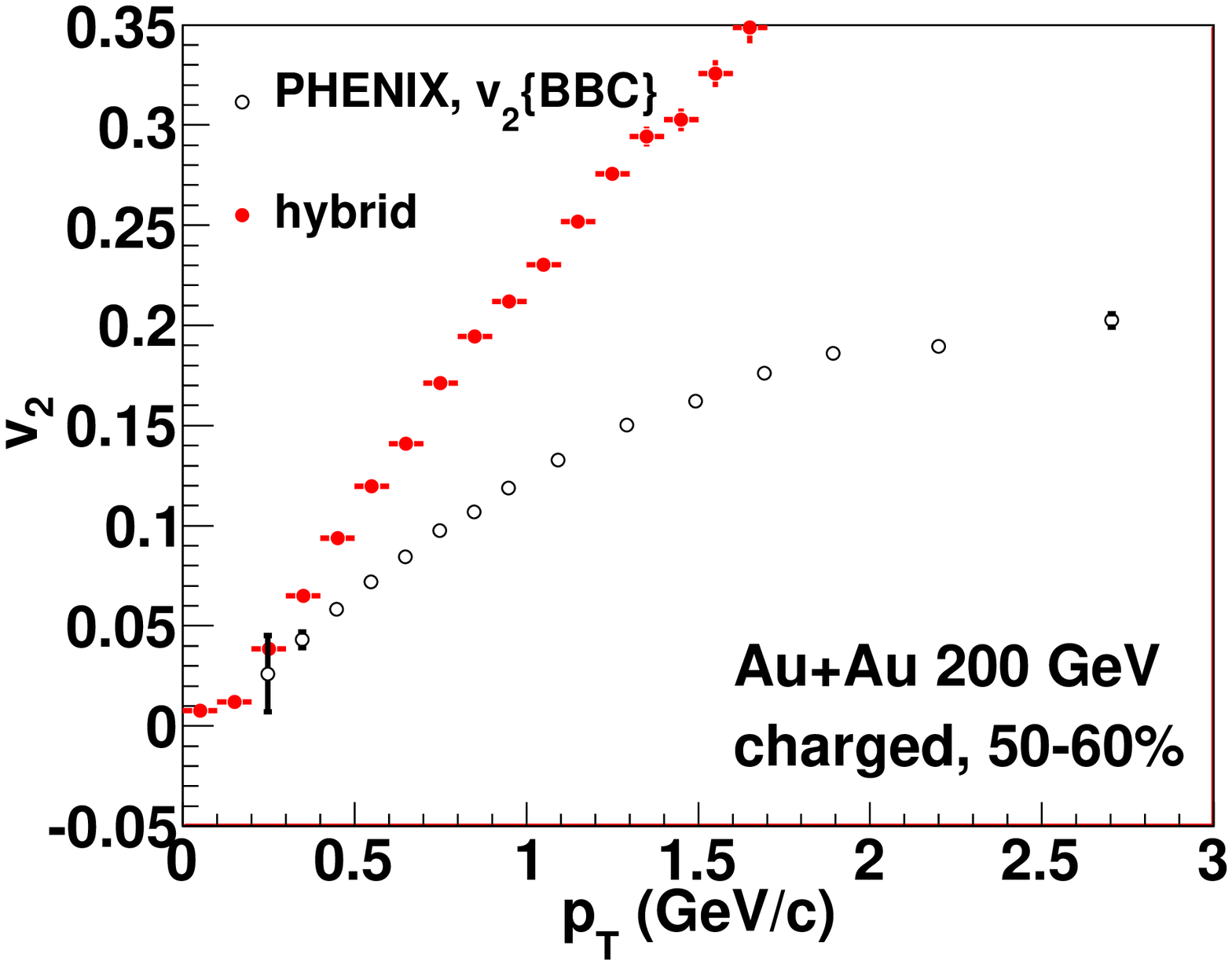,scale=0.3}
\end{minipage}
\caption{
The same as Fig.~\ref{fig:v2ptChPHENIX} but
using the KLN model initial conditions.
\label{fig:v2ptKlnChPHENIX}
}
\end{center}
\end{figure}

As already seen in the integrated $v_{2}$ in Figs.~\ref{fig:v2centPHOBOS}
and \ref{fig:v2centSTAR}, the KLN model gives
a larger $v_{2}$ than the Glauber model does.
This is again seen in $v_{2}(p_{T})$ in Figs.~\ref{fig:v2ptChPHENIX} and \ref{fig:v2ptKlnChPHENIX}:
The slope of $v_{2}(p_{T})$ from the KLN initialisation is slightly 
steeper than
that from the Glauber initialisation.

\begin{figure}[tb]
\begin{center}
\begin{minipage}[t]{6 cm}
\epsfig{file=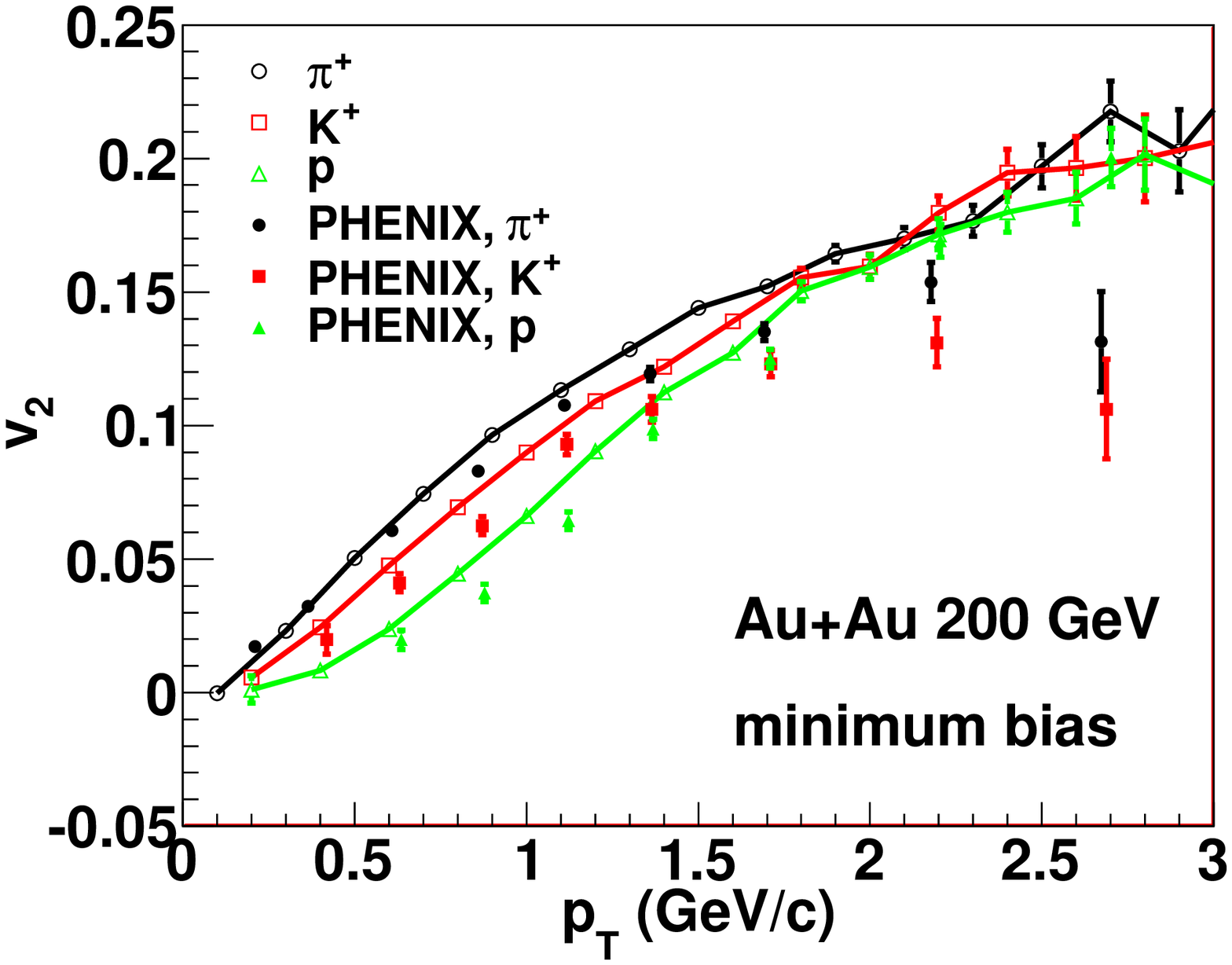,scale=0.3}
\end{minipage}
\begin{minipage}[t]{6 cm}
\epsfig{file=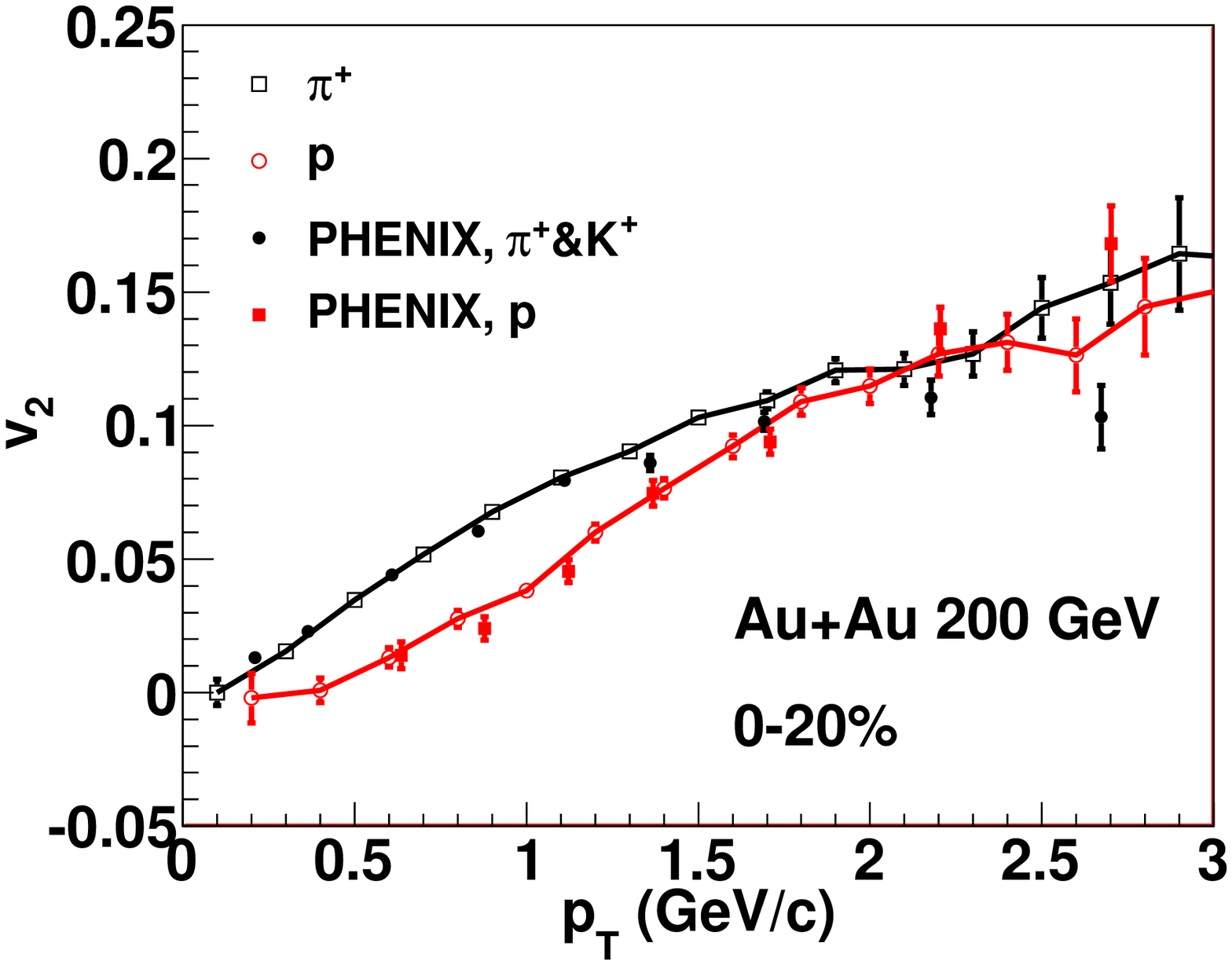,scale=0.3}
\end{minipage}
\begin{minipage}[t]{6 cm}
\epsfig{file=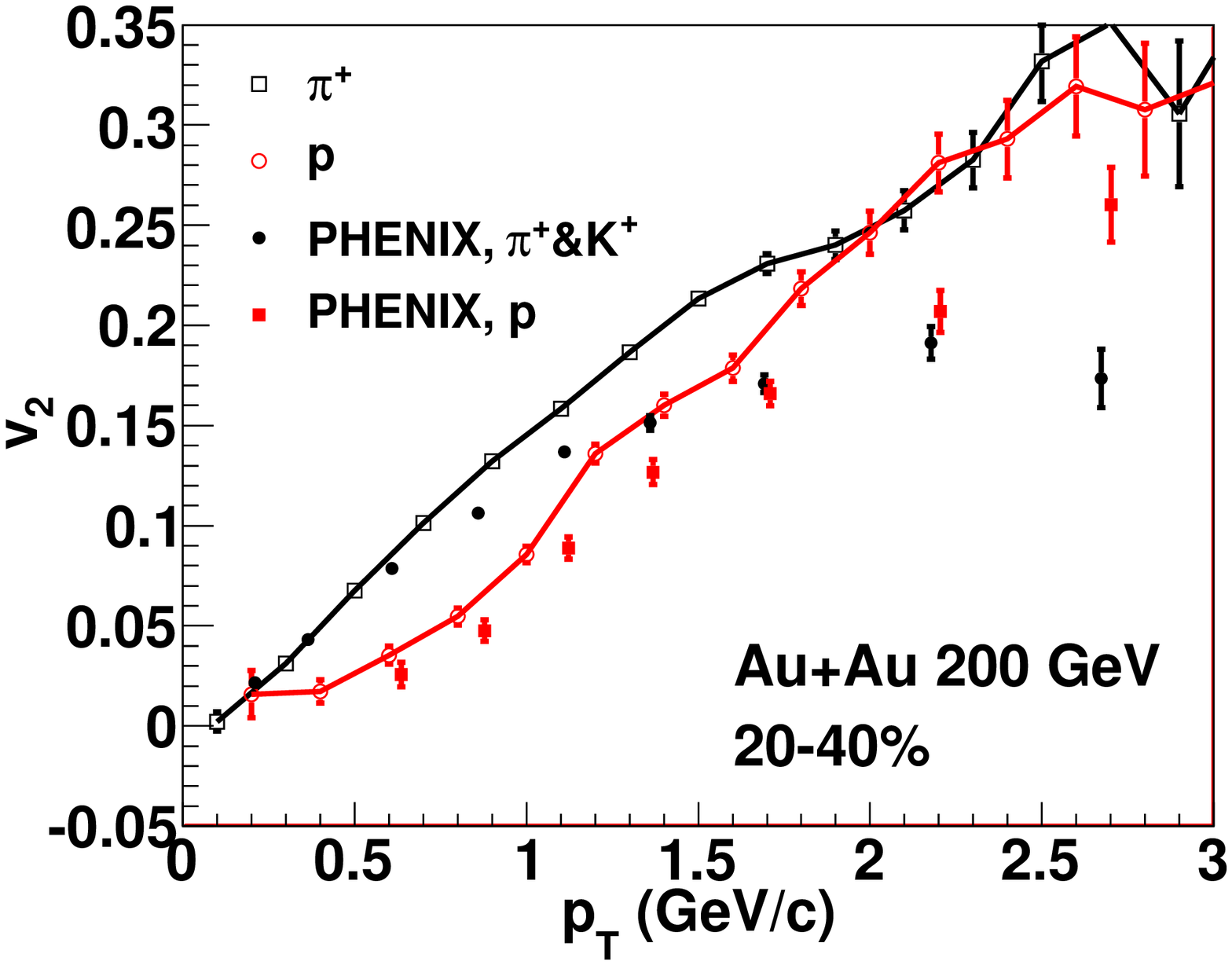,scale=0.3}
\end{minipage}
\caption{
Transverse momentum dependence of
$v_{2}$ of identified hadrons with respect to
participant plane 
using the Glauber model initial conditions
are compared with the PHENIX data \cite{PHENIXv2-2} in minimum bias 
(left), and in
0-20\% (middle) and 20-40\% (right) centralities.
\label{fig:v2ptPIDPHENIX}
}
\end{center}
\end{figure}

In Fig.~\ref{fig:v2ptPIDPHENIX},
$v_{2}(p_{T})$ for identified hadrons using the Glauber
initialisation 
are compared with the PHENIX data.
Mass splitting pattern,
which is known to come mainly from hadronic rescattering effects \cite{Hirano:2007ei},
is seen in both theoretical results and experimental data.
In low $p_{T}$ region up to $\sim 1$ GeV/$c$, we reasonably reproduce the
PHENIX data.
However, the data gradually deviate from the theoretical results above
$p_{T} \sim 1$ GeV/$c$, which suggests again the necessity of viscous corrections.
As already seen in the $v_{2}(p_{T})$ for charged hadrons,
$v_{2}$ for pions overshoots the data in semi-central collisions.

It might be also interesting to analyse $v_{2}$ for $\phi$ mesons
in low $p_{T}$ region: Since $\phi$ mesons hardly rescatter
with pions
in the late hadronic stages,  these particles
do not participate in mass splitting pattern \cite{Hirano:2007ei}.
A hint of this behaviour has been already seen in recent STAR data \cite{Zhang:2012pj}.

\subsection{\it Results at LHC\label{sec:LHCresults}}

\begin{figure}[tb]
\begin{center}
\begin{minipage}[t]{9 cm}
\epsfig{file=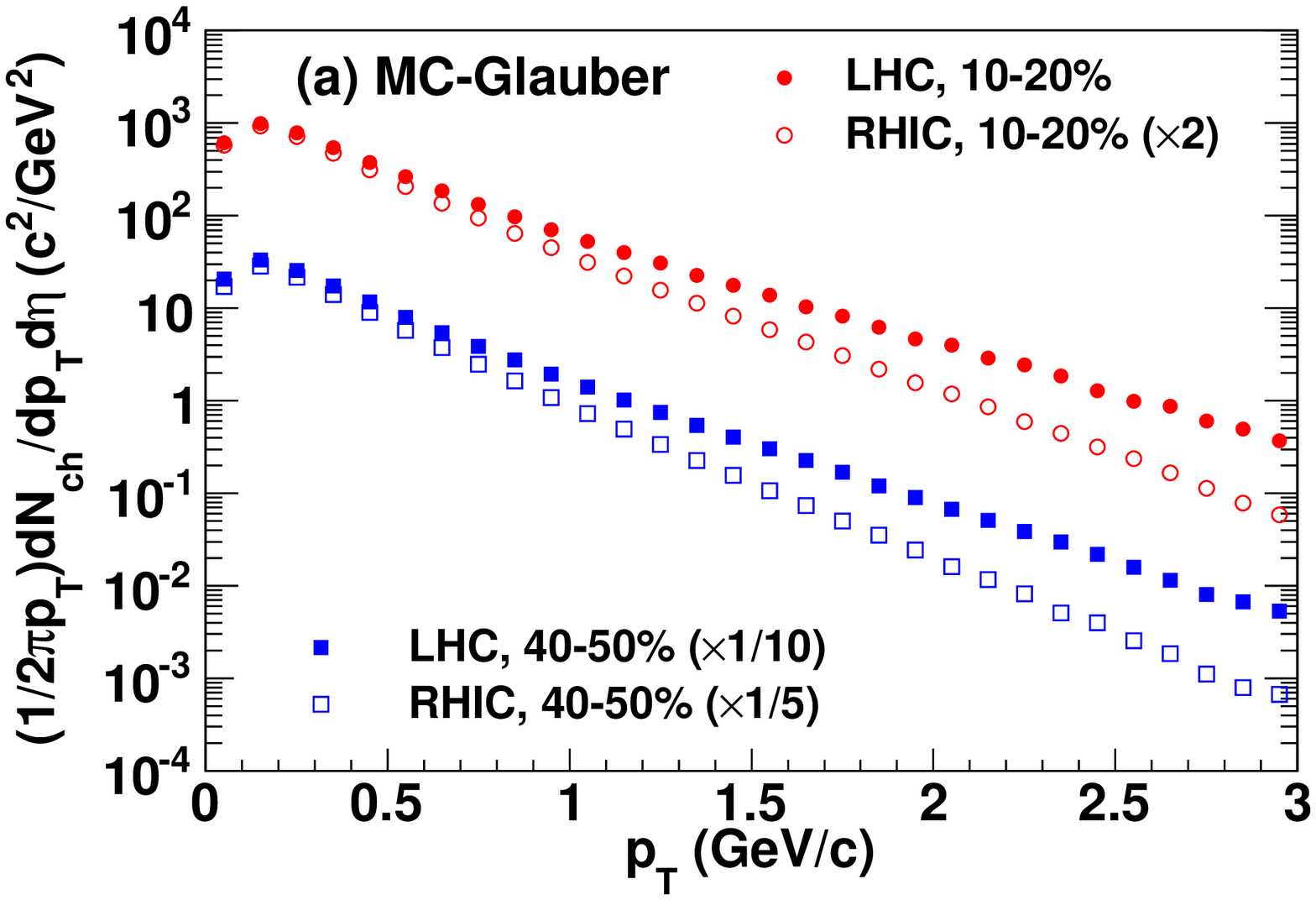,scale=0.45}
\end{minipage}
\begin{minipage}[t]{9 cm}
\epsfig{file=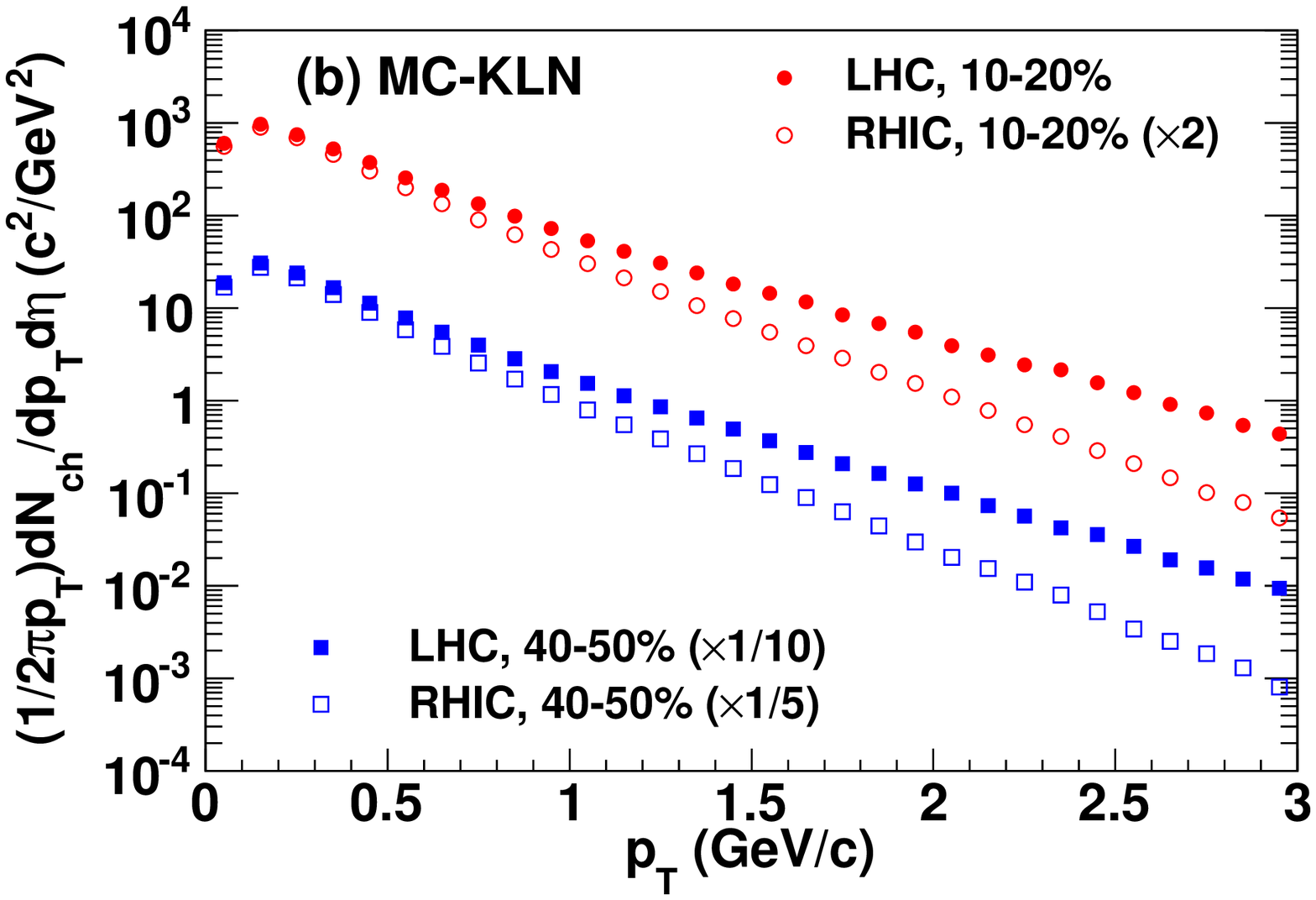,scale=0.45}
\end{minipage}
\caption{
Transverse momentum distribution of charged hadrons
at 10-20\% (circles) and 40-50\% (squares) centralities 
in Pb+Pb collisions at $\sqrt{s_{NN}}=$ 2.76 TeV (filled symbols)
and in Au+Au collisions at $\sqrt{s_{NN}}=$ 200 GeV (open symbols). 
Results were calculated using
(a) the MC-Glauber initialisation 
and (b) the MC-KLN initialisation. 
For the sake of comparison and visibility, the spectra are scaled by
2, 1/10, and 1/5 for 10-20\% at RHIC, 40-50\% at LHC and
40-50\% at RHIC, respectively. Figures are
from Ref.~\cite{Hirano:2010je}.
\label{fig:ptdistLHC}}
\end{center}
\end{figure}

Figure \ref{fig:ptdistLHC} shows a comparison of
transverse momentum distributions of charged hadrons
between RHIC and LHC energies at 10-20\% and 40-50\% centralities.
As clearly seen from figures,
the slope of the $p_{T}$ spectra becomes flatter as collision energy
and, consequently, pressure of produced matter increases.
To quantify this,
we calculate mean $p_{T}$ of charged hadrons.
In the MC-Glauber initialisation, 
mean $p_{T}$ increases from RHIC to LHC
by 21\% and 19\% in 10-20\% and 40-50\% centrality, respectively.
On the other hand, 
the corresponding relative increases are  
25\% and 24\% in the MC-KLN initialisation.
Since our calculations at RHIC were tuned to reproduce the
$p_T$-spectra, this means that at LHC the spectra calculated using
the MC-KLN initialisation are slightly flatter than those calculated using
the MC-Glauber initialisation.

\begin{figure}[tb]
\begin{center}
\begin{minipage}[t]{9 cm}
\epsfig{file=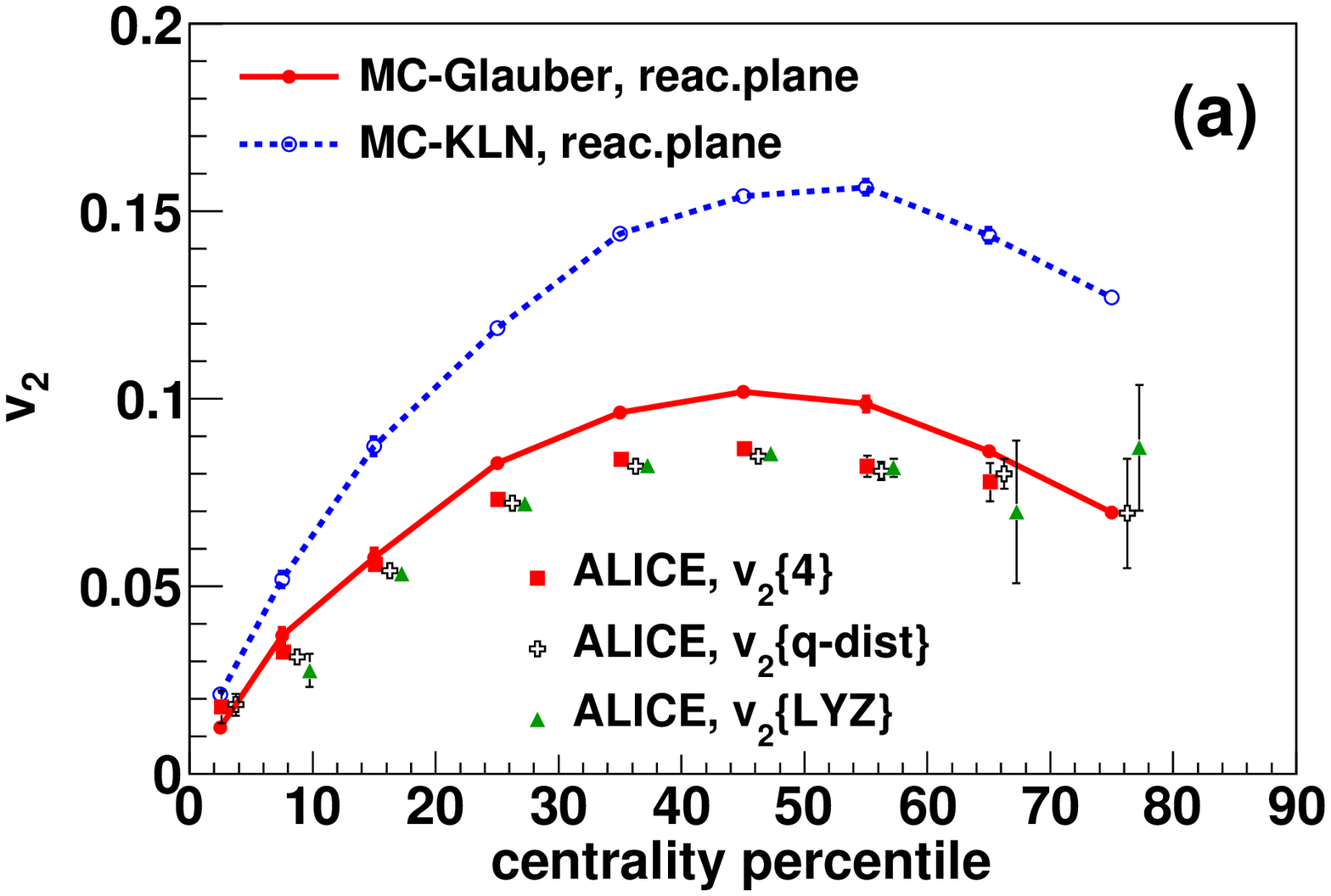,scale=0.45}
\end{minipage}
\begin{minipage}[t]{9 cm}
\epsfig{file=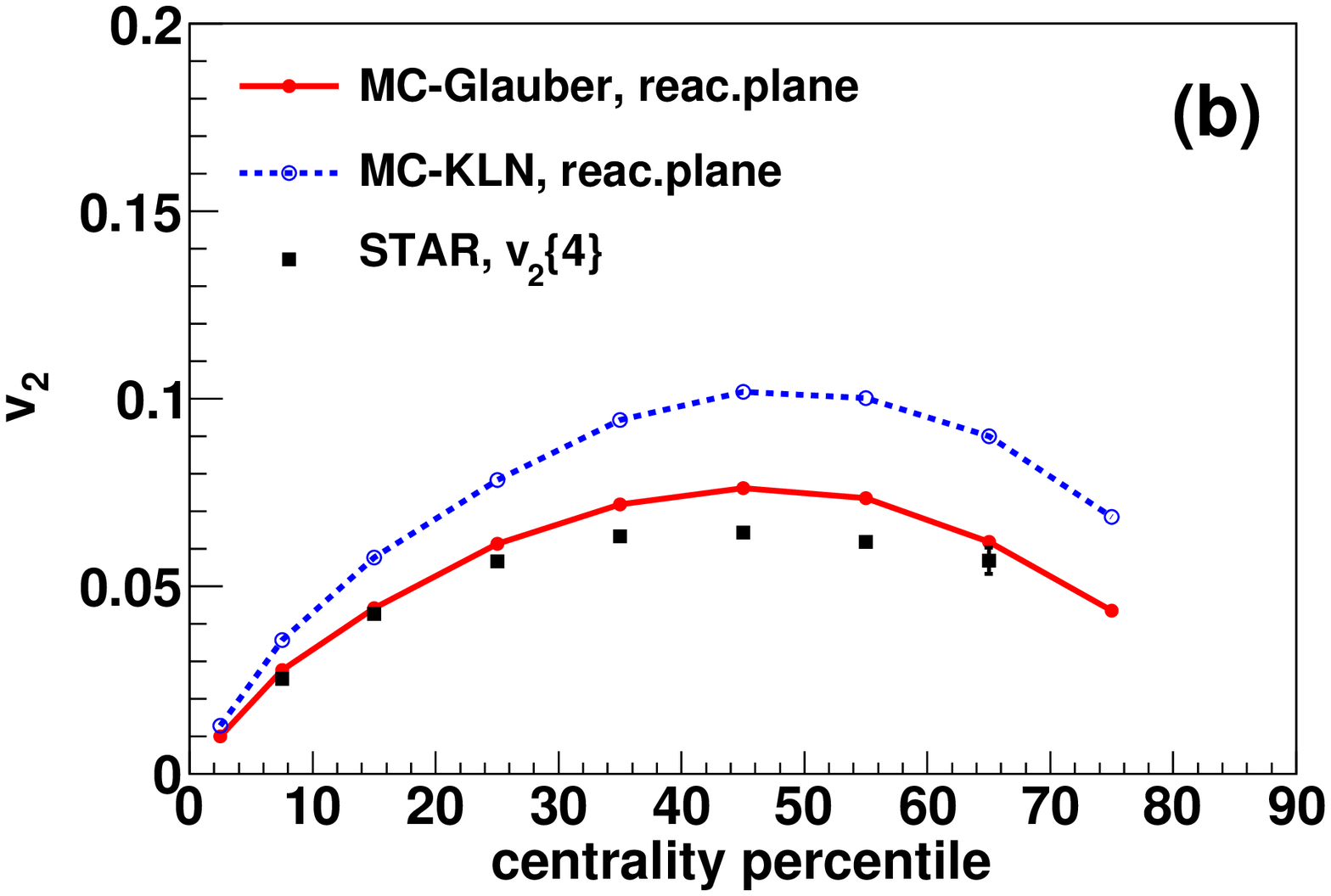,scale=0.45}
\end{minipage}
\caption{
Centrality dependence of $v_{2}$ for charged hadrons with respect to
reaction plane (model ``A'') in Pb+Pb collisions
at $\sqrt{s_{NN}}$ = 2.76 TeV  (left)  and
in Au+Au collisions
at $\sqrt{s_{NN}}$ = 200 GeV (right)
is compared with the ALICE \cite{Aamodt:2010pa}
($0.2 < p_{T} < 5$ GeV/$c$ and $\mid \eta \mid < 0.8$).
and STAR $v_{2}\{4\}$ data 
($0.15 < p_{T} < 2$ GeV/$c$ and $\mid \eta \mid < 1$), respectively.
Figures are from Ref.~\cite{Hirano:2010je}.
\label{fig:v2centALICESTAR}}
\end{center}
\end{figure}

We compare the integrated $v_2$ for charged hadrons with respect to
reaction plane with the ALICE \cite{Aamodt:2010pa} and STAR
\cite{STARv2-4} $v_{2}\{4\}$ data in Fig.~\ref{fig:v2centALICESTAR}.
When evaluating the integrated $v_2$, we take account of both
transverse momentum and pseudorapidity acceptance as done in the
experiments, \textit{i.e.} $0.2 < p_{T} < 5.0$ GeV/$c$ and 
$\mid \eta \mid < 0.8$ for ALICE, and $0.15 < p_{T} < 2.0$ GeV/$c$
and $\mid \eta \mid < 1.0$ for STAR. We want to emphasise that, not
only the $p_T$ cut~\cite{Luzum:2010ag}, but also the pseudorapidity
cut plays an important role in a consistent comparison with the
data. Due to the Jacobian for the change of variables from rapidity
$y$ to pseudorapidity $\eta$, $v_2(y=0) < v_2(\eta=0)$ for positive
elliptic flow \cite{Kolb:2001yi}.~\footnote{Notice that even if one
assumes the Bjorken scaling solution, one has to consider the
pseudorapidity acceptance since $v_{2}(\eta)$ is not constant even
if $v_2(y)$ is~\cite{Kolb:2001yi}.} In the case of the
MC-Glauber (MC-KLN) initialisation in 40-50\% centrality, $v_{2}$
integrated over the whole $p_{T}$ region is $\sim$14\%
($\sim$10\%) larger at $\eta=0$ than at $y=0$.

When the MC-Glauber model is employed
for initial profiles,
centrality dependence of integrated $v_2$
from the hybrid approach
almost agrees with both ALICE and STAR data.
Since eccentricity fluctuation contributes little and negatively
to $v_{2}\{4\}$ in a non-Gaussian distribution of
eccentricity fluctuation \cite{Ollitrault:2009ie,Voloshin:2007pc},
this indicates there is only
little room for the QGP viscosity in the model calculation.
On the other hand, apparent discrepancy
between the results from the MC-KLN initialisation
and the ALICE and STAR data
means that 
viscous corrections during the hydrodynamic evolution are required.

From RHIC to LHC, the $p_{T}$-integrated $v_2(\mid \eta \mid < 0.8)$
increases by 24\% and 25\% in 10-20\% and 40-50\% centrality, respectively,
in the MC-Glauber initialisation.
On the other hand, in the MC-KLN initialisation, the increase
reaches 42\% and 44\% 
 in 10-20\% and 40-50\% centrality, respectively.
Since eccentricity does not change significantly (at most $\pm 6$\%
in 40-50\% centrality)
from RHIC to LHC as shown in Fig.~\ref{fig:epspart} (right),
the significant increase of integrated $v_{2}$
must be attributed to a change in transverse 
dynamics.

\begin{figure}[tb]
\begin{center}
\begin{minipage}[t]{9 cm}
\epsfig{file=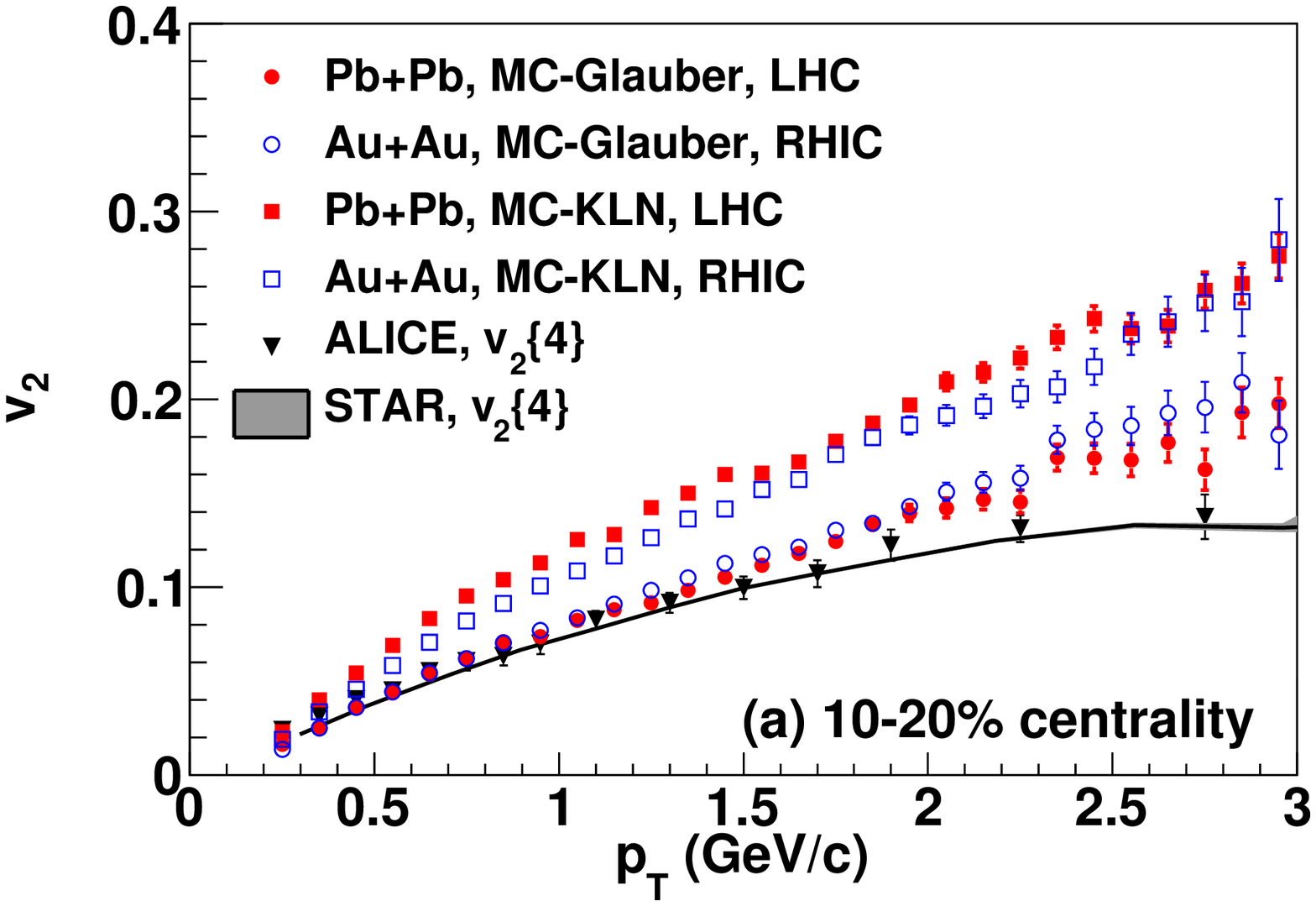,scale=0.45}
\end{minipage}
\begin{minipage}[t]{9 cm}
\epsfig{file=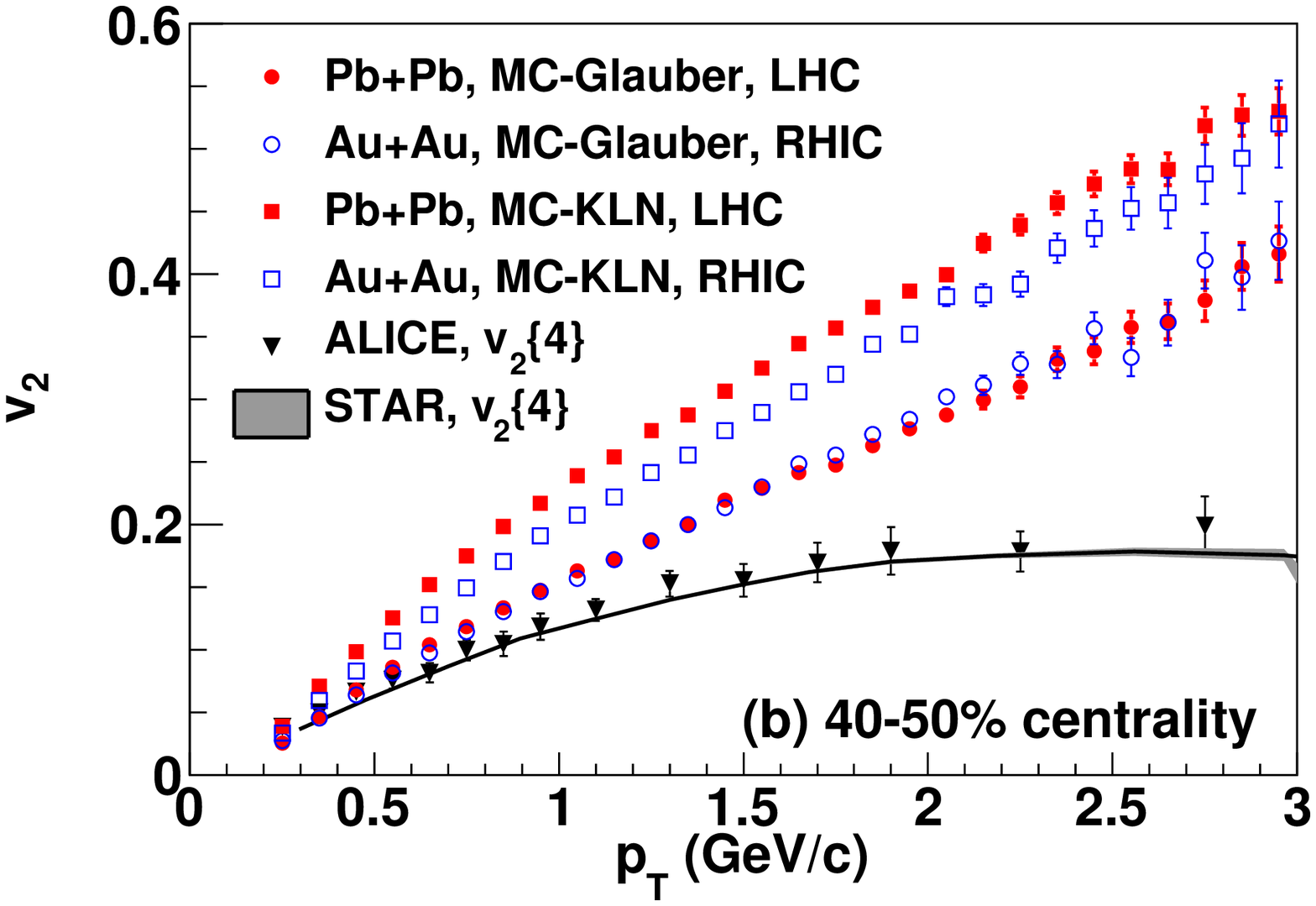,scale=0.45}
\end{minipage}
\caption{
Transverse momentum dependence of $v_{2}$ of
charged hadrons with respect to
reaction plane 
in $\sqrt{s_{NN}}$ = 2.76 TeV Pb+Pb collisions in
10-20\% (left) and 40-50\% (right) centralities.
 Figures are from Ref.~\cite{Hirano:2010je}.
\label{fig:v2ptALICESTAR}}
\end{center}
\end{figure}

Finally, we compare $v_{2}(p_{T})$ of charged hadrons
with ALICE \cite{Aamodt:2010pa} and STAR \cite{STARv2-4} data in
10-20\% (Fig.~\ref{fig:v2ptALICESTAR} (left))
 and 40-50\% (Fig.~\ref{fig:v2ptALICESTAR} (right)) centrality.
 Interestingly, the data at LHC agrees with the data
at RHIC within errors. The calculated $v_2(p_T)$ shows similar
independence of collision energy when MC-Glauber initialisation is
used, whereas MC-KLN initialisation leads to a slightly larger
$v_2(p_T)$ at the larger energy. For MC-Glauber results, the fit to
the data is fair below $p_{T} \sim$ 1.5 GeV/$c$ and $p_{T} \sim$ 0.8
GeV/$c$ momenta in the 10-20\% and 40-50\% centralities,
respectively. Results from the MC-KLN initialisation at both
energies are significantly larger than experimental data in the
whole $p_{T}$ region, which again indicates the necessity of viscous
corrections in hydrodynamic evolution. For both initialisations the
difference between the data and the calculated $v_2(p_T)$ is larger
in more peripheral collisions. This too can be understood as an
indication of viscosity, since the more peripheral the collision,
the smaller the system and the more anisotropic its shape, and both
of these qualities enhance the dissipative effects.

Due to the relationships among the $p_T$ spectrum, $p_T$
averaged $v_2$, and $p_T$ differential $v_2(p_T)$, the flatter the
$p_T$ spectrum, the larger the $v_2$ even if $v_2(p_T)$ stays the
same. It is also worth noticing that the steeper the slope of
$v_2(p_T)$, the larger the increase in $v_2$ for the same increase
in mean $p_T$. This is the main reason why quite a similar increase
of mean $p_T$ for both MC-Glauber and MC-KLN initialisations leads
to much larger increase of $v_2$ for MC-KLN than for MC-Glauber
initialisation.

Song {\it et al.}\ calculated $v_{2}$ as a function of centrality and
$p_{T}$ using viscous hydrodynamics by employing almost the same
initial conditions as we
did~\cite{Song:2010mg,Song:2010aq,Song:2011hk,Song:2011qa} (see also
Sec.~\ref{sec:e-by-e_hydro}).  So it is interesting to see how much
viscosity is required to reproduce these data in
their analyses. As shown in Fig.~\ref{fig:hybrid8}, 
$\eta/s \sim 0.08$, which is almost identical to the conjectured
minimum value \cite{Kovtun:2004de}, leads to reproduction of the
$v_{2}/\varepsilon$ data at RHIC in the MC-Glauber initialisation,
whereas $\eta/s \sim 0.16$ in the MC-KLN initialisation
\cite{Song:2010mg}.  Within the MC-KLN initialisation, they also
calculated $v_{2}$ at the LHC energy and found that 
$\eta/s \sim 0.20$-0.24 is required to describe the data.  This
increase tendency of the specific shear viscosity with temperature is
qualitatively consistent with expectation from results based on finite
temperature QCD \cite{AMY}.  Temperature dependent shear viscosity was
also discussed in \cite{Song:2010mg,Niemi:2011ix,Niemi:2012ry}.
It turned out that extracting the temperature dependence of
  $\eta/s$ from the data is demanding, since at RHIC the value of
  $\eta/s$ in the plasma phase does not affect the observed
  anisotropies. Only the minimum value reached in the transition
  region and viscosity in the hadronic phase do. At LHC the situation
  is better, but even there the hadronic viscosity affects the results
  as much as the plasma viscosity
  does~\cite{Niemi:2011ix,Niemi:2012ry}.

\section{Results from event-by-event 
hybrid simulations \label{sec:result2}}

We perform hydrodynamic simulations on an event-by-event basis
and calculate observables using $\sim$10$^5$  
``minimum bias"  events 
(events with $N_{\mathrm{part}}\ge2$ in our theoretical definition)
for each initial parameter set.
In this section, we especially focus on higher order harmonics using
several flow analysis methods.
We first overview the flow analysis methods employed in this study.
Using these methods, we analyse final particle distributions
to obtain azimuthal anisotropy coefficients.
Most of the results are obtained using
the MC-KLN initialisation at the LHC energy.
We also compare these results with the ones obtained using
the MC-Glauber initialisation.

\subsection{\it Event plane method\label{sec:eventplane}}

In the event plane method \cite{Poskanzer:1998yz},
the event plane is 
first determined using the anisotropy of the emitted particles.
The event plane is in general defined by
\begin{eqnarray}
n\Psi_{n}^{\mathrm{EP}} = \arg \sum_{j\neq i} \omega_j e^{i n \phi_j },
\end{eqnarray}
where 
$\omega_i$ is an weight and
$\phi_i$ is an azimuthal angle for each particle.
The sum is taken over an ensemble of particles
which do not coincide with particles
which $v_n$ one wants to obtain.

In fact, there have been various event plane methods 
for the different detector setups.
For example,
one can randomly divide measured particles
in a pseudorapidity region into two subgroups.
Then when one wants to obtain flow parameters for particles in one subgroup,
particles in the other subgroup are used to determine the event plane.
One can also choose two groups of particles
separated in pseudorapidity to determine the event plane,
which can eliminate short range correlations.
	
In what follows, we demonstrate the flow analysis according to the event
plane method by the ATLAS Collaboration \cite{ATLAS:2011ah}.\footnote{
The method employed here is categorised in the event plane method
in a broad sense. However, this is sometimes
called the ``$\eta$-subevent" method.}
In this analysis method, the event planes are determined using particles
in the two regions, A: $3.2 <  \eta  < 4.8$ and B: $-4.8 <  \eta  < -3.2$.
 $\omega_i$ is taken as transverse mass $m_{T}^{i}$ for each particle.
When the harmonics $v_{n}$ is calculated
in positive (negative) rapidity region,  
particles in the region B (the region A) are used to determine
the event plane $\Psi_{n}^{\mathrm{B}}$ ($\Psi_{n}^{\mathrm{A}}$)
to avoid the non-flow effect from the autocorrelation.
Centrality is defined using the total transverse energy of charged particles
deposited in these rapidity regions, 
as prescribed 
by the ATLAS Collaboration \cite{ATLAS:2011ah}.

In the ``$\eta$-subevent" method,
non-flow effects can be eliminated at midrapidity when event plane angle
is determined away from midrapidity.
On the other hand,  one naively
anticipates
the correlation between 
the event plane angle determined
in the large rapidity region and the one in the whole
rapidity region gets weaker~\cite{Petersen:2011fp}.
As will be shown, this can be corrected by taking account of
event plane resolution in the ``$\eta$-subevent" method.

Event plane angles for the $n$-th harmonics is thus
calculated using the particles
in the regions A and B as
\begin{eqnarray}
\label{eq:EPATLAS1}
n \Psi_{n}^{\mathrm{A}} & = &\arg \sum_{\mathrm{A}} m_{T}^{i} e^{i n \phi_i },\\
\label{eq:EPATLAS2}
n \Psi_{n}^{\mathrm{B}} & = &\arg \sum_{\mathrm{B}} m_{T}^{i} e^{i n \phi_i },
\end{eqnarray}
Using these angles, one obtains $n$-th harmonics
\begin{eqnarray}
v_{n}^{\mathrm{obs}} = \frac{1}{\langle N^{\mathrm{P}} + N^{\mathrm{N}} \rangle} \left\langle \sum_{\mathrm{P}} \cos n (\phi_i -\Psi_{n}^{\mathrm{B}}) +\sum_{\mathrm{N}} \cos n (\phi_i -\Psi_{n}^{\mathrm{A}})\right\rangle
\end{eqnarray}
where $\langle \cdots \rangle$ denotes the event average.
$N^{\mathrm{P}}$ ($N^{\mathrm{N}}$) is the number of particles in the positive
(negative) rapidity region.
Since the finite number of measured particles limits
resolution in estimating the event plane angle $\Psi_{n}$,
one has to consider the so-called resolution parameter $\mathcal{R}_{n}$
\begin{eqnarray}
\mathcal{R}_{n} & = & \langle \cos n(\Psi_{n} - \Psi_{n}^{\mathrm{true}}) \rangle,
\end{eqnarray}
where $\Psi_{n}$ is the event angle estimated using the measured particle
and $\Psi_{n}^{\mathrm{true}}$ is an ideal event plane angle corresponding to the infinite
number of measured particles.
In this event plane method, the resolution factor can be estimated as
\begin{eqnarray}
 \langle \cos n(\Psi_{n}^{\mathrm{A}} - \Psi_{n}^{\mathrm{B}}) \rangle
& = & \langle \cos n(\Psi_{n}^{\mathrm{A}} - \Psi_{n}^{\mathrm{B}}) \rangle \nonumber\\
& = & \langle \cos n(\Psi_{n}^{\mathrm{A}} - \Psi_{n}^{\mathrm{true}} +  \Psi_{n}^{\mathrm{true}}-\Psi_{n}^{\mathrm{B}}) \rangle \nonumber\\
& = & \langle \cos n(\Psi_{n}^{\mathrm{A}} - \Psi_{n}^{\mathrm{true}}) \cos n(\Psi_{n}^{\mathrm{B}}-\Psi_{n}^{\mathrm{true}}) \rangle  - \langle \sin n(\Psi_{n}^{\mathrm{A}} - \Psi_{n}^{\mathrm{true}}) \sin n(\Psi_{n}^{\mathrm{B}}-\Psi_{n}^{\mathrm{true}})\rangle \nonumber\\
& = & \langle \cos n(\Psi_{n}^{\mathrm{A}} - \Psi_{n}^{\mathrm{true}})\rangle\langle \cos n(\Psi_{n}^{\mathrm{B}}-\Psi_{n}^{\mathrm{true}}) \rangle \nonumber\\
& = & \mathcal{R}_{n}^{2}
\end{eqnarray}
Here we have assumed that the two groups are symmetrically 
located with respect to midrapidity
like the pseudorapidity regions A and B
so that multiplicities in the two groups are almost equal
and independent
in a sense that two-particle correlation function between
a particle from region A and the one from region B can
be factorised into two one-particle distribution functions. 
Thus,
\begin{eqnarray}
\langle \cos n(\Psi_{n}^{\mathrm{A}} - \Psi_{n}^{\mathrm{true}})\rangle=
\langle \cos n(\Psi_{n}^{\mathrm{B}} - \Psi_{n}^{\mathrm{true}})\rangle
\equiv \mathcal{R}_{n} 
\end{eqnarray}
Thus the anisotropic parameters using the event plane method become
\begin{eqnarray}
v_{n}\text{\{EP\}} & = & \frac{v_{n}^{\mathrm{obs}}}{\mathcal{R}_{n}},\\
\mathcal{R}_{n} & = & \sqrt{ \langle \cos n \left(\Psi_n^{\mathrm{A}} - \Psi_{n}^{\mathrm{B}} \right)\rangle}.
\label{eq:resoATLAS}
\end{eqnarray}

\begin{figure}[tb]
\begin{center}
\begin{minipage}[t]{9 cm}
\epsfig{file=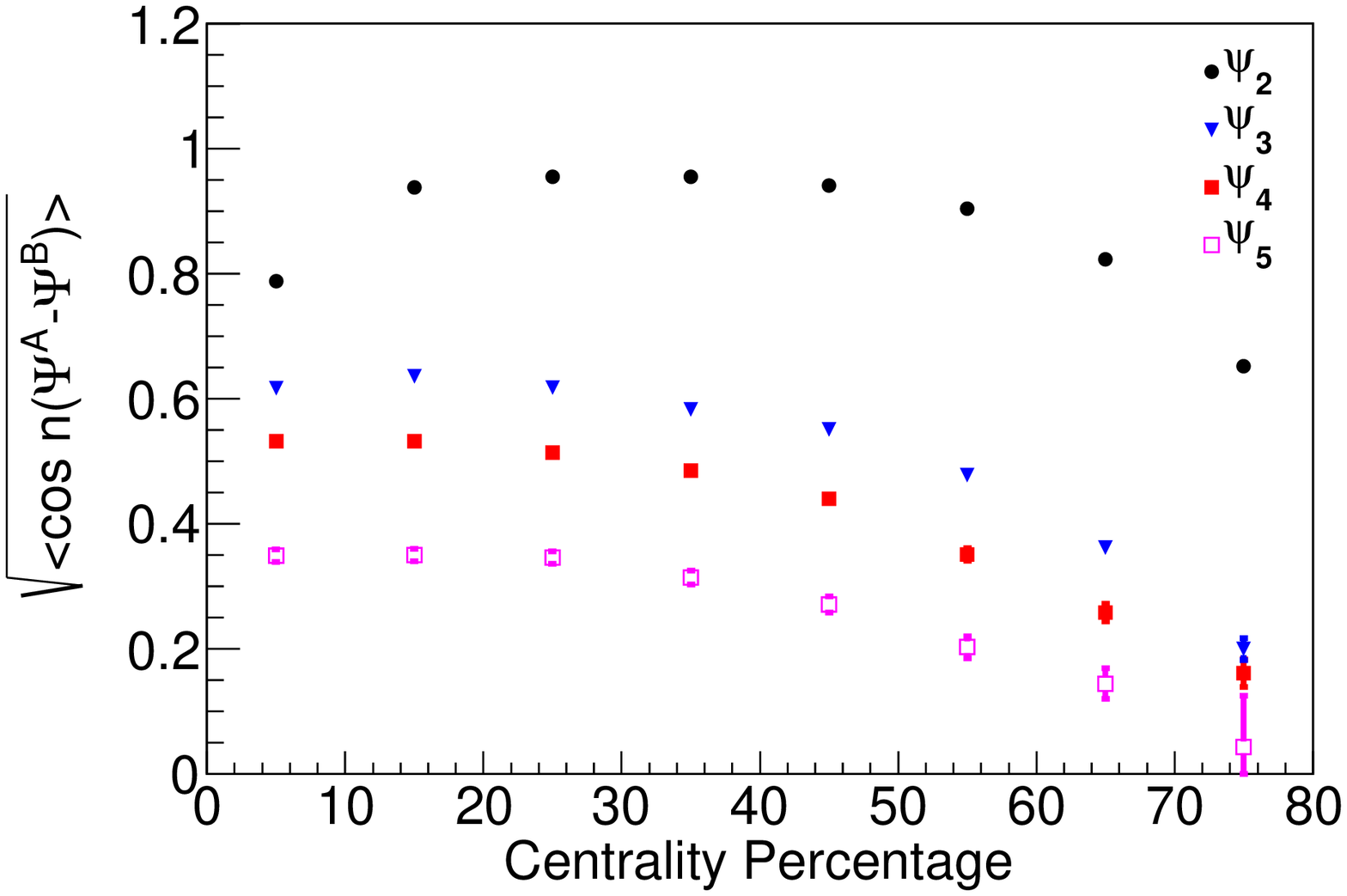,scale=0.45}
\end{minipage}
\begin{minipage}[t]{9 cm}
\epsfig{file=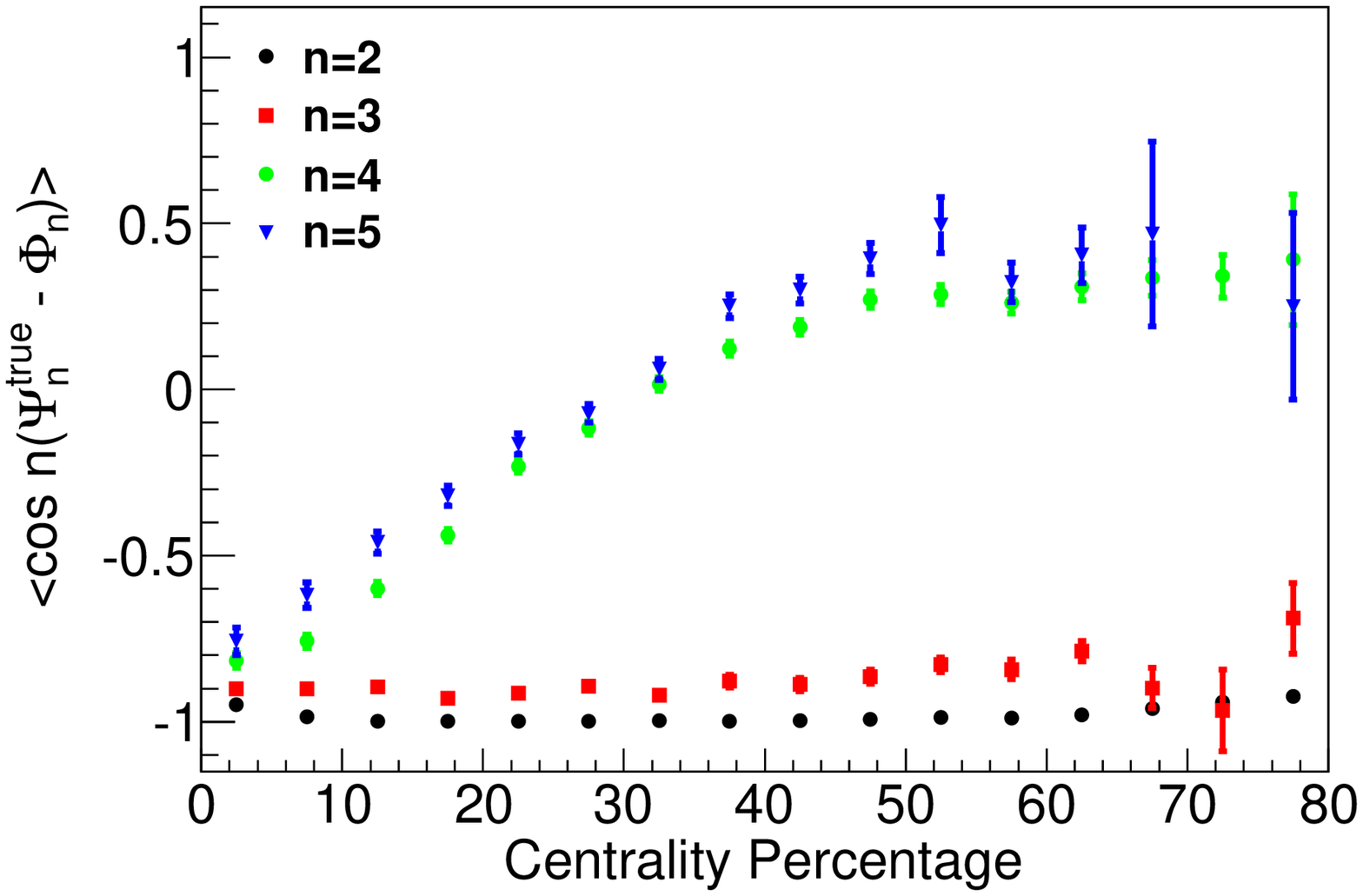,scale=0.45}
\end{minipage}
\caption{
(Left) Resolution parameter defined in Eq.~(\ref{eq:resoATLAS})
 for $n=2$, 3, 4 and 5 as a function of centrality.
(Right) Correlation $\langle \cos n (\Psi_{n}^{\mathrm{true}} - \Phi_{n}) \rangle$
between the true event plane and the participant plane as a function of centrality
in Pb+Pb collisions at $\sqrt{s_{NN}}$ = 2.76 TeV.
\label{fig:reso}}
\end{center}
\end{figure}

Figure \ref{fig:reso} (left) shows the resolution parameter $\mathcal{R}_{n}$
 ($n=2$, 3, 4 and 5) as a function of centrality.
Event plane resolution for the second harmonics
reaches almost unity in mid central collisions (20-30 \%) and
is relatively better than the others as expected from
the almond-like geometry on average.
The other event plane resolutions 
become worse as decreasing multiplicity and increasing the order $n$.
To understand the origin of the event plane, we also calculate
correlation of angles between 
the orientation angle $\Phi_n$ (see Eg.~(\ref{eq:Phin})) and the true event plane
\begin{eqnarray}
\mathcal{C}_{n}(\Psi_{n}^{\mathrm{true}}, \Phi_n) & = & \langle \cos n (\Psi_{n}^{\mathrm{true}} - \Phi_{n}) \rangle \nonumber \\
& = & \frac{\langle \cos n (\Psi_{n}^{\mathrm{A/B}} - \Phi_{n})\rangle}{\mathcal{R}_{n}}.
\end{eqnarray}
Since the number of produced particles is finite, the resolution parameter
to correct $\Psi_{n}^{A/B}$ plays again an essential role in evaluating $\mathcal{C}_{n}$.
When the anisotropic flow is generated by anisotropic pressure gradient
in the transverse plane, the correlation becomes $\sim -1$: In the case
of positive elliptic flow generated from a conventional smooth almond-like profile
elongated in $y$ direction,
$\Psi_2^{\mathrm{true}} \sim 0$ and $\Phi_2 \sim \pi/2$ and, consequently,
$\mathcal{C}_{2} \sim -1$.  
Figure \ref{fig:reso} (right)
shows $\mathcal{C}_{n}$ ($n$=2, 3, 4 and 5)
as a function of centrality.
$\mathcal{C}_{n}$ for $n$=2 and 3 are close to $-1$.
So one can interpret anisotropic flow $v_{n}$ ($n=$2 or 3)
as generated by the corresponding anisotropy $\varepsilon_{n}$ of
  the pressure gradient.
While $\mathcal{C}_{4}$ and $\mathcal{C}_{5}$
start from $-1$ in very central collisions,
increase with centrality percentile and
even become positive above 30\% centrality.
This indicates the possibility of having 
finite $v_{n}$ owing to $\varepsilon_{m}$ $(m\neq n)$.

\begin{figure}[tb]
\begin{center}
\epsfig{file=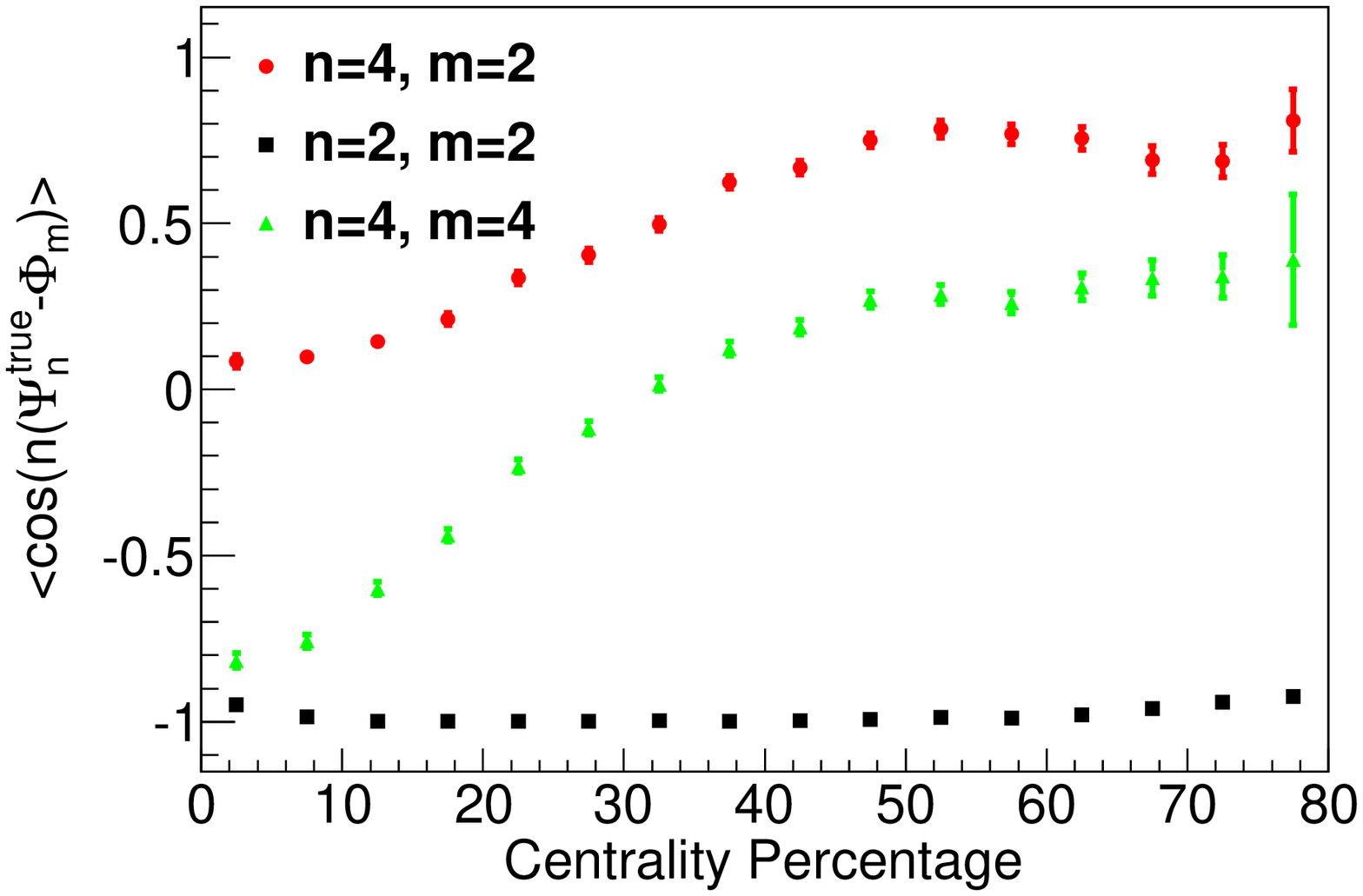,scale=0.45}
\caption{
Mixed correlation
between the true event plane $\Psi_{n}^{\mathrm{true}}$ and 
the orientation angle
$\Phi_{m}$ as a function of centrality
in Pb+Pb collisions at $\sqrt{s_{NN}}$ = 2.76 TeV.
\label{fig:mixedcorr}}
\end{center}
\end{figure}

To further investigate the origin of the event plane,
we also calculate mixed correlations
\begin{eqnarray}
\mathcal{C}_{nm}(\Psi_{n}^{\mathrm{true}}, \Phi_m) & = & \langle \cos n
 (\Psi_{n}^{\mathrm{true}} - \Phi_{m}) \rangle, \nonumber \\
\mathcal{C}_{nn}& = &\mathcal{C}_{n}.
\end{eqnarray}
Figure \ref{fig:mixedcorr} shows 
$\mathcal{C}_{42}$ together with $\mathcal{C}_{2}$ and $\mathcal{C}_{4}$
in Pb+Pb collisions at $\sqrt{s_{NN}}$ = 2.76 TeV.
If $v_{4}$ is generated solely by
$\varepsilon_{2}$, 
$|\Psi_{4}-\Phi_{2}|$ is expected to be
either 0 or $\pi/4$ due to symmetry
in the transverse plane and, consequently,
$\mathcal{C}_{42} = 1$ or $-1$.
In central collisions where the reaction zone
is almost cylindrical on average, 
there is almost no correlation between
$\Psi_{4}$ and $\Phi_{2}$.
However, the correlation between them
increases towards unity as increasing centrality percentage,
which indicates $\Psi_{4} \approx \Phi_{2} \pm n\pi/2$ ($n$: integer)
and $v_{4}$ is partly generated by $\varepsilon_{2}$ like positive
elliptic flow in non central 
collisions~\cite{Qin:2010pf,Qiu:2011iv,Borghini:2005kd}.

Relations between initial geometry fluctuation and final anisotropic
flow were discussed in Ref.~\cite{Qin:2010pf}.  They found that
$v_{4}$ decreases with increasing $\varepsilon_{2}$ and changes its
sign from positive to negative (see Fig.~17 in
Ref.~\cite{Qin:2010pf}).  The behaviour is quite similar to our
finding that the centrality dependence of $\mathcal{C}_{42}$ changes
also its sign.  Distributions of $\Psi_n - \Phi_n$ for $n=4$ and 5 at
the RHIC energy were investigated in Fig.~8 in Ref.~\cite{Qiu:2011iv}.
Contrary to the cases for $n=2$ and 3, the distributions for $n=4$ and
5 have a peak at $\Psi_n - \Phi_n =0$ only in central collisions.
Peak structure gradually disappears as moving to peripheral collisions
and eventually another peak appears at $\Psi_n - \Phi_n =\pi/n$ in the
peripheral collisions (50-60\% centrality).  These results also
suggest that $v_{n}$s at least $n=4$ and 5 are generated by other
order of initial geometry $\varepsilon_{m}$ ($m\neq n$).

Before showing detailed results from event-by-event hydrodynamic
simulations, we make two remarks here:
Since the event plane is often determined using particles in forward
and backward rapidity regions (also known as 
``$\eta$-subevent method''),
 full three dimensional dynamical simulations are essential if
  one wants to do the flow analysis in the same way than in
  experiments.
For example, the PHENIX Collaboration at RHIC 
utilises the Beam Beam Counter (BBC) or the reaction plane detector (RxNP) in forward and
backward rapidity regions to determine
event planes.
If one assumes boost invariance in dynamical calculations, one cannot perform
this kind of flow analysis exactly.

Second, the effect of experimental event plane resolution cannot
  be properly evaluated unless one samples a finite number of particles at
  particlisation.
The Cooper-Frye formula
\cite{Cooper:1974mv} has been conventionally used to calculate
particle spectra in hydrodynamic calculations, but it gives
smooth and
continuous function of momentum distributions. This corresponds to
the $N$ times over-sampling of particles from a fluid element with $N
\rightarrow \infty$ limit, which is, however, not adequate if
event-by-event hydrodynamic simulations are supposed to describe
actual events.

\subsection{\it Multi-particle cumulants\label{sec:cumulant}}

In the previous subsection, we described the method
to calculate anisotropic parameters with respect to
the event plane.
In this subsection, we discuss how to calculate
anisotropic parameter using the
particle ensemble itself, namely, the multi-particle cumulant
 method \cite{Borghini:2000sa,Borghini:2001vi}.

We first define a $2p$-particle correlation to calculate the $n$-th order
of higher harmonics as
\begin{eqnarray}
c_{n}\{2p\} & = & 
\langle \exp [i n (\phi_{1} + \phi_{2} + \cdots \phi_{p} - \phi_{p+1} - \cdots - \phi_{2p} )]\rangle 
\end{eqnarray}
where $\phi_{i}$ is azimuthal angle of the $i$-th particle under consideration.
In this subsection, $\langle \cdots \rangle$ means an average taken in two steps: First,
one averages over all possible permutations
of $p$ particles in the same event, then one averages over all events.
For example, two-particle correlations can be written as 
$c_{n}\{2\} = \langle e^{in(\phi_1-\phi_2)} \rangle$.

We next define the corresponding cumulant.
Correlations among $2p$ particles can be in general decomposed into
a sum of correlations among smaller number of particles and cumulants.
The simplest example is two-particle cumulant
\begin{eqnarray}
c_{n}\{2\} & = & d_{n}\{2\} + \langle e^{in\phi_{1}} \rangle 
\langle e^{-in\phi_{2}}\rangle \nonumber \\
& = & \langle e^{in(\phi_{1}-\phi_{2})}\rangle.
\end{eqnarray}
Due to symmetry, a term $\langle e^{in\phi_{1}} \rangle$ should vanish.
As we see, a two-particle cumulant reduces to the corresponding two-particle correlation.
This quantity contains correlation from collective flow as well as
the so-called non-flow effects such as two-body decays of
resonance particles.
The next non-trivial and important example is
four-particle cumulant:
\begin{eqnarray}
c_{n}\{4\} & = & 
\langle \exp [i n (\phi_{1} + \phi_{2}  - \phi_{3} -  \phi_{4} )]\rangle \nonumber \\
& = & \langle e^{in(\phi_{1}-\phi_{3})}\rangle \langle e^{in(\phi_{2}-\phi_{4})}\rangle
+\langle e^{in(\phi_{1}-\phi_{4})}\rangle \langle e^{in(\phi_{2}-\phi_{3})}\rangle
+ d_{n}\{4\} \nonumber \\
& = & 2c_{n}\{2\}^2 + d_{n}\{4\}.
\end{eqnarray}
Although $d_{n}\{4\}$ still contains 
non-flow effects of correlations among four particles,
this does not contain non-flow effects from two-particle correlations
and, consequently, is expected to contain much information about the
anisotropic flow.
In fact, it was shown in Ref.~\cite{Borghini:2001vi} that
these cumulants are related with higher order anisotropic flow.
So one can define higher order anisotropic parameter using $2p$-particle
cumulants as
\begin{eqnarray}
v_{n}\{2\}^{2} & = & d_{n}\{2\}, \\
v_{n}\{4\}^{4} & = & -d_{n}\{4\}. 
\end{eqnarray}
Since this method requires many particles to take correlations/cumulates,
the statistical errors tend to be large. 
Even 10$^{5}$ events are not enough:
It happens that error bars are too large to show the results 
in our statistics, which indicates the necessity of 
massive numerical simulations.

In the actual calculations, we evaluate the above correlations as follows:
We first calculate the following two quantities:
\begin{eqnarray}
q_{n} & = & \frac{1}{\sqrt{M}}\sum_{i=1}^{M} e^{in \phi_{i}},\\
u_{n} & = &  \frac{1}{M}\sum_{i=1}^{M} e^{in \phi_{i}}.
\end{eqnarray}
$q_{n}$ is a flow vector and $u_{n}$ is the one with different normalisation.
When the non-flow effects can be neglected,
$v_{n} \approx \sqrt{u_{n}^{*}u_{n}}$.
In the case of $2p$-particle correlations,  one needs to average over all possible
permutations 
excluding self-correlation terms
and take a sum such as
\begin{eqnarray}
\sum_{(i,j)} & = &\sum_{i=1}^{M} \sum_{j=1}^{M} (1-\delta_{ij})\nonumber \\
 & = & \sum_{i=1}^{M} \sum_{j=1}^{M} - \sum_{i=j=1}^{M}, \\
\sum_{(i,j,k)} & = & \sum_{(i,j)}\sum_{k=1}^{M}(1-\delta_{ki}-\delta_{kj})\nonumber \\
& = & \sum_{i}^{M}\sum_{j}^{M}\sum_{k}^{M}-\sum_{i=j=1}^{M}\sum_{k=1}^{M}
-\sum_{j=k=1}^{M}\sum_{i=1}^{M}-\sum_{k=i=1}^{M}\sum_{j=1}^{M}+2\sum_{i=j=k=1}^{M}, \\
\sum_{(i,j,k,l)} & = &\sum_{(i,j,k)}\sum_{l=1}^{M} (1-\delta_{li}-\delta_{lj}-\delta_{lk}).
\end{eqnarray}
In this way, correlation functions reduce to \cite{murase}
\begin{eqnarray}
c_{n}\{2\} & = & \frac{\left\langle \sum_{i=1}^{M} e^{in\phi_{i}}
\sum_{j=1}^{M} e^{in\phi_{j}}-\sum_{i=j=1}^{M}e^{in(\phi_{i} - \phi_{j})} \right\rangle}
{\langle M(M-1) \rangle}\nonumber \\
& = & \frac{\langle |u_n|^{2} M^{2} - M \rangle}{\langle M^2-M \rangle},\\
c_{n}\{4\} & = & \frac{\langle U_{4}M^4 -6 U_{3}M^{3} +11 U_{2} M^{2} -6M\rangle}
{\langle M(M-1)(M-2)(M-3) \rangle}, \nonumber \\
& = & \frac{\langle U_{4}M^4 -6 U_{3}M^{3} +11 U_{2} M^{2} -6M\rangle}
{\langle M^4 -6M^3 +11M^2-6M\rangle},
\end{eqnarray}
where
\begin{eqnarray}
U_{4} & = & |u_n|^{4}, \\
6U_{3} & = & 4 |u_n|^2 +u_{2n}^{*}u_{n}^2 + u_{2n}u_{n}^{*2},\\
11U_{2} & = & 8|u_n|^2 + |u_{2n}|^2+2.
\end{eqnarray}

\subsection{\it Results\label{sec:LHCresultEbyE}}

\begin{figure}[tb]
\begin{center}
\epsfig{file=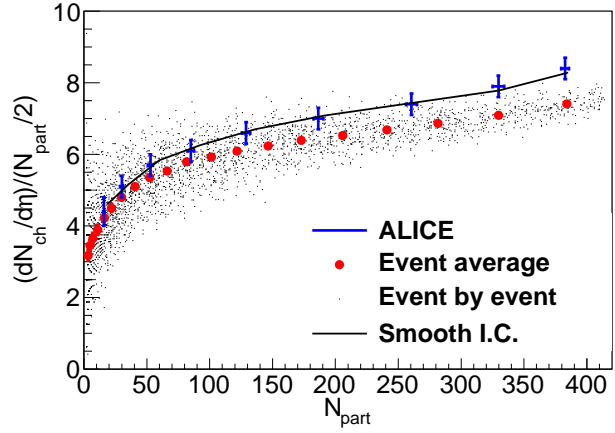,scale=0.45}
\caption{
$(dN_{\mathrm{ch}}/d\eta)/(N_{\mathrm{part}}/2)$
in $|\eta|<0.8$
as a function of $N_{\mathrm{part}}$
in Pb+Pb collisions at $\sqrt{s_{NN}} = 2.76$ TeV
for each event
and for events averaged over every 5\% centrality.
Here we use impact parameter to categorise the centrality
for simplicity
and plot results at $\langle N_{\mathrm{part}} \rangle$
for each centrality.
To exhibit how this quantity distributes event-by-event,
results from 3000 minimum bias events are shown.
Result from the smooth initial condition with the same initial parameter set
is also shown in thin solid line.
Experimental data are from ALICE Collaboration \cite{ALICEdNdeta2}.
\label{fig:dndetanpart2}}
\end{center}
\end{figure}

First, in Fig.~\ref{fig:dndetanpart2}, 
we show $(dN_{\mathrm{ch}}/d\eta)/(N_{\mathrm{part}}/2)$
as a function of $N_{\mathrm{part}}$
from event-by-event hydrodynamic simulations 
employing MC-KLN initialisation and compare
them with ALICE data \cite{ALICEdNdeta2}.
For comparison, we average over these events at every 5 \% centrality.
Although we used the same, adjusted parameter set as 
in the smooth initial conditions using
the MC-KLN initialisation which reproduces the ALICE data reasonably well,
results are systematically smaller than the data, in particular,
in central events.
In this paper, we do not try to fine-tune
initial parameters.
It would be interesting to see how the effects of the
  fluctuations of the gluon production itself would change this
  behaviour. In principle these fluctuations can be included using the
  negative binomial distribution (N.B.D.)
  ~\cite{Qin:2010pf,Tribedy:2010ab,Tribedy,Dumitru:2012yr}, but since
  the rapidity dependence of the required N.B.D. is not known, 
  it is not clear
  how to implement these fluctuations in a fully three dimensional
  calculations.

\begin{figure}[tb]
\begin{center}
\begin{minipage}[t]{6 cm}
\epsfig{file=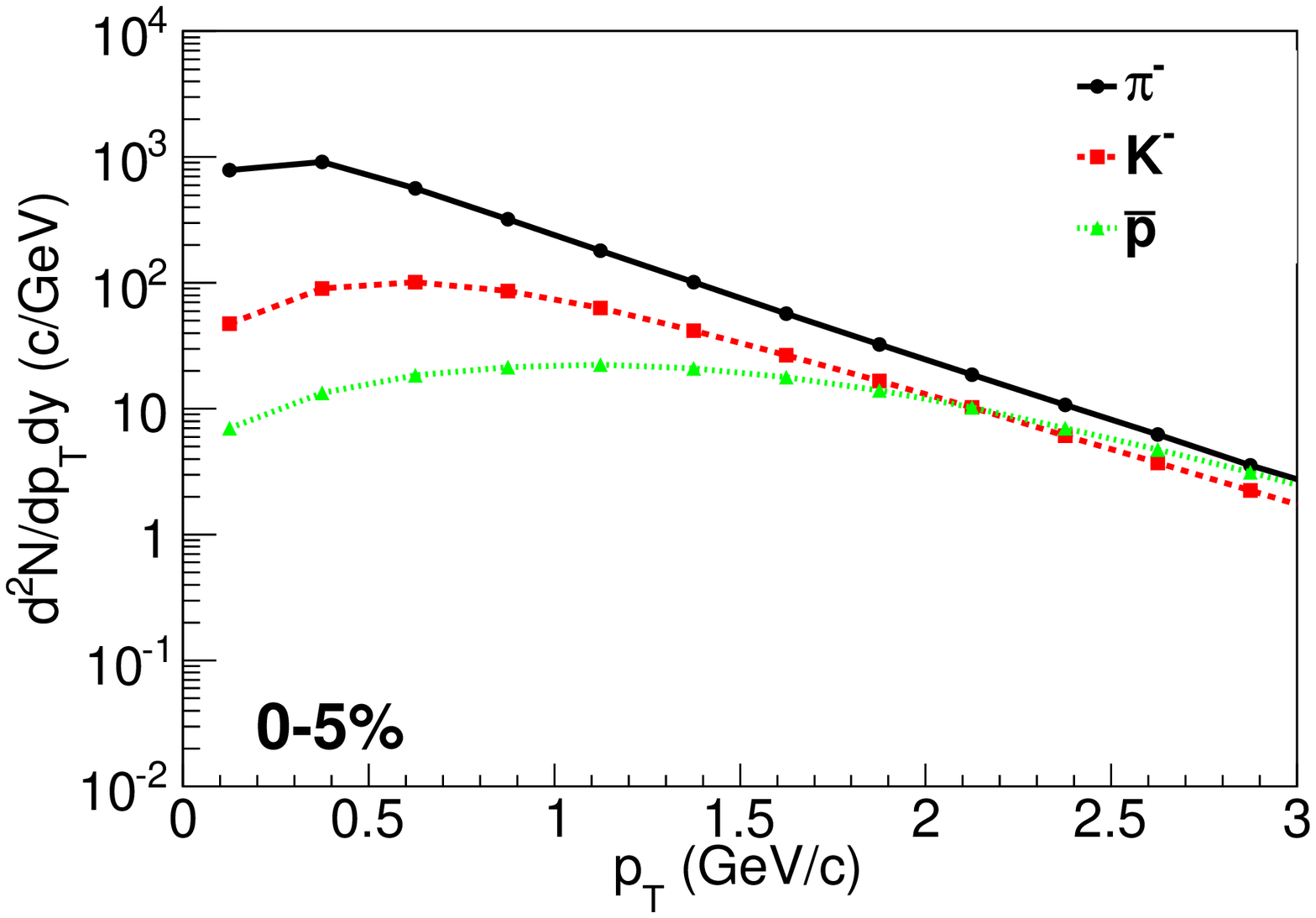,scale=0.3}
\end{minipage}
\begin{minipage}[t]{6 cm}
\epsfig{file=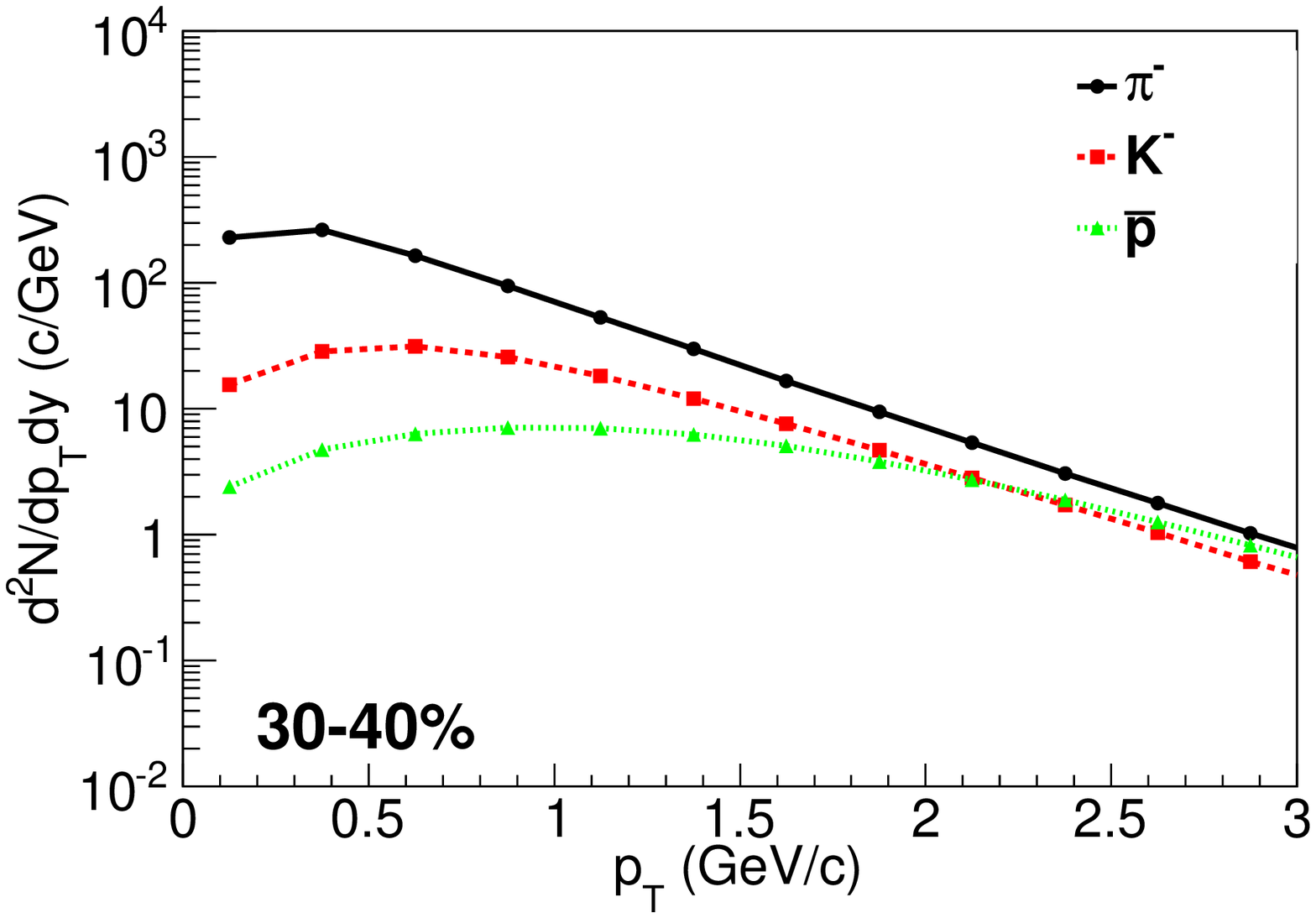,scale=0.3}
\end{minipage}
\begin{minipage}[t]{6 cm}
\epsfig{file=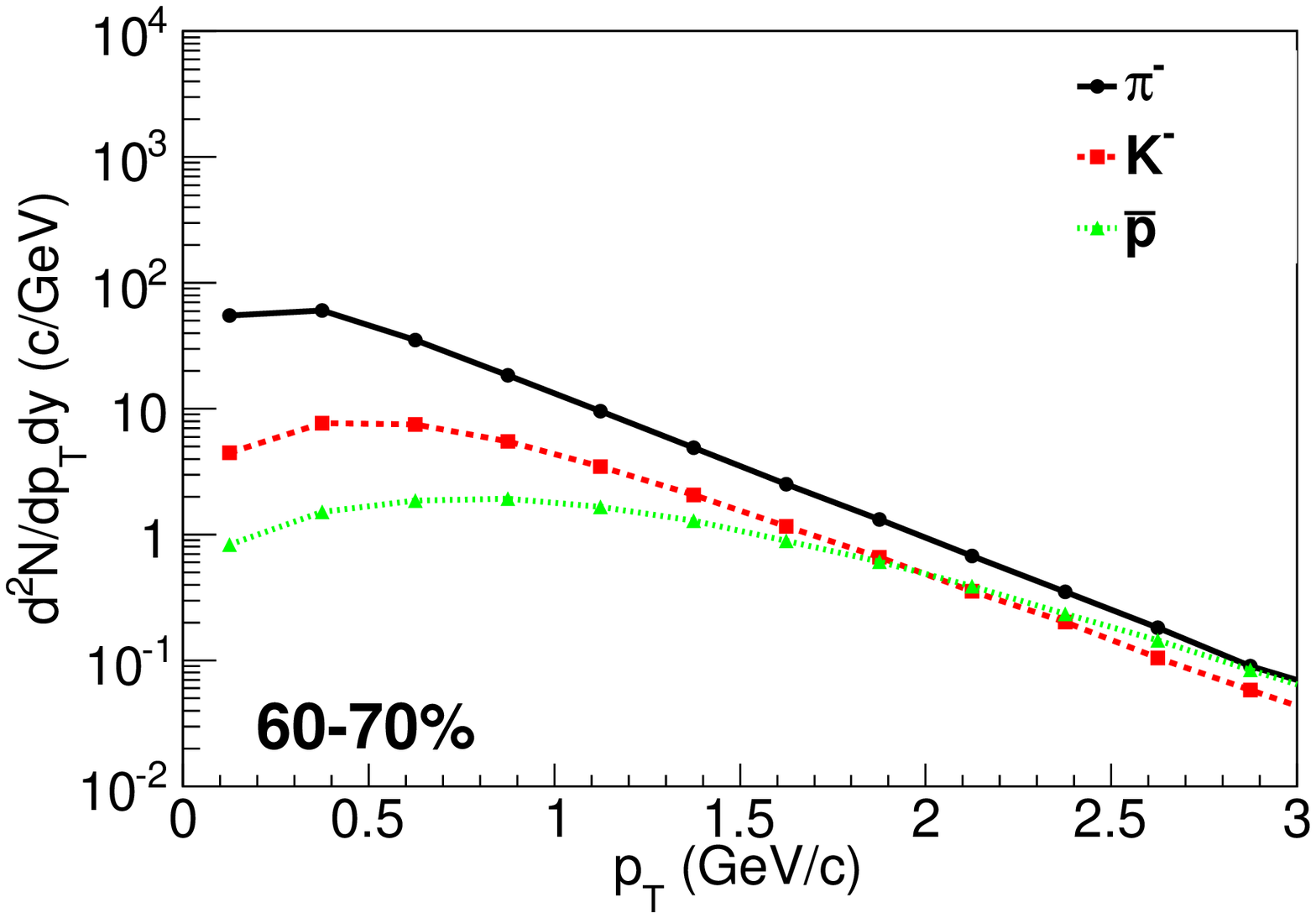,scale=0.3}
\end{minipage}
\caption{
$p_{T}$ distribution of identified hadrons
at midrapidity in Pb+Pb collisions at $\sqrt{s_{NN}}$ = 2.76 TeV
at 0-5\% (left), 30-40\% (middle) and 60-70\%  (right) centralities.
\label{fig:dndptPIDLHC}}
\end{center}
\end{figure}

In Fig.~\ref{fig:dndptPIDLHC},
$p_{T}$ distributions of $\pi^{-}$, $K^{-}$ and $\bar{p}$
using the MC-KLN initialisation in Pb+Pb collisions at the LHC energy
are shown at 0-5\%, 30-40\% and 60-70\% centralities.
Anti-proton yield becomes comparable with negative pion yields
at $p_{T}\sim$ 3 GeV/$c$ in all these results,
which is consistent with the preliminary ALICE data \cite{Floris:2011ru}.
Thus the switching temperature $T_{\mathrm{sw}} = 155$ MeV, which
  affects both particles yields and the slopes of the $p_{T}$ spectra,
  and was adjusted to reproduce particle ratios of identified hadrons
  at RHIC, works reasonably well at the LHC energy too.
Detailed comparison of these results with the data
would give more precise information about the switching temperature.

\begin{figure}[tb]
\begin{center}
\begin{minipage}[t]{9 cm}
\epsfig{file=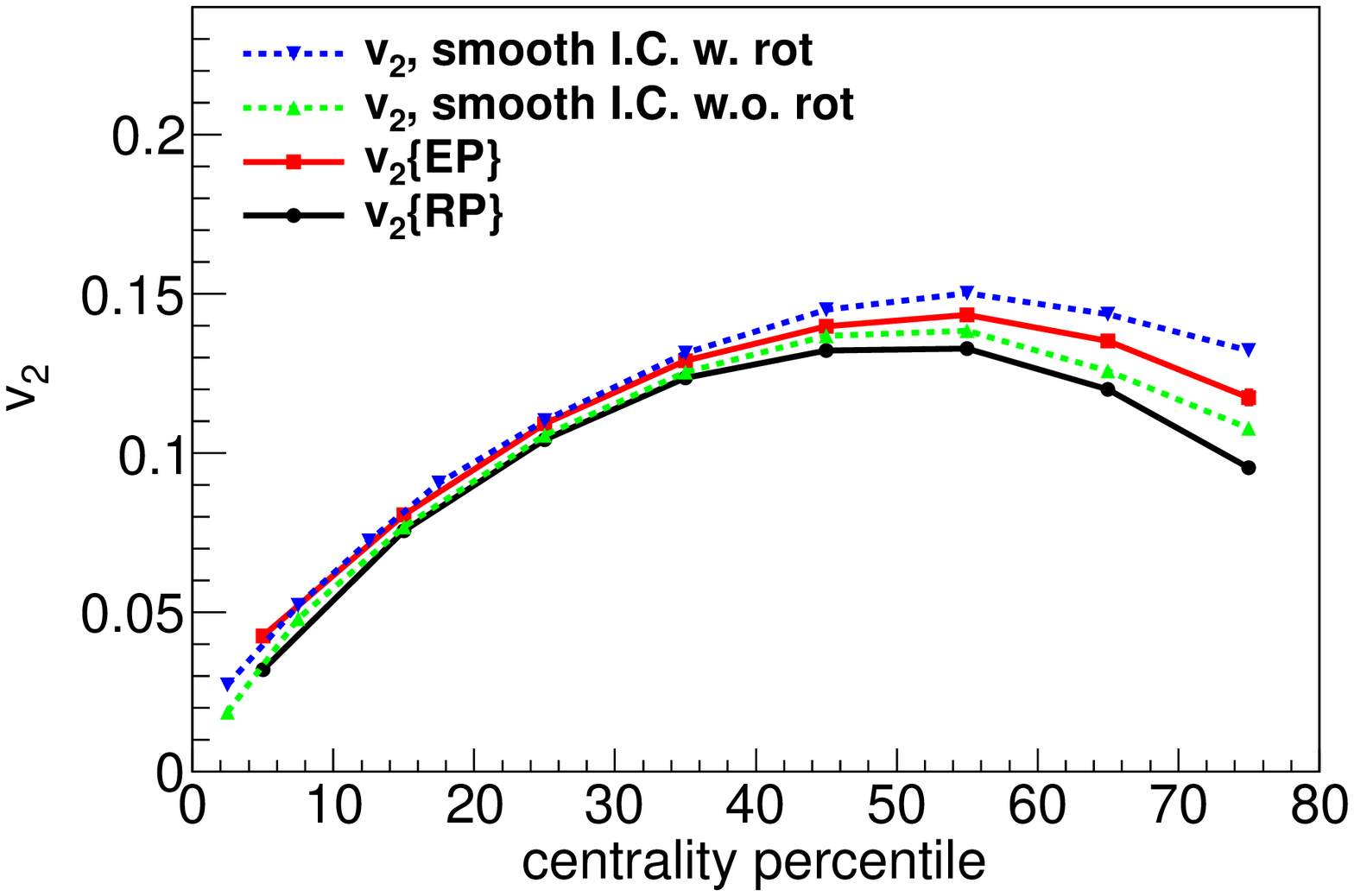,scale=0.45}
\end{minipage}
\begin{minipage}[t]{9 cm}
\epsfig{file=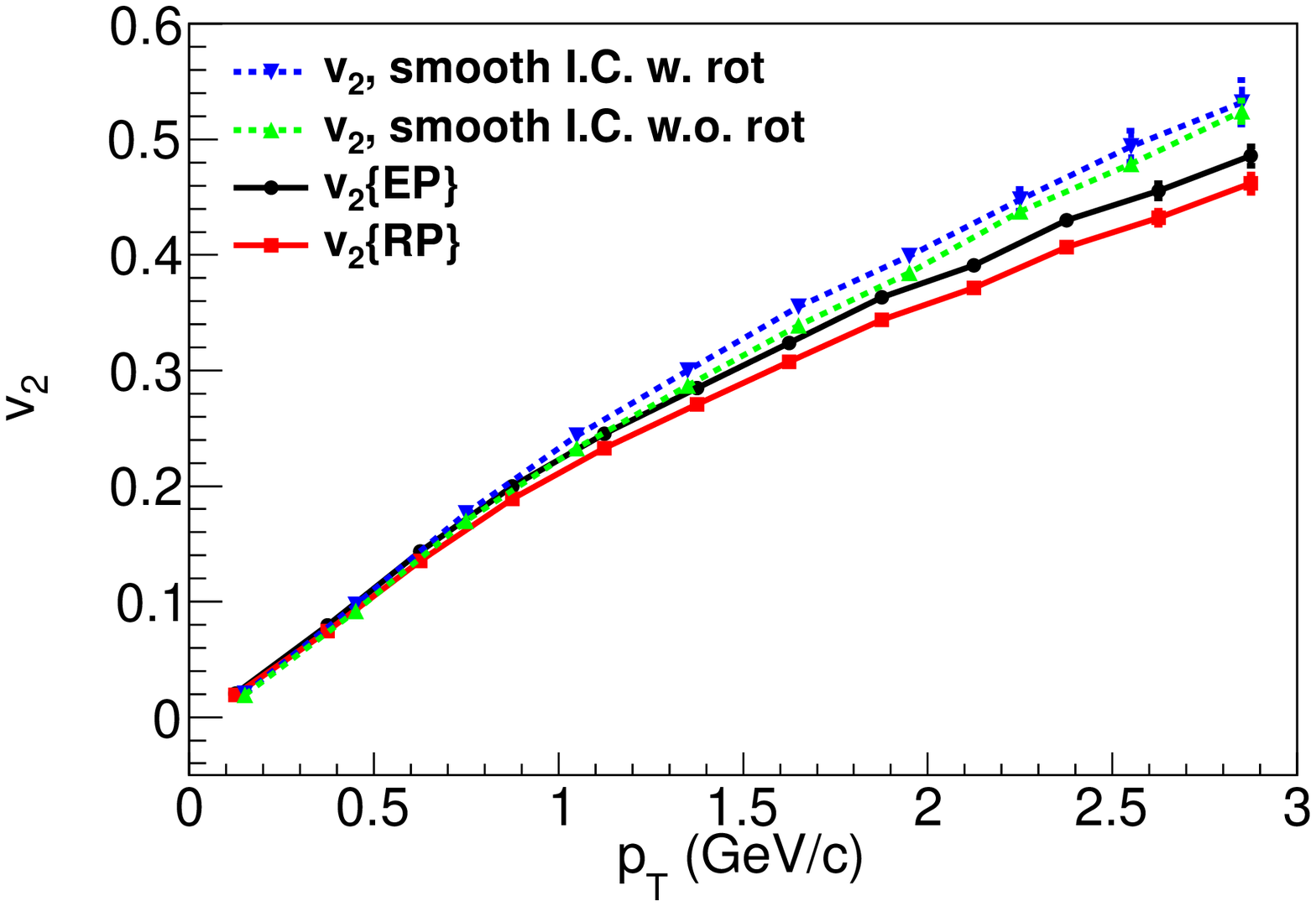,scale=0.45}
\end{minipage}
\caption{
Centrality (left) and transverse momentum (right) dependences
 of $v_{2}$ of charged hadrons 
in Pb+Pb collisions at $\sqrt{s_{NN}}$ = 2.76 TeV
using smooth initial conditions with (inverse triangle) 
or without (triangle) shifting the centre of mass and rotation to match
participant plane angle.
These are compared with $v_{2}$\{EP\} (square) and $v_{2}$\{RP\} (circle)
(see the text).
Transverse momentum dependence is shown at 40-50\% centrality.
\label{fig:comp}}
\end{center}
\end{figure}

We compare the results of $v_{2}$ from event-by-event hydrodynamic simulations
with those from hydrodynamic simulations with
conventional, smooth initial conditions.
Figure \ref{fig:comp} (left) shows centrality dependence of 
$p_{T}$-integrated $v_2$ ($0<\eta<1$) 
of 
charged hadrons in Pb+Pb collisions at $\sqrt{s_{NN}}$ = 2.76 TeV
using MC-KLN initialisation.
Although the difference between $v_{2}$\{EP\} and $v_{2}$
from event-averaged initial conditions with
shift of the centre of mass and rotation of participant plane
discussed in the previous section
is hardly seen
in central to semi-central collisions (0 - 40\%),
the two results deviate from each other above $\sim 40$\%:
$v_{2}$\{RP\}, which is
$v_{2}$ with respect to the reaction plane in event-by-event simulations 
(reaction plane method),
 is slightly smaller than $v_{2}$ from event averaged
initial conditions and the difference increases with increasing centrality
percentile.
This is also the case for the difference between $v_{2}$\{RP\}
and $v_{2}$ from event-averaged initial conditions without
shift of the centre of mass or rotation of participant plane.
Figure \ref{fig:comp} (right) shows
$p_{T}$ dependence of $v_{2}$ for charged hadrons
in the same collision system at 40-50\% centrality.
In both cases, there is almost no difference
between event-by-event simulations and simulations with event-averaged
initial conditions in the low $p_{T}$ region.
However, results deviate from each other gradually
with increasing $p_{T}$ above $\sim 1$ GeV/$c$ \cite{Andrade:2008xh}.
From a point of view of conventional hydrodynamic simulations with smooth
initial conditions, event-by-event simulations mimic shear viscous effects
since shear viscosity reduces $p_{T}$-integrated $v_{2}$ and $v_{2}$ at high 
$p_{T}$~\cite{Teaney:2003kp}.

\begin{figure}[tb]
\begin{center}
\begin{minipage}[t]{9 cm}
\epsfig{file=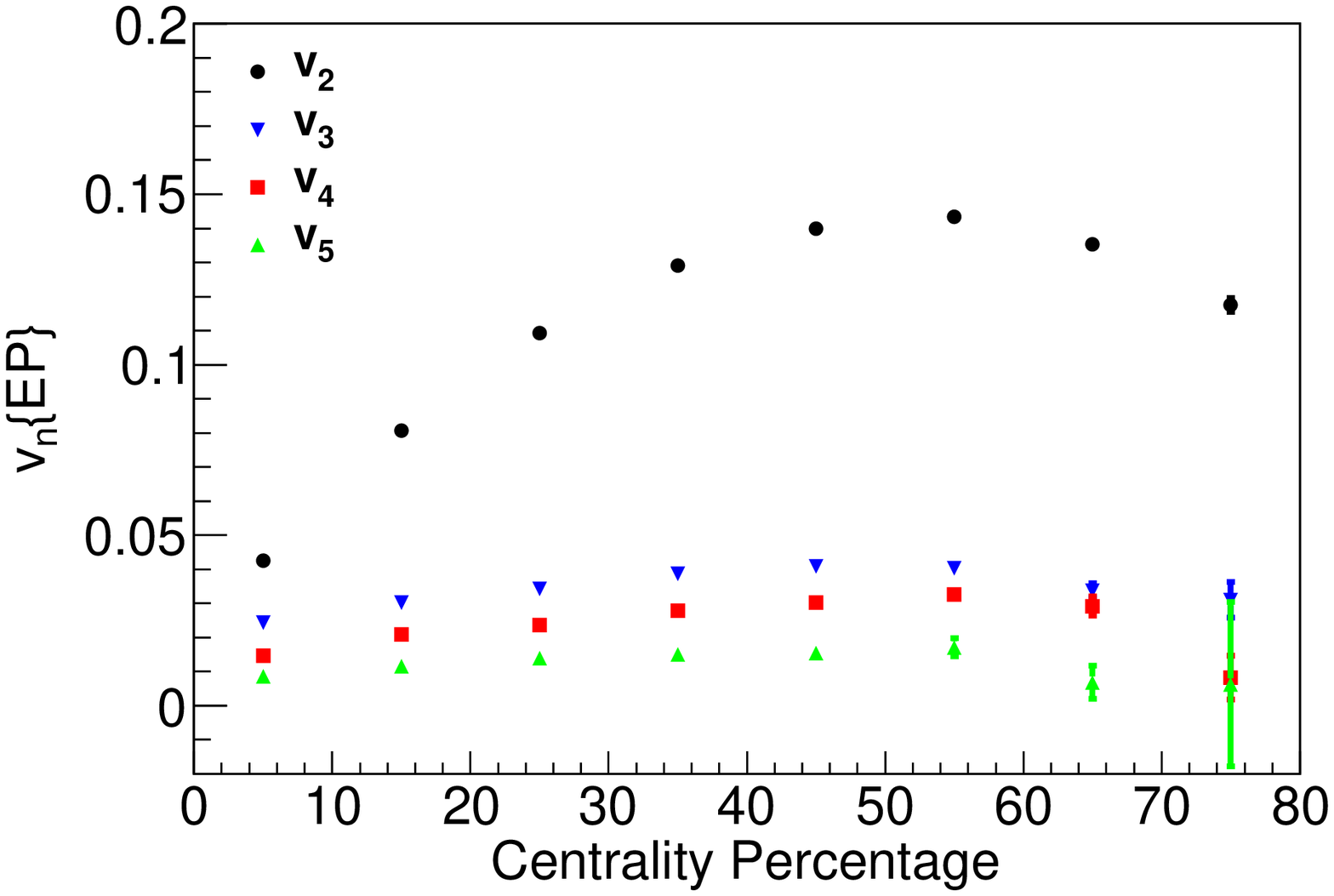,scale=0.45}
\end{minipage}
\begin{minipage}[t]{9 cm}
\epsfig{file=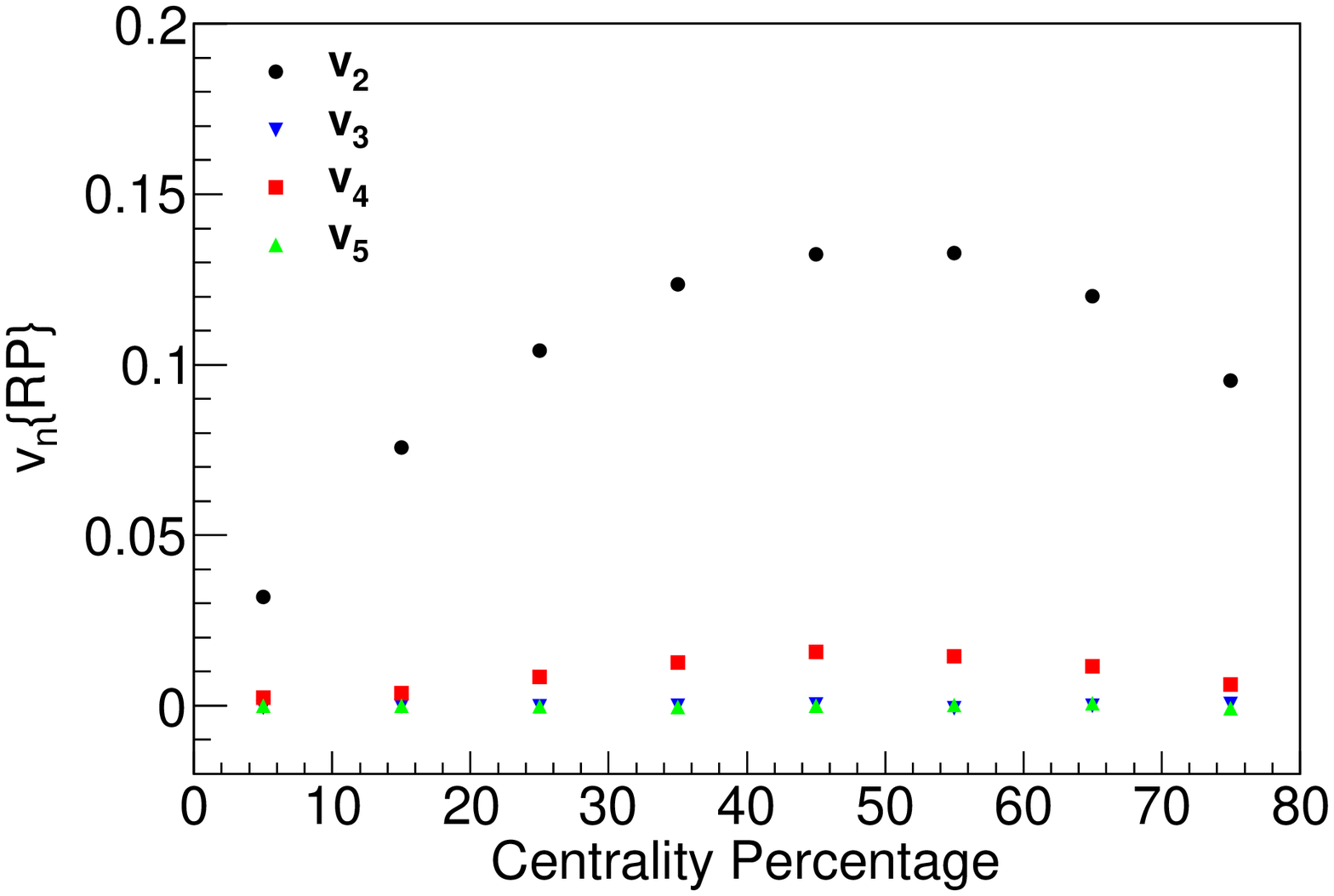,scale=0.45}
\end{minipage}
\begin{minipage}[t]{9 cm}
\epsfig{file=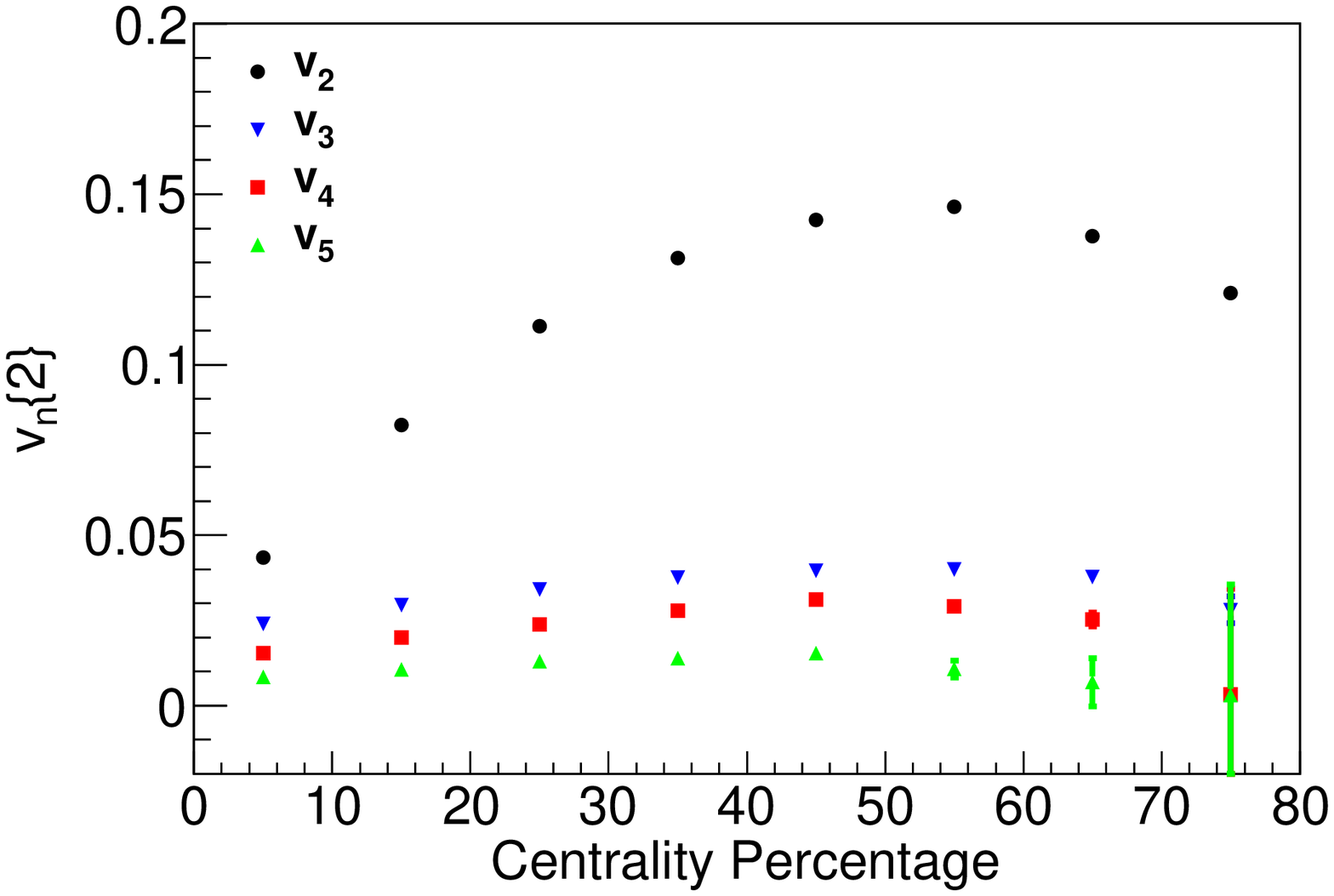,scale=0.45}
\end{minipage}
\begin{minipage}[t]{9 cm}
\epsfig{file=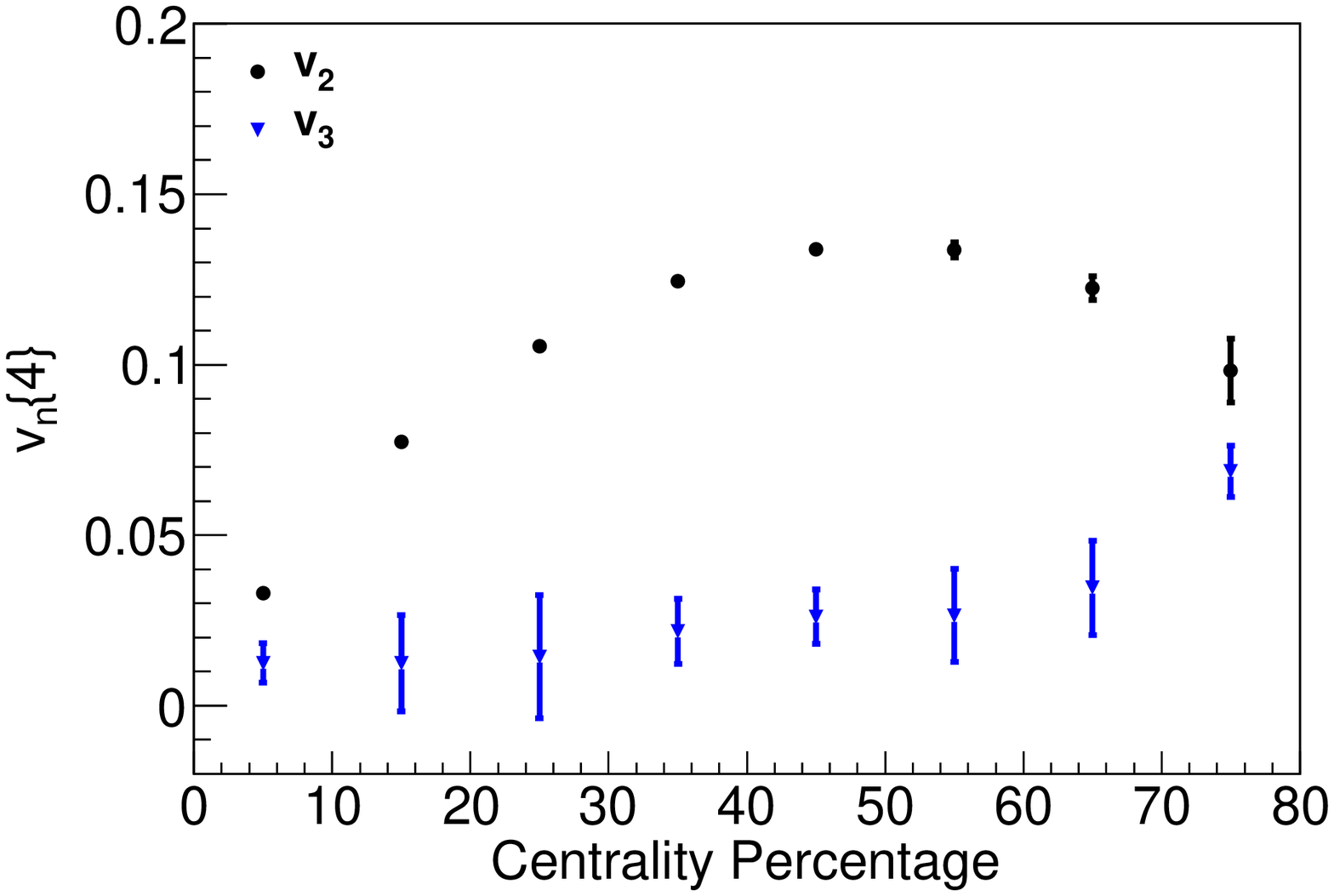,scale=0.45}
\end{minipage}
\caption{
$v_{n}$ using the event plane method (upper left),
$v_{n}$ using the reaction plane method (upper right)
$v_{n}$ using the two particle cumulant method (lower right) and
$v_{n}$ using the four particle cumulant method (lower right).
$v_4$\{4\} and $v_{5}$\{5\} are omitted due to less statistics.
\label{fig:vncent}}
\end{center}
\end{figure}

Figure \ref{fig:vncent} shows
event anisotropies $v_{n}$ ($n=$2, 3, 4 and 5)
by various flow analysis methods employed in this study.
Centrality dependences of $v_{n}$\{EP\} are almost identical to
those of $v_{n}$\{2\}.
$v_{3}$\{RP\} and $v_{5}$\{RP\}
vanish as they should.
$v_{4}$\{RP\} is non-zero
only in mid-central collisions (20-70\% centrality).
Since $v_{n}$ using the four particle cumulant method
demands higher statistics,
we only show $v_{2}$\{4\} and $v_{3}$\{4\}.
Although $v_{3}$\{4\} has large error bars,
 it seems to increase with increasing centrality percentage.
This is contrary to $v_{3}$\{EP\} and $v_{3}$\{2\}
which first increase up to 50-60\% centrality, but begin to
  decrease in more peripheral collisions.
This emphasises the importance of employing the same flow
  analysis method both in the theory calculations and in the
  experimental data analysis.

\begin{figure}[tb]
\begin{center}
\begin{minipage}[t]{9 cm}
\epsfig{file=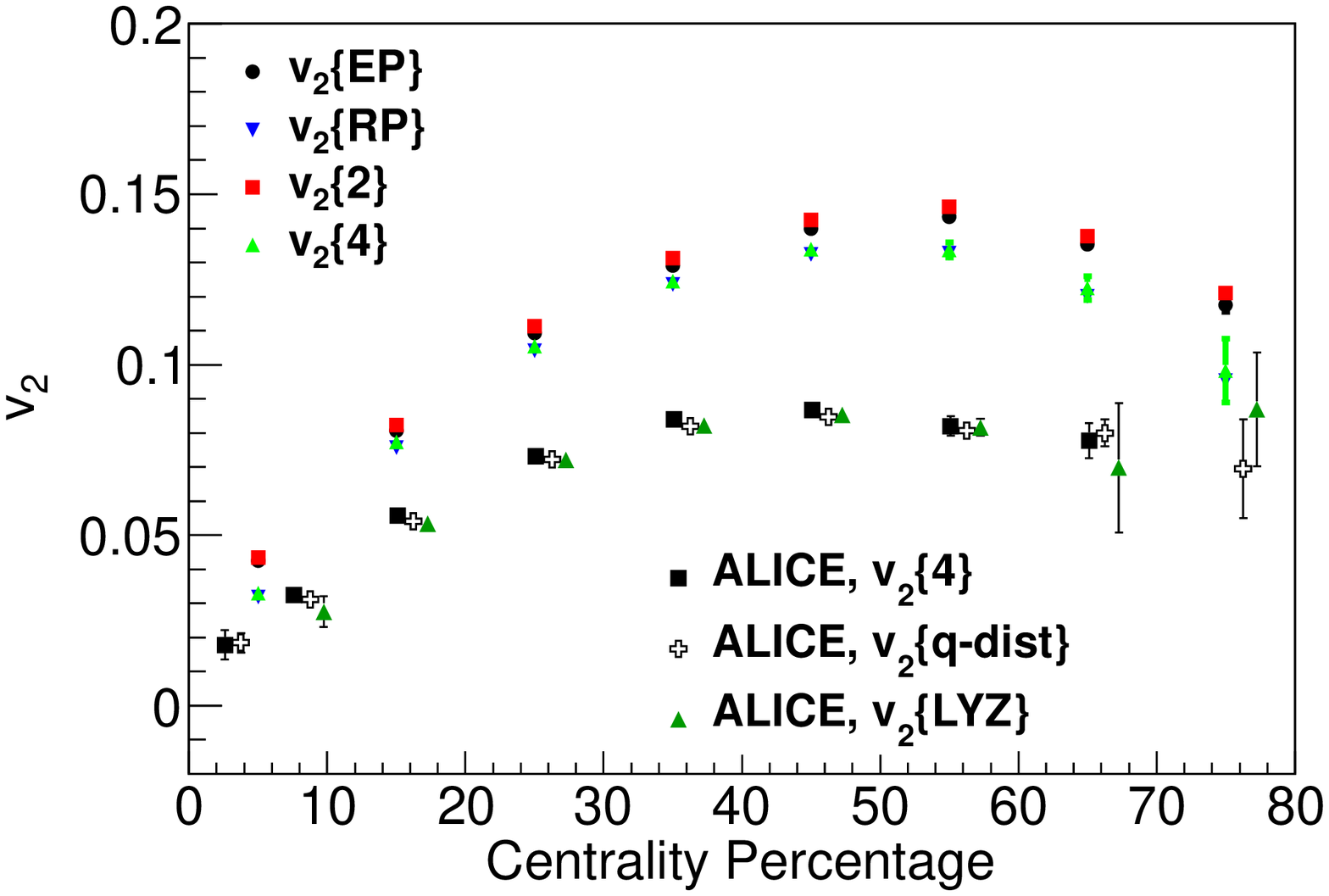,scale=0.45}
\end{minipage}
\begin{minipage}[t]{9 cm}
\epsfig{file=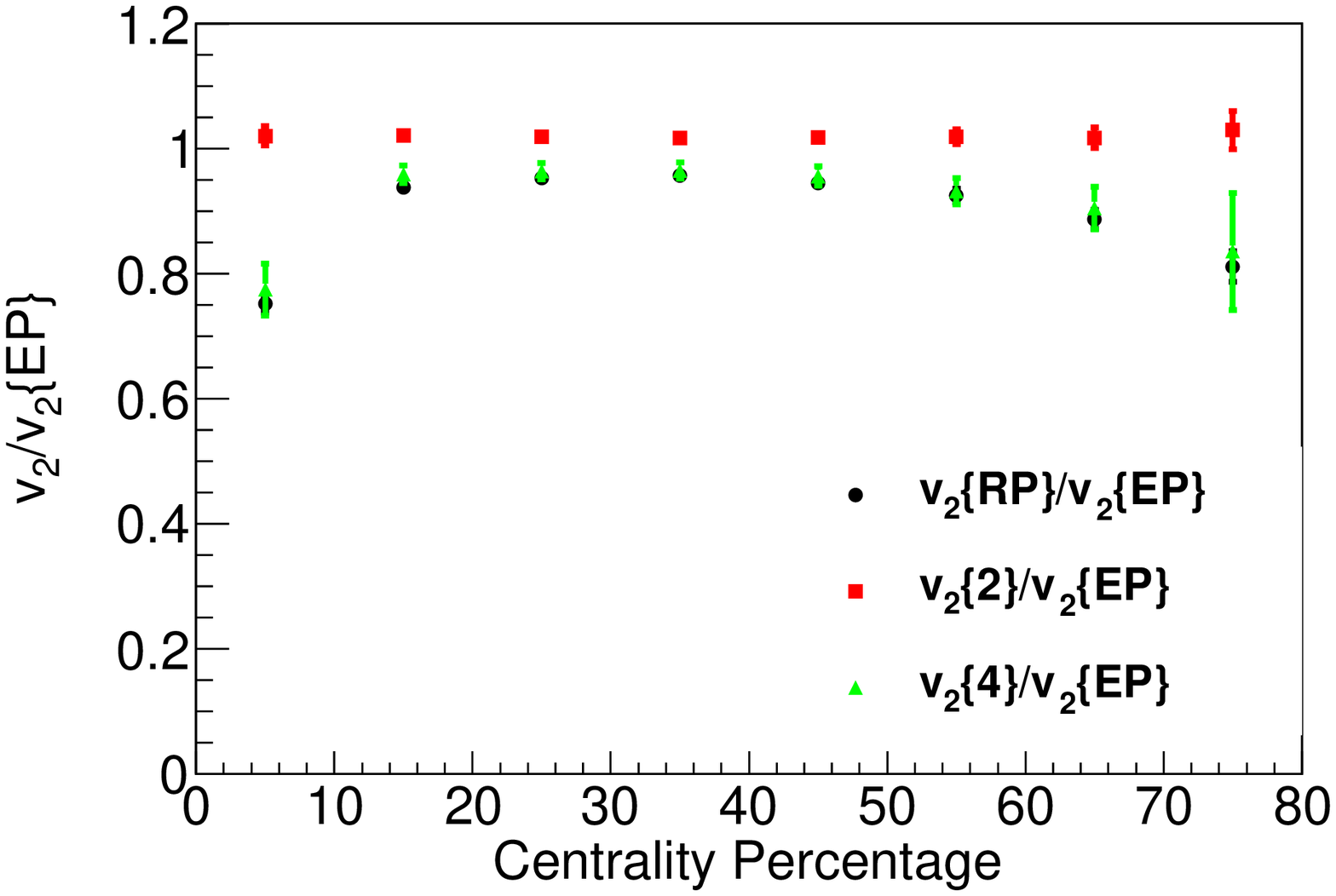,scale=0.45}
\end{minipage}
\caption{
(Left) Centrality dependence of  $v_{2}$ using
the event plane method, the reaction plane method,
the two particle cumulant and the four particle cumulant
in MC-KLN initialisation compared with the ALICE data \cite{Aamodt:2010pa}.
(Right) Ratio of $v_{2}$ to $v_{2}\{$EP$\}$ as a function of centrality.
\label{fig:v2}
}
\end{center}
\end{figure}

Figure \ref{fig:v2} (left)
shows the centrality dependence of $v_{2}$
compared with the ALICE data \cite{Aamodt:2010pa}.
It has been known that ideal hydrodynamics with the CGC based 
initial state  overshoots
the data by 50-60\%.
The deviation could have been understood as viscous effects,
which we do not discuss in this paper.
It should be noted that the data do not contain charged hadrons below $p_{T} = 0.15$
GeV/$c$ but that the hydrodynamic results do.
If we took account of momentum cut of the ALICE setup,
the calculated results would have become larger.
Ratios of $v_{2}$
to $v_{2}$\{EP\} are shown to see 
the difference among the various flow analysis methods
in Fig.~\ref{fig:v2} (right).
$v_{2}$\{2\}
is almost identical to $v_{2}$\{EP\}.
On the other hand, $v_{2}$\{RP\} almost traces $v_{2}$\{4\}.
In central (0-10\%) and peripheral (60-80\%) collisions,
$v_{2}$\{RP\} and $v_{2}$\{4\} are 10-20\% smaller than $v_{2}$\{EP\}
and $v_{2}$\{2\},
which can be understood as a consequence of eccentricity
fluctuations shown 
in Fig.~\ref{fig:epspart}.
On the other hand, the difference between $v_{2}$\{EP\} and $v_{2}$\{RP\}
is $\sim$5 \% in semi-central collisions.
Correspondences of between $v_{2}$\{2\} and $v_{2}$\{EP\}
and between $v_{2}$\{4\} and $v_{2}$\{RP\}
are also discussed using UrQMD in Ref.~\cite{Zhu:2005qa}.

\begin{figure}[tb]
\begin{center}
\begin{minipage}[t]{9 cm}
\epsfig{file=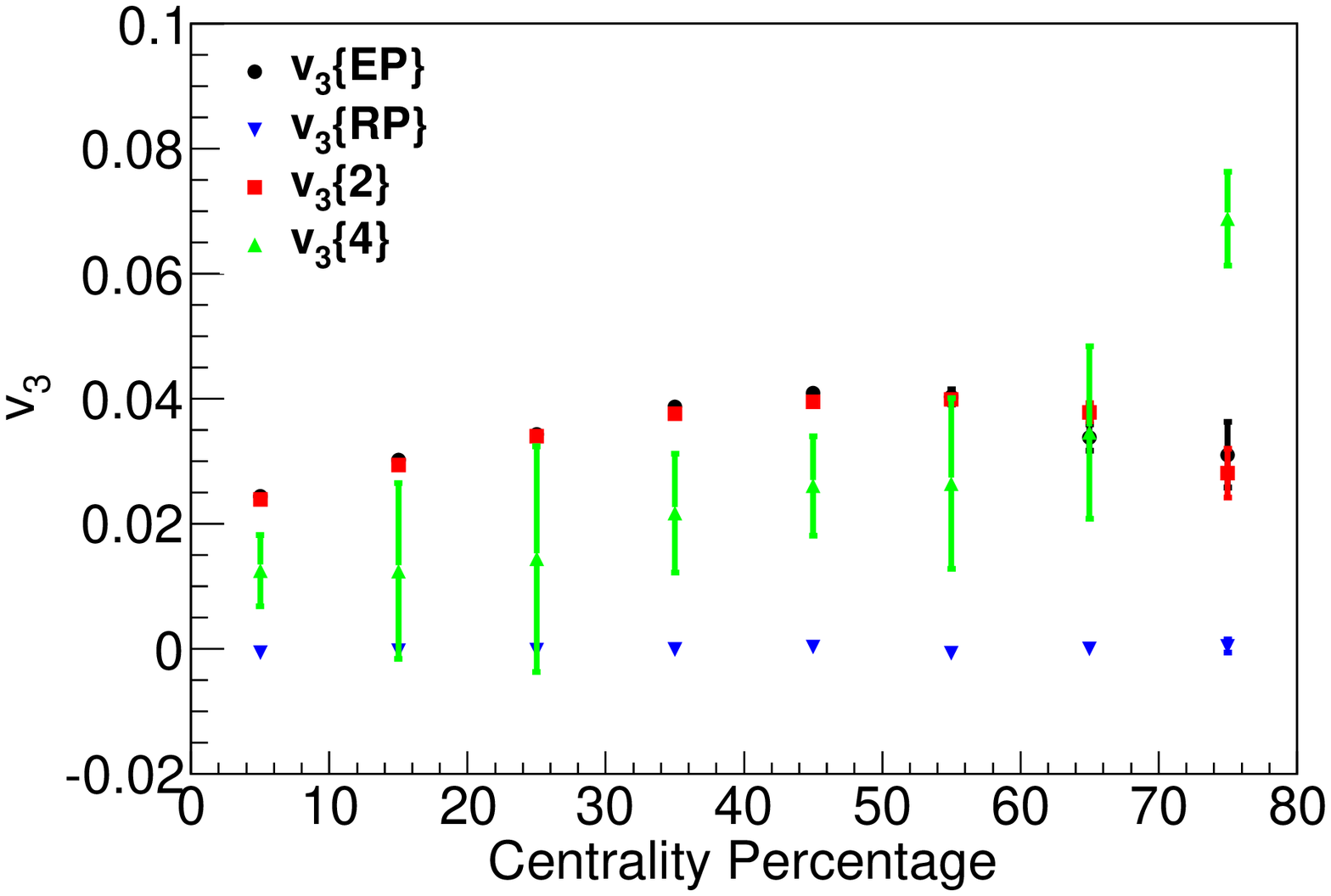,scale=0.45}
\end{minipage}
\begin{minipage}[t]{9 cm}
\epsfig{file=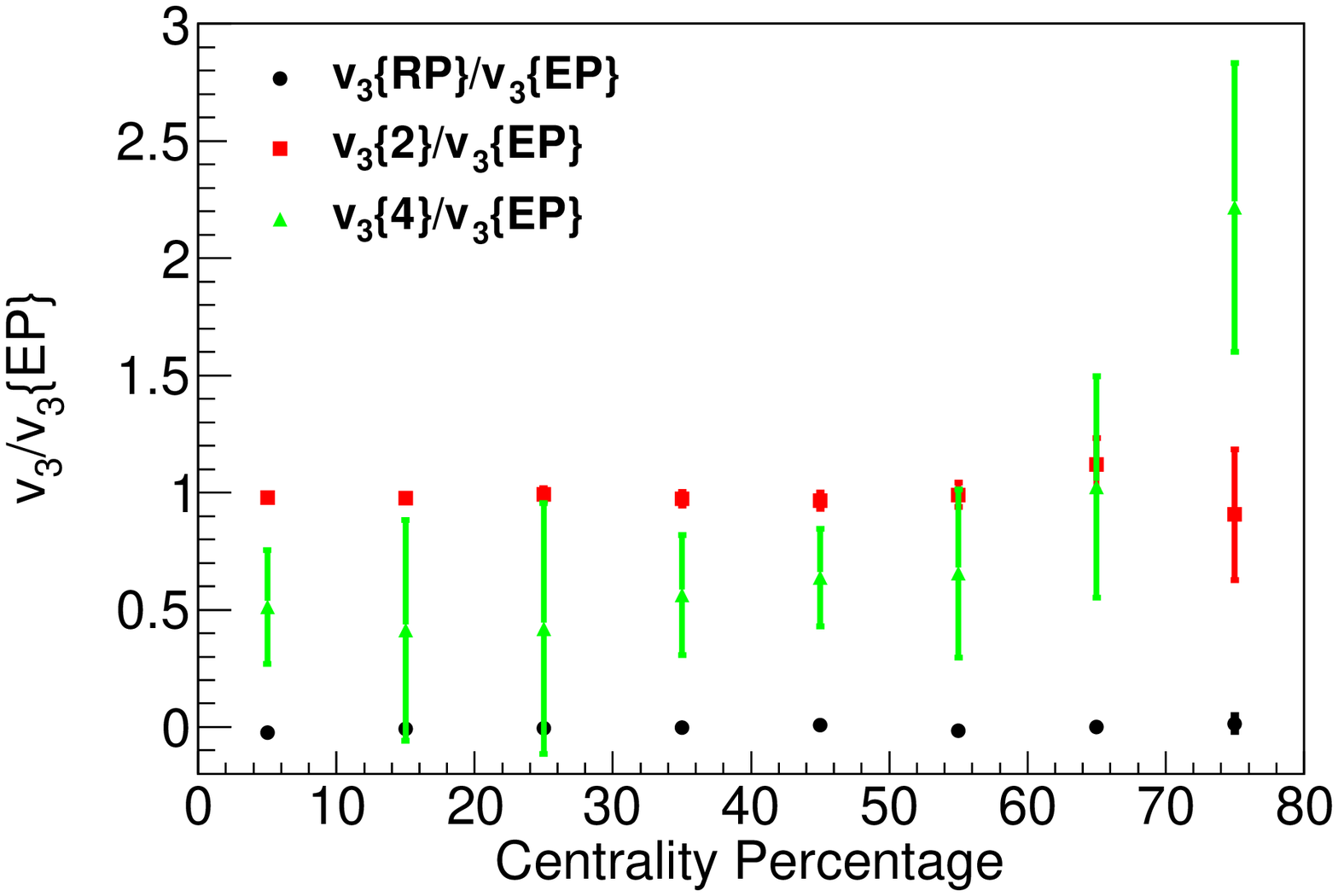,scale=0.45}
\end{minipage}
\caption{
(Left) Centrality dependence of $v_{3}$ using the event plane method, the reaction plane method,
the two particle cumulant and the four particle cumulant.
(Right) Ratio of $v_{3}$ to $v_{3}$\{EP\} as a function of centrality.
\label{fig:v3}}
\end{center}
\end{figure}
        
Figure \ref{fig:v3} shows
centrality dependence of $v_{3}$ (left) 
and its ratio to $v_{3}$\{EP\} (right).
If odd harmonics are generated solely by fluctuation in initial
transverse profiles, $v_{3}$\{RP\} should vanish since the
initial fluctuation does not correlate
with the reaction plane.
This is, in fact, seen in Fig.~\ref{fig:v3}.
As seen in $v_{2}$, $v_{3}$\{EP\} 
is almost identical to $v_{3}$\{2\}.
Due to poor statistics, $v_{3}$\{4\} has  large errors.
Nevertheless, it seems to be finite and smaller than $v_{3}$\{EP\} and $v_{3}$\{2\}
up to $\sim$60 \% centrality:
$v_{3}$\{4\} is roughly half of $v_{3}$\{2\} and
$v_{3}$\{EP\} \cite{Bhalerao:2011yg,Bhalerao:2011bp}.
Ratios of $v_{3}$ to $v_{3}$\{EP\} as functions of
centrality are shown in Fig.~\ref{fig:v3}
to see the dependence of $v_3$ on flow analysis methods 
 more clearly.
It would be interesting to gain more statistics to confirm
whether $v_{3}$\{4\} differs from the other $v_{3}$.

\begin{figure}[tb]
\begin{center}
\begin{minipage}[t]{9 cm}
\epsfig{file=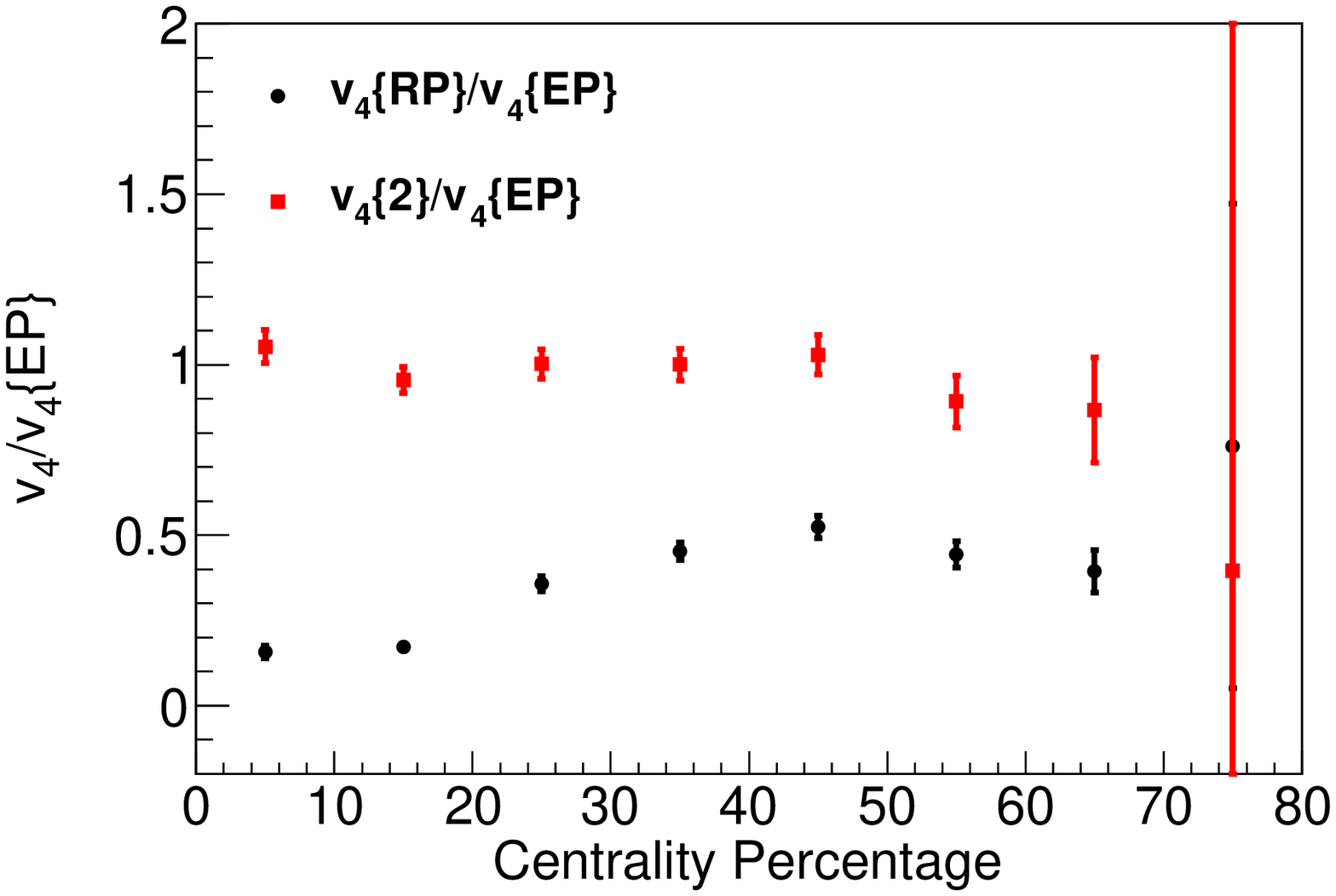,scale=0.45}
\end{minipage}
\begin{minipage}[t]{9 cm}
\epsfig{file=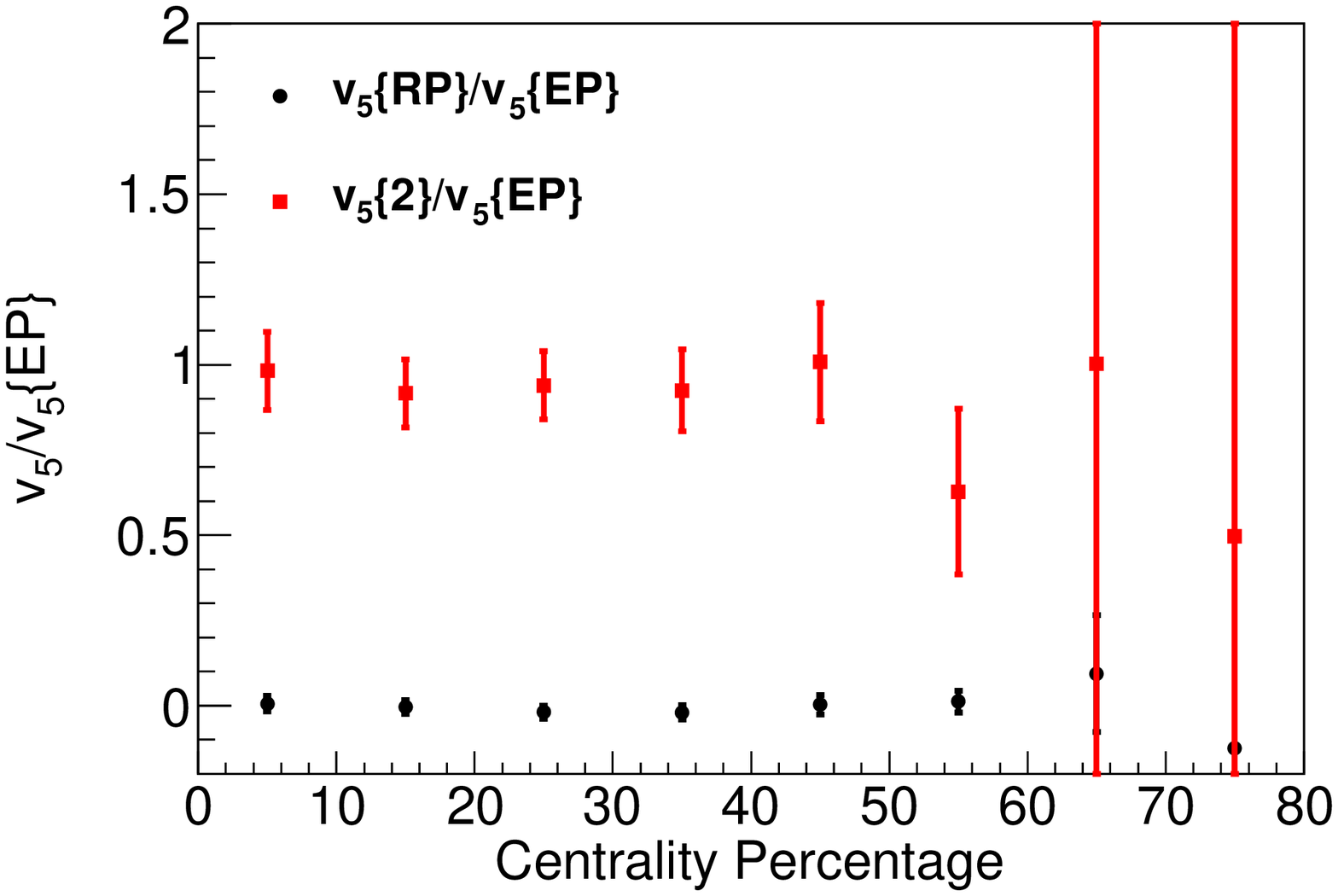,scale=0.45}
\end{minipage}
\caption{
(Left) Ratio of $v_{4}$ to $v_{4}$\{EP\} as a function of centrality.
(Right) Ratio of $v_{5}$ to $v_{5}$\{EP\} as a function of centrality.
\label{fig:v45}}
\end{center}
\end{figure}

Figure \ref{fig:v45} shows centrality dependence of ratios of
fourth (left)
and fifth (right) harmonics. Due to poor statistics, harmonics
using the four particle cumulant method are omitted.
Although $v_{4}$\{2\} and $v_{5}$\{2\} have large error bars,
they seem to agree with the harmonics evaluated using the event plane
  method, in the same way $v_2$\{2\} and $v_3$\{2\} do.
$v_{4}$\{RP\} is finite, while $v_{5}$\{RP\} vanishes
 as already seen in Fig.~\ref{fig:vncent}.

\begin{figure}[tb]
\begin{center}
\begin{minipage}[t]{9 cm}
\epsfig{file=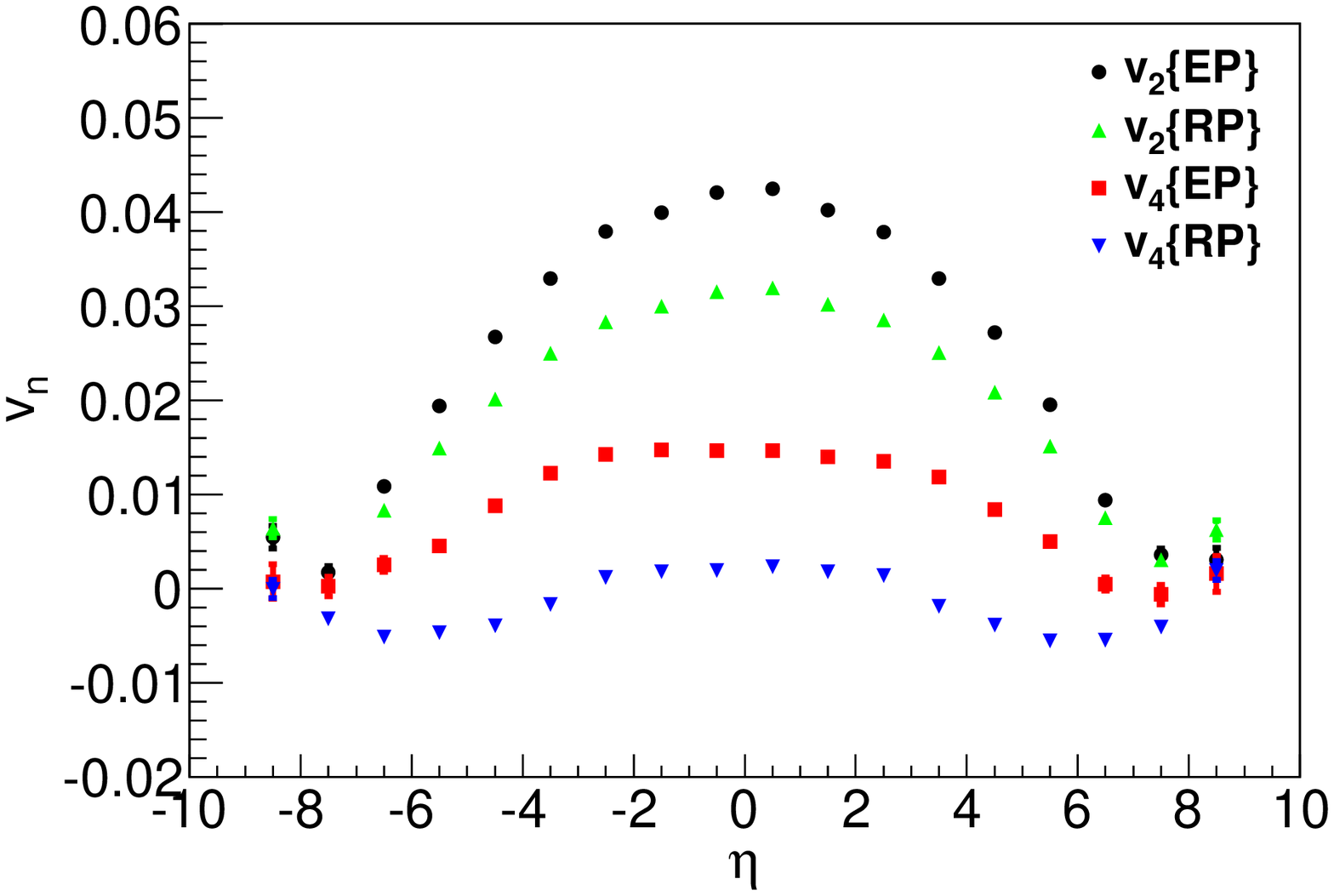,scale=0.45}
\end{minipage}
\begin{minipage}[t]{9 cm}
\epsfig{file=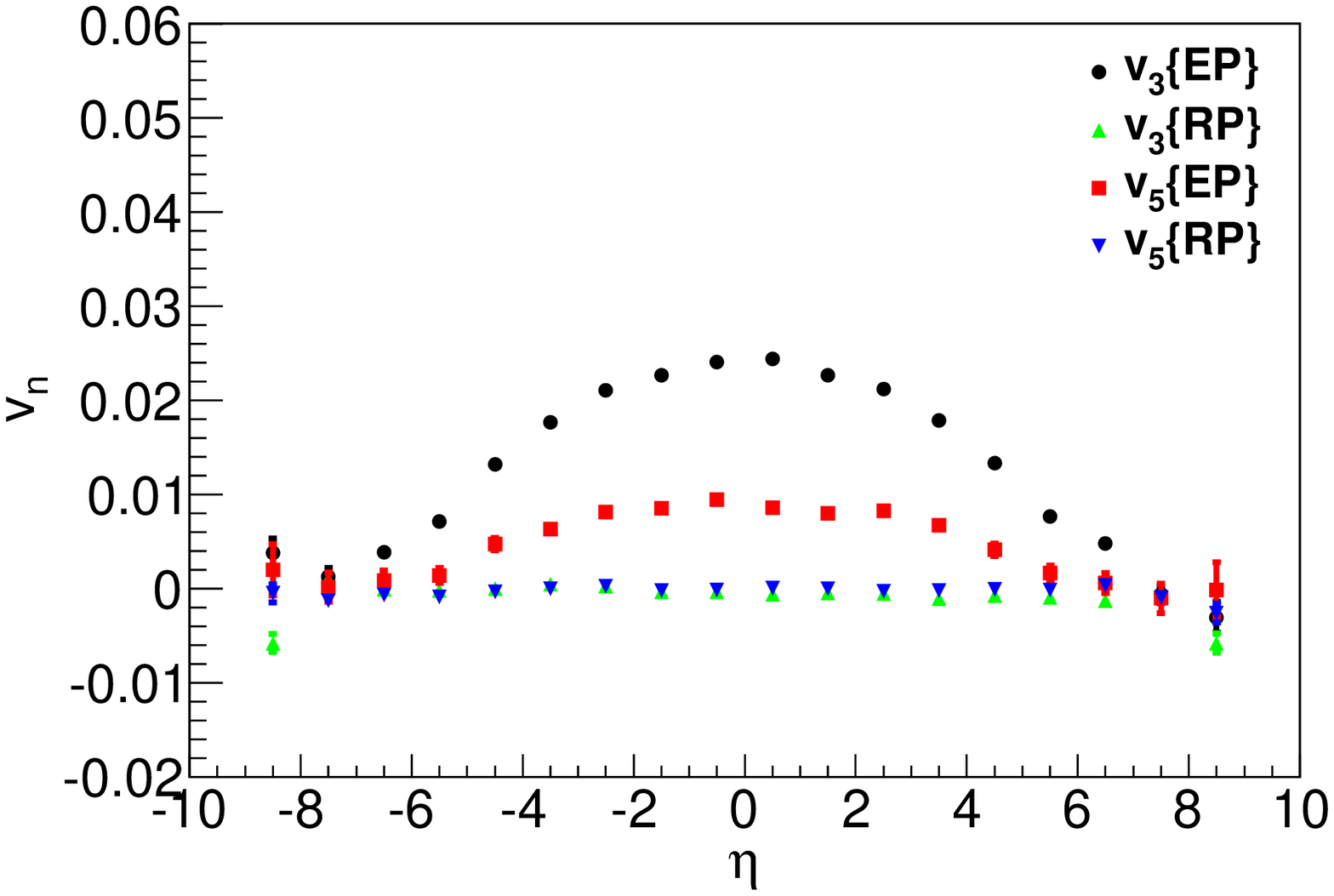,scale=0.45}
\end{minipage}
\caption{
Pseudorapidity dependence of $v_{2}$ and $v_{4}$ (left)
and $v_{3}$ and $v_{5}$ (right) at 0-10\% centrality. Harmonics evaluated 
using the event plane method are compared with those evaluated
with respect to the reaction plane.
\label{fig:vmevmr0}}
\end{center}
\end{figure}

\begin{figure}[tb]
\begin{center}
\begin{minipage}[t]{9 cm}
\epsfig{file=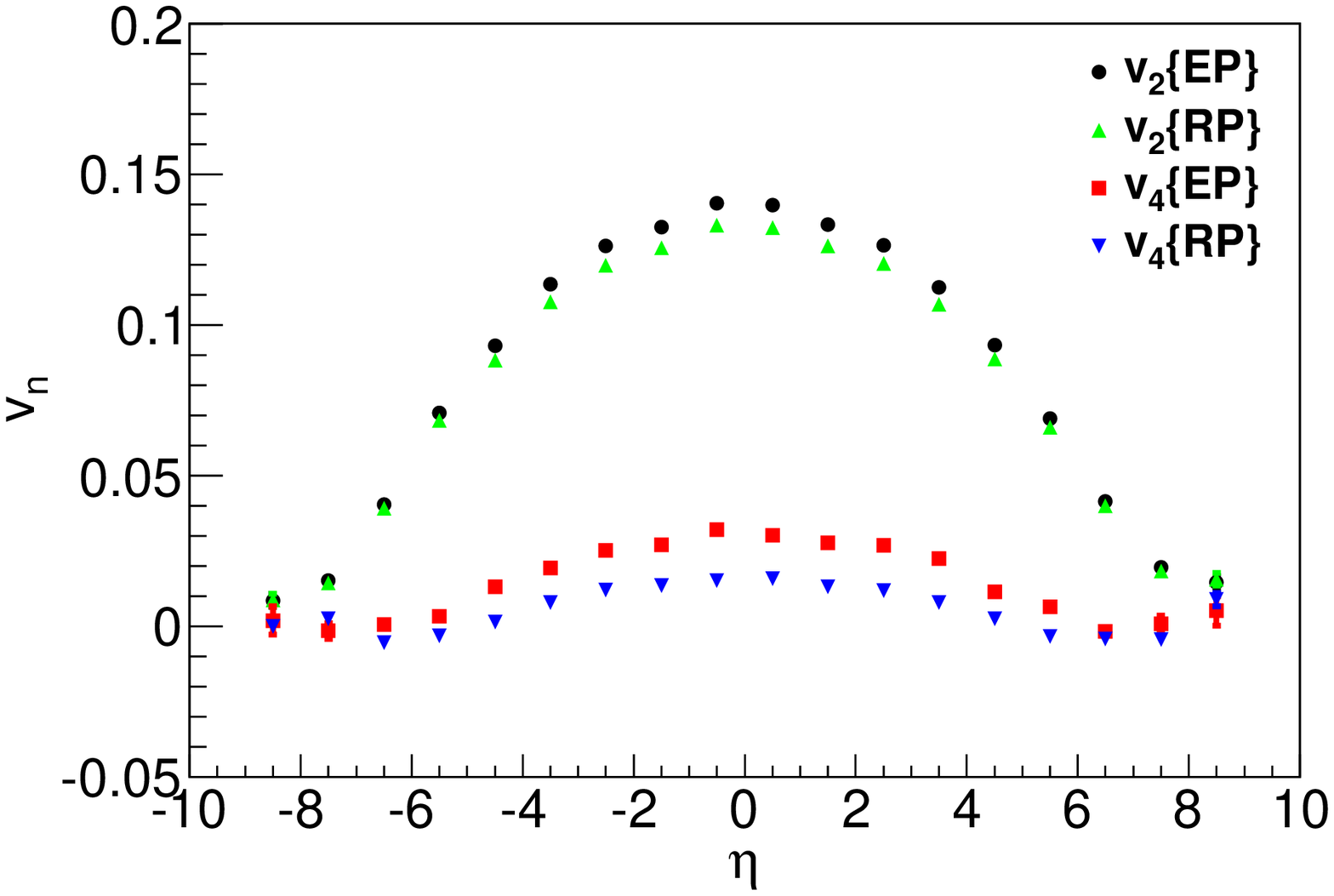,scale=0.45}
\end{minipage}
\begin{minipage}[t]{9 cm}
\epsfig{file=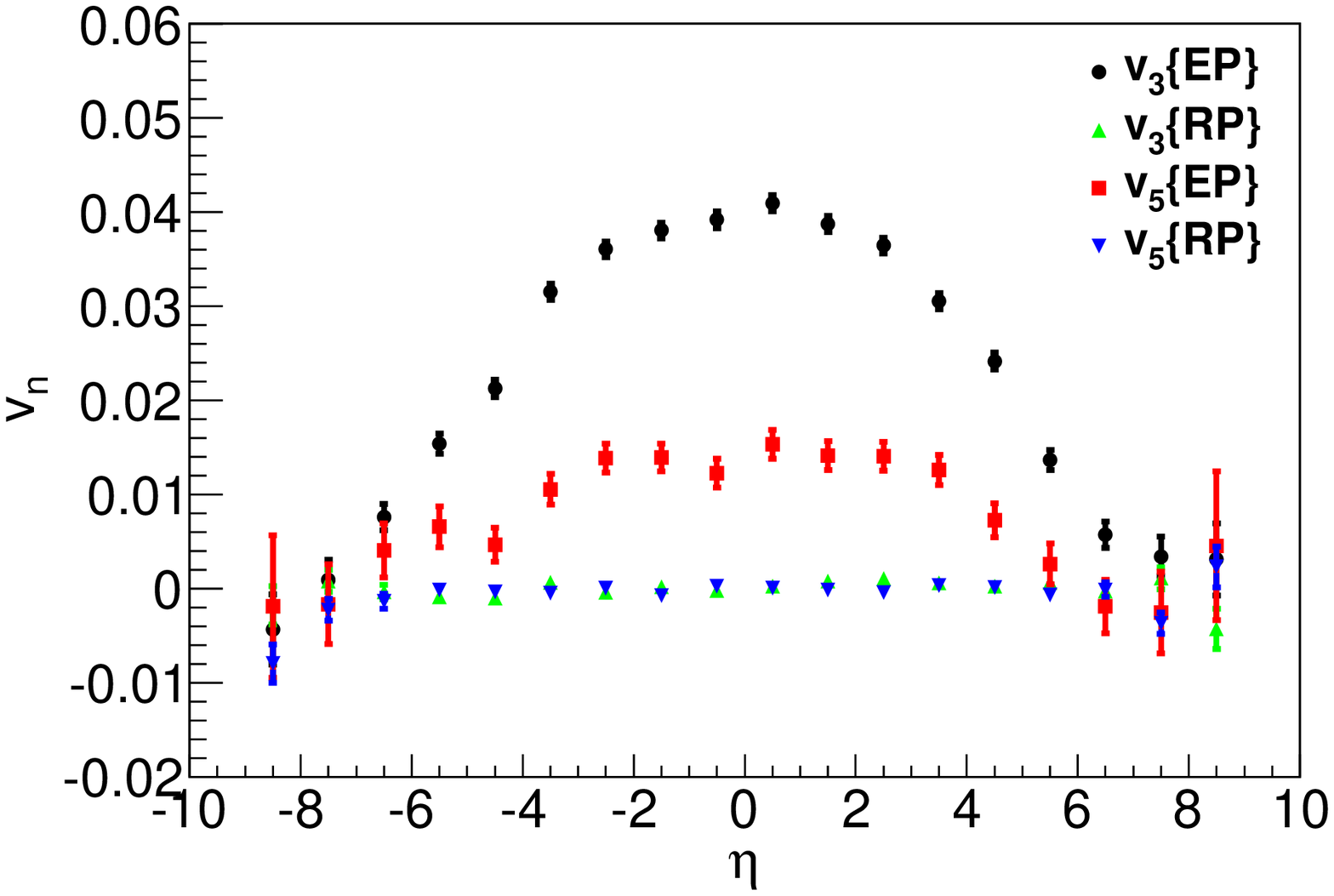,scale=0.45}
\end{minipage}
\caption{
The same as Fig.~\ref{fig:vmevmr0} but at 40-50\% centrality.
\label{fig:vmevmr4}}
\end{center}
\end{figure}

\begin{figure}[tb]
\begin{center}
\begin{minipage}[t]{9 cm}
\epsfig{file=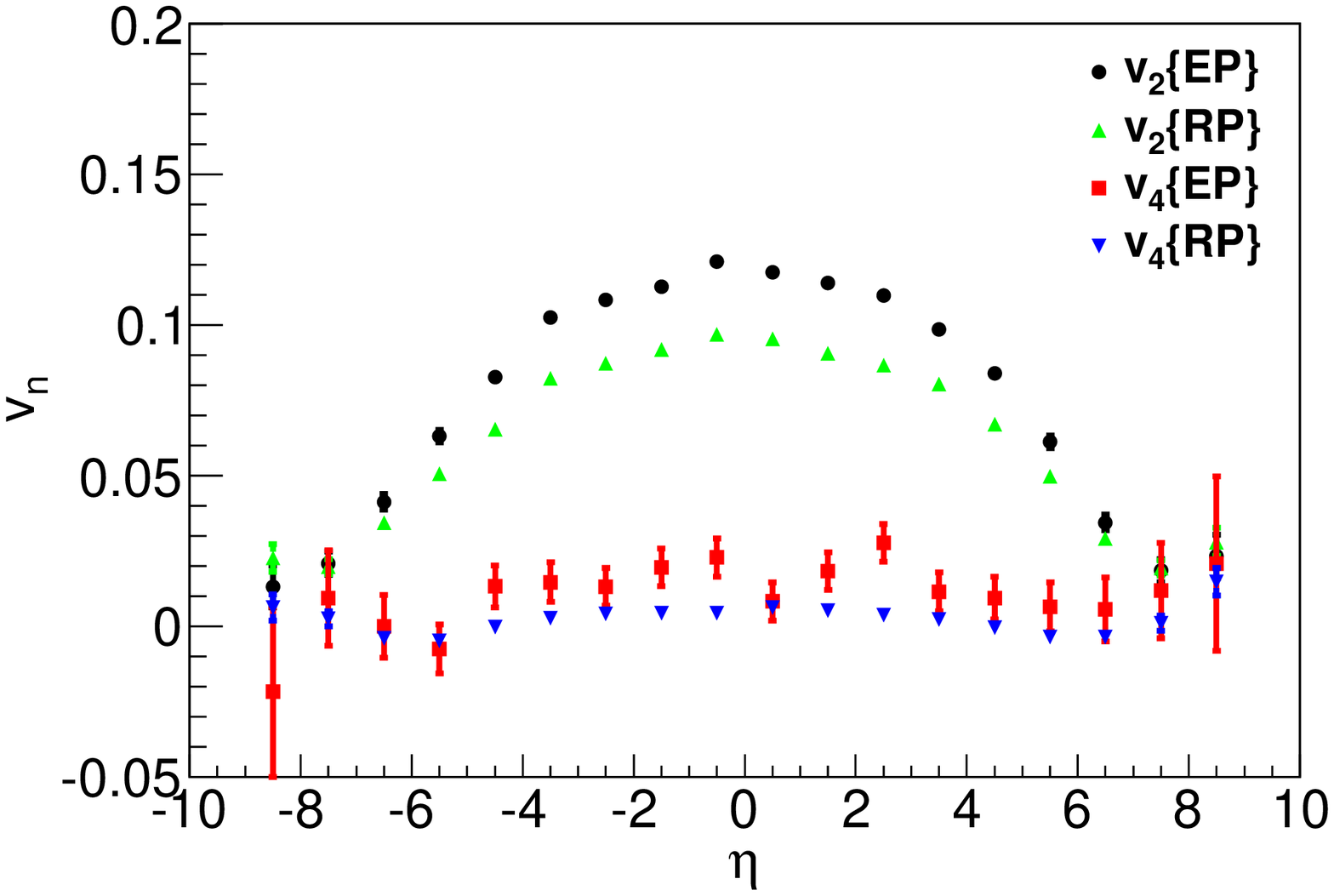,scale=0.45}
\end{minipage}
\begin{minipage}[t]{9 cm}
\epsfig{file=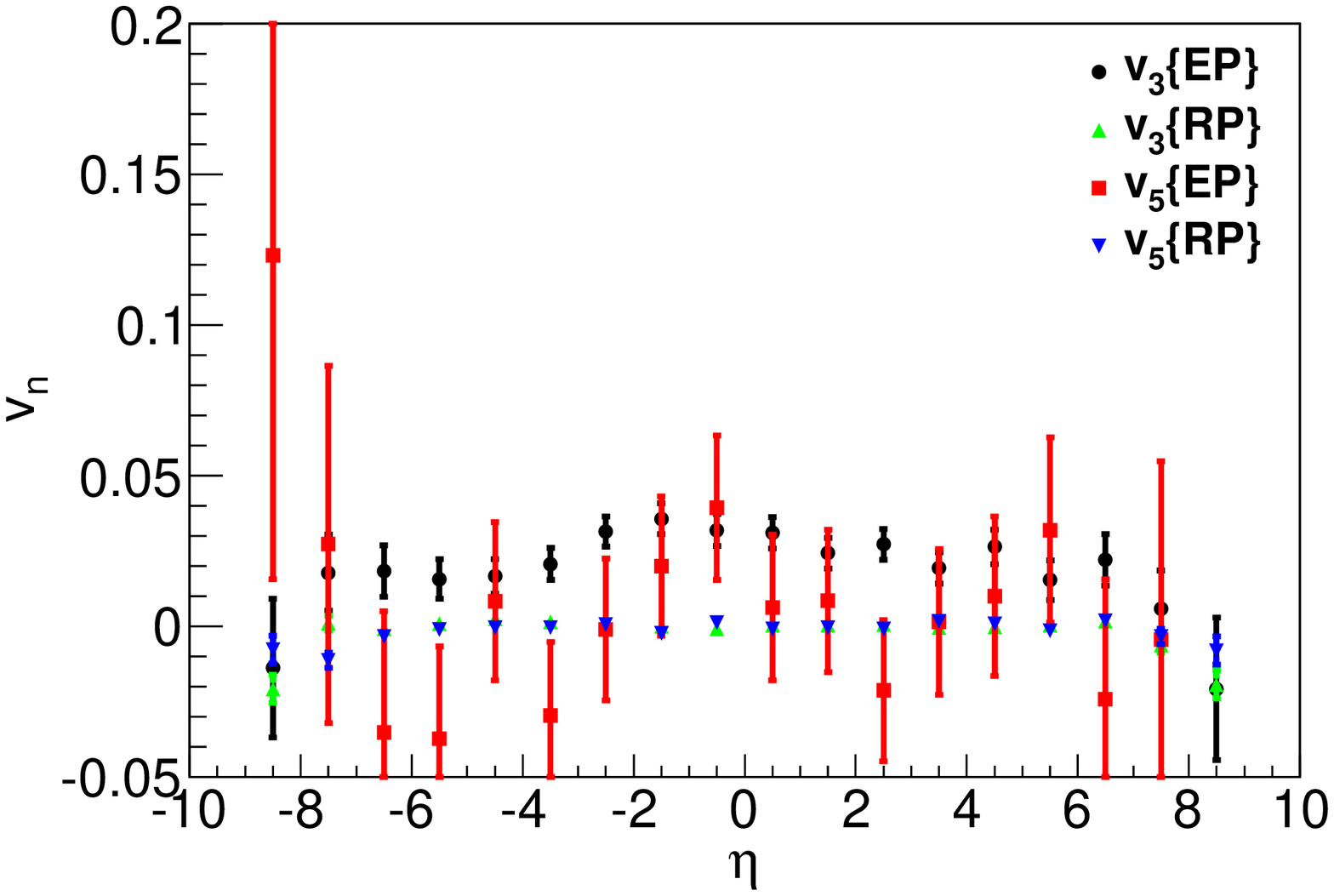,scale=0.45}
\end{minipage}
\caption{
The same as Fig.~\ref{fig:vmevmr0} but at 70-80\% centrality.
\label{fig:vmevmr7}}
\end{center}
\end{figure}

Figures \ref{fig:vmevmr0}, \ref{fig:vmevmr4} and \ref{fig:vmevmr7}
show pseudorapidity dependence of higher order harmonics in 0-10\%,
40-50\% and 70-80\% centrality, respectively.  The second harmonics
$v_{2}$, which are shown in the left panels, always have a maximum at
midrapidity and decrease as moving away from midrapidity although
$\varepsilon_{2}$ is almost constant as a function of $\eta_{s}$.
This triangular shape was also measured at the RHIC energy.  
As we discussed in Sec.~\ref{sec:afterburner} the rapidity dependence
  is easy to understand as a consequence of the space-time rapidity
  dependence of the initial energy density even if the initial
  eccentricity hardly depends on $\eta_s$
  (Fig.~\ref{fig:epsilonpp-eta}). The larger the initial density, the
  longer the lifetime of the low viscosity QGP phase where $v_2$ is
  built up much more efficiently than in the highly dissipative
  hadronic phase, see Fig.~\ref{fig:v2eta} and related discussion on
  page~\pageref{discuss:v2eta}.

Difference between $v_{2}$\{EP\} and $v_{2}$\{RP\}
is relatively large in central (0-10\%) and peripheral (70-80\%) 
collisions, whereas
the difference is very small in semi-central collisions (30-40\%).
This difference at midrapidity was already shown in
  Fig.~\ref{fig:v2}, but it persists up to $\mid \eta \mid \sim 5$.
$v_{4}\{\mathrm{EP}\}$ also depends on $\eta$ 
even if $\varepsilon_{4}\{\mathrm{PP}\}$ is almost independent of
  $\eta_{s}$ as shown in Fig.~\ref{fig:epsilonpp-eta}.

Odd harmonics ($n=$3 and 5) shown in the right panel are finite
  near midrapidity when event plane method is used to evaluate them,
  whereas they vanish when the reaction plane method is used. The
  $v_{3}$\{EP\} and $v_{5}$\{EP\} have a broad peak at midrapidity,
decrease as moving away from midrapidity
and eventually vanish near the beam rapidity.
Note that $v_{5}$ has large error bars in the 70-80\% centrality
due to small multiplicity.
It should be noted that all $v_{n}$\{EP\}$(\eta)$ 
have almost no boost invariant region
which is in good contrast to $\varepsilon_{n} (\eta_{s})$ shown in 
Fig.~\ref{fig:epsilonpp-eta}.
As mentioned, this 
can be understood as the space-time rapidity dependence
of the lifetime of the QGP fluid.

\begin{figure}[tb]
\begin{center}
\begin{minipage}[t]{9 cm}
\epsfig{file=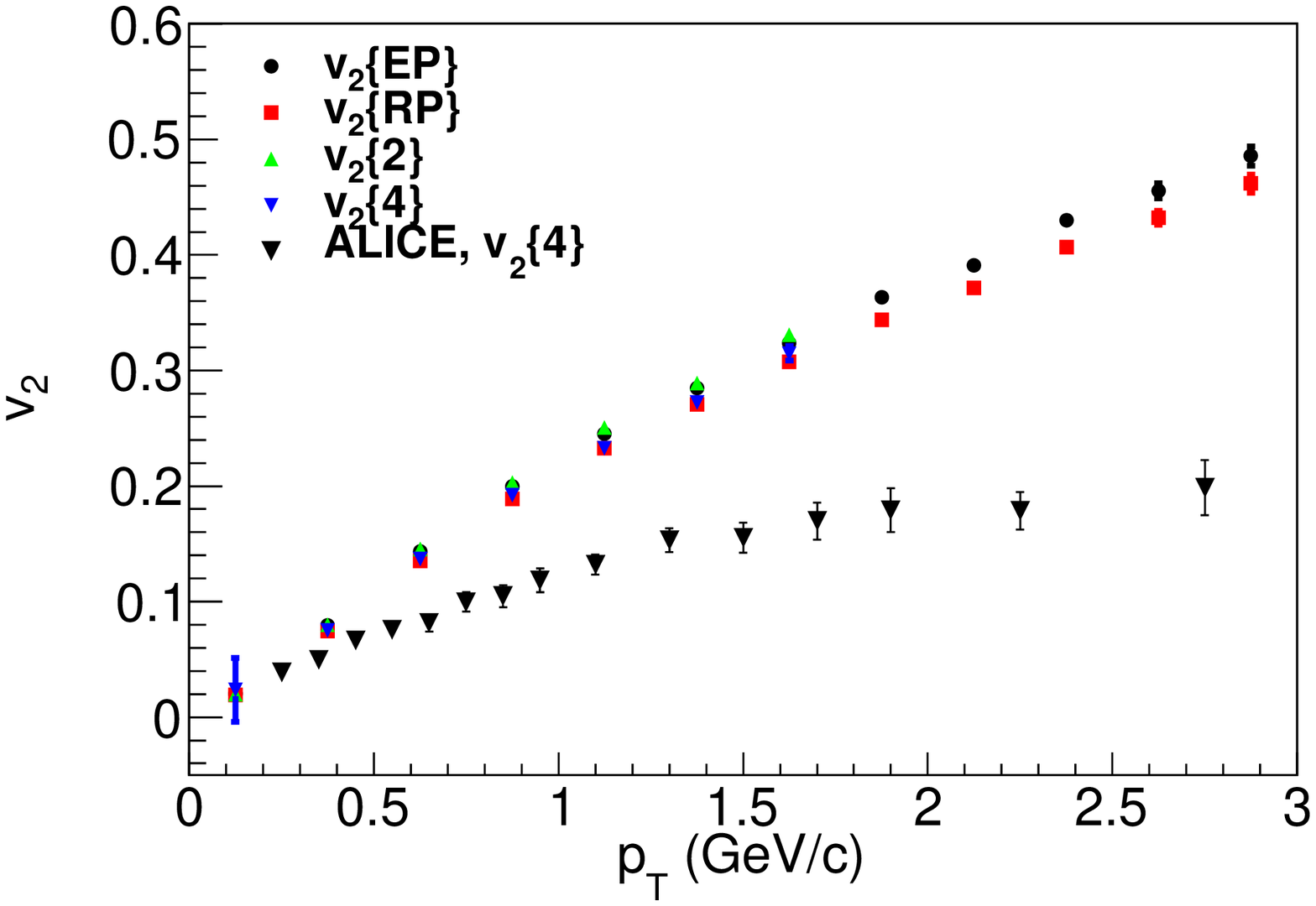,scale=0.45}
\end{minipage}
\begin{minipage}[t]{9 cm}
\end{minipage}
\caption{
Transverse momentum dependence of $v_{2}$ using
the event plane method, the reaction plane method, the two particle cumulant
and the four particle cumulant 
calculated using MC-KLN initialisation compared with the ALICE
$v_{2}\{4\}$ data at 40-50\% centrality \cite{Aamodt:2010pa}.
Due to the lack of statistics,
results are shown up to 1.625 GeV/$c$ for the two particle
cumulant and the four particle cumulant methods.
\label{fig:vmpt}}
\end{center}
\end{figure}

Transverse momentum dependences of $v_{2}$ evaluated using
the four methods described
in this study are compared with the ALICE $v_{2}$\{4\} data
\cite{Aamodt:2010pa}
at 40-50\% centrality in Pb+Pb collisions at $\sqrt{s_{NN}}$ = 2.76 TeV
in Fig.~\ref{fig:vmpt}.
Up to $p_{T}\sim$ 1.5 GeV/$c$ where, in the $M$-particle cumulant methods,
 at least $M$ particles 
are binned in all events in this centrality,
the difference among the four methods is very small.
Above this, the difference between $v_{2}$\{EP\} and $v_{2}$\{RP\}
is visible but still not so significant.

\begin{figure}[tb]
\begin{center}
\begin{minipage}[t]{6 cm}
\epsfig{file=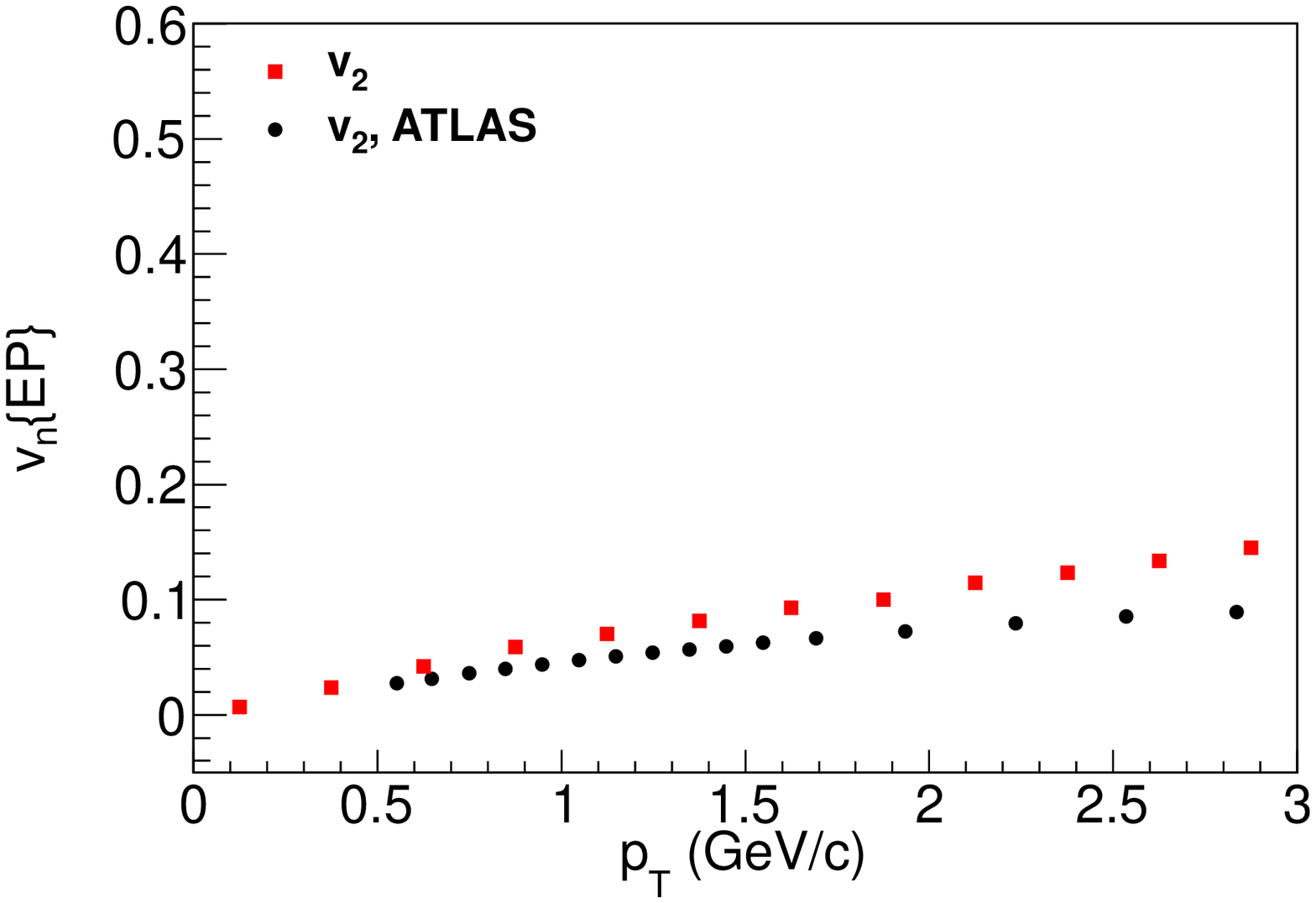,scale=0.3}
\end{minipage}
\begin{minipage}[t]{6 cm}
\epsfig{file=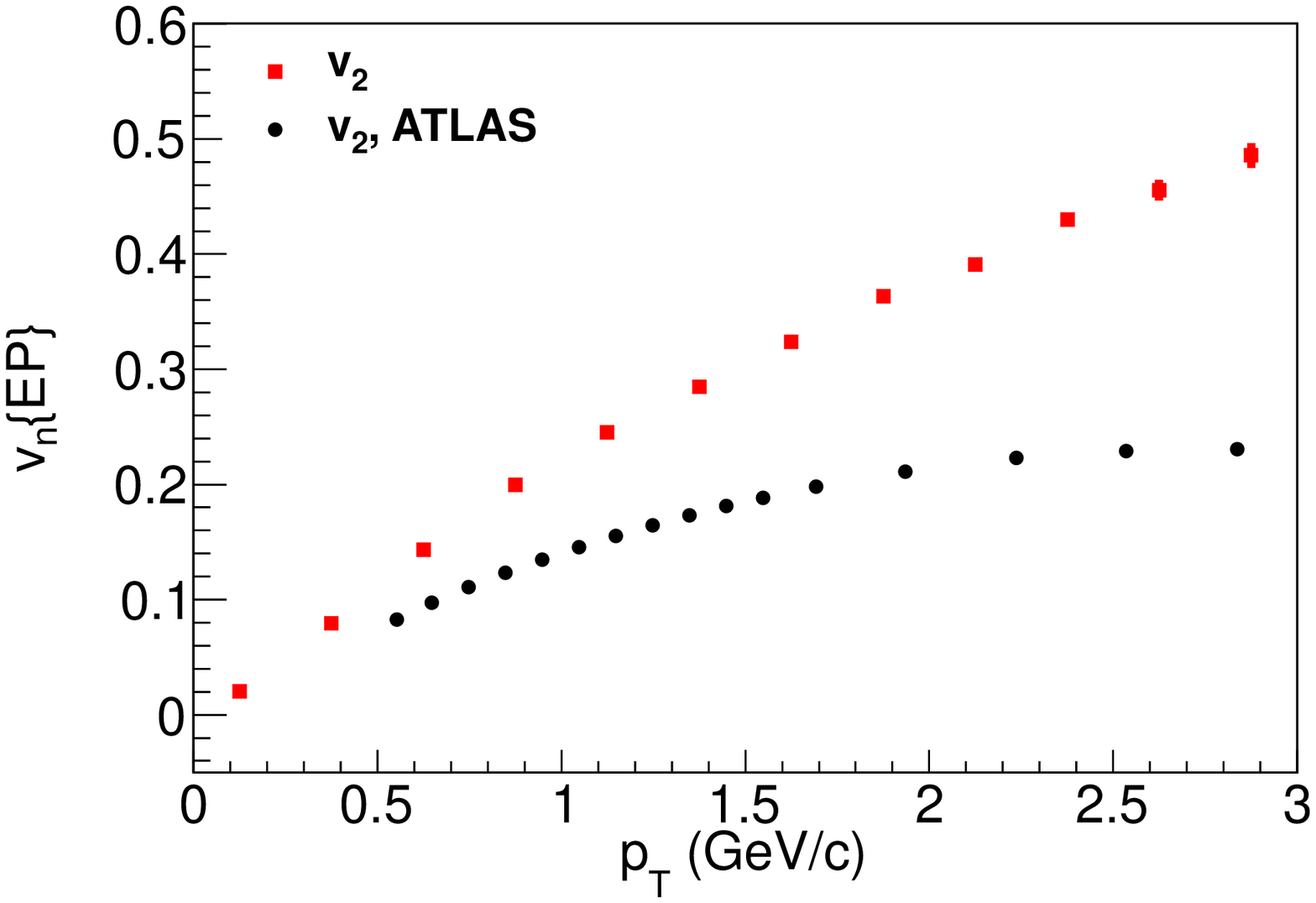,scale=0.3}
\end{minipage}
\begin{minipage}[t]{6 cm}
\epsfig{file=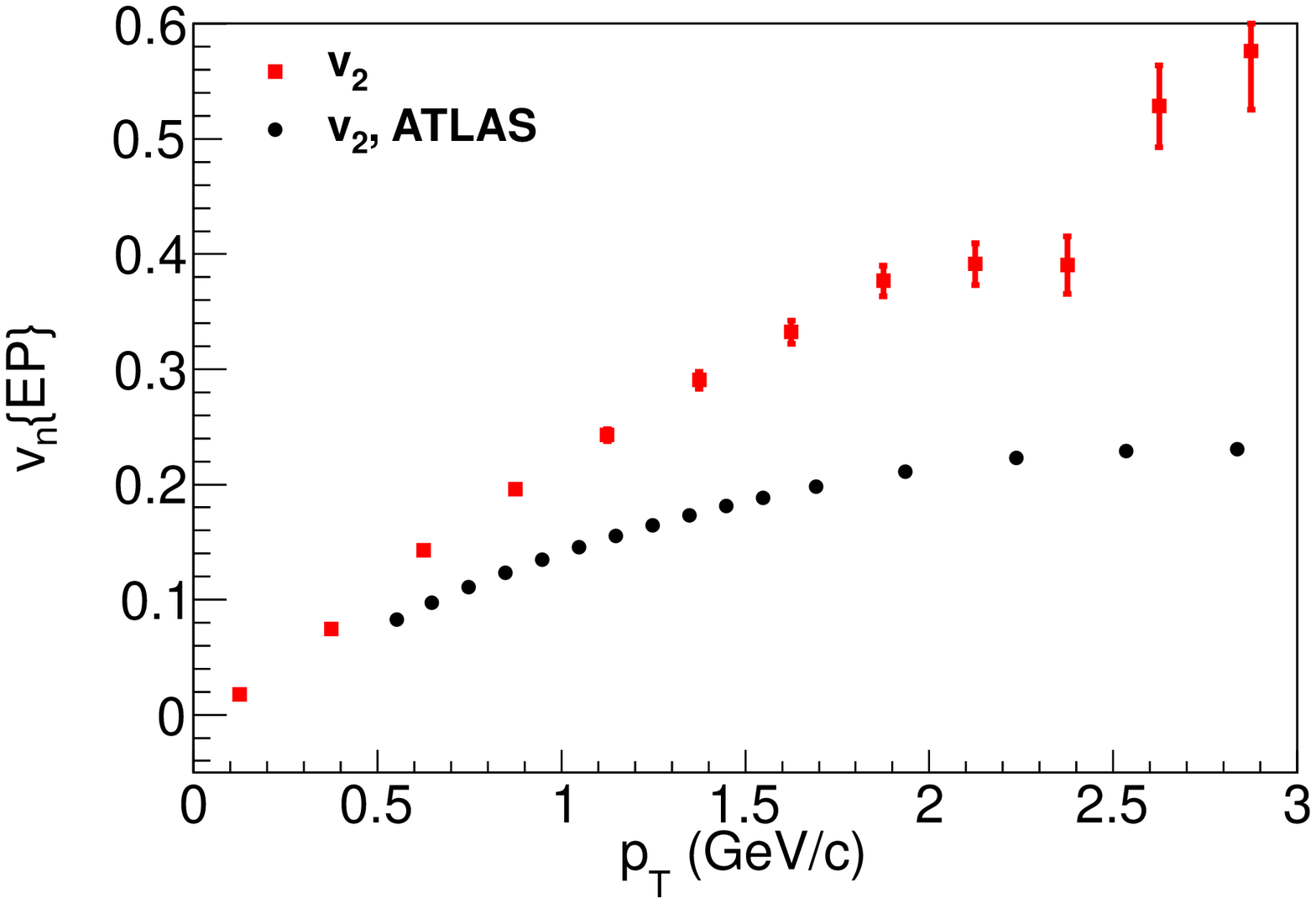,scale=0.3}
\end{minipage}
\begin{minipage}[t]{6 cm}
\epsfig{file=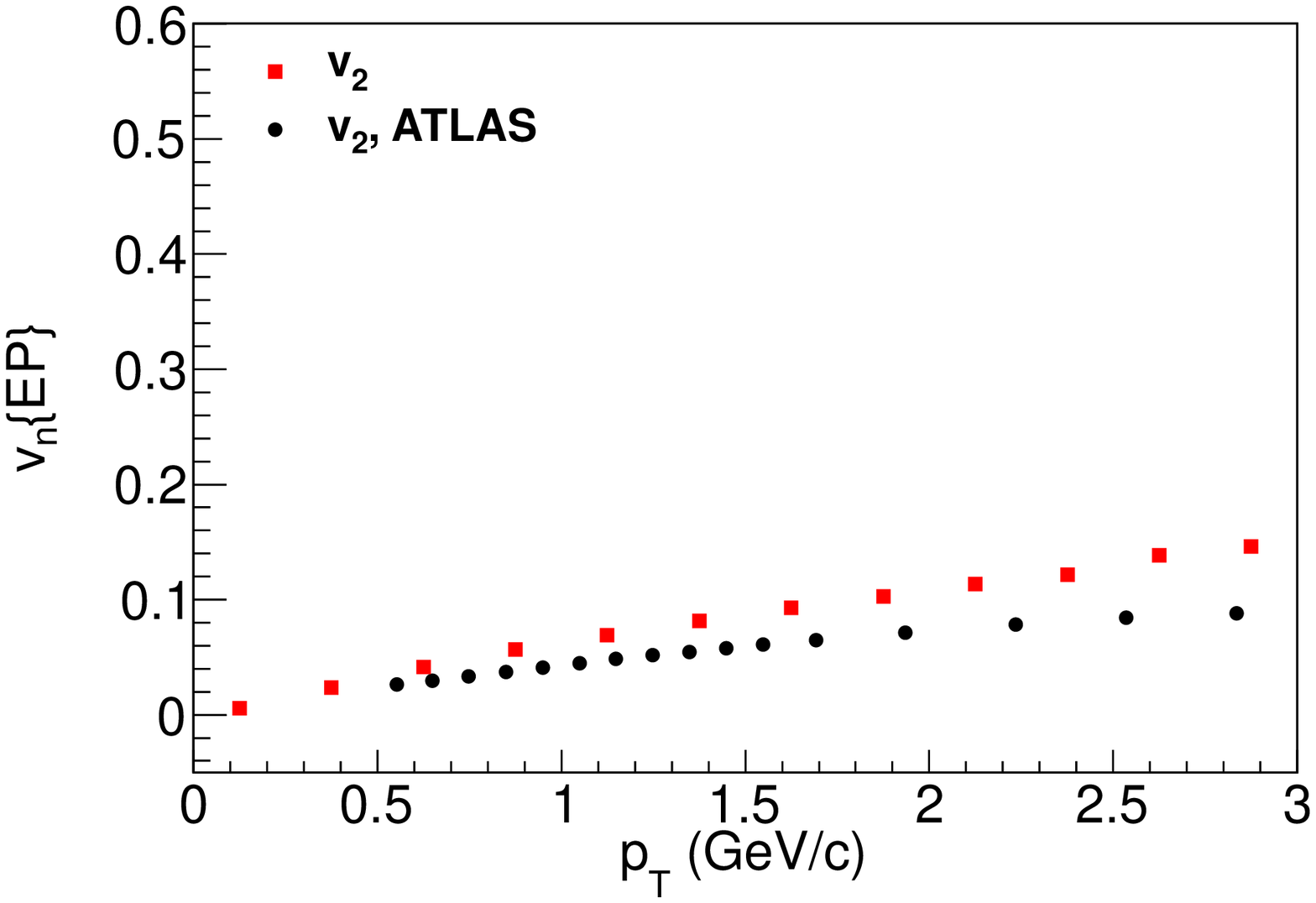,scale=0.3}
\end{minipage}
\begin{minipage}[t]{6 cm}
\epsfig{file=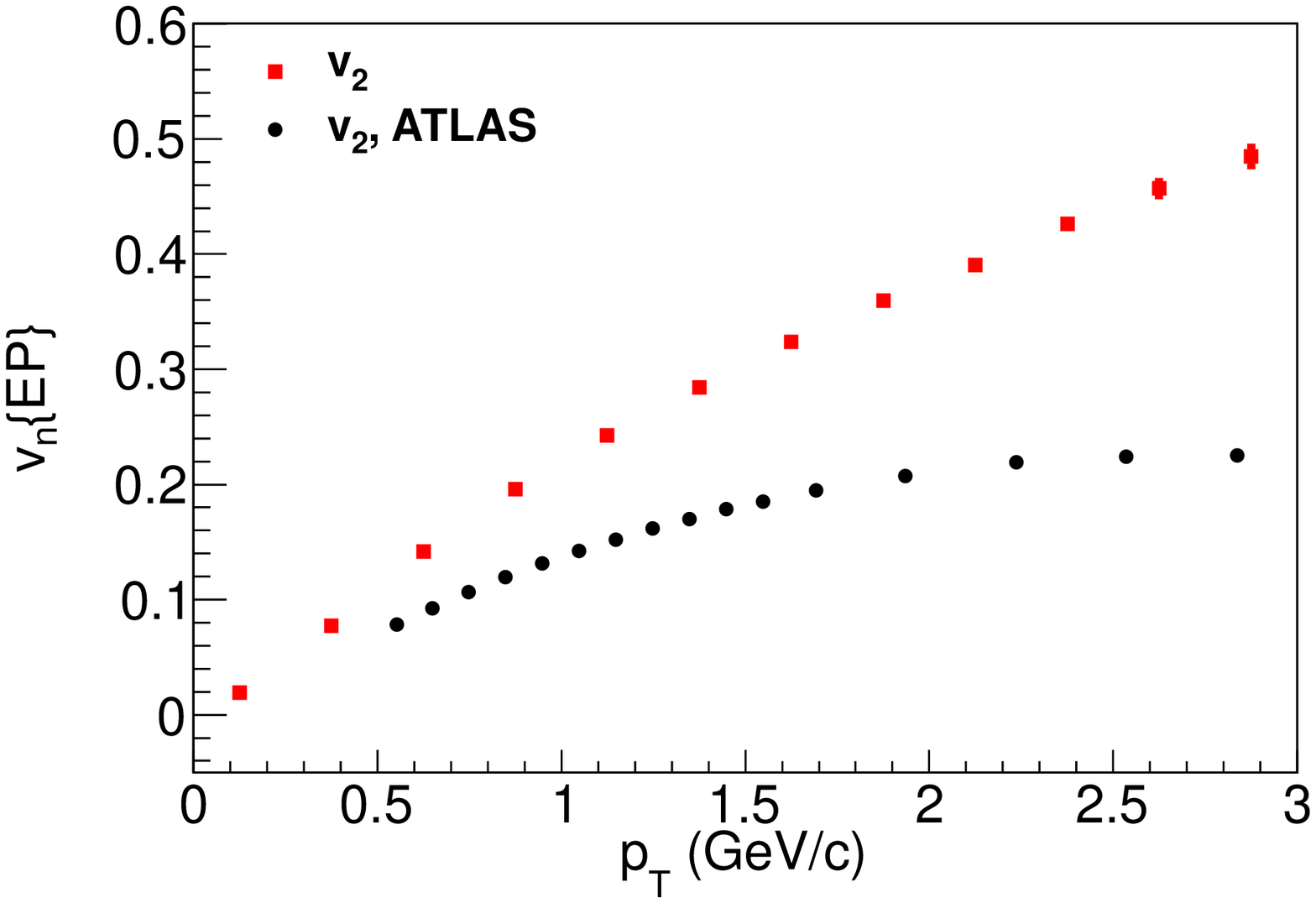,scale=0.3}
\end{minipage}
\begin{minipage}[t]{6 cm}
\epsfig{file=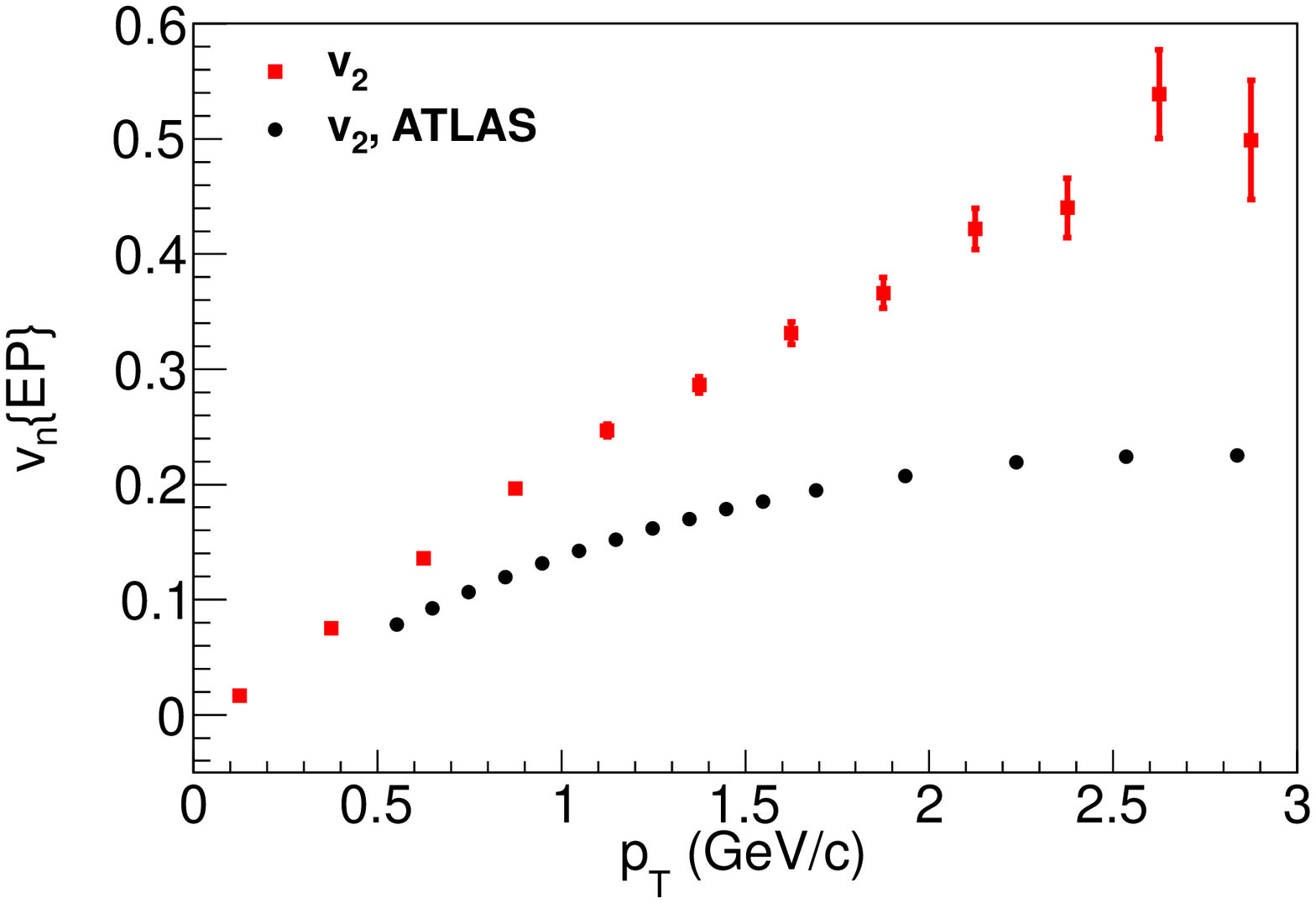,scale=0.3}
\end{minipage}
\begin{minipage}[t]{6 cm}
\epsfig{file=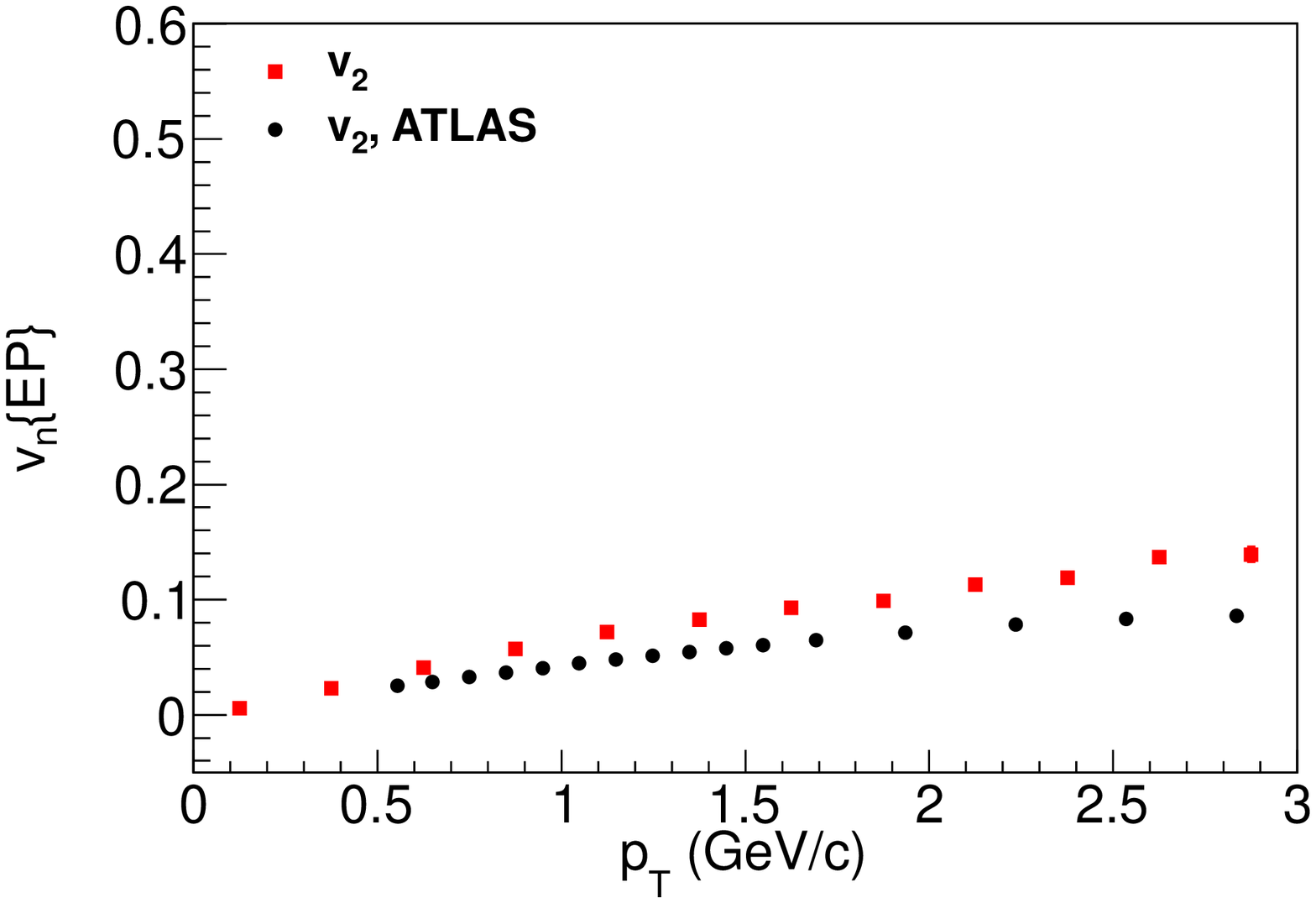,scale=0.3}
\end{minipage}
\begin{minipage}[t]{6 cm}
\epsfig{file=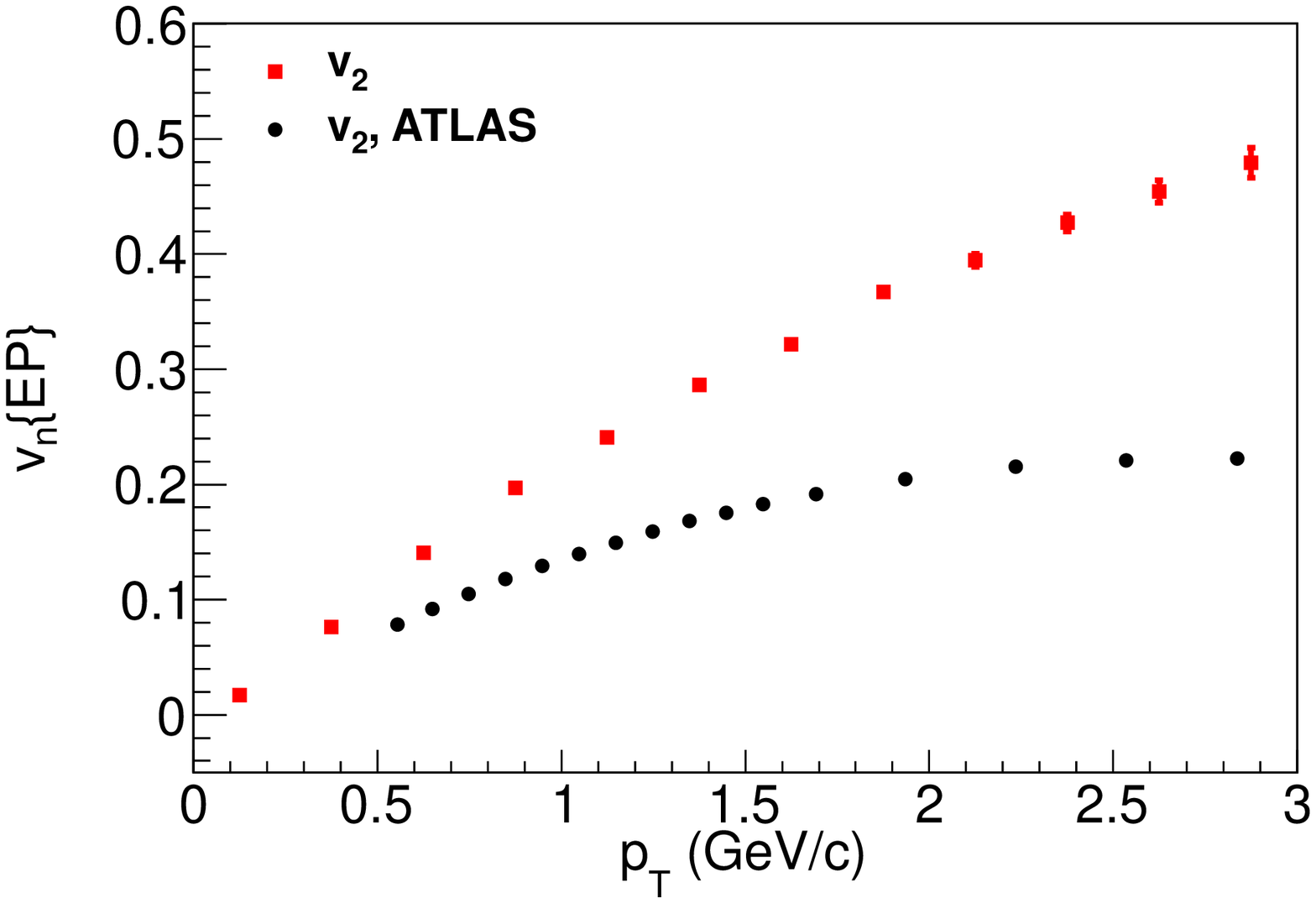,scale=0.3}
\end{minipage}
\begin{minipage}[t]{6 cm}
\epsfig{file=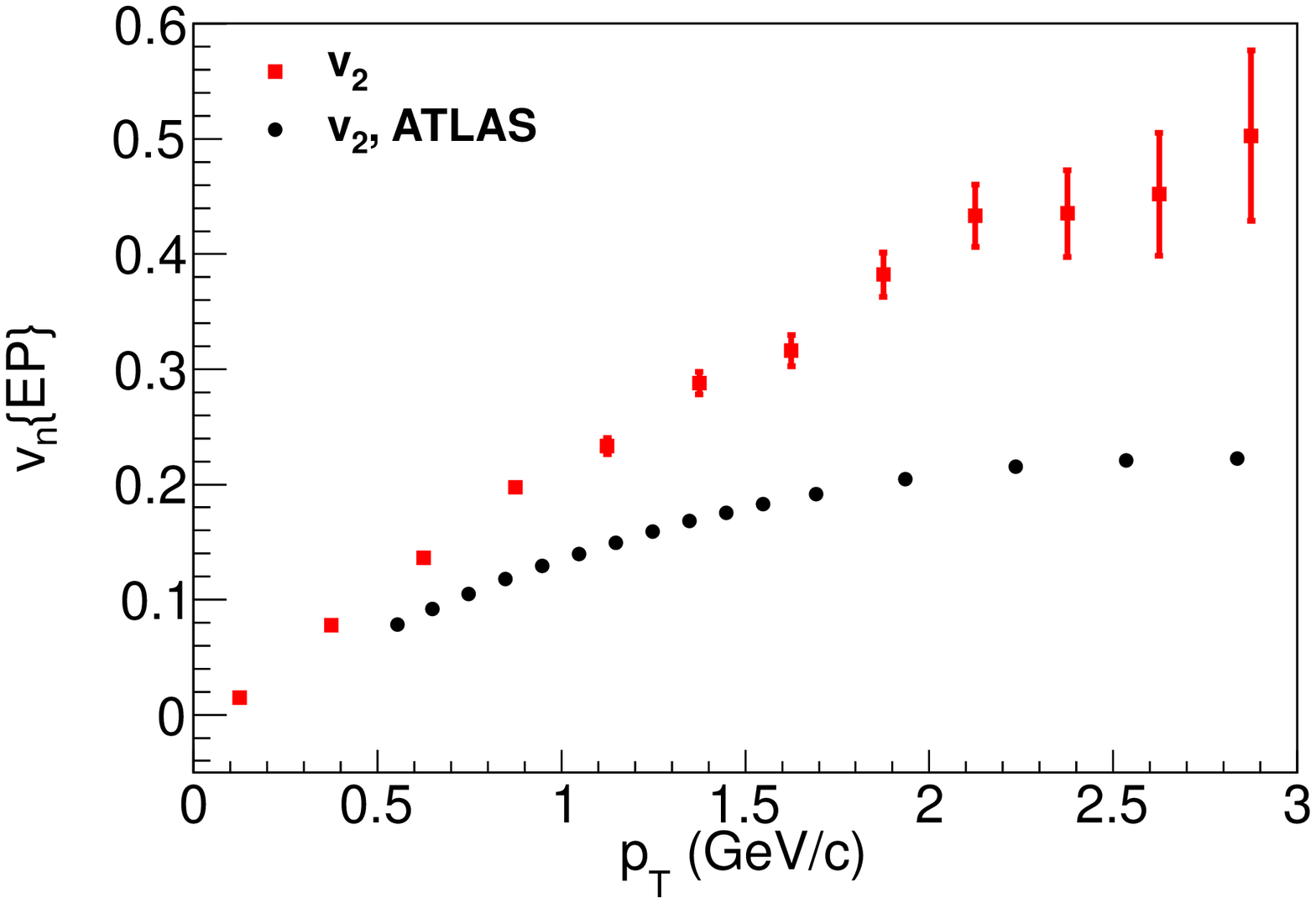,scale=0.3}
\end{minipage}
\caption{
Transverse momentum dependence of $v_{2}$ of
charged hadrons
in $0<\mid \eta \mid < 1$ (top),
$1<\mid \eta \mid < 2$ (middle)
and $2<\mid \eta \mid < 2.5$ (bottom)
 using the event plane method, and
 compared with the ATLAS data \cite{ATLAS:2011ah}
 at 0-10\% (left), 40-50\% (middle) and 70-80\% (right) centralities. 
\label{fig:v2ept}}
\end{center}
\end{figure}

Transverse momentum dependence of $v_{2}$ using
the event plane method
is compared with the ATLAS $v_{2}$ data \cite{ATLAS:2011ah}
in  Pb+Pb
central (0-10\%), semi-central (40-50\%) and peripheral (70-80\%)  collisions 
in $0<\mid \eta \mid < 1$ (top),
$1<\mid \eta \mid < 2$ (middle)
and $2<\mid \eta \mid < 2.5$ (bottom)
at $\sqrt{s_{NN}}$ = 2.76 TeV 
in Fig.~\ref{fig:v2ept}.
We emphasise here that we employ the same flow analysis method
as the ATLAS
Collaboration \cite{ATLAS:2011ah}.
$v_{2}$ from event-by-event ideal hydrodynamic simulations 
overshoots the ATLAS data at
all centralities regardless of the pseudorapidity regions.

\begin{figure}[tb]
\begin{center}
\begin{minipage}[t]{6 cm}
\epsfig{file=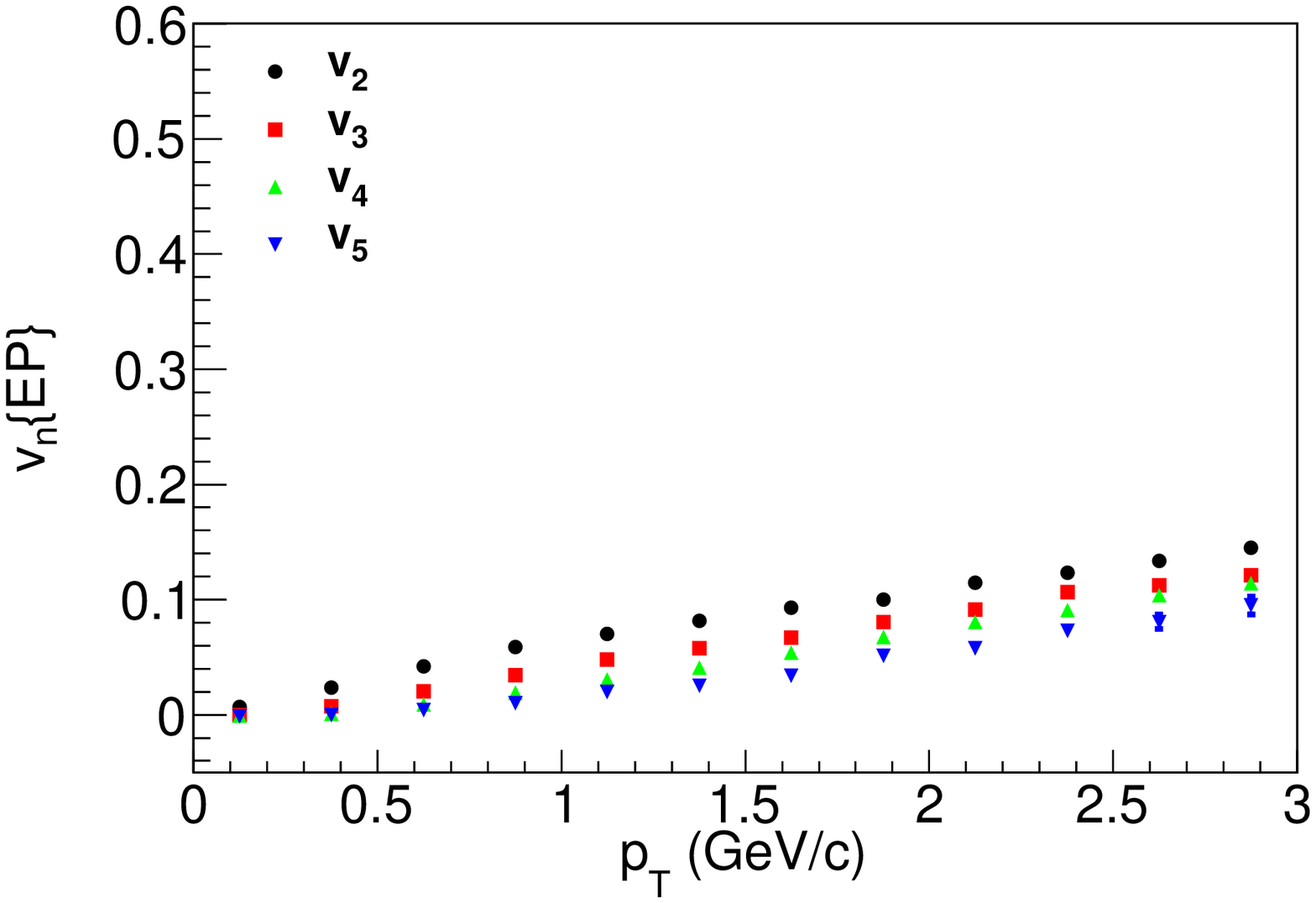,scale=0.3}
\end{minipage}
\begin{minipage}[t]{6 cm}
\epsfig{file=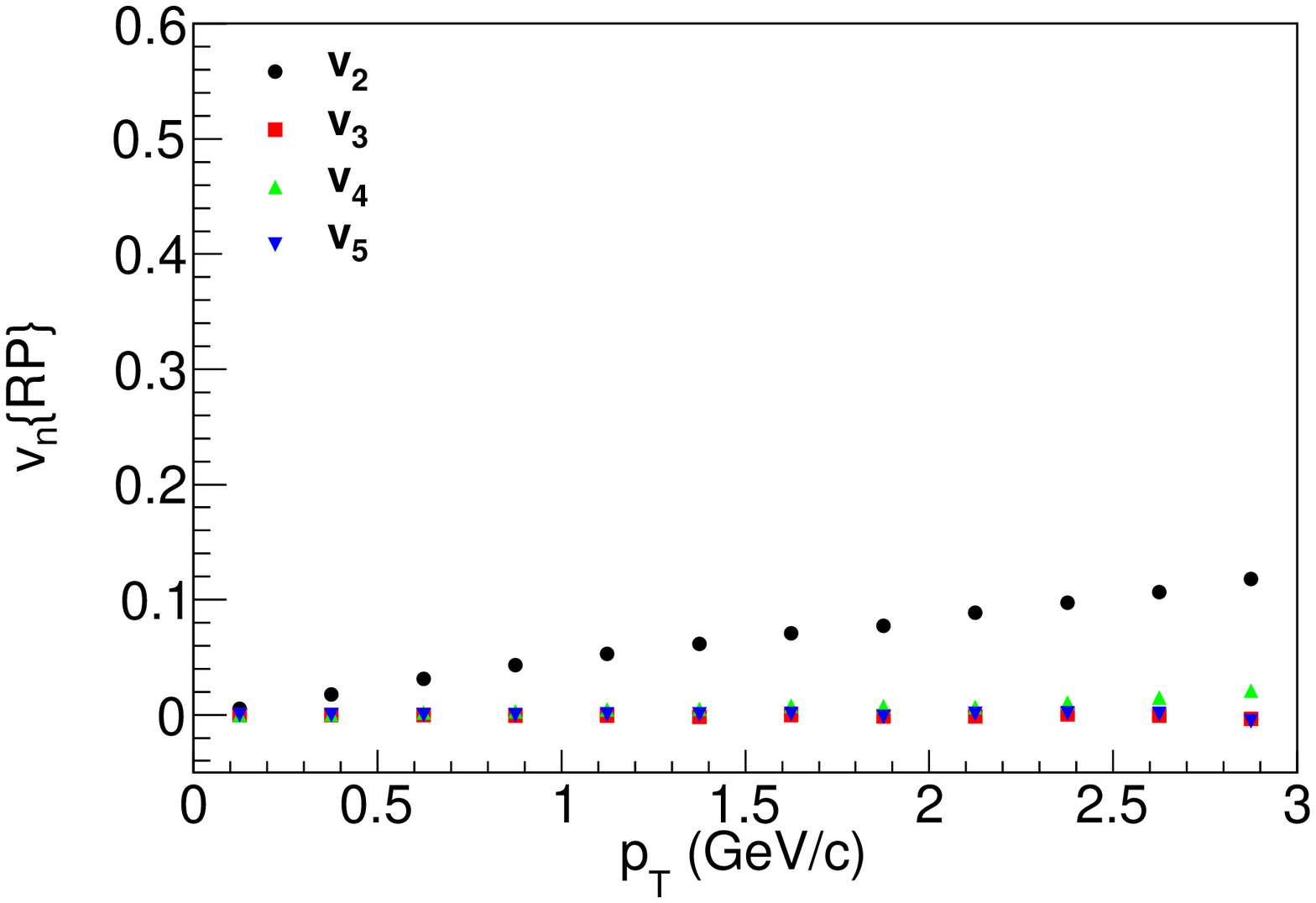,scale=0.3}
\end{minipage}
\begin{minipage}[t]{6 cm}
\epsfig{file=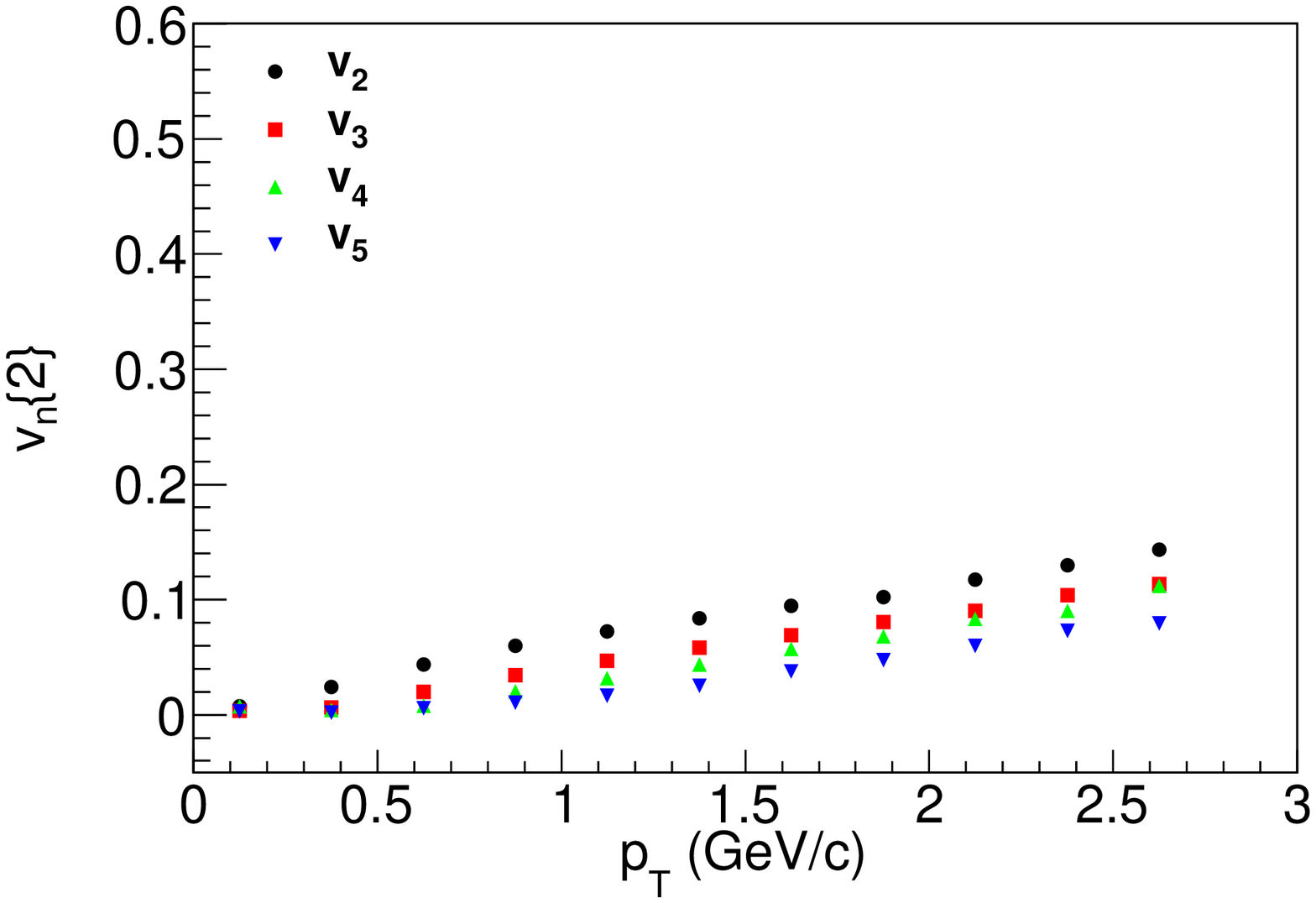,scale=0.3}
\end{minipage}
\caption{
Transverse momentum dependence of 
harmonics, $v_{n}$ ($n$= 2, 3, 4 and 5), using the event plane method (left), 
the reaction plane method (middle), and
the two particle cumulant method (right) at 0-10\% centrality.
\label{fig:vmpt0}}
\end{center}
\end{figure}

\begin{figure}[tb]
\begin{center}
\begin{minipage}[t]{6 cm}
\epsfig{file=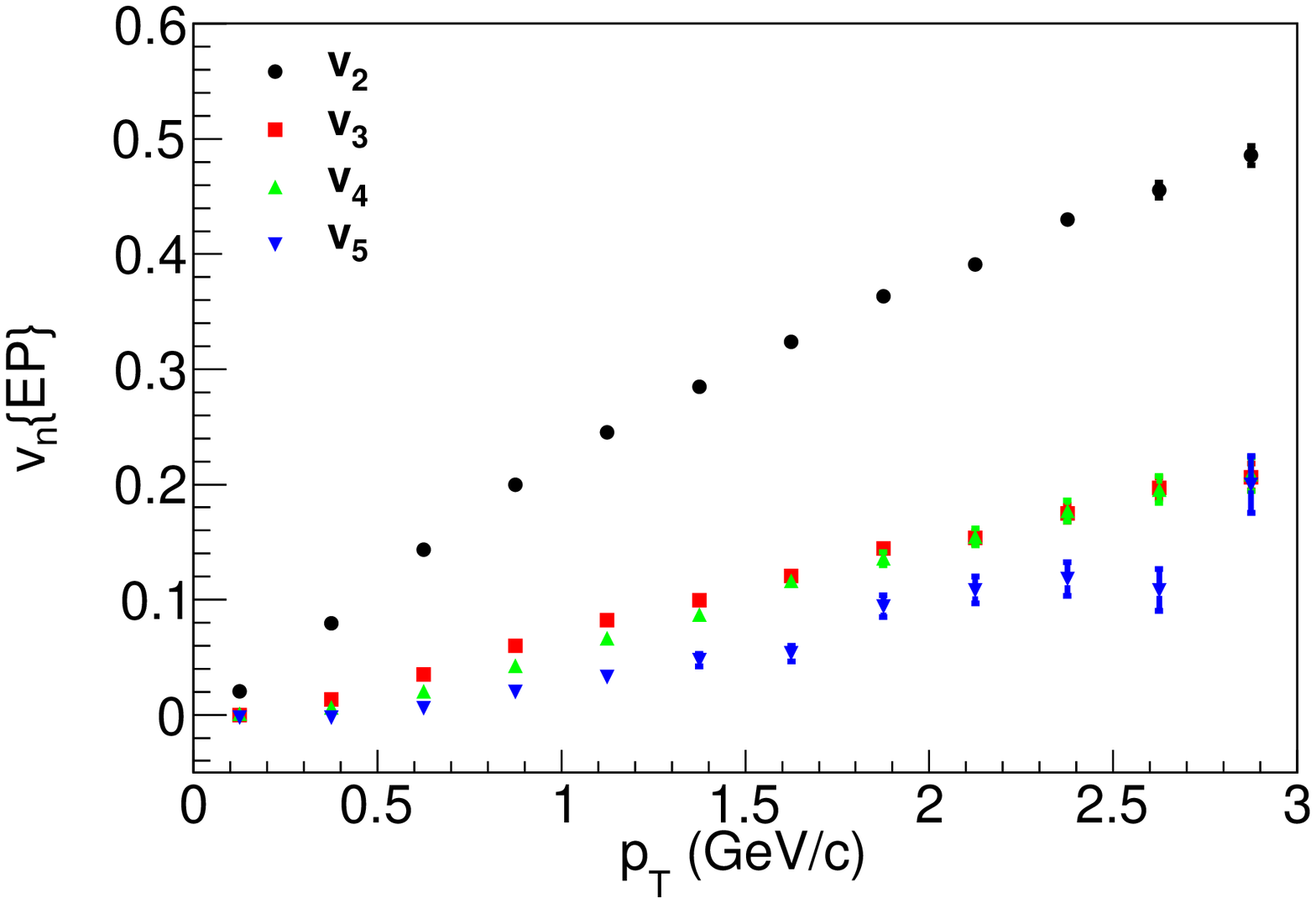,scale=0.3}
\end{minipage}
\begin{minipage}[t]{6 cm}
\epsfig{file=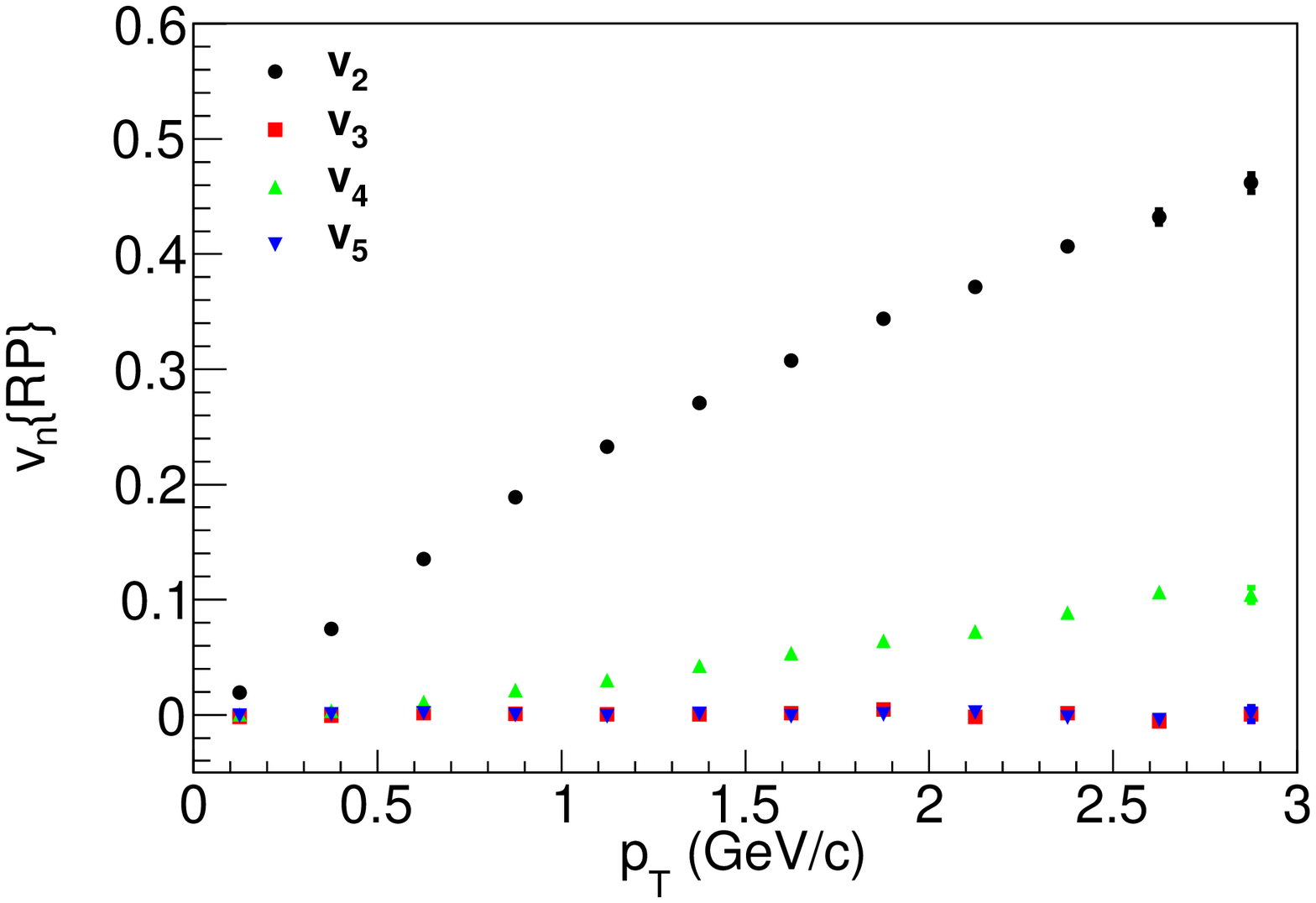,scale=0.3}
\end{minipage}
\begin{minipage}[t]{6 cm}
\epsfig{file=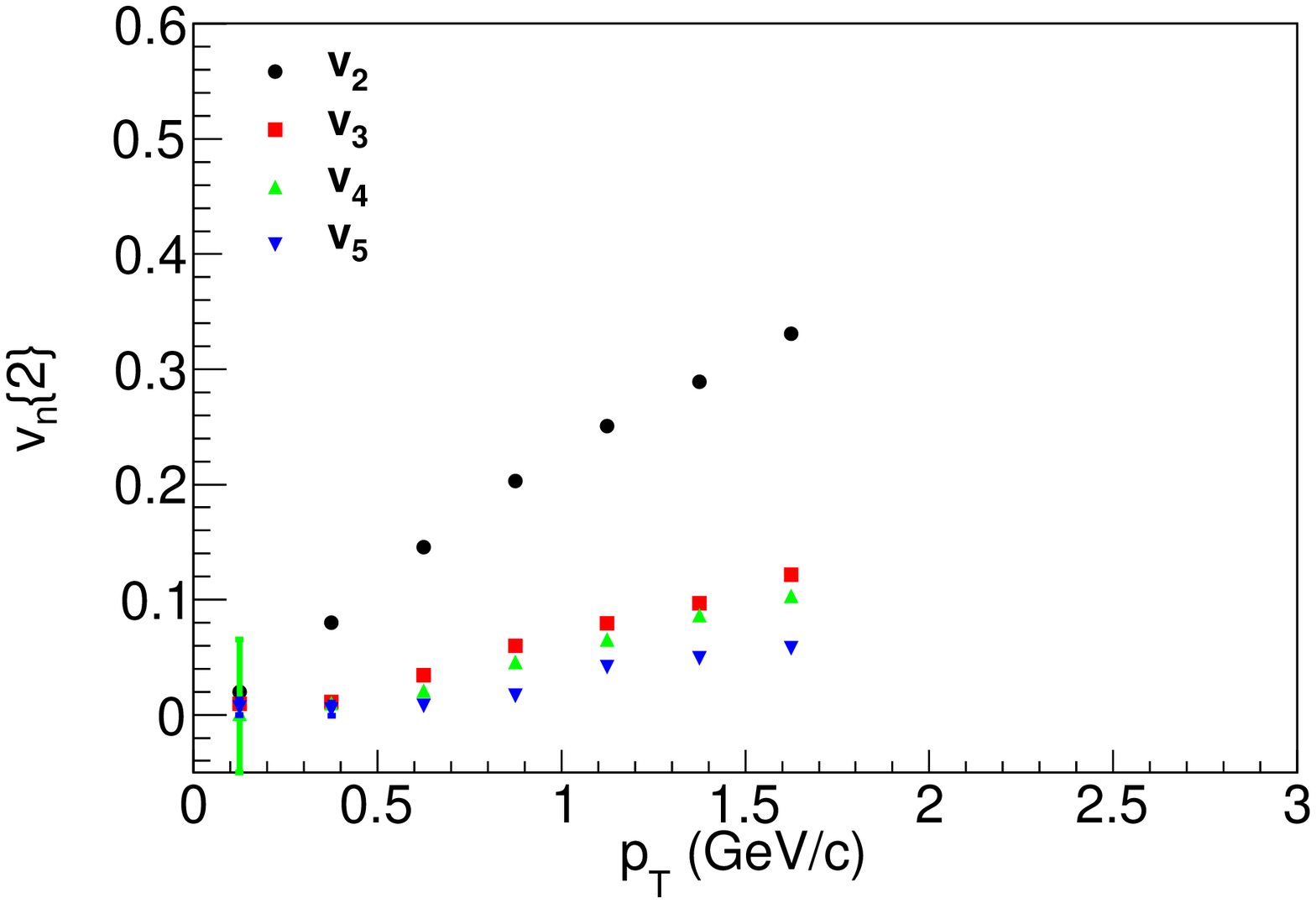,scale=0.3}
\end{minipage}
\caption{
The same as Fig.~\ref{fig:vmpt0} but at 40-50\% centrality.
\label{fig:vmpt4}}
\end{center}
\end{figure}

\begin{figure}[tb]
\begin{center}
\begin{minipage}[t]{6 cm}
\epsfig{file=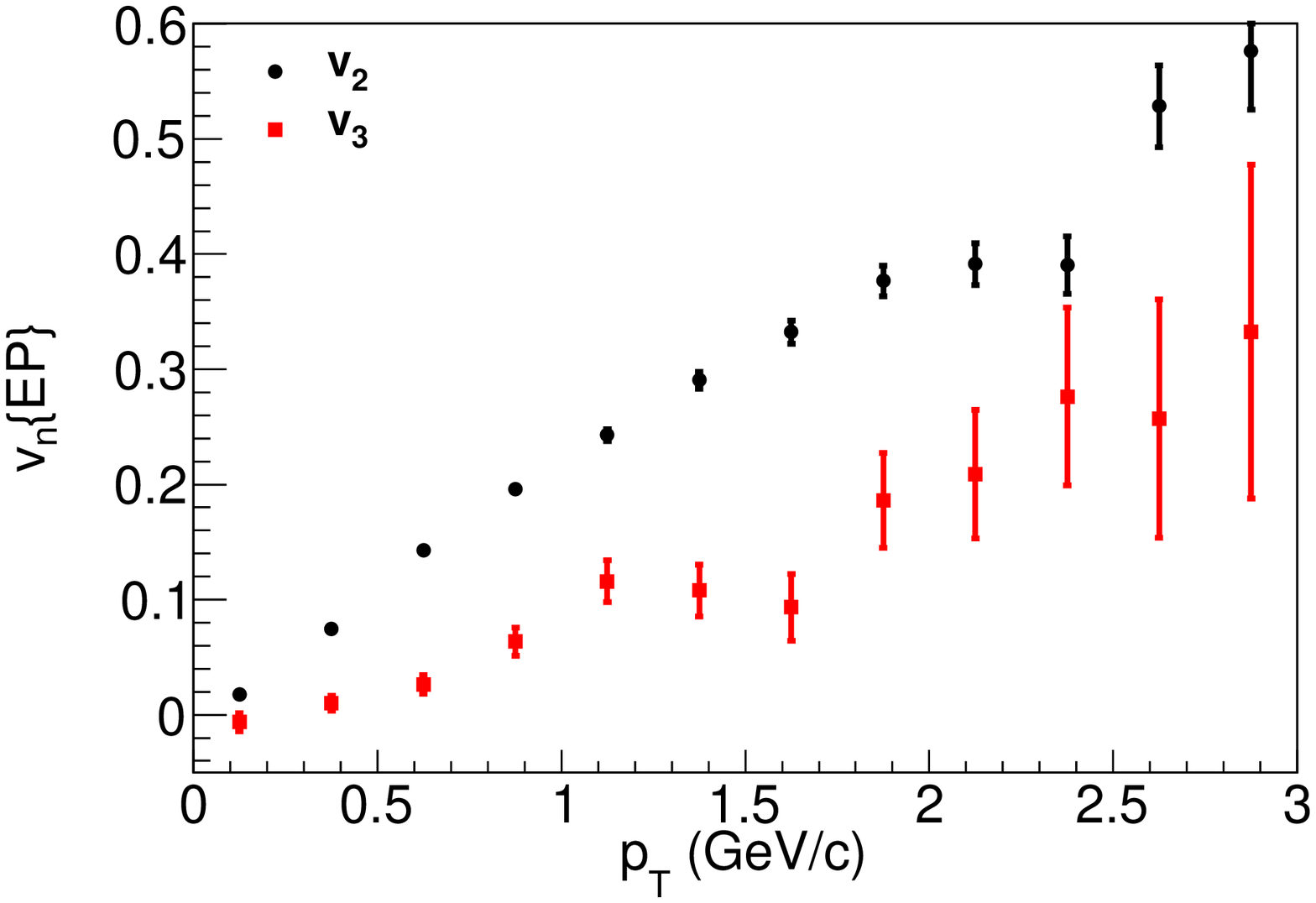,scale=0.3}
\end{minipage}
\begin{minipage}[t]{6 cm}
\epsfig{file=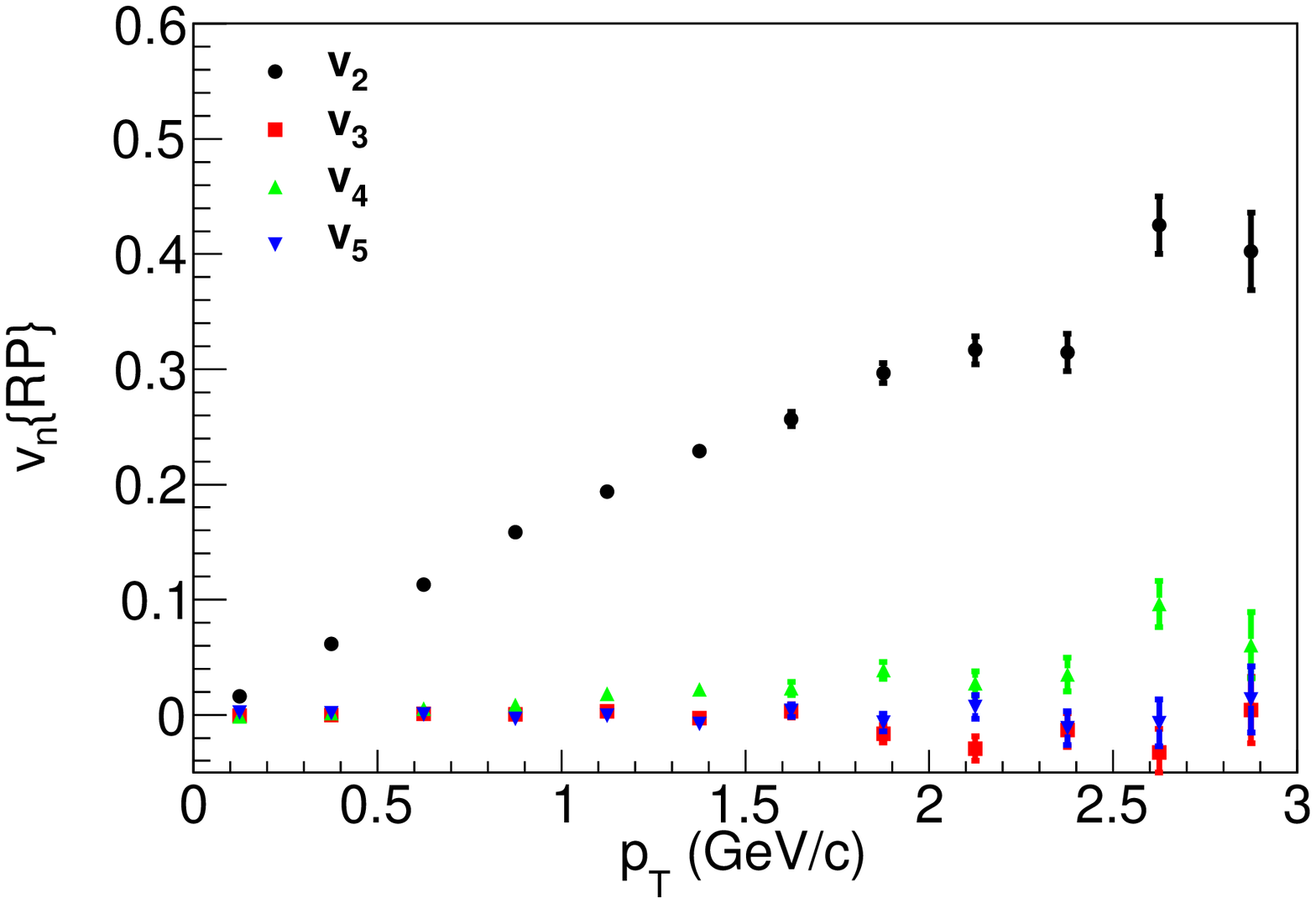,scale=0.3}
\end{minipage}
\begin{minipage}[t]{6 cm}
\epsfig{file=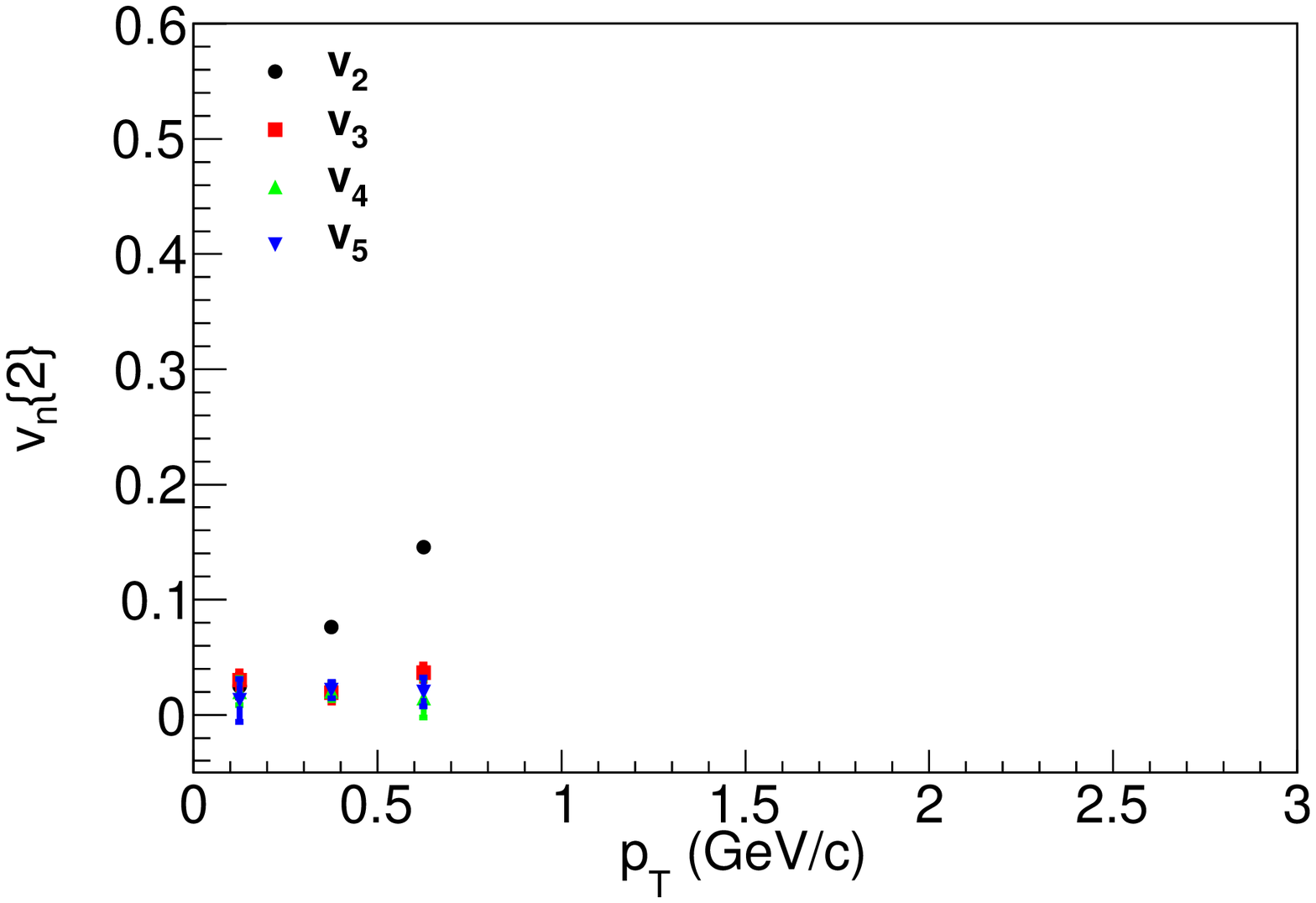,scale=0.3}
\end{minipage}
\caption{
The same as Fig.~\ref{fig:vmpt0} but at 70-80\% centrality.
$v_{4}$ and $v_{5}$ are omitted in the event plane method since
the errors are too large due to poor resolution
of the event plane angle.
\label{fig:vmpt7}}
\end{center}
\end{figure}

Transverse momentum dependences of $v_{n}$ ($n$=2, 3, 4 and 5) 
of charged hadrons
at midrapidity are shown at 0-10\% (Fig.~\ref{fig:vmpt0}),
40-50\% (Fig.~\ref{fig:vmpt4}) and 70-80\% (Fig.~\ref{fig:vmpt7}) centralities
in Pb+Pb collisions at $\sqrt{s_{NN}}$=2.76 TeV.
In each figure, results using the event plane method,
the reaction plane method
and the two-particle cumulant method
are compared with each other.
Due to the poor statistics, we omit the results from the four-particle cumulant method.
In central collisions (0-10\% centrality),
all $v_{n}$\{EP\} are close to each other.
The magnitude and the order of $v_{n}$\{2\}
are almost identical 
to those of $v_{n}$\{EP\}.
However, in the case of the reaction plane method,
only $v_{2}$ is finite and the other harmonics vanish
in $0<p_{T} < 3$ GeV/$c$.
In semi-central collisions (40-50\% centrality),
$v_{2}$ deviates from other harmonics, which reflects the almond like 
average geometry 
in non-central collisions.
Again, the pattern of $v_{n}$\{2\} is quite similar to that of $v_{n}$\{EP\}.
$v_{2}$ using these three methods is
almost the same,
$v_{3}$ and $v_{5}$ vanish when the reaction plane method
is used
as expected from an argument of fluctuating initial 
conditions,\footnote{
Regarding $v_{5}$, a more precise reason is still not clear
since the correlation between $\Psi_{5}$ and $\Phi_{5}$
is non-trivial as shown in Fig.~\ref{fig:reso}.}
whereas
$v_{4}$\{RP\} becomes finite although its magnitude is rather 
smaller than $v_{4}$\{EP\} and $v_{4}$\{2\}.
The sensitivity of $v_4$ on the analysis methods indicates that 
measured $v_{4}$ contains both fluctuation and geometry effects.
In peripheral collisions, it is quite hard to obtain higher order
harmonics with small errors
for the given number of  events since
not only the number of events but also the number of measured particles
do matter.
In the event plane method, resolution parameter
becomes worse with decreasing centrality, which, in turn,
increases the errors of $v_{n}^{\mathrm{obs}}$.

\begin{figure}[tb]
\begin{center}
\begin{minipage}[t]{6 cm}
\epsfig{file=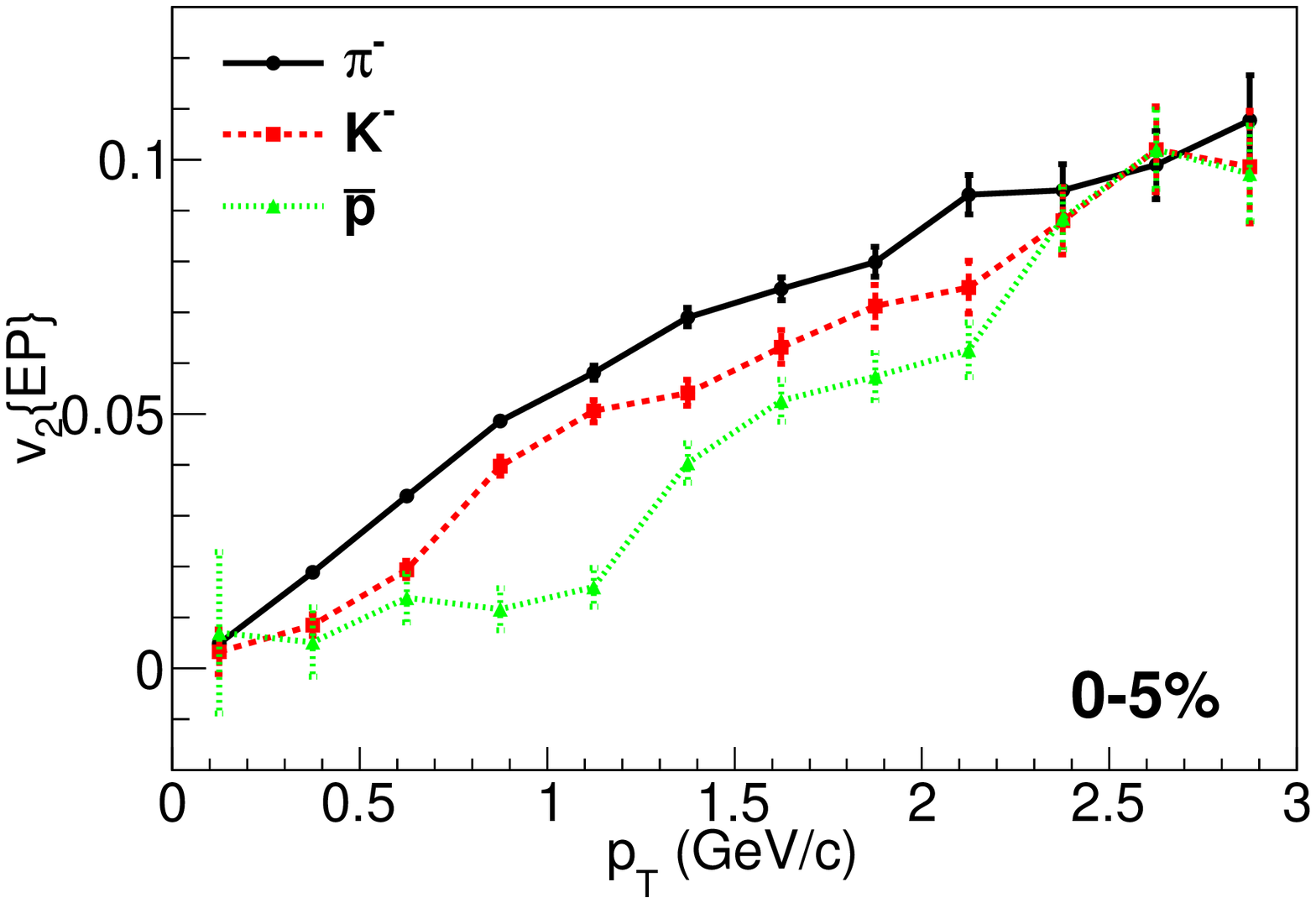,scale=0.3}
\end{minipage}
\begin{minipage}[t]{6 cm}
\epsfig{file=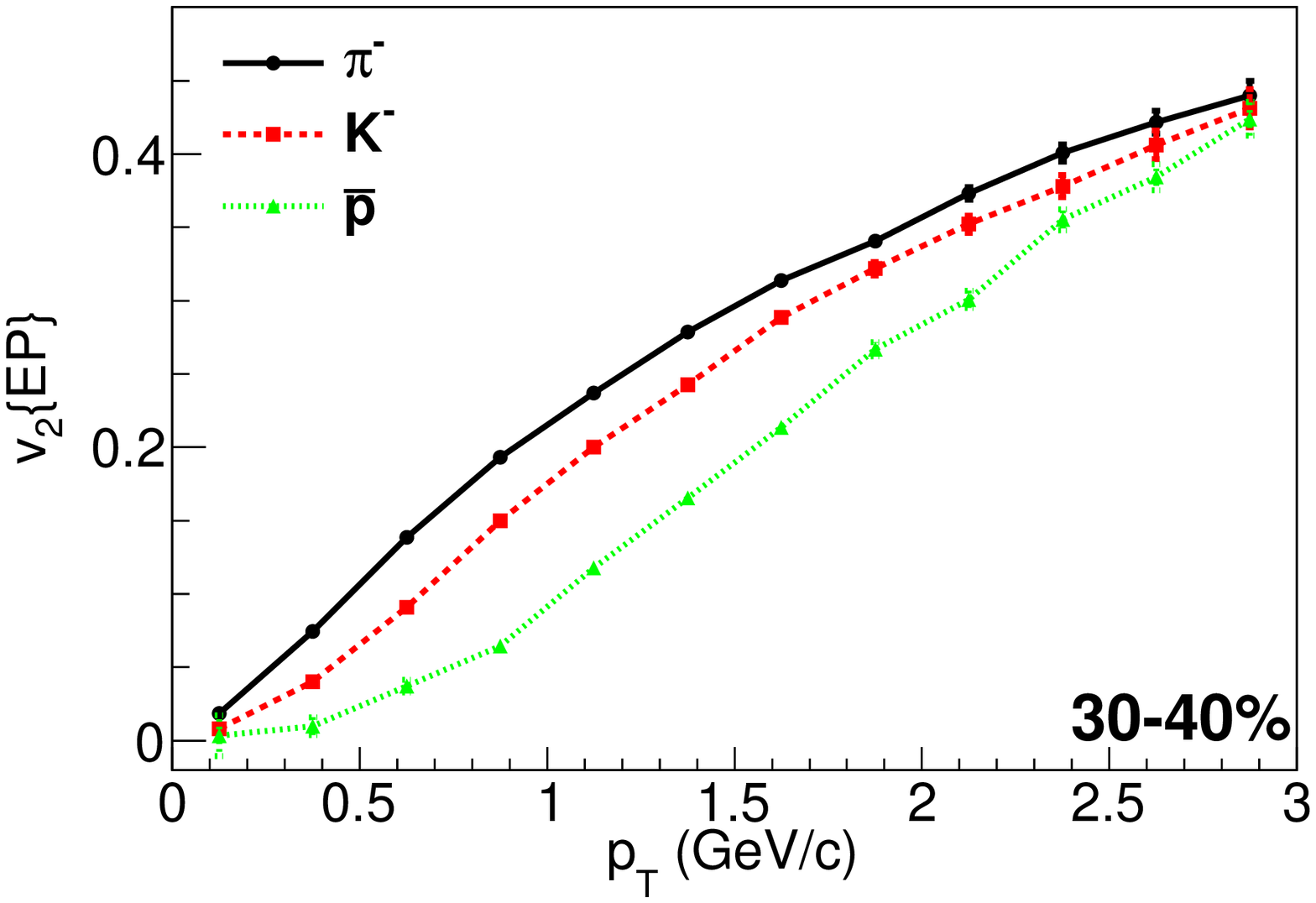,scale=0.3}
\end{minipage}
\begin{minipage}[t]{6 cm}
\epsfig{file=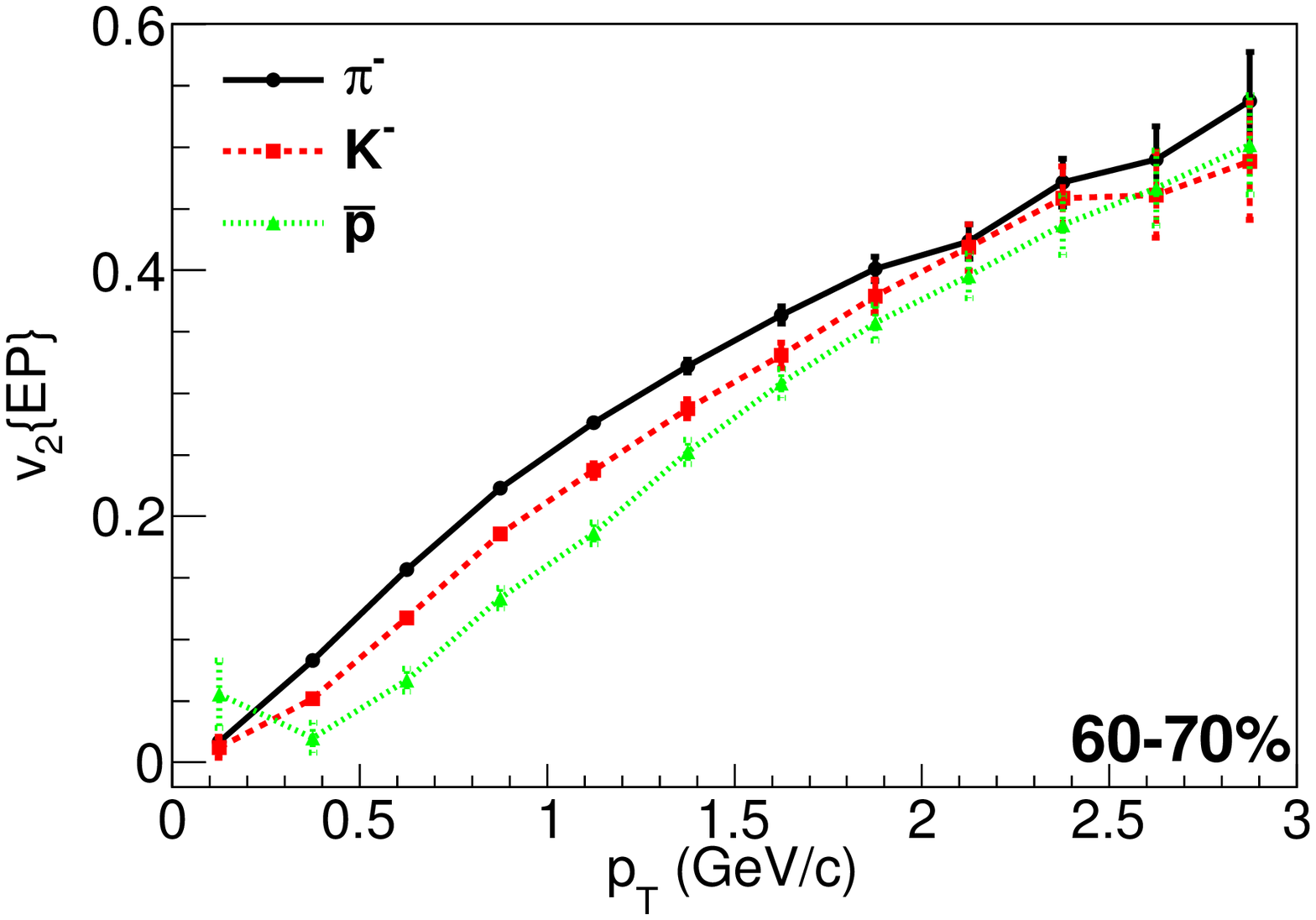,scale=0.3}
\end{minipage}
\caption{
$v_{2}$\{EP\} of identified hadrons
in $|\eta|<0.8$ in Pb+Pb collisions at $\sqrt{s_{NN}}$ = 2.76 TeV.
Results at 0-5\% (left), 30-40\% (middle) and 60-70\%  (right) centralities
are shown.
\label{fig:v2ptPIDLHC}}
\end{center}
\end{figure}

\begin{figure}[tb]
\begin{center}
\begin{minipage}[t]{6 cm}
\epsfig{file=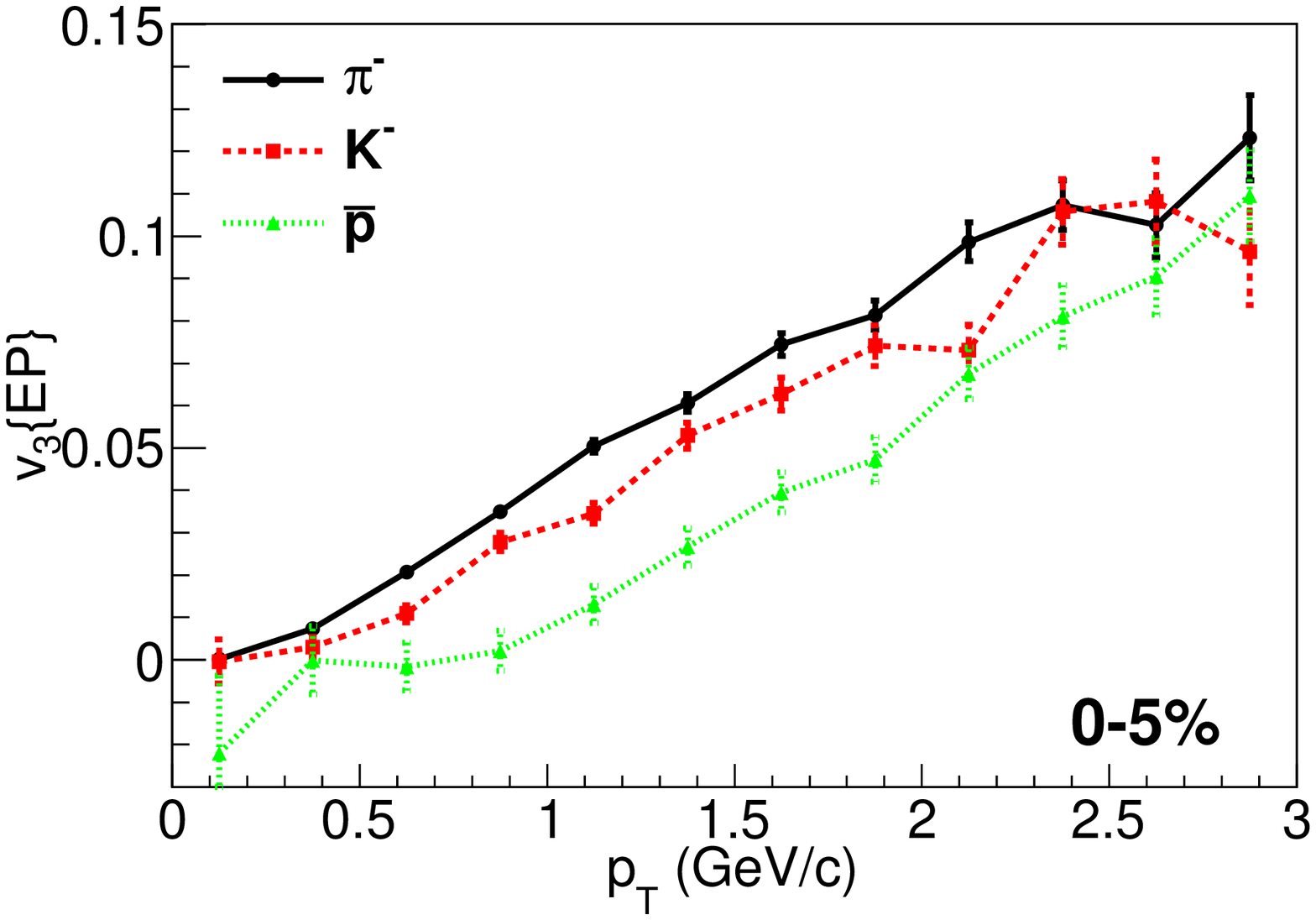,scale=0.3}
\end{minipage}
\begin{minipage}[t]{6 cm}
\epsfig{file=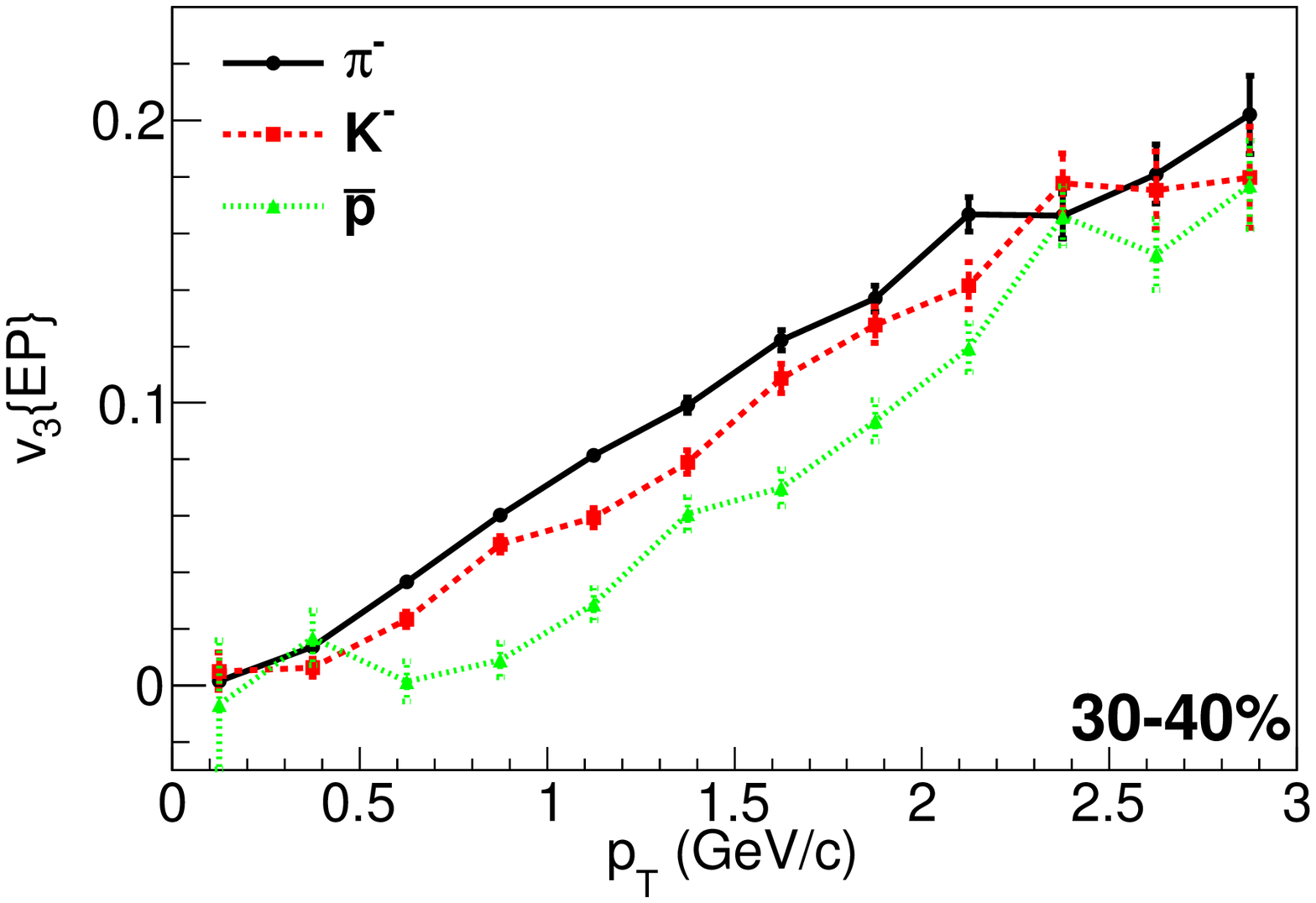,scale=0.3}
\end{minipage}
\begin{minipage}[t]{6 cm}
\epsfig{file=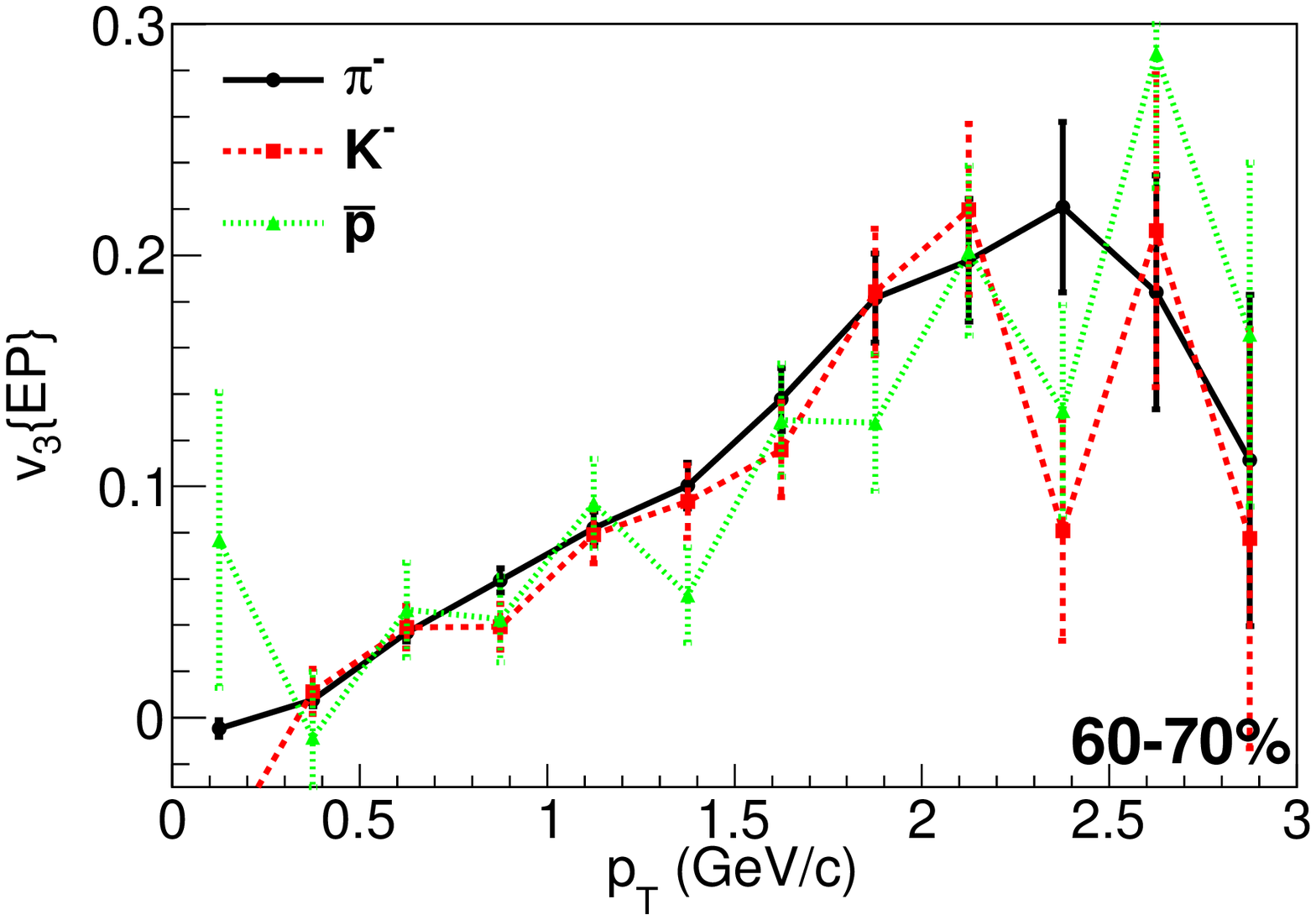,scale=0.3}
\end{minipage}
\caption{
The same as Fig.~\ref{fig:v2ptPIDLHC}
but for $v_{3}$\{EP\}.
\label{fig:v3ptPIDLHC}}
\end{center}
\end{figure}

We also show $v_{2}$\{EP\} and
$v_{3}$\{EP\} of identified hadrons ($\pi^{-}$, $K^{-}$
and $\bar{p}$) in Figs.~\ref{fig:v2ptPIDLHC} and \ref{fig:v3ptPIDLHC},
respectively,  at midrapidity ($|\eta|<0.8$)
at three centralities (0-5\%, 30-40\% and 60-70\%) 
in Pb+Pb collisions at the LHC energy.
Mass splitting pattern, namely $v_{n}^{\pi}>v_{n}^{K}>v_{n}^{p}$
for $n=2$ and 3, is clearly seen in all results except 
in $v_{3}$ at
60-70\% centrality where, due to smaller multiplicity,
resolution is poor and, consequently, errors are very large.
Similar mass splitting pattern
was also found in Ref.~\cite{Petersen:2011sb},
where it was pointed out that the mass splitting pattern is not
  necessarily a consequence of the existence of the QGP.

When radial flow is sufficiently large, $v_{2}$ of protons
could be negative in the low $p_{T}$ region \cite{Huovinen:2001cy}.
However, this cannot be clearly seen even though
radial flow at the LHC energy is expected to be 
larger than that at the RHIC energy.
$v_{n}(p_{T})$ for identified hadrons
would give further constraints for the dynamical modelling
at the LHC energy
as well as at the RHIC energy.

\begin{figure}[tb]
\begin{center}
\begin{minipage}[t]{9 cm}
\epsfig{file=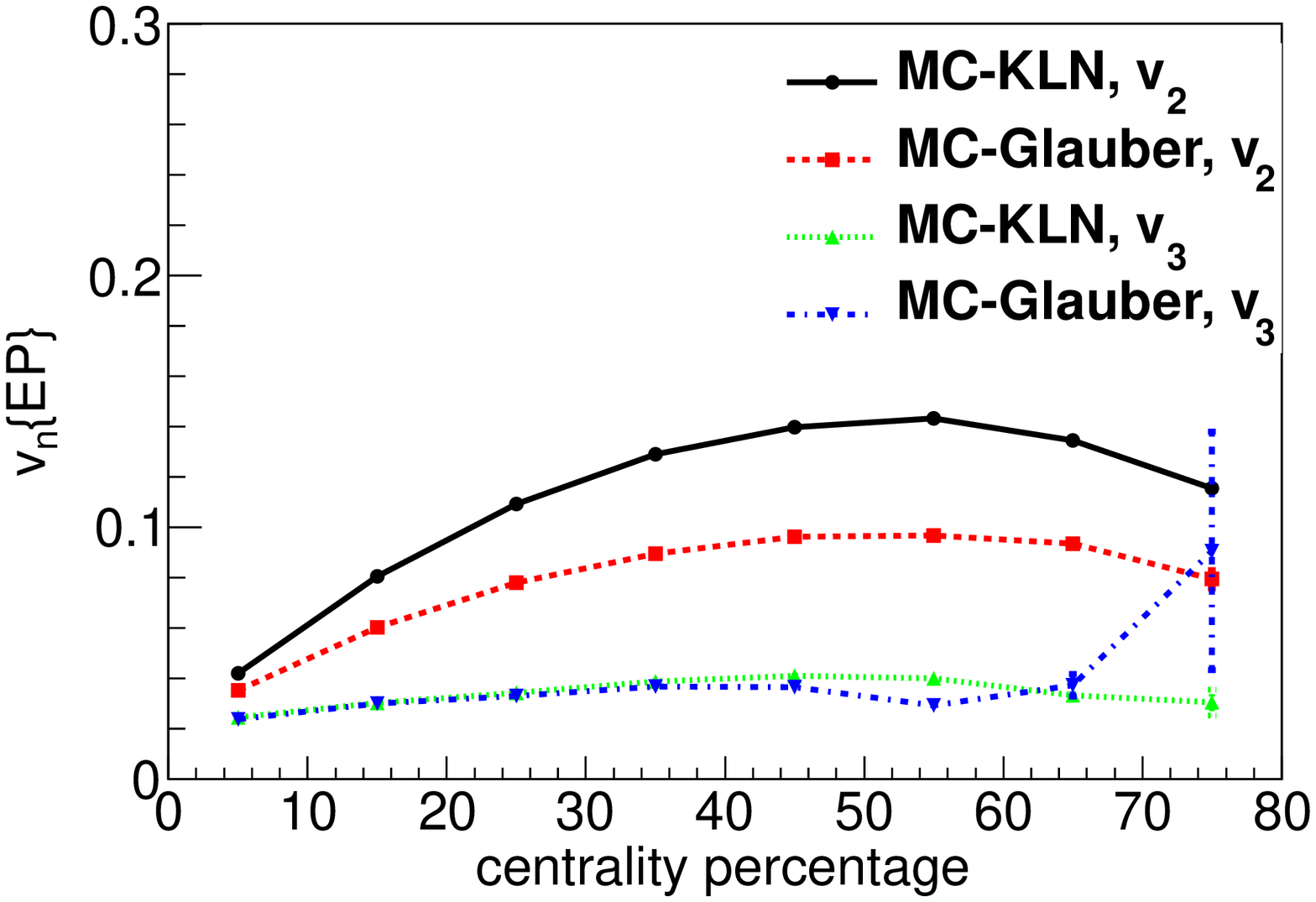,scale=0.45}
\end{minipage}
\begin{minipage}[t]{9 cm}
\epsfig{file=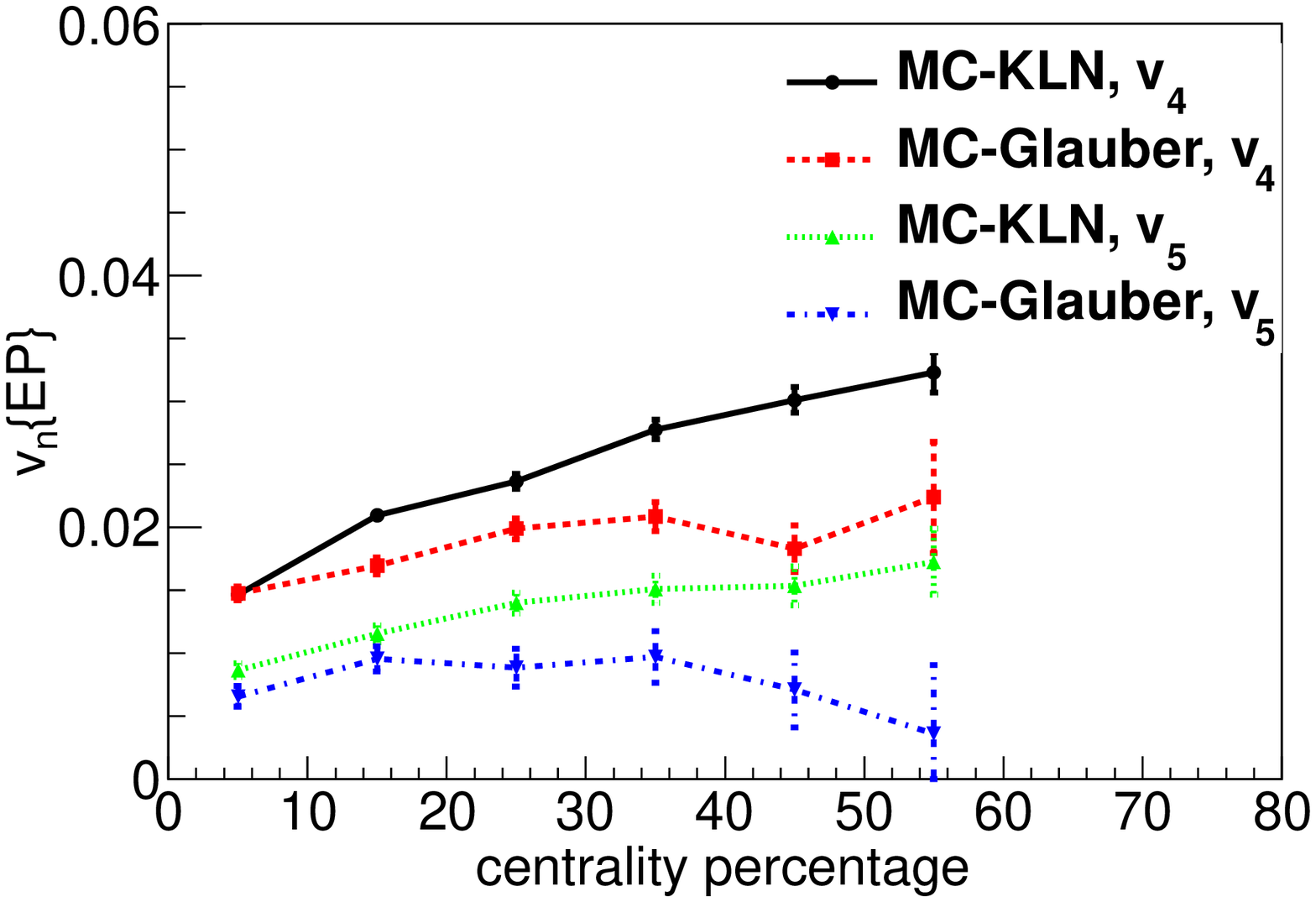,scale=0.45}
\end{minipage}
\caption{
Centrality dependence of $v_{2}$ and $v_{3}$ (left)
and $v_{4}$ and $v_{5}$ (left) of charged hadrons
at midrapidity ($0<\eta<1$)
in Pb+Pb collisions at $\sqrt{s_{NN}}=2.76$ TeV.
Results obtained using the MC-KLN model are compared
with the ones obtained using the MC-Glauber model.
\label{fig:KlnVsGlauber}}
\end{center}
\end{figure}

So far we have shown the results 
obtained using the MC-KLN initialisation.
We next compare these results with the ones 
obtained using the MC-Glauber initialisation.
As already seen in the previous section,
the MC-KLN model
gives a larger  eccentricity $\varepsilon_{2}$ 
than the MC-Glauber model
and, consequently, leads to larger $v_{2}$.
This is again seen in Fig.~\ref{fig:KlnVsGlauber} (left).
Regardless of different models, $v_{3}$ using the MC-KLN model
is almost identical to $v_{3}$ using the Glauber model.
This is due to the fact that, in both models, 
$v_{3}$ is generated by triangular
anisotropic shape of the reaction zone, $\varepsilon_{3}$
and that $\varepsilon_{3}$ is determined by the granular structure of
the colliding nuclei as shown in Fig.~\ref{fig:KlnVsGlauberEps}.
On the other hand, 
one cannot see the similarity of $v_{n}$ ($n=4, 5$) between the MC-KLN model
and the MC-Glauber model in Fig.~\ref{fig:KlnVsGlauber} (right).
In semi-central collisions (10-60\% centrality)
$v_{n}$ ($n=4, 5$) using the MC-KLN model are
systematically larger than the ones using the MC-Glauber model.
In spite of a fact that higher order flow coefficients
such as $v_{4}$ and $v_{5}$
are correlated with lower order flow coefficients 
as demonstrated in Fig.~\ref{fig:mixedcorr},
simultaneous analyses of $v_{n}$ ($n\lsim 5$) 
can be used to discriminate between the CGC and
Glauber pictures for the initial conditions and would
thus help to provide useful information about the transport
properties of the QGP.
Although 
the simultaneous analysis of $v_{2}$ and $v_{3}$ revealed the MC-KLN model
 apparently contradicts to the PHENIX data \cite{Adare:2011tg},
these calculations lack multiplicity fluctuations
which would affect $\varepsilon_{n}$~\cite{Dumitru:2012yr}
and thus improve the fit to the data.

\section{Conclusion \label{sec:conclusion}}
 
In this paper, we have performed simulations of relativistic heavy ion collisions
at RHIC and LHC energies
using an integrated approach of dynamical modelling,
in which Monte-Carlo calculations of initial collisions
based on the colour glass condensate or the Glauber pictures,
a fully (3+1)-dimensional ideal hydrodynamic model
with the state-of-the-art lattice QCD equation of state,
and a hadronic cascade are combined.

We employed the Monte-Carlo
versions of the Kharzeev-Levin-Nardi model
and the Glauber model
and, using them, generated many events to obtain
initial conditions for the subsequent hydrodynamic
evolution.
We had two options for initial conditions:
One is the single smooth initial condition for a given centrality
as we 
take an average of the entropy profiles in each centrality class.
By rotating each initial profile to match its participant plane
with the reaction plane,
we obtained initial conditions containing some fluctuation effects.
It turned out that these initial conditions
are necessary to obtain large elliptic flow parameter
in small systems, such as the matter created in Cu+Cu collisions,
and in 
systems which are almost cylindrical on average, such as
central events in Au+Au collisions.
However, these initial conditions
did not contain higher odd harmonic components
and, therefore,
could not lead to reproduction of anisotropic flow parameters with odd harmonics.
We neglected the longitudinal structure of the created matter in this option
and assumed boost invariance. Therefore, we could not
discuss (pseudo-)rapidity dependence of observables.

The other option is to utilise each initial profile from the Monte-Carlo
calculations on event-by-event basis.
Since the entropy production in the primary 
nucleon-nucleon collision happens
when the two nucleons in the colliding nuclei
are sufficiently close in the transverse plane,
the entropy density profile contains
fluctuations from configuration of nucleons
in the colliding nuclei.
The resultant example of initial conditions exhibits
a bumpy structure and contains also odd spatial anisotropies
unlike in the first option.
In this option, we also considered space-time rapidity dependence
of initial conditions. The KLN model gives rapidity
distribution of produced gluons through the 
$k_{T}$ factorisation formulation.
We simply identified the momentum rapidity distribution with
the space-time rapidity distribution in the transverse plane
to obtain initial conditions in the whole configuration space.
The Glauber model does not tell us anything about the longitudinal structure
of entropy production.
We modelled initial longitudinal structure of
entropy profile based on a string picture, which we called
a modified BGK (Brodsky-Gunion-K\"uhn) model.
In this model, the space-time rapidity dependence of entropy density
was calculated using the numbers of participants and that of
binary collisions as given by the Glauber model.

Using these initial conditions, we performed
ideal hydrodynamic simulations
solving the expansion numerically in all three dimensions using
the Milne coordinates ($\tau$, $x$, $y$, $\eta_{s}$).
For the equation of state, we used a model
in which the lattice QCD results in high temperature regions
are smoothly connected with
a resonance gas 
equation of state with the same hadrons than the event generator JAM.
The resultant equation of states exhibits
a crossover behaviour rather than a phase transition.
We employed the Piecewise Parabolic Method
to solve the hydrodynamic equations.
This method has been known as a robust algorithm against
strong shock waves, and is thus
ideal for  simulating fluid evolution starting from bumpy initial conditions
on an event-by-event basis.
We described the space-time evolution of the QGP fluids
all the way down to the switching temperature.
The switching temperature was chosen to reproduce
the final particle ratios of  pions, kaons and protons
at the full RHIC energy.
With the subsequent
hadronic evolution, we found $T_{\mathrm{sw}} = 155$ MeV
leads to a good description of yields and slopes in transverse momentum
distribution for identified hadrons at RHIC.
From a macroscopic hydrodynamic picture to
a microscopic particle picture at $T_{\mathrm{sw}}$,
we employed the Cooper-Frye formula
rejecting in-coming particles which contribute
as a negative number in the phase space.
By sampling hadrons on the particlisation hypersurface
according to this prescription,
we obtained ``initial" phase space distribution
for the subsequent hadronic cascading
on an event-by-event basis.
We simulated the space-time evolution
of hadron gas utilising a hadronic cascade, JAM.
Transport models can naturally treat the late low density phase
of the system in the heavy ion collision where the
system is not expected to be in a local equilibrium
any more, and provide a realistic freeze-out 
which depends on the particle species.

We simulated $\sim$10$^{5}$ ``minimum bias" events with $N_{\mathrm{part}}\ge 2$
in the event-by-event option for each parameter set.
We confirmed the average multiplicity in event-by-event
 simulations is smaller than the multiplicity using the smooth
averaged initial conditions.
We also obtained $p_{T}$ spectra for identified hadrons and
found anti-proton yields become comparable with negative pion
yield at $p_{T}\sim 3$ GeV/c.
So far comparisons of theoretical results using a flow analysis method
have been made with experimental data
obtained using the different flow analysis method.
We analysed
the final particle distribution in a spirit of performing
(almost) the same procedure
as the experimentalists have done.
We used particle multiplicity in forward rapidity
to perform centrality cuts.
When we calculated the anisotropic flow parameters,
we employed several typical flow analysis methods such as
reaction plane method, event plane method and
two- and four-particle cumulant methods.

We found that some of the different flow analysis
methods lead to almost the same results. For instance,
$v_{n}$ using the event plane method
is almost the same as $v_{n}$ using the two-particle
cumulant method within errors.
Notice that contributions from jet fragmentation
were missing in our dynamical simulations,
which omits
one of the major non-flow effects in the two-particle cumulant method.
On the other hand,
the four-particle cumulant
method gives different results.
$v_{2}$ using the four-particle cumulant method
is more like theoretically calculated $v_{2}$
with respect to the reaction plane.
Although we did not have enough statistics to draw firm conclusions,
in our calculations,
$v_{3}$ using the four particle
cumulant method was roughly half
of $v_{3}$ obtained using the other two methods.
$v_{2}$ using the event plane method
contains effects of participant plane fluctuation:
In the limit of vanishing centrality percentage,
$v_{2}$ using the event plane method 
stays finite.
$v_{2n+1}$ are expected to vanish if measured with respect to
the reaction plane.
However, $v_{2n+1}$ are finite  due to fluctuating initial transverse profiles
in the event plane method and the two-particle
cumulant method.

We evaluated the correlations 
between the event plane and the orientation angles
to see a possible mechanism of generating  higher order anisotropic flow.
For $v_{2}$ and $v_{3}$,
these two angles are strongly correlated:
Anisotropic flow is generated mainly by pressure gradient
in the direction of
short axis in each participant anisotropy.
On the other hand, non-trivial correlations are seen
in $v_{4}$ and $v_{5}$.
$v_{4}$ is first driven by the fourth spatial anisotropy
in central collisions, but
could be driven by \textit{elliptic flow} with respect to
reaction plane in mid-central to peripheral collisions.
This could be seen in the mixed correlation
between $\Phi_{2}$ and $\Psi_{4}$.

We used two models of initial conditions, namely MC-KLN model
and MC-Glauber model,
and made a systematic comparison between them.
Although entropy density profile is different in
these two models,
both models can lead to almost identical
particle yields and $p_{T}$ spectra.
On the other hand, the spatial anisotropy $\varepsilon_{2}$
from the MC-KLN model is larger than
the one from the MC-Glauber model,
which leads to discrepancy of $v_{2}$ between
the MC-KLN model and the MC-Glauber model.
Within ideal hydrodynamic simulations,
the difference can be seen in all $v_{2}$ results.
However,
$\varepsilon_{3}$ from the MC-KLN model 
is almost the same as the one from the MC-Glauber model
because
it originates from initial
fluctuation of transverse profiles, which is treated in a similar
  fashion in both models.
Consequently, there is almost no difference of $v_{3}$ between these two models.
As claimed by the PHENIX Collaboration \cite{Adare:2011tg},
one can use this fact
to discriminate the MC-KLN model from the MC-Glauber model.
Simultaneous analyses of higher order anisotropies
such as $v_{4}$ and $v_{5}$,
in addition to $v_{2}$ and $v_{3}$,
would also provide more information about the initial conditions.

In this paper, we restricted our discussion to ideal hydrodynamic
description of the QGP fluids.
However, to understand the
transport properties of the QGP,
viscous corrections to both dynamics and particle spectra
are mandatory.
So far, there have been several viscous hydrodynamic
simulations to analyse anisotropic parameters at RHIC and LHC
energies. Some of the simulations have been performed
in (2+1) dimensional space with an assumption of
boost invariance. However, given a fact that
event planes are determined in forward rapidity regions
in some flow analyses experimentally,
fully three dimensional simulations are necessary 
to describe the actual dynamics.

Final higher harmonics were turned out to be sensitive to
the flow analysis method.
A good example was the triangular flow $v_{3}$:
Although $v_{3}$ with respect to the event plane
is almost similar to
$v_{3}$ from the two-particle cumulant method,
$v_{3}$ from the four-particle cumulant method
is roughly a half of them.
This has already been confirmed in the experimental data \cite{Aamodt:2011by}.
In the hydrodynamic simulations,
one usually obtains smooth distributions using the Cooper-Frye
formula. However,
Monte-Carlo sampling from these smooth momentum distributions
is necessary to perform particle-based analysis method, \textit{i.e.},
event-plane method or multi-particle cumulants methods.
Of course, the subsequent hadronic cascading is also 
important and required to describe the
gradual freezeout in the hadronic rescattering stage.

We also showed that,
for higher order harmonics ($n \ge 4$),
the participant plane defined using initial profiles
does not correlate with
the event plane defined using final particle samples.
This means that  calculations
using event-averaged initial conditions
where the orientation angles of the anisotropy $\varepsilon_n$ are
  matched, are questionable.

The PHENIX Collaboration found \cite{Adare:2011tg} that the Glauber initialisation
followed by viscous hydrodynamic simulations
with $\eta/s = 0.08$ simultaneously reproduced
$v_{2}$ and $v_{3}$ as functions of centrality,
while the KLN initialisation with $\eta/s=0.16$
leads to a reasonable reproduction only of $v_{2}$.
After that, it was found that fluctuation of particle production
obeying KNO scaling at midrapidity enhances initial
spatial anisotropy
$\varepsilon_{2n+1}$ ($n\ge1$) 
at all centralities \cite{Dumitru:2012yr}.
The initial conditions
including this feature
 could lead to enhancement of  $v_{2n+1}$ with keeping $v_{2}$.
This would change our understanding of
$v_{2}$ and $v_{3}$ as functions of centrality above.
Although it would be interesting to
consider this KNO scaling idea in the integrated dynamical model,
how to formulate it in rapidity direction is an open question.

We  assumed that initial entropy production
can directly be used as initial conditions in hydrodynamic simulations
and neglected the description of
any thermalisation processes. 
As known, thermalisation is one of the outstanding
open questions
in the physics of relativistic heavy ion collisions.

In the event-by-event hydrodynamic simulations,
thermal fluctuations during evolutions might have been
important \cite{Kapusta:2011gt}. In this case, hydrodynamic equations
(more specifically, constitutive equations)
are no longer the deterministic equations, but 
become the stochastic equations.
Since the fluctuation-dissipation relation
tells us thermal fluctuation of the energy momentum tensor
is intimately related with viscosity, one should
include fluctuation in the dynamical evolution, in particular,
in the event-by-event simulations.

Switching from a hydrodynamic picture to a particle picture
has several open issues.
Some procedures
adopted in the present study
do not respect energy-momentum conservation.
We neglected in-coming particles which contribute
as negative number in the phase space distribution.
This could have been resolved partly by simulating
hydrodynamics and hadron cascade simultaneously
and by explicitly treating absorption of particles
coming inside fluid regions.
The negative contributions are an issue for the space-like
  hypersurface elements, but even for the time-like elements, the
  sampling of the finite number of particles violates the
  energy-momentum conservation in an individual event.
The conservation recovers only when over-sampling
of particles is made.

Regarding a switch from a fluid to a gas, one natural question
would be when and how hydrodynamic picture breaks down.
In the present study,
the switching temperature
is just an adjustable parameter to reproduce
particle ratios among hadrons.
It is interesting to note that
the switching temperature $T_{\mathrm{sw}}=155$ MeV
 obtained in this study
is very close to 
the pseudocritical temperature
of the chiral phase transition.
In fact, it has been claimed \cite{Torrieri:2007fb,Rajagopal:2009yw} that bulk viscosity
enhanced in the vicinity of cross-over region
would trigger this transition from thermalised fluid to individual particles.

\section*{Acknowledgment}

The authors acknowledge the fruitful discussion with
S.~Esumi and Y.~Hori about
the flow analysis methods.
T.H. acknowledge a kind hospitality
from the nuclear theory group
at Lawrence Berkeley National Laboratory 
where a part of work was done when he visited
there on sabbatical.
The work of T.H. was supported by
Grant-in-Aid for Scientific Research
Nos.~22740151 and 22340052,
 the work of
P.H.\ by BMBF under contract no.\ 06FY9092,
the work of K.M.\ by JSPS Research Fellowships for Young Scientists,
and the work of Y.N.\  by
Grant-in-Aid for Scientific Research
No.~20540276.

\section*{Appendix}

In this Appendix, we show some technical details about momentum integration of 
the Cooper-Frye formula \cite{Cooper:1974mv} for the purpose of
less numerical costs \cite{murase}.

Since the number is Lorentz invariant,
Equation (\ref{eq:CFpm}) can be Lorentz-transformed to the local rest frame
using four flow velocity $u^{\mu}$ and written as
\begin{eqnarray}
\Delta N_{\pm} &=& \int \frac{d^3p}{E}\frac{[p\cdot \Delta \sigma]_{\pm}}{\exp[(p\cdot u-\mu)/T]-\epsilon } \nonumber \\
& = & \int \frac{d^3\bar{p}}{\bar{E}}\frac{[\bar{p}\cdot \Delta \bar{\sigma}]_{\pm}}
{\exp[(\bar{E} -\mu)/T] -\epsilon }
\end{eqnarray}
where $[\cdots]_{\pm} = \Theta(\pm\cdots) |\cdots |$.
Hereafter integral variables $\bar{p}=(\bar{E}, \bar{\bm{p}})$ can be renamed as $p$.
In the following, we discuss only about out-going particles $[\cdots]_{+}$. Results
for in-coming particles $[\cdots]_{-}$ can be easily obtained by replacing $\Delta \bar{\sigma}$
with $-\Delta \bar{\sigma}$.
Taking the $z$ axis being parallel to $\Delta \bar{\boldsymbol \sigma}$,
\begin{eqnarray}
[p \cdot \Delta \bar{\sigma}]_{+} & = & [E \Delta \bar{\sigma}^{0}-p_{z}\mid \Delta
 \bar{\boldsymbol \sigma} \mid ]_{+}\nonumber\\
& = & [E \Delta \bar{\sigma}^{0}-p \cos \theta \mid \Delta \bar{\boldsymbol \sigma} \mid ]_{+}.
\end{eqnarray}
Thus, the above integration becomes
\begin{eqnarray}
\Delta N_{+} & = & \int \frac{dp}{\exp[(E -\mu)/T] -\epsilon }\frac{2\pi p^2 d\cos\theta}{E}
[E \Delta \bar{\sigma}^{0}-p \cos \theta \mid \Delta \bar{\boldsymbol \sigma} \mid ]_{+}.
\end{eqnarray}

When $\mid \Delta \bar{\boldsymbol \sigma}\mid = 0$,
we easily integrate the above equation and the results becomes
\begin{eqnarray}
\Delta N_{+} & = & 4 \pi \int \frac{p^2 dp}{\exp[(E -\mu)/T] -\epsilon }
[\Delta \bar{\sigma}^{0}]_{+}.
\end{eqnarray}

When $\mid \Delta \bar{\boldsymbol \sigma}\mid > 0$,
a short calculation leads to
\begin{eqnarray}
[E \Delta \bar{\sigma}^{0}-p \cos \theta | \Delta \bar{\boldsymbol \sigma} |]_{+}
& = & p | \Delta \bar{\boldsymbol \sigma}| 
\left[\frac{E \Delta \bar{\sigma}^{0}}{p |\Delta \bar{\boldsymbol \sigma}|}
- \cos \theta \right]_{+}\nonumber \\
& = & p | \Delta \bar{\boldsymbol \sigma} | [A-\cos \theta]_{+}\\
A & = & \frac{E}{|\bm{p} |}\frac{\Delta \bar{\sigma}^{0}}{| \Delta \bar{\boldsymbol \sigma}|}
\end{eqnarray}
First, we integrate the above equation with respect to $\cos \theta$
\begin{eqnarray}
\int_{-1}^{1} d\cos\theta [A-\cos \theta]_{+} & = & \int_{-1}^{1}d\cos\theta (A-\cos \theta)
\Theta(A-\cos \theta) \nonumber \\
& = & 2A \Theta(A-1) + \frac{(A+1)^2}{2} \Theta(1-|A|) ,
\end{eqnarray}
where $\Theta$ is the Heaviside function.
Let us define ``velocity" and ``momentum" 
of surface vector as $v_{\bar{\sigma}} = \Delta \bar{\sigma}^{0}/| \Delta \bar{\boldsymbol \sigma}|$
and $p_{\bar{\sigma}} = m| v_{\bar{\sigma}}|/\sqrt{1-v_{\bar{\sigma}}^{2}}$, respectively.

Finally we obtain
\begin{eqnarray}
\Delta N_{+} & = & 4\pi \int_{0}^{\infty} \frac{p^2dp}{\exp[(E -\mu)/T] -\epsilon}  
[\Delta \bar{\sigma}^{0}]_{+}\nonumber \\
&+& \Theta(1- | v_{\bar{\sigma}}| ) \left(\pi | \Delta \bar{\boldsymbol \sigma} | 
 \int_{p_{\bar{\sigma}}}^{\infty}\frac{p(E^2 v_{\bar{\sigma}}^2 + p^2)dp}
{\exp[(E -\mu)/T] -\epsilon} 
-2 \pi | \Delta \bar{\sigma}^{0}| \int_{p_{\bar{\sigma}}}^{\infty} 
\frac{p^2dp}{\exp[(E -\mu)/T] -\epsilon} \right). \nonumber \\
\end{eqnarray}
Remembering that the hypersurface elements have been Lorentz-transformed, 
we have to Lorentz-transform back to the laboratory frame:
\begin{eqnarray}
\Delta \bar{\sigma}^{0} & = & u \cdot \Delta \sigma \\
\mid \Delta \bar{\boldsymbol \sigma} \mid^{2} & = & 
(\Delta \bar{\sigma}^{0})^2-\Delta \bar{\sigma}\cdot\Delta \bar{\sigma} \nonumber\\
& = & -(g_{\mu \nu}-u_{\mu} u_{\nu})\Delta \sigma^{\mu} \Delta \sigma^{\nu} 
\end{eqnarray}

Now the three-dimensional integral with a complicated integrand such as 
$[\cdots]_{\pm}$ reduces to the one-dimensional one and it is easily
done numerically.

\end{document}